\documentclass{aa}  
\usepackage[table, dvipsnames]{xcolor}
\usepackage{lscape}
\usepackage{setspace}
\usepackage{chngpage}
\usepackage[english]{varioref}
\usepackage{pythonhighlight}
\usepackage{booktabs}
\usepackage{placeins}
\usepackage{graphicx}
\usepackage{txfonts}
\usepackage{hyperref}
\usepackage{caption}

\begin{document}

   \title{The VMC Survey -- L. Type II Cepheids in the Magellanic Clouds}

   \subtitle{Period-Luminosity relations in the near-infrared bands.}

   \author{T. Sicignano \inst{1,2}
          \and 
          V. Ripepi \inst{2}
          \and 
          M. Marconi \inst{2}
          \and 
          R. Molinaro \inst{2}
          \and
          A. Bhardwaj \inst{2}
          \and
          M.-R. L. Cioni \inst{3}
          \and
          R. de Grijs \inst{4,5,6}
          \and
          J. Storm \inst{3}
          \and
          ~~~M. A. T. Groenewegen \inst{7}
          \and 
          V. D. Ivanov \inst{8}
          \and 
          G. De Somma \inst{2,9,10}
          }

   \institute{Scuola Superiore Meridionale, Largo S. Marcellino 10, 80138 Napoli, Italy 
   \and
   INAF-Osservatorio Astronomico di Capodimonte, Salita Moiariello 16, 80131, Napoli, Italy\\  \email{correponding authors: teresa.sicignano@inaf.it, vincenzo.ripepi@inaf.it}
   \and
   Leibniz-Institut f\"ur Astrophysik Potsdam (AIP), An der Sternwarte 16, D-14482 Potsdam, Germany
   \and
   School of Mathematical and Physical Sciences, Macquarie University, Balaclava Road, Sydney, NSW 2109, Australia 
   \and
   Astrophysics and Space Technologies Research Centre, Macquarie University, Balaclava Road, Sydney, NSW 2109, Australia 
   \and 
   International Space Science Institute--Beijing, 1 Nanertiao, Zhongguancun, Hai Dian District, Beijing 100190, China
   \and 
   Koninklijke Sterrenwacht van België, Ringlaan 3, 1180, Brussels, Belgium
   \and 
   European Southern Observatory, Karl-Schwarzschild-Str. 2, 85748 Garching bei München, Germany
   \and
   Istituto Nazionale di Fisica Nucleare (INFN), Sezione di Napoli, Via Cinthia 21, 80126 Napoli, Italy
   \and
   INAF-Osservatorio Astronomico d'Abruzzo, Via Maggini sn, 64100 Teramo, Italy
}

   \date{Received November 17, 2023; accepted January 16, 2024}

 
  \abstract
   {
    Type II Cepheids (T2Cs) are less frequently used counterparts of classical or Type I Cepheids which provide the primary calibration of the distance ladder for measuring the Hubble constant in the local Universe. In the era of the Hubble Tension, T2C variables together with the RR Lyrae stars and the tip of the red giant branch (TRGB) can potentially provide classical Cepheid independent calibration of the cosmic distance ladder.
   }
  {Our goal is to provide an absolute calibration of the Period-Luminosity, Period-Luminosity-Color and Period-Wesenheit relations($PL$, $PLC$ and $PW$, respectively) of T2Cs in the Large Magellanic Cloud (LMC), which traditionally serves as a crucial first anchor of the extragalactic distance ladder. 
  }
  {We exploited time-series photometry in the near-infrared (NIR,  $Y,\,J$ and $K_s$) bands for a sample of $\sim 320$ T2Cs in the Magellanic Clouds (MCs). These observations were acquired during 2009--2018 in the context of the VMC ESO public survey (``The VISTA near-infrared $YJK_s$ survey of the Magellanic System''). The NIR photometry from the VMC survey was supplemented with well-sampled optical light curves and accurate pulsation periods from the OGLE IV survey (Optical Gravitational Lensing Experiment) and the $Gaia$ mission. We used the best-quality NIR light curves to generate custom templates for modelling sparsely sampled light curves in $YJK_s$ bands. 
   }
   {The best-fitting $YJK_s$ template light curves were used to derive accurate and precise intensity-averaged mean magnitudes and pulsation amplitudes of 277 and 62 T2Cs in the LMC and SMC, respectively. We used optical and NIR mean magnitudes for different T2C subclasses (BLHer, WVir and RVTau) to derive $PL/PLC/PW$ relations in multiple bands, which were calibrated with the geometric distance to the LMC from Eclipsing Binaries and with the $Gaia$ parallaxes. 
   We used our new empirical calibrations of $PL/PW$ relations to obtain distances to 22 T2C-host Galactic globular clusters, which were found to be systematically smaller by $\sim$0.1 mag and 0.03--0.06 mag compared with the literature, when the zero points are calibrated with the distance of LMC or $Gaia$ parallaxes, respectively. A better agreement is found between our distances and those based on RR Lyrae stars in globular clusters, providing strong support for using these population II stars together with the TRGB for future distance scale studies.}
    {}
 

   \maketitle
%

\section{Introduction} \label{intro}
 The determination of the Hubble constant ($H_0$), which parametrizes the expansion rate of the Universe, is highly debated given the ongoing tension between the $H_0$ values derived from the extragalactic distance scale, traditionally adopting classical Cepheids (CC) and Type Ia Supernovae, and those inferred by the {\it Planck} mission based on the $\Lambda$ Cold Dark Matter theory applied to the Cosmic Microwave Background measurements \citep[see e.g.][]{Riess2022,Planck2020}. Given the ongoing Hubble tension, CC-independent calibration of the cosmic distance ladder using stellar standard candles of different ages and metallicities is being explored to investigate possible sources of systematic uncertainties in $H_0$ determinations. One of the most promising alternative calibrations based on the tip of the red giant branch (TRGB) stars provides an $H_0$ value that is intermediate between Cepheid-Supernovae and {\it Planck} measurements \citep{Freedman2021}. Population II stars, such as Type II Cepheids (T2Cs) and RR Lyrae variables (RRL), can be used independently or together with the TRGB stars providing another alternative calibration for the local $H_0$ measurements based on old stellar standard candles.

T2Cs are probably old, low-mass, and metal-poor stars. Based on their pulsation periods, they are separated into 3 types: i) \textit{BL Herculis} stars (BLHer) with periods between 1 and 4 days;
ii) \textit{W Virginis} stars (WVir)  with periods between 4 and 20 days;
iii) \textit{RV Tauri} stars (RVTau) with periods between 20 and 150 days. However, these period boundaries are only approximate, in particular, between WVir and RVTau stars. \citet{soszynski2008optical} noticed a subsample of WVir stars showing peculiar light curves and brighter magnitudes than normal WVir stars of a given period. These peculiar WVir (pWVir) are hypothesized to have a binary origin. 


From the evolutionary point of view, BLHer, WVir, and RVTau are thought to be in different evolutionary phases going from post Horizontal Branch (HB) to post Asymptotic Giant Branch \citep[AGB,][]{gingold1976evolutionary,gingold1985evolutionary}. BLHer are low-mass stars having exhausted their core helium on the portion of the HB bluer than the classical instability strip (IS) populated by RRL. They are ascending towards the AGB phase, becoming increasingly brighter and redder. In this process, BLHer crosses the classical IS at luminosities higher than that of the RRL \citep[see e.g.][and references therein]{Dicriscienzo2007,mdc07}. 
The WVir stars exhibit a more advanced evolutionary phase than BLHer stars. Indeed, they are thought to be AGB stars crossing the IS at higher luminosities (which explains the longer periods) compared to BLHer on the blueward and redward path of the blue loop during the thermal pulse phase \citep[see e.g.][and references therein]{bono2020evolutionary,2002PASP..114..689W}.  
Historically, RVTau stars have been considered to be the post-AGB stars which cross the IS at even higher luminosities compared to WVir, but recent studies have also suggested a connection to binary evolution \citep[see e.g.][]{Groenewegen17}.

T2Cs can be used as distance indicators because they follow very tight Period-Luminosity, Period-Wesenheit\footnote{The Wesenheit magnitudes are constructed to be reddening free by definition \citep[][]{madore1982period}}, and Period-Luminosity-Colour relationships \citep[$PL/PW/PLC$, see e.g.][]{feast2008luminosities,Matsunaga2011,ripepi2015vmc,Bhardwaj2017_LMC,bhardwaj2020}. These empirical relations based on T2Cs are almost parallel to those of CCs but at significantly fainter (i.e. 1--1.5 mag) magnitudes. \citet{ripepi2015vmc}, \citet{Bhardwaj2017_LMC}, and \citet[][and references therein]{bhardwaj2022rr} suggested that the RRL $PL$ relations are an extension of T2Cs $PL$ relations at shorter periods in the near-infrared (NIR), in agreement with theoretical predictions \citep[see e.g.][]{caputo04why,mdc07}. In particular, BLHer and WVir follow common $PL$ relations in the NIR bands but different $PL$ relations in the optical \citep[e.g.][and references therein]{ripepi2015vmc, Bhardwaj2017_Bulge}. The NIR PL relations of short-period T2Cs exhibit less scatter than the RVTau stars, especially in the Large Magellanic Cloud (LMC). The NIR $PL$ relations have several advantages compared to the optical bands due to smaller extinction, lower variability amplitudes, and smaller temperature variations, which lead to tighter $PL$ relations. 

The T2Cs are generally found in Galactic globular clusters (GGCs) having blue HB morphology, in all components of the Milky Way, namely, the bulge, disc and halo. However, these stars are rare in dwarf galaxies in the Local Group, except for a few hundred stars in the Magellanic Clouds \citep[MCs; ][]{soszynski2018ogle} and 4 in the Sagittarius dwarf Spheroidal galaxy \citep{Hamanowicz2016}. The MCs also host standard candles such as CCs ($\sim11,000$) and  RRL stars ($\sim45,000$) \citep[e.g.][]{soszynski2018ogle,clementini2019gaia,Clementini2022,ripepi2022gaia}. The absolute calibration of NIR $PL$, $PW$, and $PLC$ relations for T2Cs in the MCs will be useful for the distance scale because the LMC and (to a lesser extent) the SMC, have traditionally served as an anchor of the distance ladder \citep[e.g.][and references therein]{Riess2022}. 
The simultaneous presence of different standard candles in the same galaxy also allows us to compare the distances obtained with the different indicators, thus verifying the possible presence of systematic errors. Particularly important is the homogeneity of the RRL and T2C distance scales, as both types of pulsators can be used to calibrate the TRGB method in an alternative route to calibrate $H_0$ \citep[]{Beaton2016}.

The LMC and the Small Magellanic Cloud (SMC) comprise the Magellanic system together with the Bridge connecting them and the Stream, a \ion{H}{I} feature covering more than 100 deg on the sky \citep[e.g.][]{mathewson1974magellanic}. The LMC and the SMC lie at an average distance of $\sim$ 50 and 60 kpc, respectively \citep[e.g.][]{pietrzynski2019distance,graczyk2020}. 
In the context of NIR survey of the MCs, a great contribution has been provided by the ESO (European Southern Observatory) public survey  "VISTA (Visible and Infrared Survey Telescope for Astronomy) survey of the Magellanic Clouds system" \citep[VMC;][]{cioni2011vmc}. The main aims of VMC are (i) to reconstruct the spatially resolved star formation history and (ii) to infer an accurate three-dimensional map of the whole Magellanic system. The survey was designed to observe in three bands $Y,J,K_s$ and obtain deep time-series photometry reaching even the faintest Cepheids of all types and the vast majority of RRLs known in the MCs. The interested readers are referred to \citet{cioni2011vmc} for details regarding the VMC survey. Time-series $YJK_s$ observations from the VMC survey are particularly useful in providing calibration of $PL/PW$ relations of primary stellar standard candles in the MCs.

The NIR $PL/PW$ relations of T2Cs have been explored in different environments: Galactic Bulge \citep[][]{Bhardwaj2017_Bulge,Braga2018}, GGCs \citep[][]{Matsunaga2006, bhardwaj2022rr, ngeow2022} and the MCs \citep[][]{Matsunaga2011,ripepi2015vmc,Bhardwaj2017_LMC}.
Several studies, in particular those from the OGLE survey and the $Gaia$ mission, have also investigated PL relations at optical wavelengths \citep[][and references therein]{Alcock1998,Iwanek2018, ripepi2022gaia}. The accurate and precise parallaxes from the $Gaia$ mission \citep{GaiaPrusti2016} have also allowed the calibration of the $PL/PW$ relations based on Galactic field T2Cs in the $Gaia$ bands \citep{ripepi2022gaia} and NIR bands \citep{Wielgorski2022}. A small, but non-negligible, effect of metallicity on the absolute NIR magnitude of T2Cs was found by \citet{Matsunaga2006} and \citet{Wielgorski2022}. However, \citet{ngeow2022} did not find any statistically significant dependence of $PL$ relations on metallicity for T2Cs in agreement with theoretically predicted relations \citep{Dicriscienzo2007, Das2021}. 
 


The scope of this work is to take advantage of Data Release 5 and 6 of the VMC survey\footnote{The ESO archive at https://www.eso.org/qi/ contains all the information on the different data products of the VMC survey.} to derive new accurate NIR $PL/PW$ relations for the T2C in both the LMC and SMC following the previous similar work by \citet{ripepi2015vmc}. Compared with this paper, we have more than doubled the sample of T2Cs in the MCs and we have at our disposal both a new accurate distance of the LMC (and SMC) and the parallaxes of the $Gaia$ mission Data Release 3 \citep[DR3][]{GaiaPrusti2016,vallenari2022gaia} for Galactic T2Cs. Overall, we will be able to calculate and calibrate $PL/PW$ relations in the MCs which will be particularly useful for the distance scale based on population II stars.

\section{Type II Cepheids in the VMC survey}

The observing strategy of the VMC survey is discussed in detail in \citet{cioni2011vmc}. In brief, the VMC surveyed an area of 184 deg$^2$ covering almost the entire LMC, SMC and Bridge in three bands $Y$, $J$ and $K_s$ (central wavelenght $\lambda_c$ = 1.02, 1.25 and 2.15 $\mu$m, respectively) using the VIRCAM \citep[VISTA InfraRed CAMera; ][]{Dalton2006,Emerson2006} instrument attached to the VISTA telescope. The VMC $K_s$-band time-series observations were planned to obtain 13 different epochs (eleven reaching a limiting magnitude of $K_s \sim $ 19.3 mag with a S/N $\sim$ 5, plus two shallower obtained with half of the exposure time)  executed over several consecutive months with a specific observing cadence to obtain well-sampled light curves for pulsating variables \citep[for details see][]{cioni2011vmc,Ripepi2012,Moretti2014}. For the \textit{Y} and \textit{J} bands, four epochs were planned, of which two were shallower. In practice, the actual number of epochs is larger, as many observations were repeated because of lower quality (i.e. data not matching the observational constraints), even if good enough to build light curves \citep[see e.g.][]{ripepi2016vmc}. 

The raw VISTA images acquired for the VMC survey were reduced by the VISTA Data Flow System \citep[VDFS;][]{irwin2004vista} pipeline at the Cambridge Astronomical Survey Unit (CASU)\footnote{http://casu.ast.cam.ac.uk/}. The data used in this work were reduced using version 1.5 of the VDFS pipeline. The photometry is in the VISTA photometric system, which is described in detail by \citet{Gonzalez-Fernandez2018}.   
The data reduced by the VDFS pipeline at CASU are ingested into the VISTA Science Archive \citep[VSA;][]{Cross2012}.

T2Cs belonging to the MCs were identified and fully characterised in terms of period determination and epoch of maximum light by the OGLE IV survey \citep[Optical Gravitational Lensing Experiment,][]{soszynski2018ogle} and the $Gaia$ mission \citep{GaiaPrusti2016,vallenari2022gaia,clementini2016gaia,clementini2019gaia,Ripepi2019_reclassificaiton,ripepi2022gaia}.

In total, we have a list of 345 T2Cs from OGLE IV and 30 additional objects from $Gaia$ DR3 catalogue. We used the information (including classification) from OGLE IV, when avaible, and from the $Gaia$ DR3 for the remaining cases. 
We cross-matched coordinates of our T2C sample with the VMC sources in the VSA database with 1 arcsec tolerance, preserving in the sample even objects with only two epochs of observation in a single band. We thus downloaded the $Y$, $J$ and $K_s$ time-series photometry for 318 and 21 T2Cs from the OGLE IV and $Gaia$ DR3 lists, respectively (see Sect.~\ref{opticalData}). 

\begin{figure}
\centering
\sidecaption
    \includegraphics[width=9cm]{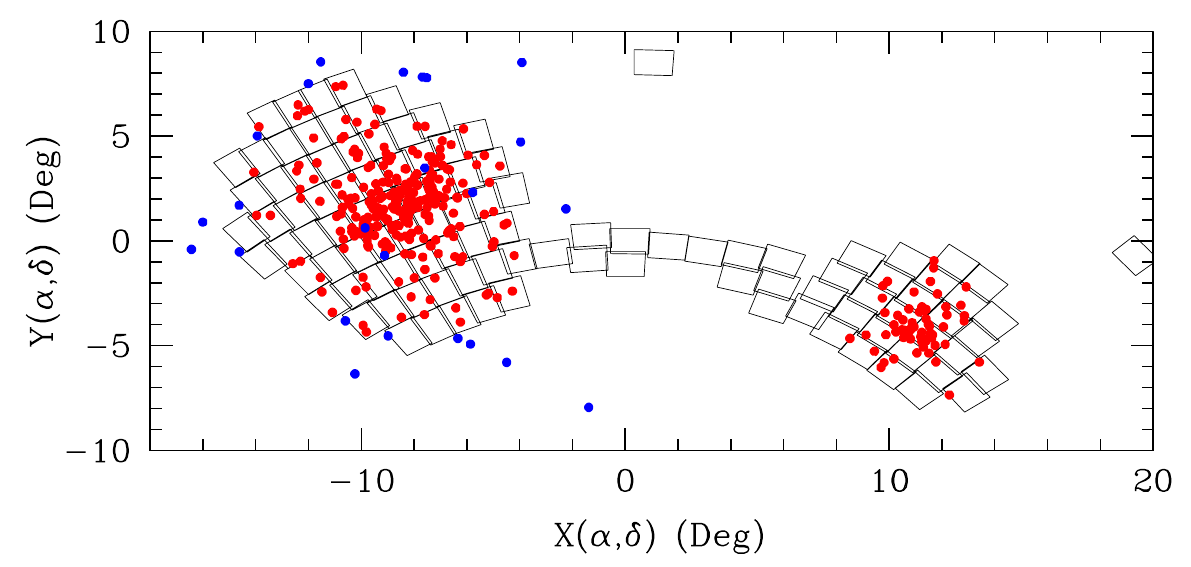}
    \caption{\label{stars}  Distribution of the T2Cs in the MCs. The red and blue filled circles show the stars with and without a match in the VMC database at the VSA (see text). The projection (zenithal equidistant) is centred at RA= 55.0 deg DEC= $-$73.0 deg. The empty rectangles trace the footprint of the VMC survey.} 
	\end{figure}

T2Cs are distributed in the two Clouds as follows: 62 T2Cs (20 BLHer, 7 WVir, 17 pWVir, and 18 RVTau) belong to the SMC and 277 (86 BLHer, 104 WVir, 26 pWVir, and 61 RVTau) to the LMC. The spatial distribution of T2Cs in the MCs is shown in Fig.~\ref{stars} in comparison with the VMC footprint. While most of the missing stars are placed outside the VMC footprint, there are four stars from the OGLE IV survey which have no VMC counterpart within 1 arcsec, namely OGLE-LMC-T2CEP-005, 030, 169, 173. A visual inspection of the OGLE and VMC images revealed that the centroids and photometry of these objects were affected by close companion stars not resolved in OGLE and/or VMC data. To be conservative, these four problematic objects were excluded from our sample.

In the end, typically, we have 5-6 epochs in $Y$ and $J$ and 14-15 in $K_s$, as shown in Fig.~\ref{epochs}. For some stars, the number of epochs is much larger. Indeed, if a variable falls into a region of overlap between two pawprints, the number of epochs is increased.
Note that not all the T2Cs have VMC observations in all the filters: 315, 329 and 339 stars have photometry in the $Y$, $J$, and $K_s$ bands, respectively. These samples represent 84\%, 87\% and 90\% of the known T2Cs in the area surveyed by VMC. An example of a representative time series photometry in the $J$ band is shown in Table~\ref{dati}. The full version of the table will be published online at the CDS \footnote{Centre de Données astronomiques de Strasbourg, https://cds.u-strasbg.fr/}.

\begin{table}
\begin{center}
\caption{Example of time series photometry for the star OGLE-SMC-T2CEP-46 in the \rm{J} band. The machine-readable version of the full table will be published online at the CDS (Centre de Données astronomiques de Strasbourg, https://cds.u-strasbg.fr/). 
\label{dati}} 
\begin{tabular}{c c c } 
 \hline
 \noalign{\smallskip} 
    HJD & $J$& $\sigma_{J}$ \\ [0.5ex] 
    d & mag & mag \\
 \noalign{\smallskip} 
    \hline
   55808.778293   &  15.3155   &   0.0039   \\
   55808.817253   &  15.3006   &   0.0038   \\
   55810.805700  &  15.2809   &   0.0039   \\
   55824.680427  &   15.2455   &   0.0036   \\
   56102.881943   &  16.0358  &    0.0057   \\
   56155.714828  &   15.3134 &     0.0043   \\
   56155.807728   &  15.3377  &    0.0044   \\
   56173.674063   &  15.3698  &    0.0043   \\
   56175.665638  &   15.6619  &    0.0049   \\
   56190.670667   &  15.5947    &  0.0044  \\
   56229.653334   &  15.5178  &    0.0050  \\
   56262.603381  &   15.2783    &  0.0045  \\
   56509.822565  &   15.9970   &   0.0058  \\
   \hline
\end{tabular}
\end{center}
\end{table}

\section{Data analysis}

\begin{figure}
\sidecaption
    \vbox{\includegraphics[width=\hsize]{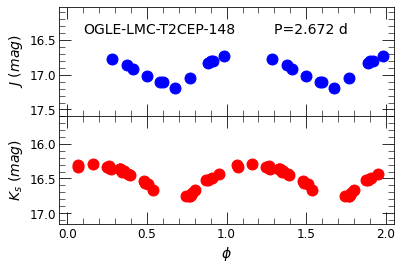}
    \includegraphics[width=\hsize]{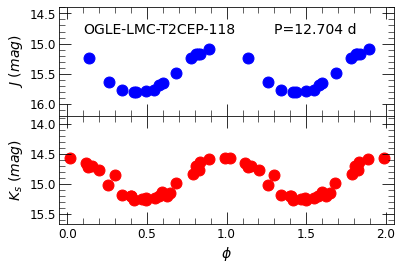}
    \includegraphics[width=\hsize]{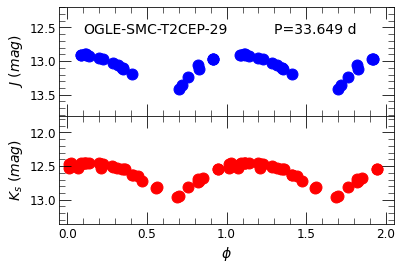}
    }
    \caption{\label{lightcurves} Examples of light curves for different variability types. From top to bottom: BLHer, WVir, and RVTau, respectively. Both $J$ and $K_s$ bands data are shown. }	
	\end{figure}

The time series light curves for the targets T2Cs were folded in phase using the periods and epoch of maximum from the literature (OGLE IV survey and $Gaia$ mission DR3). The example light curves for the different T2C sub-classes are shown in  Fig.~\ref{lightcurves}.
The following subsections define the procedure to determine our targets' intensity-averaged magnitudes in each band. We followed the same methodology already adopted in our previous works, consisting of the use of templates to fit sparsely sampled light curves \citep[see][]{ripepi2016vmc,ripepi2017vmc,ripepi2022vmc}.  

\subsection{Template derivation}
\begin{figure}
\sidecaption
\centering
    \includegraphics[width=\hsize]{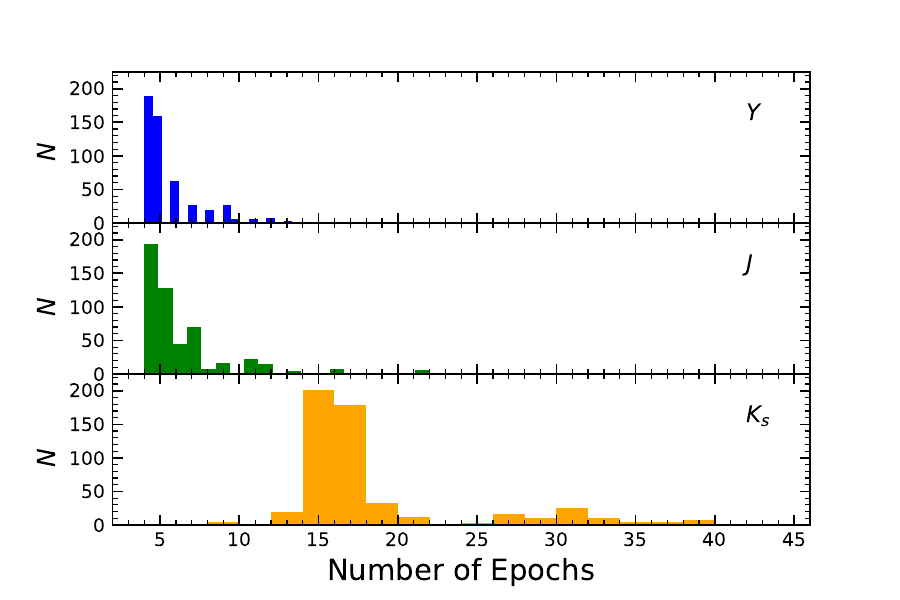}
    \caption{\label{epochs} Number of epochs for the T2C sample analysed in this work for the different VMC bands.}
\end{figure}

The first step of creating templates was to choose the best light curves among our sample of stars. For this purpose, we selected a subsample of stars with an adequate number of epochs ($>$ 10)~ to ensure a good probability of having well-sampled light curves. As shown in Fig.~\ref{epochs}, this first selection reduced significantly the number of objects to be considered in $Y$ and $J$, while several tens of stars are available in $K_s$.

Then, we proceeded with a visual inspection of the light curves, considering only those showing the least scatter and ensuring the entire period range spanned by the data. We also retained the variety of light curve shapes representing the four different T2C types investigated in this work (BLHer, WVir, pWVir, RVTau).  

Following previous works on templates of CCs in the VMC survey \citep[i.e.][]{ripepi2022vmc}, the template-fitting procedure is carried out in two subsequent stages: i) fitting the light curve with a spline function, and ii) using a Fourier truncated series with at least 10 terms to fit the continuous spline curve obtained in the previous step. This two-step procedure was very successful in avoiding numerical ringing in the light curve fitting due to the small number of epochs and large phase gaps.

The first step was carried out with a custom build {\tt Python} code, which adopts the {\tt splrep} function to fit smooth spline curves to the data. The best smoothing factors were found using visual inspection of each light curve. The fitting spline curves have been subsequently transformed into templates by subtracting their intensity-averaged mean magnitude and re-scaling the amplitude to unity.

\begin{figure}
\sidecaption
\centering
    \includegraphics[width=\hsize]{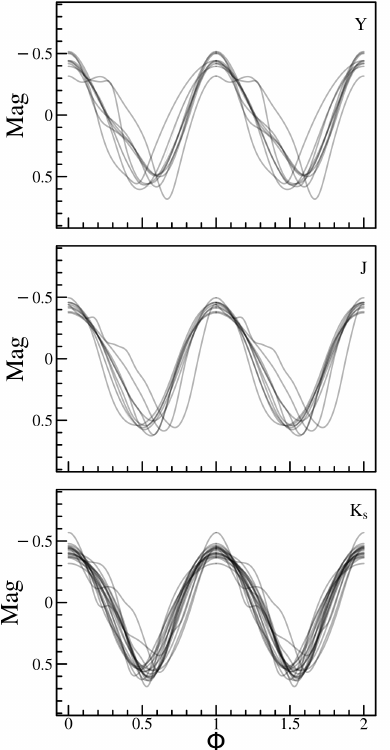}
    \caption{\label{figtemplates} Templates created in each band (labelled in the figures).}
\end{figure}

The second step consisted of fitting these normalised templates with a truncated Fourier series with 10 terms to ensure a perfect correspondence of the analytical function to the spline curve. The fitting equation is:
\begin{equation}
    m(\phi)= A_0 + \sum^N_1 A_i\, {\rm cos}\,(2\pi i \phi + \Phi_i) 
\end{equation}
\noindent
where N is the number of harmonics, $A_0$ the average magnitude (which in our case is zero), $\phi$ is the phase (intended as a dimensional substitute of time, it is, therefore, an observational independent variable); while $A_i$ and $\Phi_i$ are the unknown coefficients of the series, i.e. the amplitude and the phase of each term.  
In the end, each template was transformed into a series of Fourier coefficients which are available in App.~\ref{fittemplates}.

The final template sample, shown in Fig.~\ref{figtemplates},~ contains 16 models for the $Y$ band, 15 for the $J$ band, and 31 for the $K_s$ band. For each band, the templates are further divided by star type (BLHer, WVir, pWvir, RVTau).

\subsection{Template fitting to the observed light curves}


We used the procedure discussed in \citet[][]{ripepi2022vmc} to fit our light curves with the templates described above. Since they have zero average and amplitude one, each template is adjusted to the real light curves by varying 3 parameters: i) a magnitude shift $\delta M$; ii) a scale factor, $a$, that increases or reduces the template amplitude to adapt it to that of the observed light curves; and iii) a phase shift $\delta \phi$ which takes into account possible differences in phase between templates and observed light curves. It is usual to impose that the maximum of the light curve is at phase 1 (or 0). To this aim, literature epochs of maximum from OGLE in the $I$ band and from $Gaia$ in the $G$ band were used. The $\delta \phi$ term also takes care of any small difference between the epoch of maximum in these two bands.

It is possible to obtain these three unknown numbers for each template by minimising the following $\chi^2$ function:

\begin{equation} \label{chi2}
    \chi^2 = \sum_l^{N_{pts}} \frac{[m_l - (a\times M_t(\phi_l + \delta\phi)+ \delta M)]^2}{\sigma_l^2}
\end{equation}

\noindent
where $N_{pts}$ is the number of epochs, $m_i$, $\phi_l$ and $\sigma_l$ are the observed magnitudes, the corresponding phases, and the uncertainties on the magnitudes, respectively. $M_t(\phi_l)$ represents the template as a function of phase. 

The fitting routine has an outlier rejection procedure, as the VMC light curves may show one or more bad measurements, due to various reasons (e.g. the star being hit by a cosmic ray or falling on a bad detector column, etc.). 
Outliers are detected by analysing the distribution of the residuals from the fit and spotting points outside the interval $\pm3.5\times DMAD $, where \textit{DMAD} is Double-Median Absolute Deviation \footnote{{\it DMAD} is calculated treating separately the values smaller and larger than the median of the considered distribution.}.

Since we have fitted all the available templates to each star, we decide the best-fitting template using the so-called G parameter \citep[introduced by ][]{ripepi2016vmc}, which is defined as follows: 

\begin{equation}
    G = \Bigl(\frac{1}{\sigma} \Bigr)^2 \times \Bigl(\frac{N_U}{N_T} \Bigr)^4
\end{equation}

\noindent
where $N_U$ is the effective number of points used in the fit (after the rejection of outliers) and $N_T$ is the initial number of points (including outliers).\\
At the first visual analysis of the values of the amplitudes in the $Y$ and $J$ bands, we noticed a problem in the choice of the template for several objects for which the few epochs we have available in these bands were not evenly distributed along the light curve. 

\begin{figure*}
\centering
    \vbox{
    \hbox{
    \includegraphics[width=0.45\textwidth]{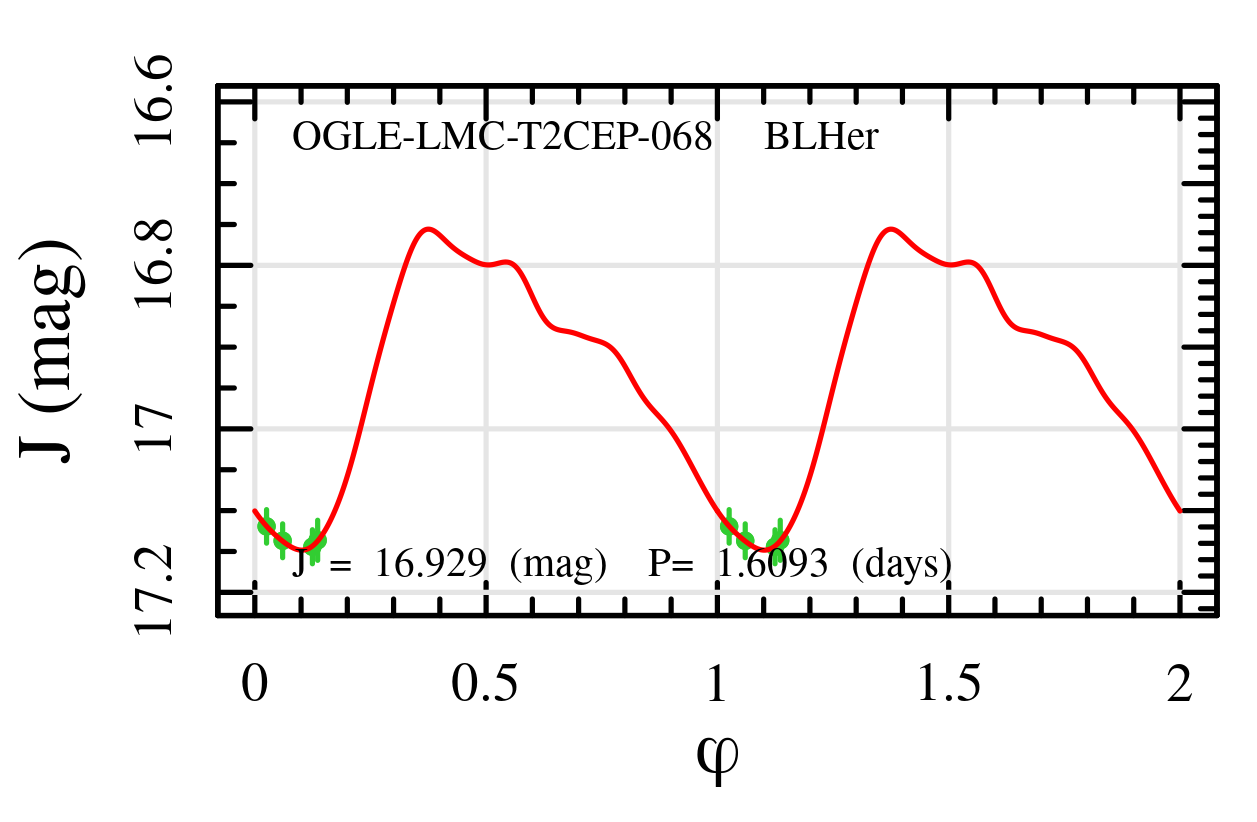}
    \includegraphics[width=0.45\textwidth]{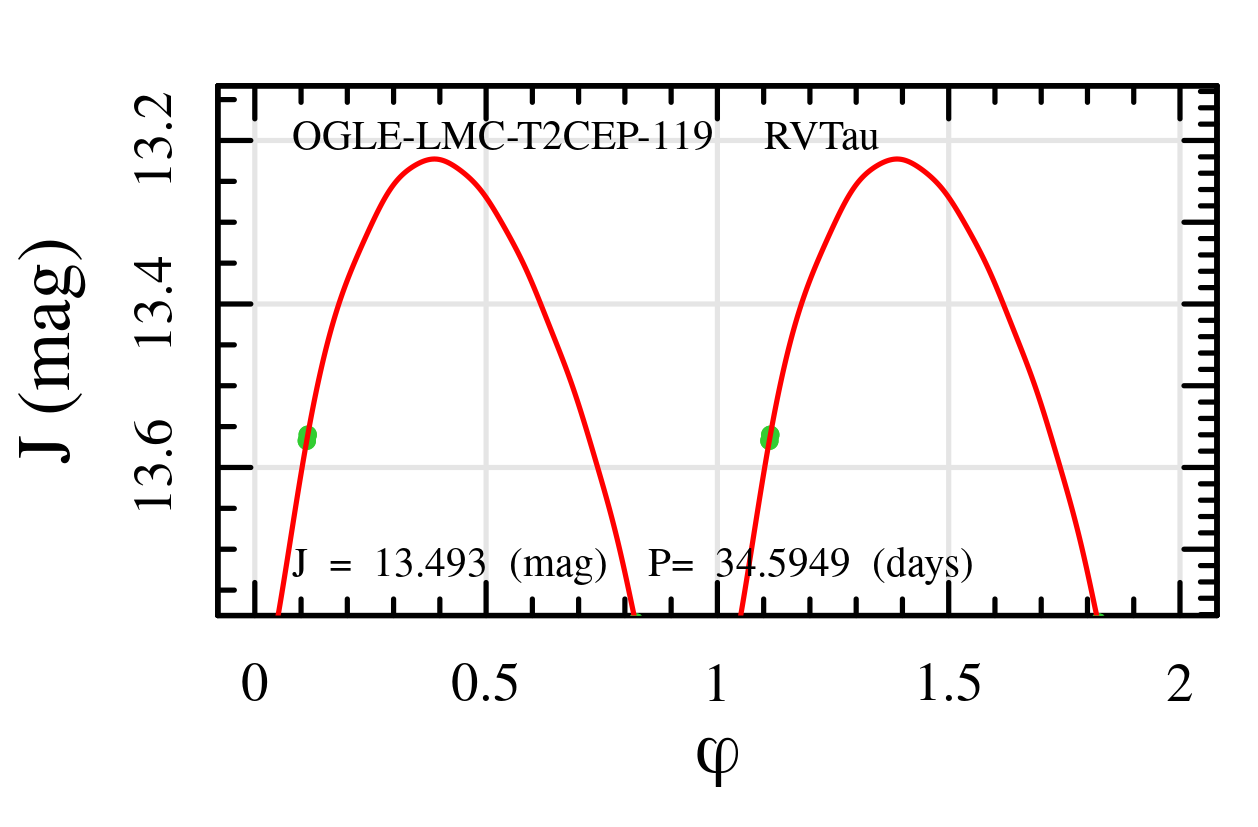}}
    \hbox{
    \includegraphics[width=0.45\textwidth]{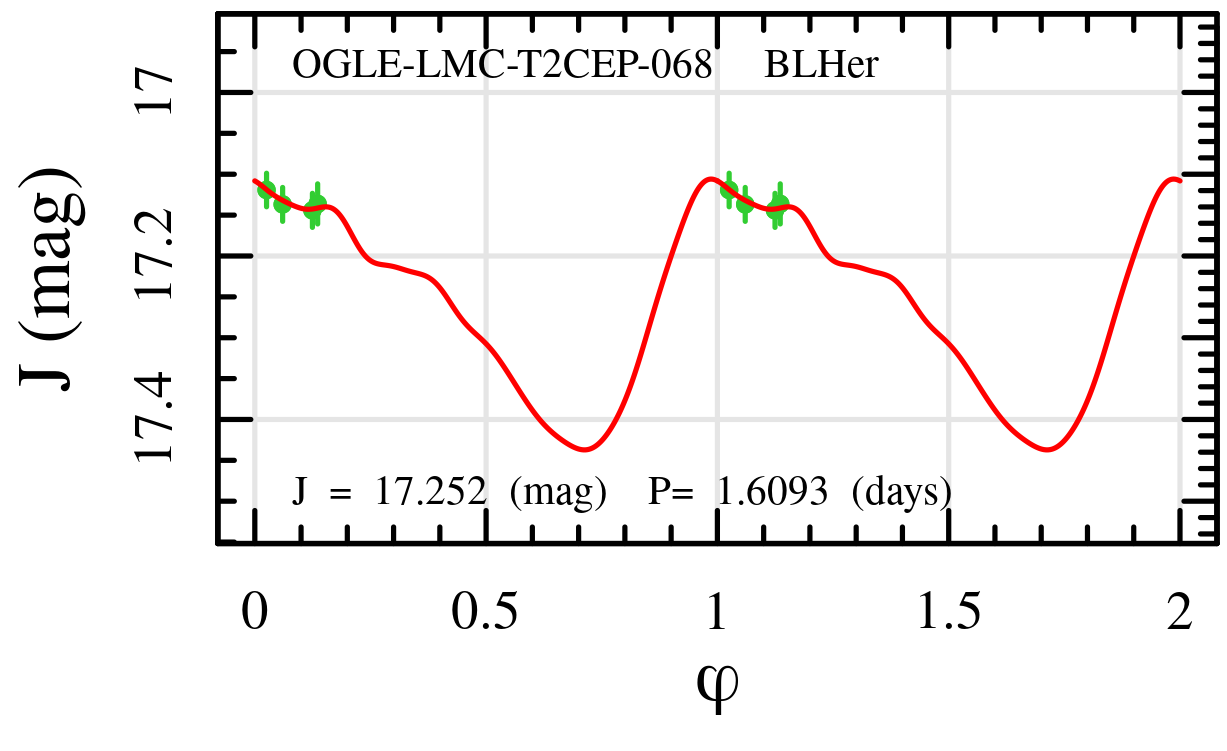}
    \includegraphics[width=0.45\textwidth]{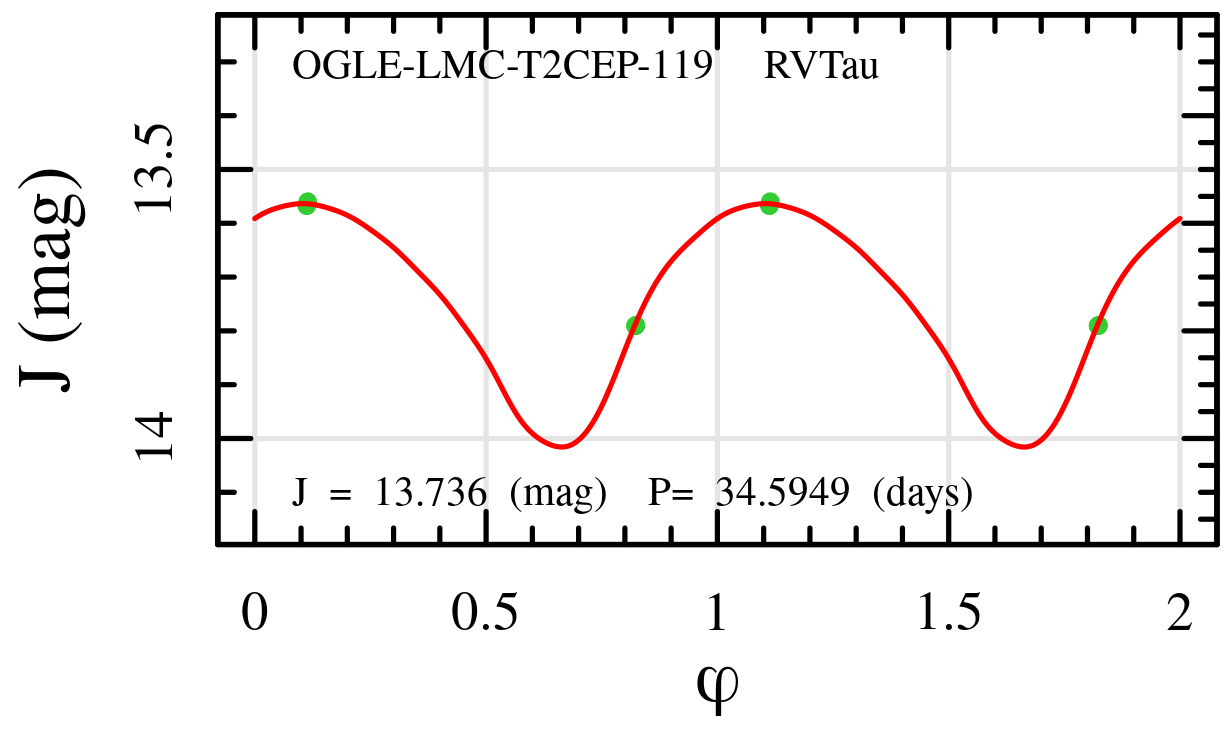} }
    }
    \caption{\label{prob} Top: Light curves with incorrect templates. Note that for the RVTau variable, two points are out of the plot due to the very wrong amplitude. Bottom: light curves the right template fitting for the same stars of the top panels.}	
	\end{figure*}
 

Indeed, as shown in Fig.~\ref{prob} (top), objects with few observations not well spaced in phase, do not allow proper constraint on three parameters of the fit of Eq.~\ref{chi2}. To overcome this problem, we introduced a parameter that evaluates the light curve's sampling. This \textit{uniformity index} was adopted according to the definition by 
 \citet{madore2005nonuniform}:
 
\begin{equation}
    U_N^2 = \frac{N}{(N-1)} \Bigl[1 - \sum^N_{l=1} (\phi_{l+1}-\phi_l)^2 \Bigr]
\end{equation}

\noindent
where $N=N_T$ and $\phi_l$ are defined as above (note that to calculate $U_N^2$, the epochs must be ordered in phase). Interpreted as a normalised variance, the $U_N^2$ statistic has a value of unity when the distribution of points is non-redundantly uniform over the light curve. After extensive visual inspection of large numbers of light curves, we verified that objects with $U_N^2<0.6$ have too scanty light curves to use Eq.~\ref{chi2} with three free parameters. Indeed, in these cases the sampling of the epochs prevented the algorithm from determining correctly the phase shift and the amplitude scaling. As a consequence i) the phasing of the templates was very different from the expected ones based on the analysis of the OGLE $I$ light curves; ii) the amplitude in $Y$ or in $J$ was larger than that in the $I$ band, while it is expected the amplitudes decrease monotonically from optical to NIR bands.
In all these cases, we decided to reduce the free parameters from three to two, by imposing a fixed amplitude scaling. To determine the proper amplitude scaling, we selected all the objects with $U_N^2>0.9$ and used their amplitudes in the $I$ band (from the OGLE IV survey) to derive simple linear relations which are shown in Table~\ref{amp}. 
The adoption of this procedure fixes the problem as shown in Fig.~\ref{prob}(bottom). 

\begin{table}
 \begin{center}
\footnotesize\setlength{\tabcolsep}{3pt}
\caption{\label{amp}\\ Coefficients of the linear relation between amplitude in $I$ band and in $Y/J$ bands.} 
\begin{tabular}{lllllllll } 
    \hline
 \noalign{\smallskip} 
    Band  &  Group & $\alpha $ &  $\sigma_{\alpha}$ & $\beta$ & $\sigma_{\beta} $ & RMS \\
    & & mag& mag & & & mag \\
 \noalign{\smallskip}
 \hline  
 \noalign{\smallskip} 
   $Y$   &  WVir & 0.005 & 0.005 & 0.980 & 0.019 & 0.04     \\
   $Y$   &  BLHer$\&$RVTau& 0.047 &  0.007 & 0.736 & 0.026 & 0.09\\
   $J$   &  WVir & 0.006 & 0.007 & 0.914 & 0.024 & 0.05   \\
   $J$   &  BLHer$\&$RVTau& 0.049 &  0.012 & 0.669 & 0.063 & 0.08\\
   \hline
\end{tabular}
\tablefoot{Relations are in the form: $Amp(Y~{\rm or}~J)= \alpha_{\lambda} + \beta_{\lambda} \times Amp(I)$. Note that the WVir subclass follows significantly different relations compared with BLHer and RVTau variables, which follow the same relations.}
\end{center}
\end{table}

We used a Monte Carlo approach similar to \citet{ripepi2016vmc} to estimate the uncertainties on the fitted parameters. In brief, for each fitted source we generated 100 bootstrap simulations of the observed time series. The template fitting procedure was repeated for each mock time series and a statistical analysis of the obtained fitted parameters was performed. Our fitted parameters error estimate is given by the robust standard deviation (1.4826 $\times$ MAD) of the distributions obtained by the quoted bootstrap simulations.

\begin{figure*}
\sidecaption
\centering
    \vbox{
    \hbox{
    \includegraphics[width=0.32\textwidth]{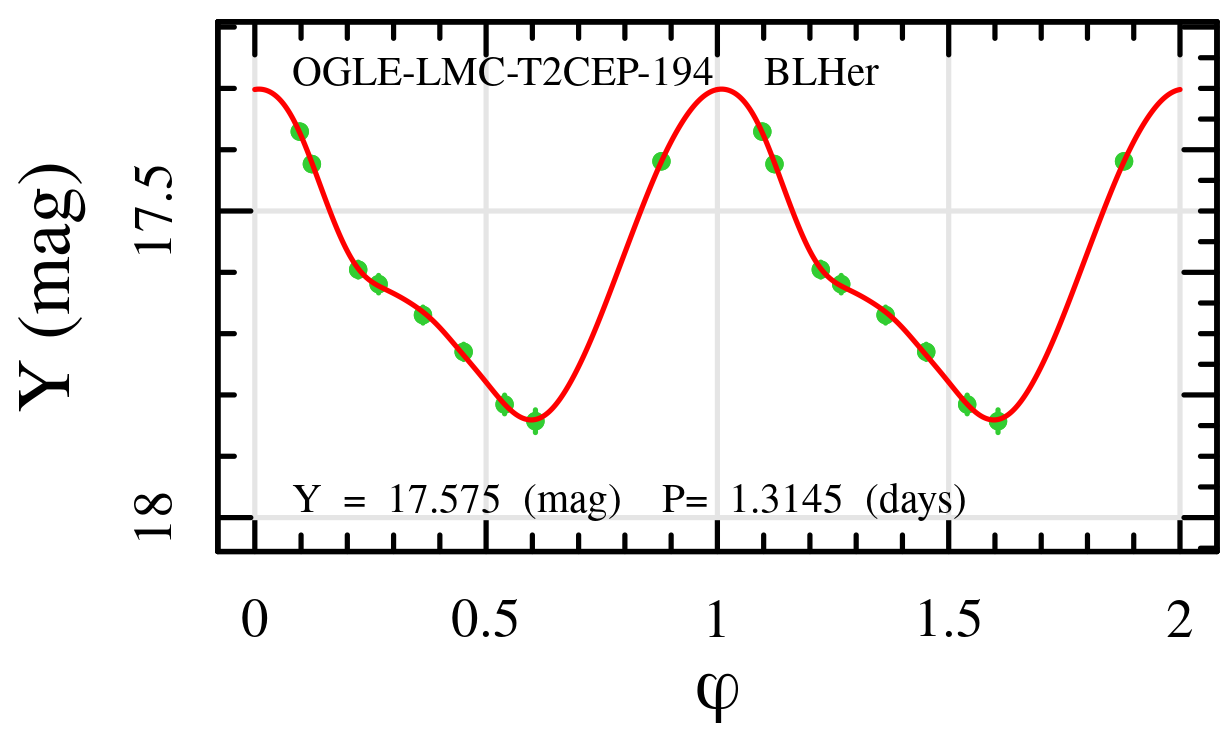}
    \includegraphics[width=0.32\textwidth]{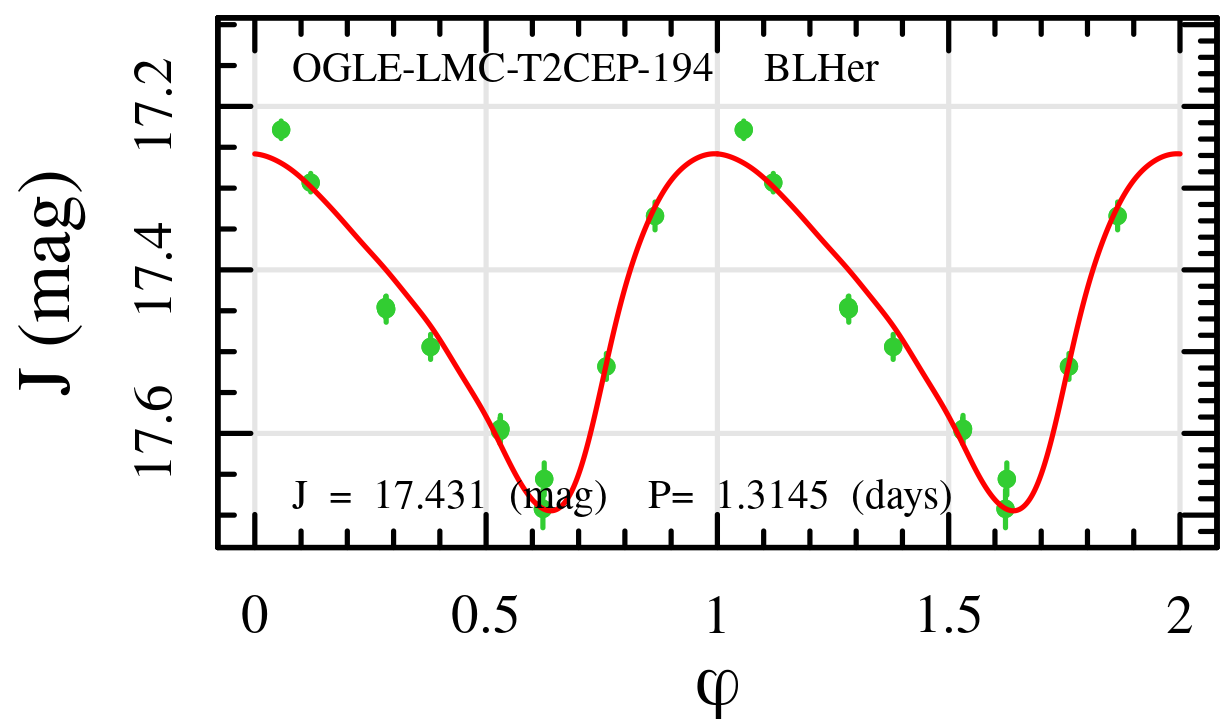}
    \includegraphics[width=0.32\textwidth]{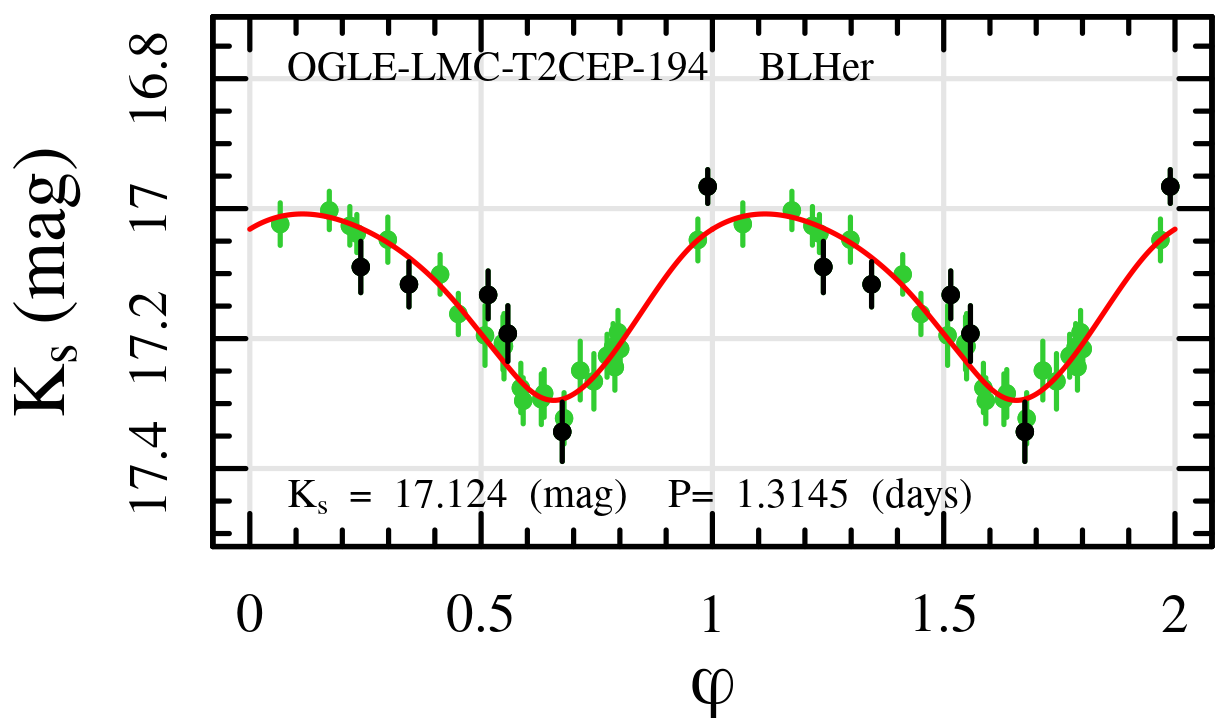}}
    \hbox{
    \includegraphics[width=0.32\textwidth]{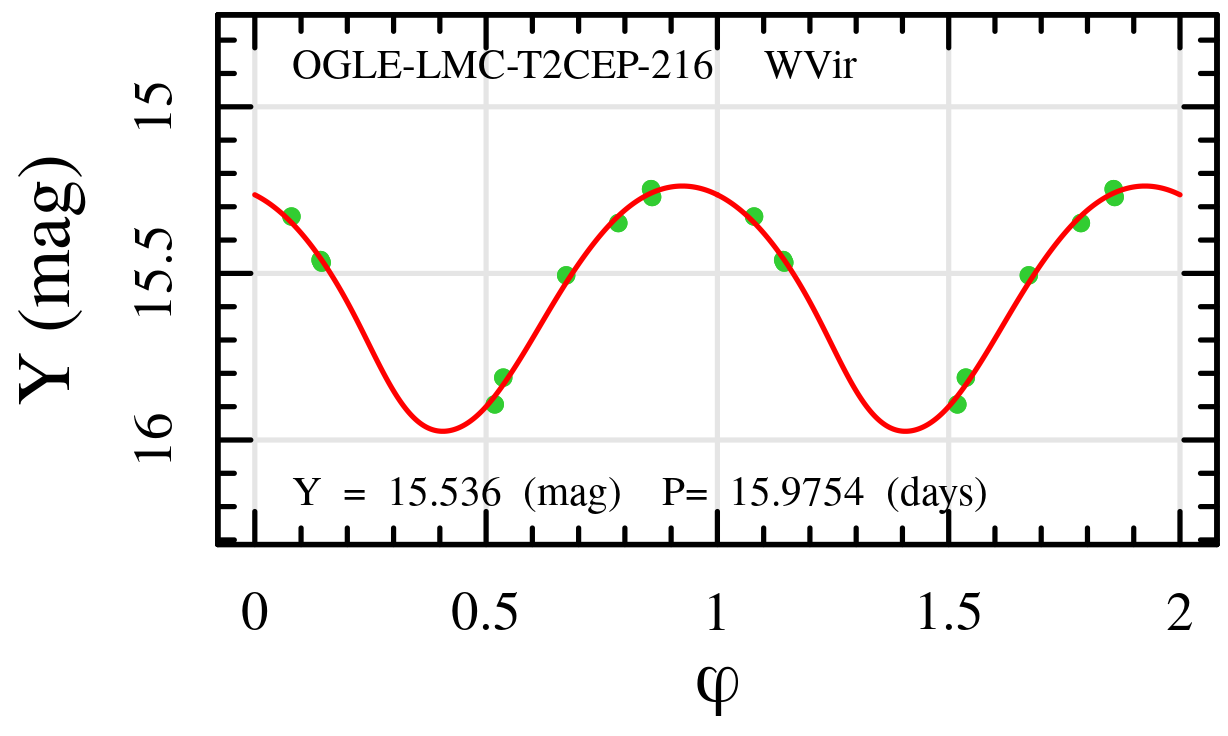}
    \includegraphics[width=0.32\textwidth]{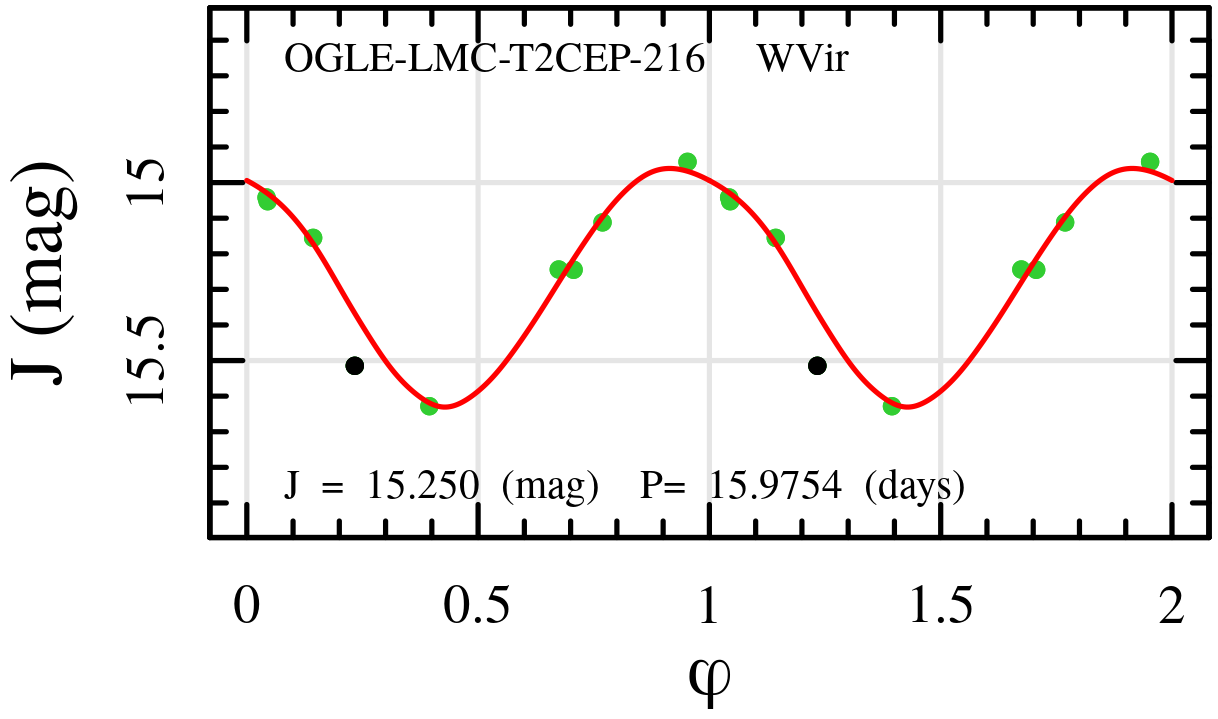}
    \includegraphics[width=0.32\textwidth]{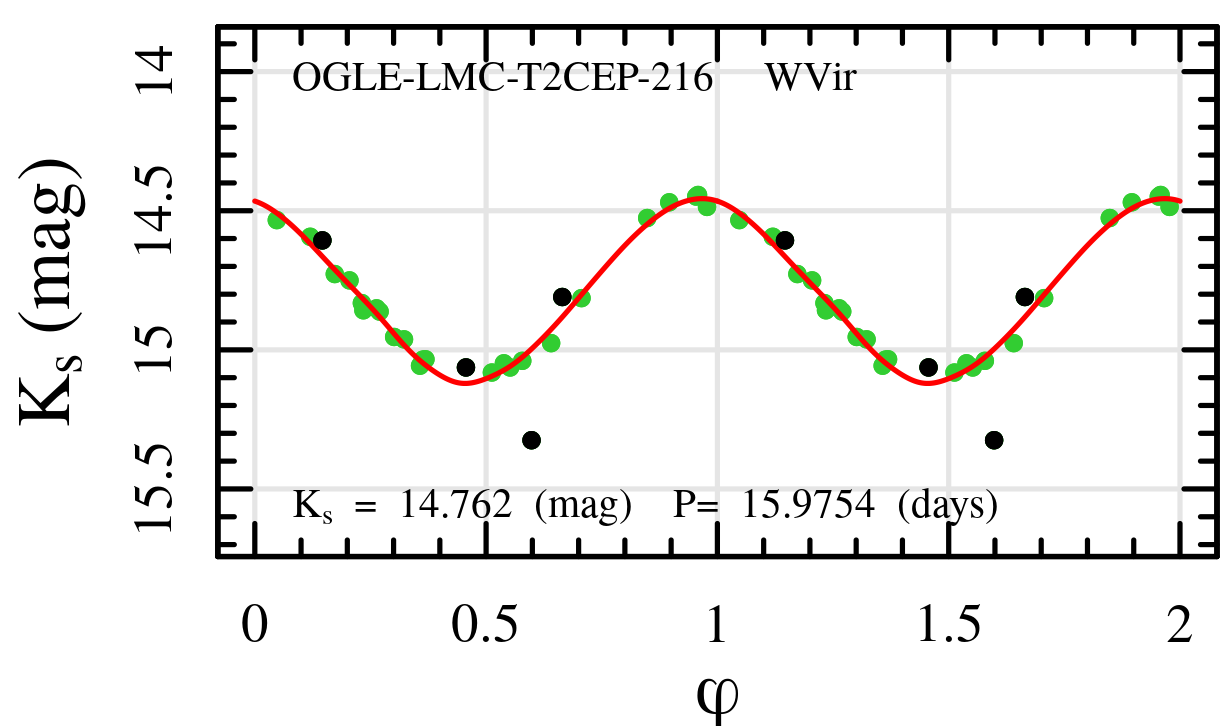}}
    \hbox{
    \includegraphics[width=0.32\textwidth]{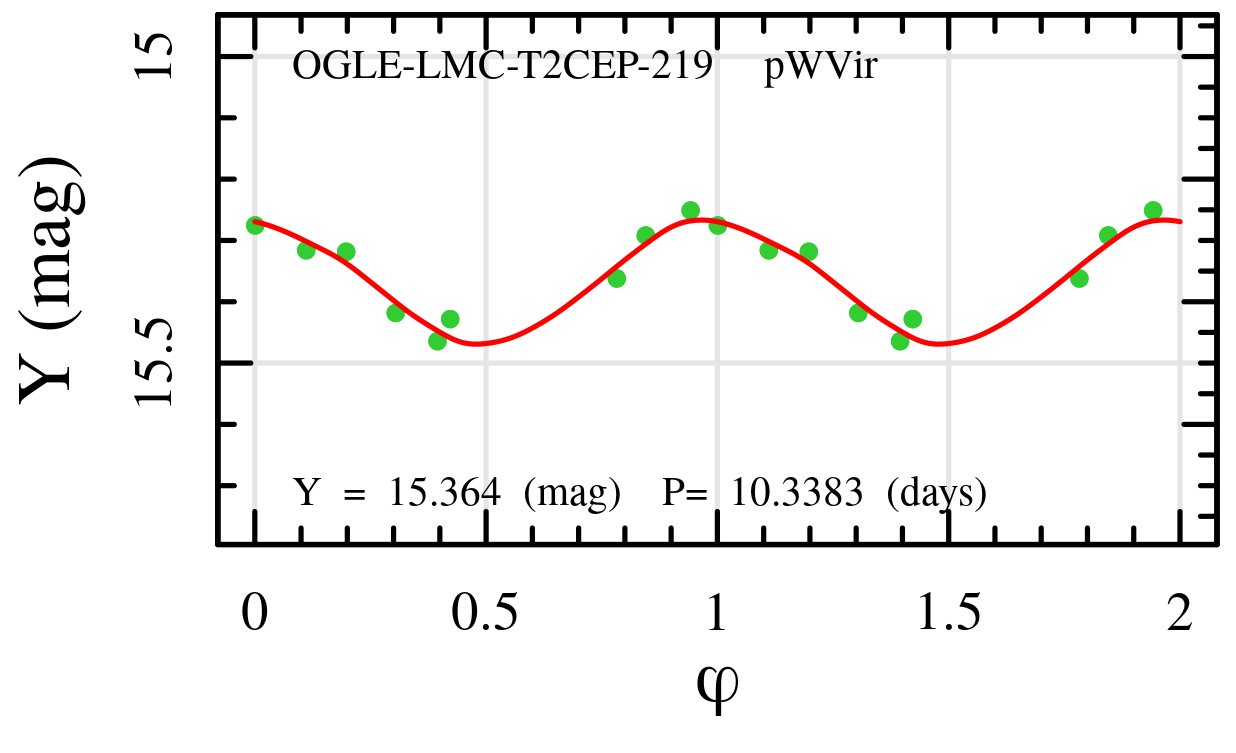}
    \includegraphics[width=0.32\textwidth]{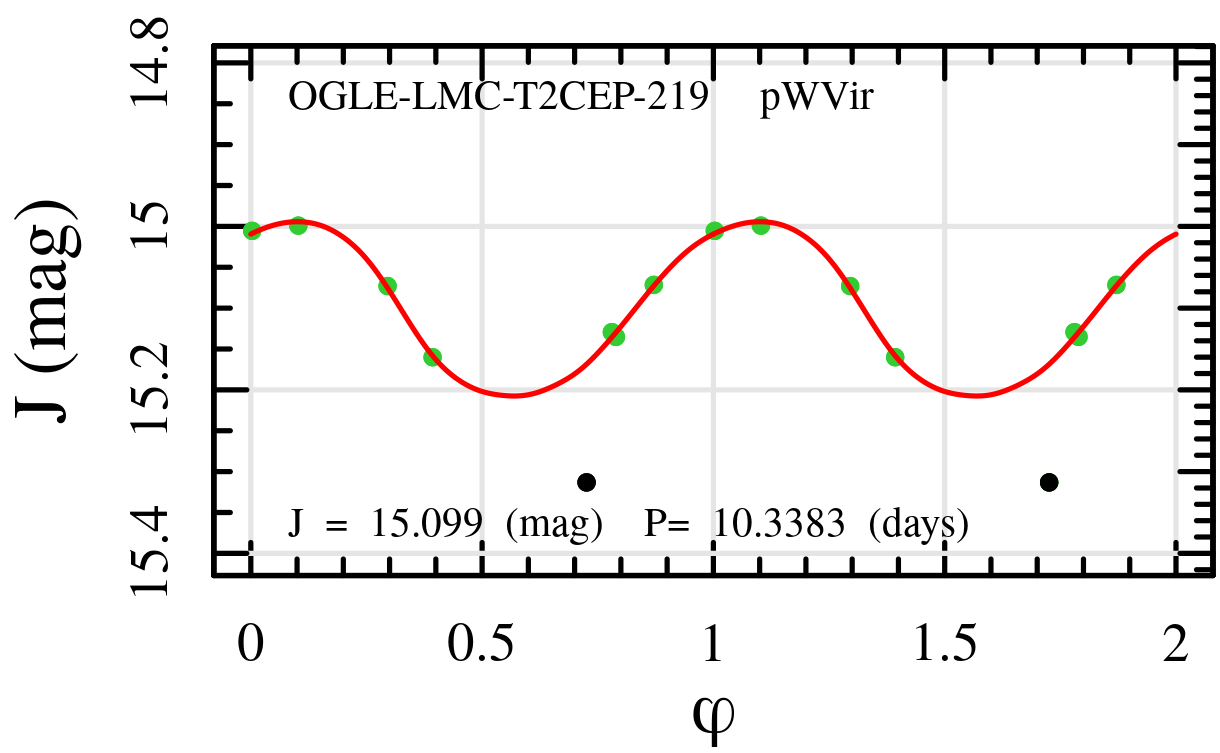}
    \includegraphics[width=0.32\textwidth]{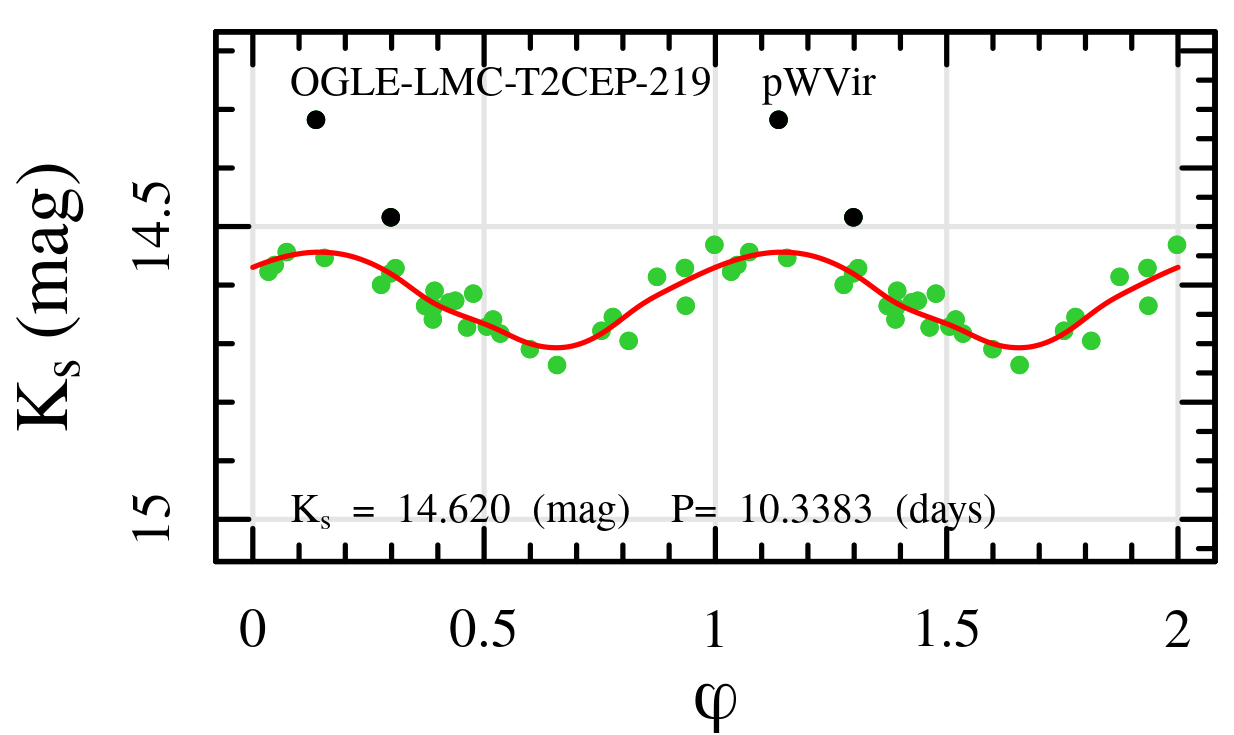}}
    \hbox{
    \includegraphics[width=0.32\textwidth]{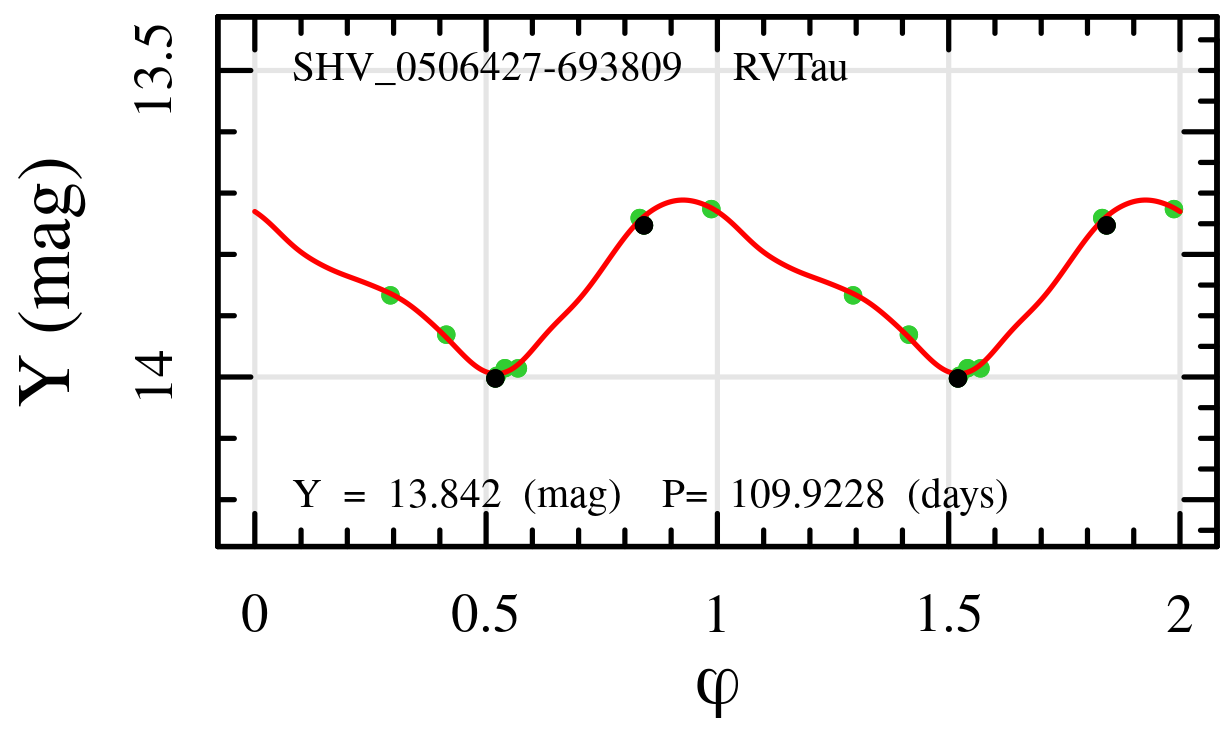}
    \includegraphics[width=0.32\textwidth]{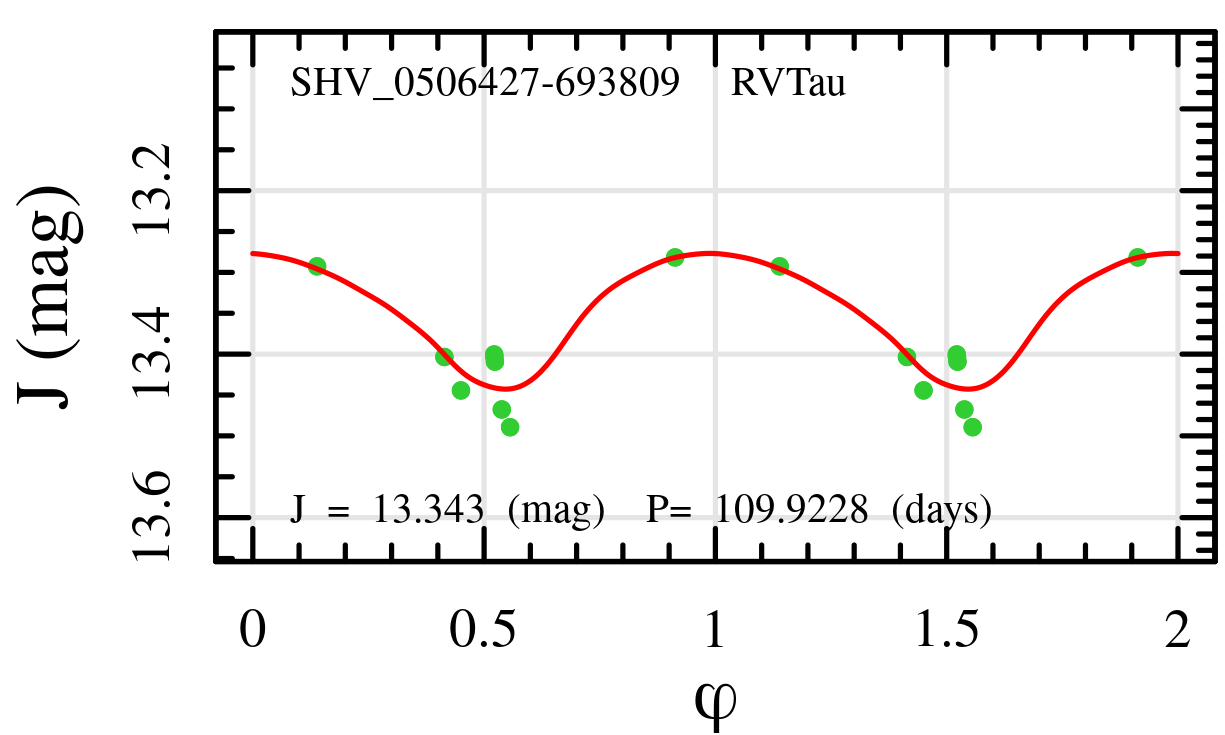}
    \includegraphics[width=0.32\textwidth]{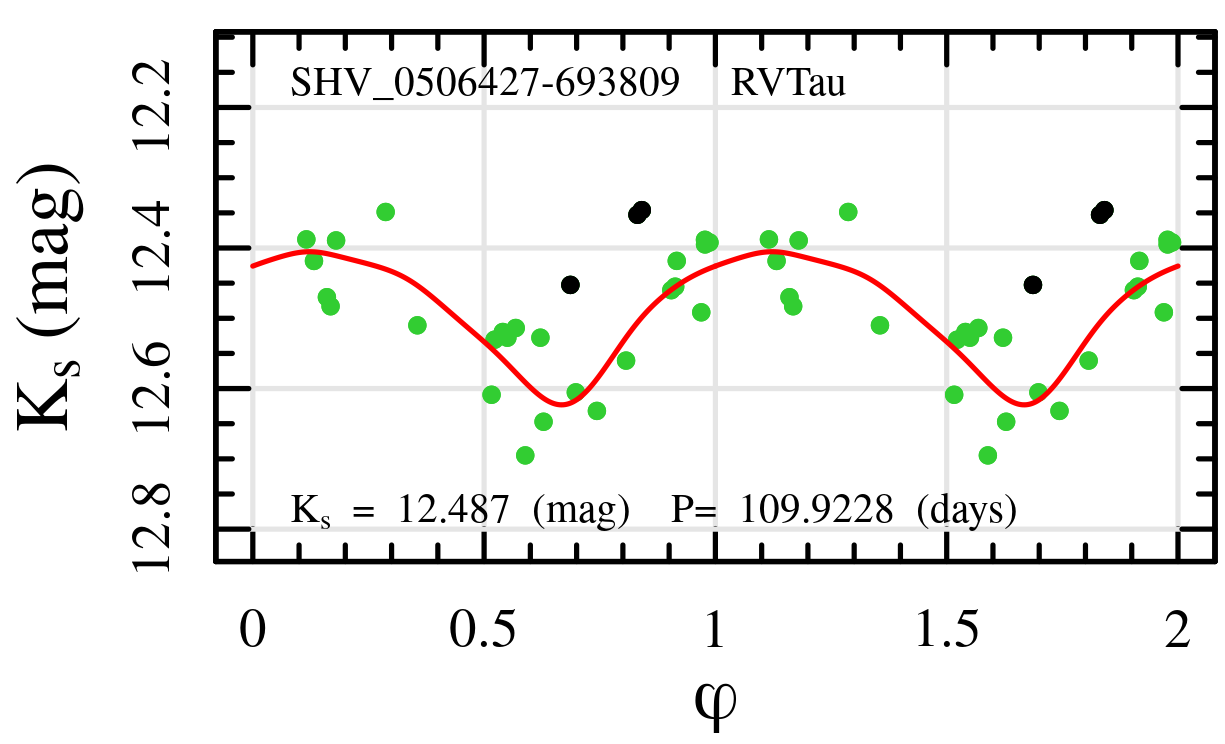}
    }
    }
    \caption{\label{templatefity} Examples of a BLHer, a WVir, a pWVir and a RVTau (from top to bottom row) light curves fitted in $Y,\,J,\,K_s$ bands (left to right
    columns).\\
    Notes: the green filled circles are the observations; the red solid line is the best template for the light curve; the black filled circles are the outliers data not used in the fit.}	
	\end{figure*}

The procedure outlined above allowed us to fit satisfactorily all the light curves using the best templates possible, eventually enabling the determination of accurate intensity-averaged magnitudes (and amplitudes) in all the bands. The final results for the whole sample of 339 targets are shown in Table~\ref{dativmc}, while examples of fitted light curves are shown in Fig.~\ref{templatefity}.

\begin{sidewaystable}
\footnotesize\setlength{\tabcolsep}{3pt}
\caption{VMC photometric parameters for all the 339 LMC and SMC T2Cs analysed in this work.}
\label{dativmc}
\begin{tabular}{lcclccccclcccclcccclccl}
\hline  
\noalign{\smallskip}   
       ID     &  RA & Dec & CLASS & $P$ & $\langle Y \rangle$ & $\sigma_{\langle Y \rangle}$& A($Y$) &  $\sigma_{{\rm A}(Y)}$ & $f_Y$ & $\langle J \rangle$ & $\sigma_{\langle J \rangle}$&   A($J$) &  $\sigma_{{\rm A}(J)}$  & $f_J$ & $\langle K_\mathrm{s} \rangle$ & $\sigma_{\langle K_\mathrm{s} \rangle}$&   A($K_\mathrm{s}$) &  $\sigma_{{\rm A}(K_\mathrm{s})}$ & $f_{K_s}$ & E(V-I) & $f_{E(V-I)}$  & Source \\
   & deg & deg &  & days & mag &mag & mag & mag & & mag & mag & mag & mag & &  mag & mag &  mag & mag &  &  &    \\                     
(1)  & (2)  & (3) & (4) & (5) & (6) & (7) &(8) & (9) & (10) & (11) & (12)  & (13) & (14) & (15) &(16)  & (17) &(18) & (19) & (20) & (21)  & (22) & (23) \\                     
\noalign{\smallskip}
\hline  
\noalign{\smallskip}
  OGLE-LMC-T2CEP-290 & 72.79843 & $-$69.00914 & 1O & 0.932 & 17.500 & 0.040 & 0.23 & 0.07 & f & 17.251 & 0.001 & 0.16 & 0.01 & t & 16.890 & 0.010 & 0.11 & 0.02 & f& 0.194 & 1 &  OGLE\\
  OGLE-LMC-T2CEP-287 & 95.17501 & $-$72.52670 & BLHer & 1.344 & 17.650 & 0.010 & 0.35 & 0.01 & t & 17.610 & 0.030 & 0.25 & 0.08 & f & 17.259 & 0.007 & 0.20 & 0.02 & f & 0.102 & 1 & OGLE\\
  OGLE-LMC-T2CEP-285 & 94.45717 & $-$69.76087 & BLHer & 1.050 & 17.900 & 0.020 & 0.44 & 0.03 & f & 17.720 & 0.010 & 0.38 & 0.05 & f & 17.430 & 0.010 & 0.30 & 0.03 & f & 0.093 & 1 & OGLE\\
  OGLE-LMC-T2CEP-283 & 94.24198 & $-$71.51516 & WVir & 9.503 & 15.858 & 0.001 & 0.39 & 0.03 & f & 15.609 & 0.002 & 0.23 & 0.01 & t & 15.131 & 0.004 & 0.23 & 0.01 & f& 0.115 & 1 &  OGLE\\
  OGLE-LMC-T2CEP-282 & 93.73022 & $-$67.06919 & pWVir & 6.304 & 16.140 & 0.009 & 0.24 & 0.06 & f & 15.787 & 0.001 & 0.86 & 0.01 & f & 15.621 & 0.005 & 0.24 & 0.03 & f & 0.072 & 1 &  OGLE\\
\noalign{\smallskip}
\hline  
\noalign{\smallskip}
\end{tabular}
\tablefoot{ Columns: $(1)$ Identification; $(2)$-$(3)$ RA and Dec; $(4)$ Class of T2C: BLHer= BLHerculis, RVTau= RV Tauri, WVir= W Virginis, pWVir= peculiar W Virginis; $(5)$ Period; $(6)$-$(7)$ Intensity-averaged magnitude in $Y$ and relative uncertainty; $(8)$-$(9)$ Peak-to-peak amplitude in $Y$ and relative uncertainty;$(10)$ tag indicating if a fixed value for the amplitude was used to calculate the averaged-intensty magnitude (f=false, t=true); $(11)$ to $(15)$ as for columns $(6)$ to $(10)$ but for the $J$ band; $(16)$ to $(20)$ as for column $(6)$ to $(10)$ but for the $K_\mathrm{s}$ band; $(21)$ $E(V-I)$ values adopted in this work; $(22)$ flag indicating the origin of the reddening value;  $(23)$ flag indicating the source (OGLE\,IV or $Gaia$) of period, identification and epoch of maximum. A portion is shown here for guidance regarding its form and content.The machine-readable version of the full table will be published at the CDS (Centre de Données astronomiques de Strasbourg, https://cds.u-strasbg.fr/).}
\end{sidewaystable}

\subsection{Reddening estimates for the target stars}

To deal with the interstellar extinction, we took advantage of the recent accurate reddening maps published by \citet{skowron2021ogle}. These maps have a varying spatial resolution, from 1.7 arcsec $\times$ 1.7 arcsec in the centre of the MCs to 27 arcsec $\times$ 27 arcsec in the peripheries, where there are fewer stars. They do not cover the entire extent of the MCs and were complemented by the same authors with the \citet[][SFD hereafter]{schlegel1998maps} maps, which have a lower spatial resolution. 
As some of our targets were very distant from the MCs centres, the derived reddening values $E(V-I)$ can be based either on \citet{skowron2021ogle} or the SFD maps. 

The coefficients used to correct the magnitudes for the extinction were calculated according to \citet{cardelli1989relationship} assuming $\rm{R_V} = A(V)/E(B-V)$ = 3.23  \citep{inno2013distance}. For the $Gaia$ bands we used the coefficients published by \citet{casagrande2018use}. The adopted extinction coefficients, corresponding to the color-excess $E(V-I)$, are reported in Table~\ref{ass}.

\begin{table}
 \begin{center}
\caption{\label{ass}\\ Extinction coefficient in different bands as a function of E($V-I$).} 
\begin{tabular}{c c } 
\\
    \hline
    Band & Interstellar absorption \\ [0.5ex] 
    \hline
   $G_{BP}$  &   2.678   \\
   $V$  & 2.564   \\
   $G$  & 2.175 \\
   $G_{BP}$  & 1.615   \\
   $I$   &  1.564   \\
   $Y$   &  1.000     \\
   $J$   &  0.743   \\
   $K_s$  & 0.307   \\
   \hline
\end{tabular}
\end{center}
\end{table}

\subsection{Complementary optical data}
\label{opticalData}

We complemented the NIR VMC data with literature optical photometry to study the variation of $PL$ relations with wavelength. In addition, we can also use Wesenheit magnitudes mixing optical and infrared bands. We added data in the $V,\,I$ bands from the OGLE IV survey and in the  $G,\,G_{BP},\,G_{RP}$ bands from the $Gaia$ mission \citep[][]{ripepi2022gaia}. The optical magnitudes from the literature were already calculated as intensity-averaged ones, so they do not need any transformation. All the optical data is listed in Table~\ref{datiottico}.

The flag "SOURCE", in Table~\ref{datiottico}, indicates whether the object was identified by OGLE IV or $Gaia$ (as in Table~\ref{dativmc}). The flag "SOS" discerns the technique used to calculate the average magnitudes in the $Gaia$ bands\footnote{SOS stands for Specific Object Studies \citep[see e.g.][]{ripepi2022gaia}}: the value 0 indicates that the magnitudes are calculated with the standard technique adopted for all the stars 
\citep[see e.g.][]{Evans2018} and not specific for pulsating variables; the value 1 means that the magnitudes in the $Gaia$ bands are calculated with the averaged-intensity technique after modelling the light curve \citep[e.g.][]{clementini2016gaia}; the value 2 specifies that there are no magnitudes in the $Gaia$ bands. 
Therefore, among the 318 T2Cs with OGLE IV identification, 316 have $Gaia$ magnitudes: 85 and 231 with flag "SOS" = 0 and 1, respectively. The two missing stars have the flag "SOS" = 2. 

Note that, while all the stars in the OGLE IV catalogue have the average $I$ band magnitudes, not all have the $V$ measurement. Moreover, the $V$, $I$ data is completely missing for the stars originating from the $Gaia$ catalogue and not present in OGLE IV. To recover the missing values in $V,\,I$ bands, we used the photometric transformations between Johnson and $Gaia$ bands provided by \citet{pancino2022gaia}. They use high-order polynomials to obtain the $V$ and $I$ band photometry from $G,\,G_{BP},\,G_{RP}$ bands with uncertainties of 0.01 and 0.03 mag, respectively.
The accuracy of transformations has been tested by \citet{2023arXiv231003603T}, by comparing the $V$ and $I$ values from the \citet{pancino2022gaia} equations and those available in the literature for a sample of Galactic CCs. The calculated $V$ magnitudes are accurate within 0.01 mag, as expected, while the $I$ magnitudes result are too bright by 0.03 mag. As a consequence, we added to the transformed $I$ magnitudes an offset of 0.03 mag. 

The flag "VI", in Table~\ref{datiottico}, marks the origin of the values of the $V\,I$ magnitudes: the first part indicates the origin of the $V$ value, the second of the $I$ one. Usually, the stars identified by the OGLE survey have the flag "VI" equal to OGLE,OGLE. However, $V$ magnitudes are missing for many T2Cs in the OGLE sample. In these cases, we adopted the $Gaia$ magnitudes to estimate this quantity and the resulting "VI" flag is "P22,OGLE". For the stars identified only by $Gaia$ both $V$ and $I$ magnitudes are estimated from $G,\,G_{BP},\,G_{RP}$ data, so that we have "VI" flag equal to "P22,P22". 
Note that the $V\,I$ magnitudes estimated from the $Gaia$ bands may have lower quality if the flag "SOS" is 0. In any case, the effect of the adoption of the standard averaged magnitudes are expected to be contained in a few per cent errors \citep[see discussions on this point in ][]{ripepi2022classical,GaiaDrimmel2023}.

\begin{sidewaystable}
\footnotesize\setlength{\tabcolsep}{3pt} \caption{Optical photometric parameters for all the 339 LMC and SMC T2Cs analysed in this work. }
\label{datiottico}
\begin{tabular}{lcccccccccccccccccc}
\hline  
\noalign{\smallskip}   
       ID     &  RA & Dec & CLASS & $P$ &  $\sigma_{{\rm P}}$ & $\langle V \rangle$ &  $\langle I \rangle$ &$epoch_I$  & A(I) & $\langle G \rangle$ & $\sigma_{\langle G \rangle}$& $\langle G_{BP} \rangle$ & $\sigma_{\langle G_{BP} \rangle}$ & $\langle G_{RP} \rangle$ & $\sigma_{\langle G_{RP} \rangle}$ & Source & SOS & VI  \\
   & deg & deg &  & days & days &mag & mag & days &  mag & mag & mag & mag & mag &  mag & mag &   &  &     \\                     
(1)  & (2)  & (3) & (4) & (5) & (6) & (7) &(8) & (9) & (10) & (11) & (12)  & (13) & (14) & (15) &(16)  & (17) &(18) & (19)  \\                     
\noalign{\smallskip}
\hline  
\noalign{\smallskip}  
  DR3\_4687428801808318208 & 16.12685 & $-$72.61781 & RVTau & 60.100 & 0.100 & --- & --- & 1529.710 & --- & 15.158 & 0.002 & 15.380 & 0.005 & 14.673 & 0.007 & $Gaia$ & 1 & P22,P22\\
  OGLE-LMC-T2CEP-016 & 73.94154 & $-$69.12951 & RVTau & 20.305 & 0.002 & 15.883 & 15.454 & 7008.381 & 0.166 & 15.794 & 0.009 & 15.930 & 0.007 & 15.488 & 0.008 & OGLE & 1 & OGLE,OGLE\\
  OGLE-LMC-T2CEP-017 & 74.06676 & $-$68.271 & WVir & 14.456 & 0.002 & --- & 15.995 & 7014.267 & 0.856 & 16.730 & 0.010 & 17.140 & 0.050 & 16.060 & 0.030 & OGLE & 1 & P22,OGLE\\
  OGLE-LMC-T2CEP-018 & 74.62422 & $-$69.40808 & BLHer & 1.380 & 0.001 & 18.634 & 17.959 & 7000.813 & 0.615 & 18.498 & 0.007 & 18.560 & 0.070 & 18.000 & 0.020 & OGLE & 1 & OGLE,OGLE\\
  OGLE-LMC-T2CEP-019 & 74.70599 & $-$68.07435 & pWVir & 8.673 & 0.001 & 16.850 & 15.989 & 7002.910 & 0.633 & 16.665 & 0.007 & 16.860 & 0.010 & 15.960 & 0.050 & OGLE & 1 & OGLE,OGLE\\
  OGLE-SMC-T2CEP-15 & 12.40383 & $-$73.16708 & BLHer & 2.570 & 0.001 & 17.189 & 16.884 & 7002.500 & 0.172 & 17.181 & 0.003 & 17.253 & 0.006 & 16.900 & 0.010 & OGLE & 1 & OGLE,OGLE\\
\noalign{\smallskip}
\hline  
\noalign{\smallskip}
\end{tabular}
\tablefoot{Columns: $(1)$ Identification from OGLE\,IV or $Gaia$; $(2)$-$(3)$ RA and Dec; $(4)$ Class: BLHer= BLHerculis (T2C); RVTau= RV Tauri(T2C); WVir= W Virginis (T2C); pWVir= peculiar W Virginis (T2C); $(5)$ Period; $(6)$ period error; $(7)$ magnitude in $I$ band from OGLE; $(8)$ Magnitude in $V$ band from OGLE; $(9)$ epoch; $(10)$ amplitude in the $I$ band;$(11)$-$(12)$ Magnitude in $G$ and relative uncertainty; $(13)$-$(14)$ As for column $(11)$ and $(12)$ but for the $G_{BP}$;$(15)$-$(16)$ As for column $(11)$ and $(12)$ but for the $G_{RP}$; $(17)$ flag indicating the source used for identification;$(18)$ flag indicating how the $Gaia$ magnitudes have been calculated; $(19)$ flag indicating what is the origin of the $V\,I$ magnitude. A portion is shown here for guidance regarding its form and content. The machine-readable version of the full table will be published at the CDS (Centre de Données astronomiques de Strasbourg, https://cds.u-strasbg.fr/).}
\end{sidewaystable}

\section{Period--Luminosity, Period--Wesenheit and Period--Luminosity--Colour relations}

The multi-band photometry obtained as explained in the previous section has allowed us to calculate a significant number of $PL$, $PLC$ and $PW$ relations for the different classes of pulsating stars of interest for this work, separately for the LMC and SMC. In particular, we adopted different combinations of types: (i) BLHer; (ii) WVir; (iii) BLHer and WVir; (iv) BLHer, WVir and pWVir; (v) BLHer, WVir, pWVir and RVTau. Next, we describe briefly the sequence of procedures followed to calculate the above-mentioned relationships.

\begin{figure*} 
    \vbox{
    \hbox{
    \includegraphics[width=0.5\textwidth]{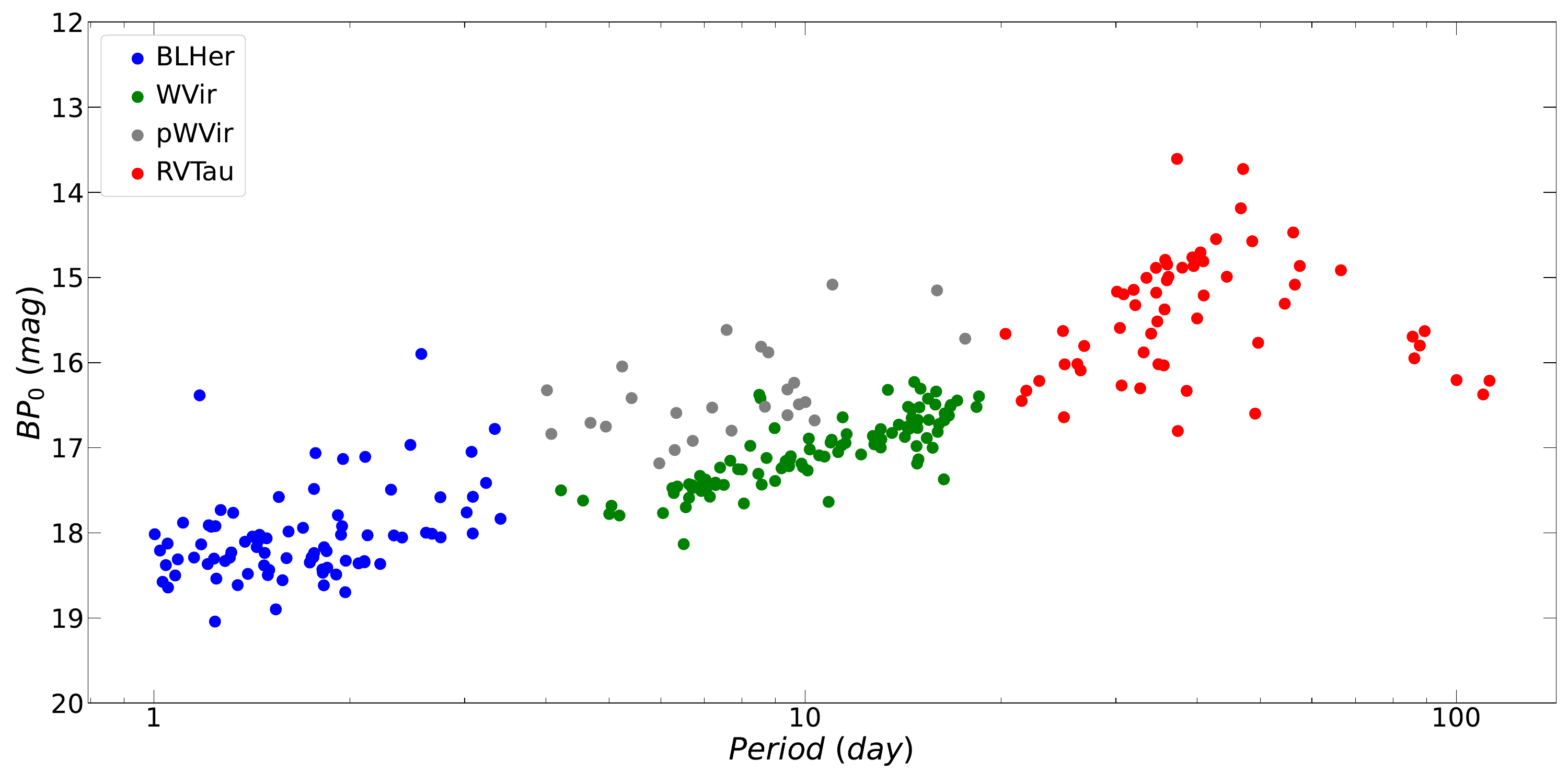}
    \includegraphics[width=0.5\textwidth]{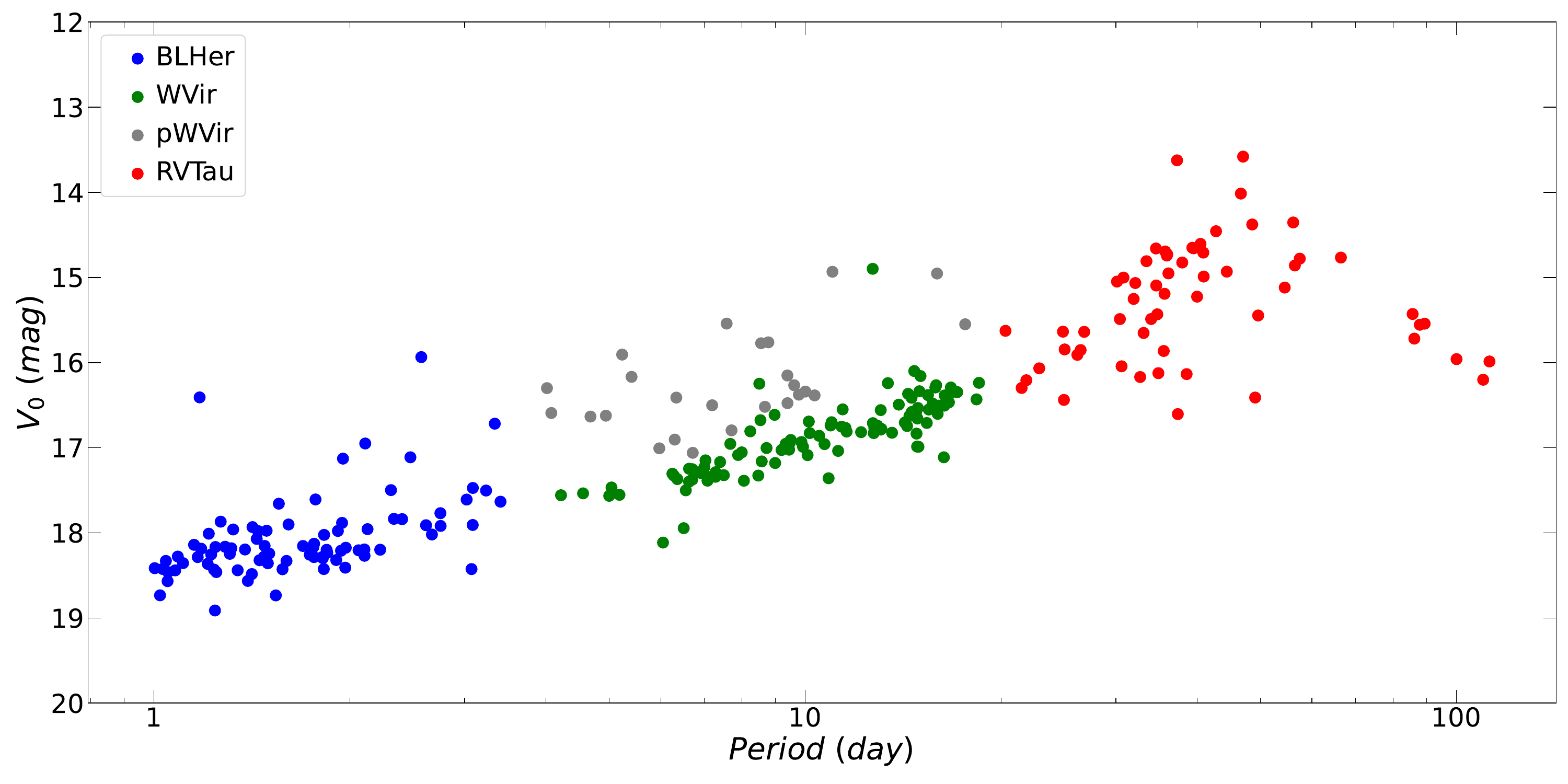}
    }
    \hbox{
    \includegraphics[width=0.5\textwidth]{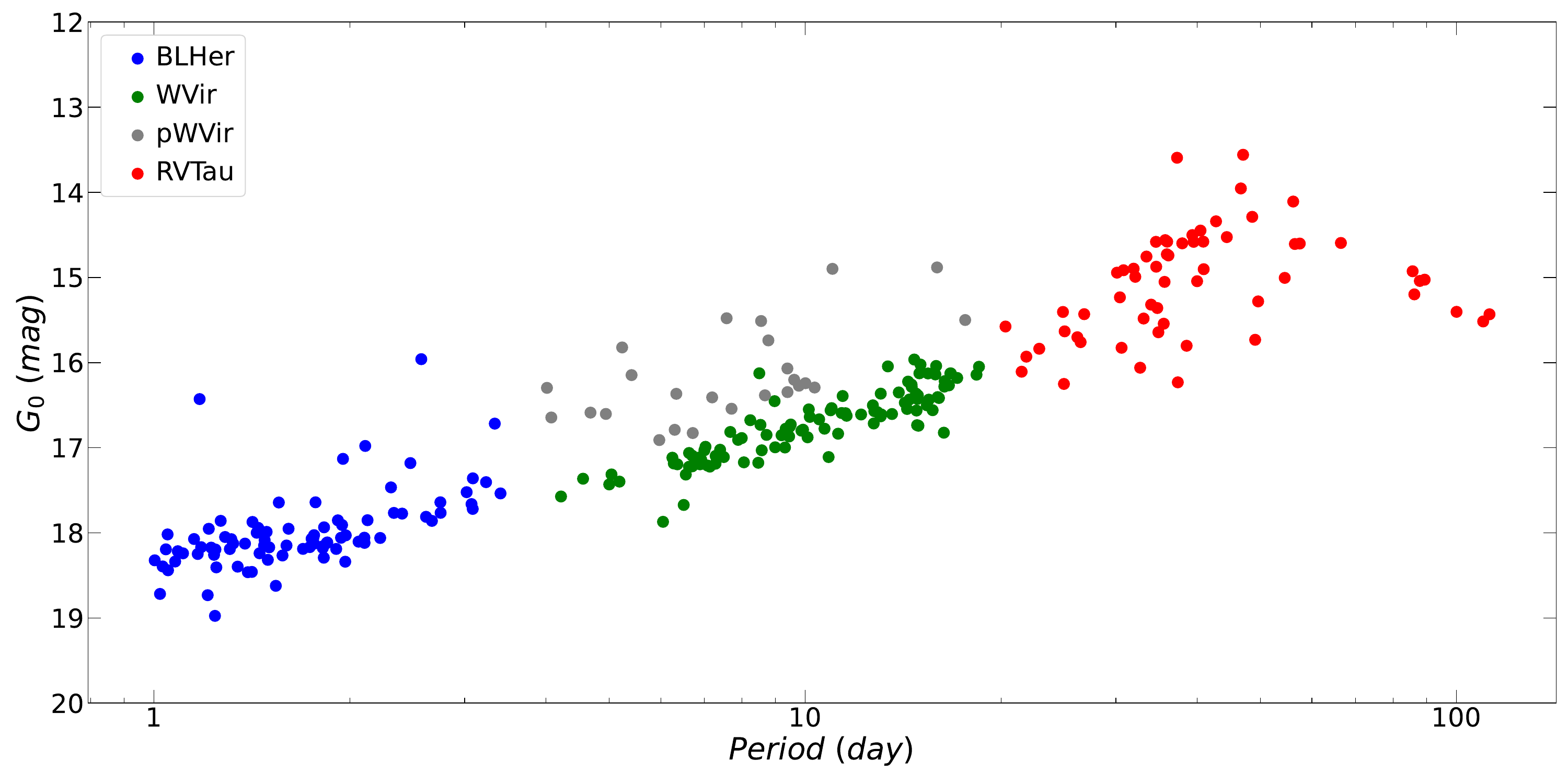}
    \includegraphics[width=0.5\textwidth]{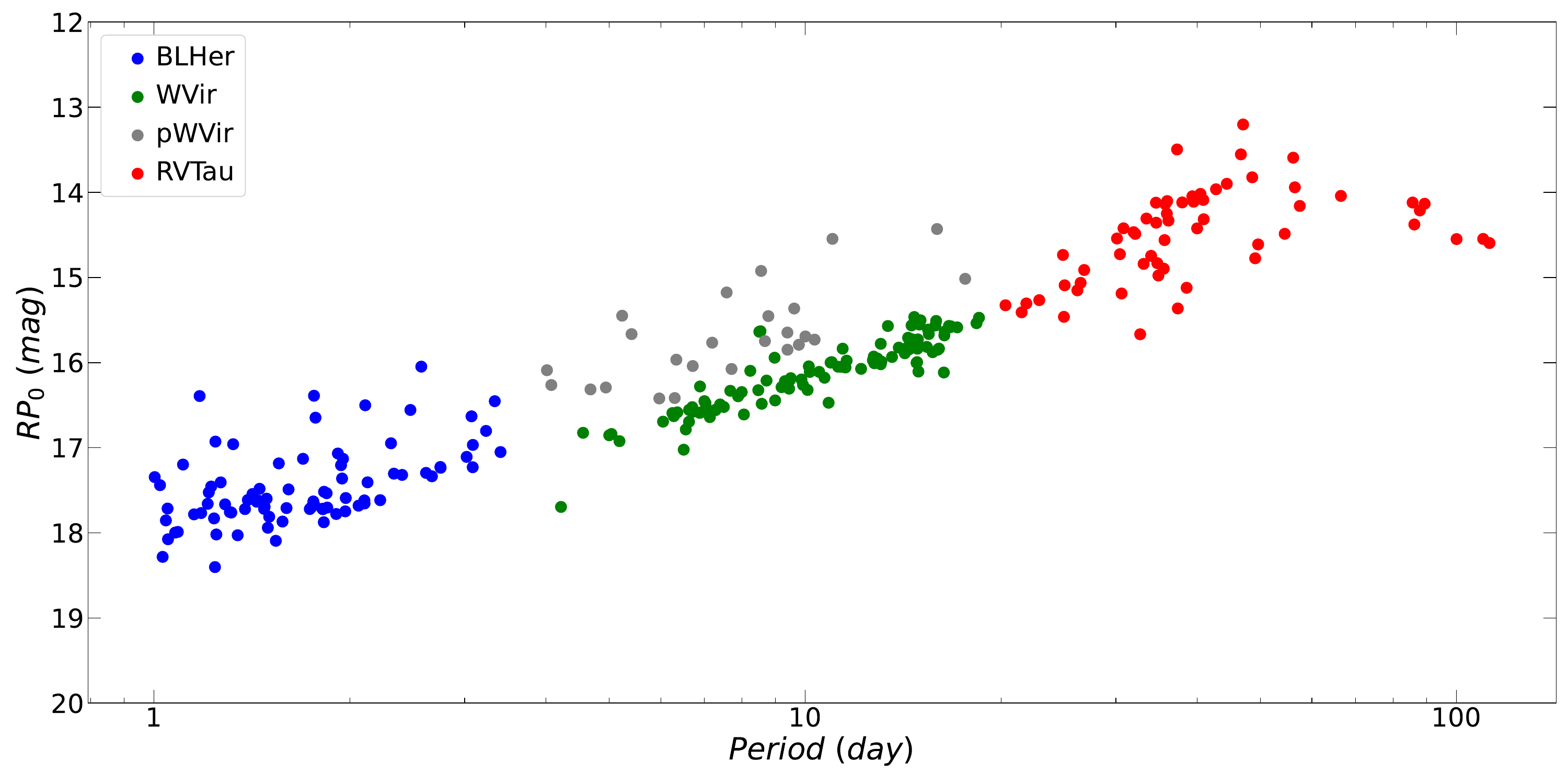}
    }
    \hbox{
    \includegraphics[width=0.5\textwidth]{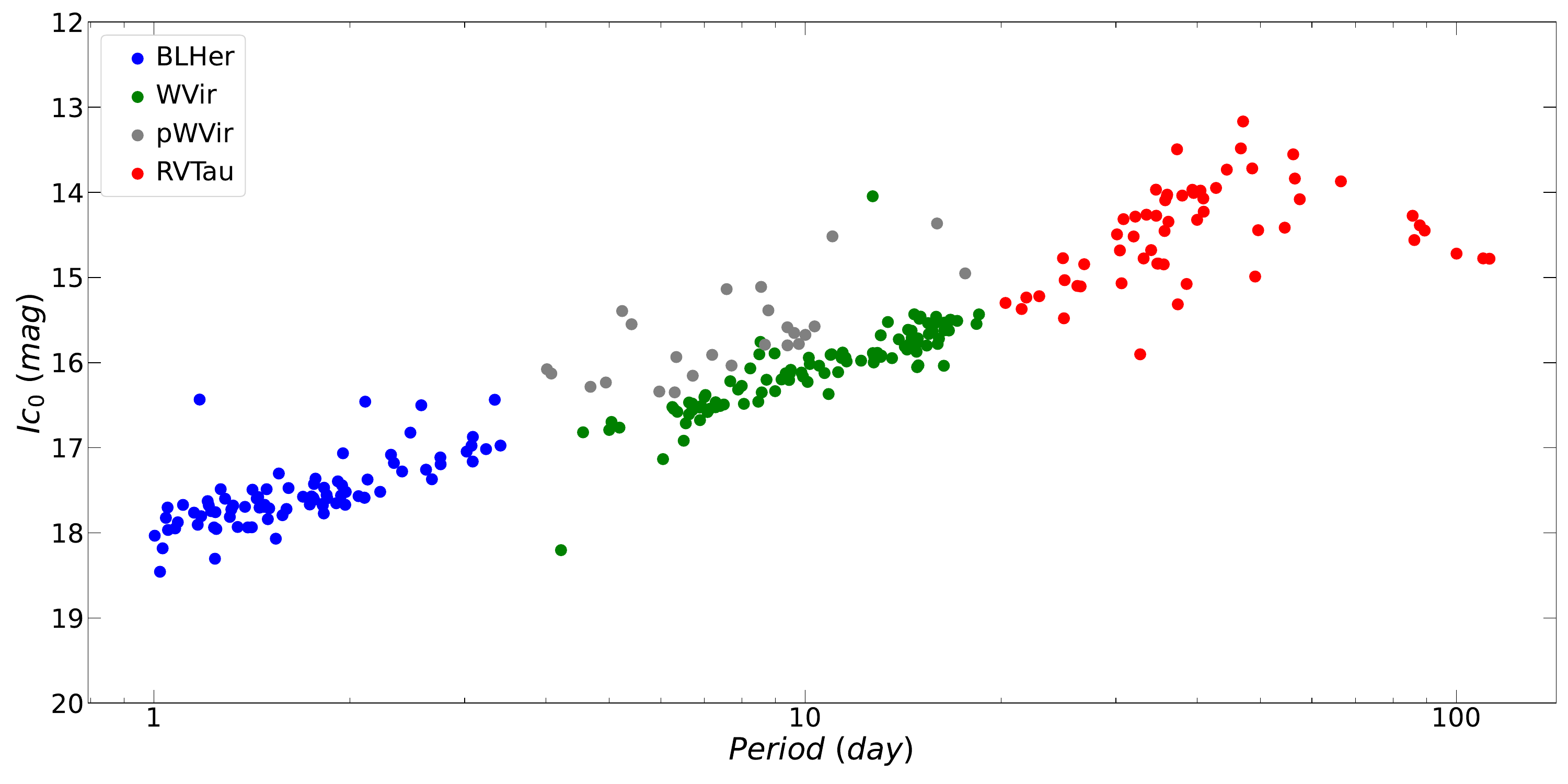}
    \includegraphics[width=0.5\textwidth]{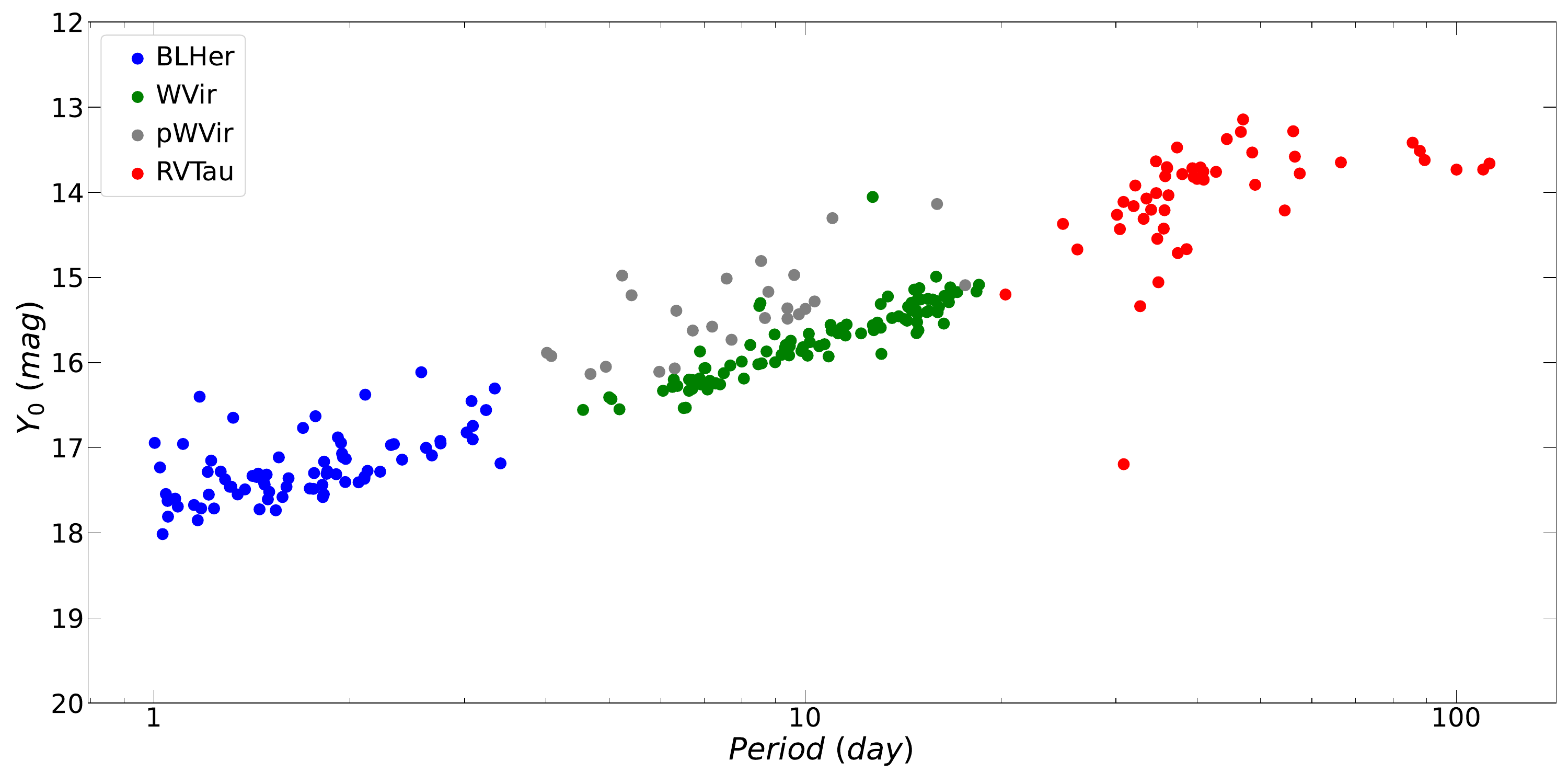}
    }
    \hbox{
    \includegraphics[width=0.5\textwidth]{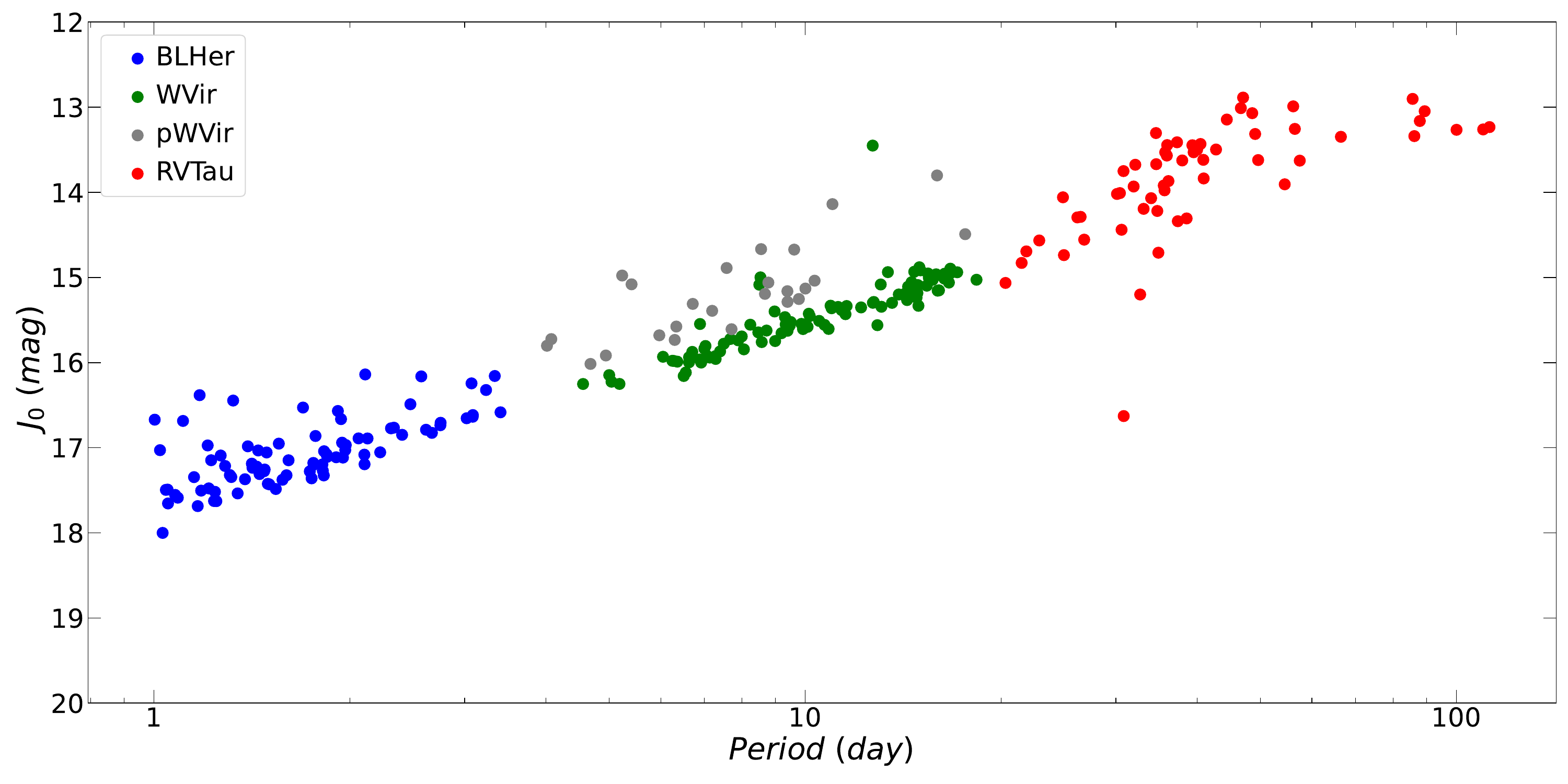}
    \includegraphics[width=0.5\textwidth]{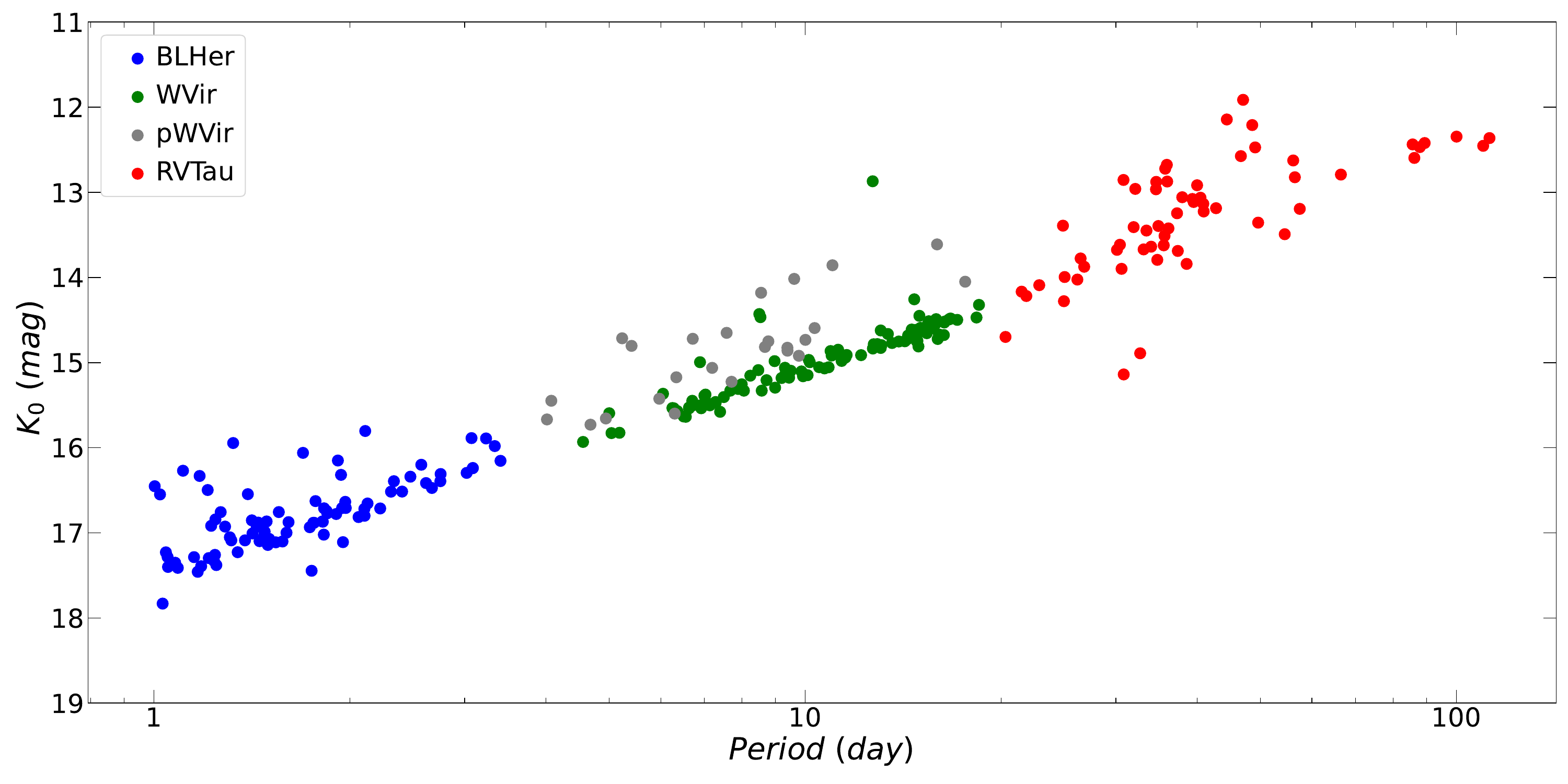}
    }
    
    }
    \caption{\label{pllmc} $PL$ relations in all the considered bands from the optical to the NIR for T2Cs in the LMC. Blue, green, grey and red filled circles represent BLHer, WVir, pWVir and RVTau pulsators, respectively. Note that the error bars are smaller than the dimensions of the dots.}	
	\end{figure*}

\begin{figure*}
    \vbox{
    \hbox{
    \includegraphics[width=0.5\textwidth]{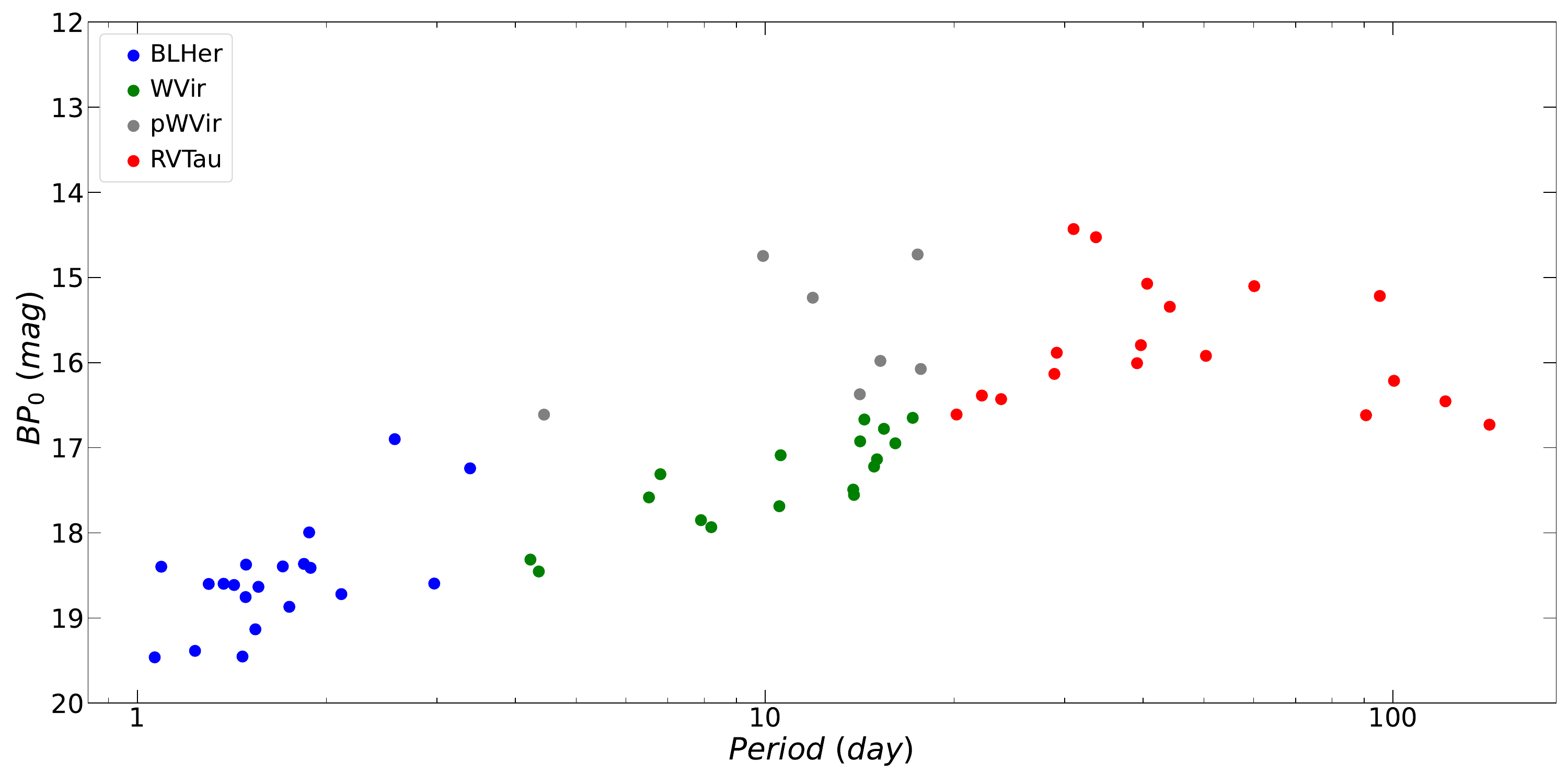}
    \includegraphics[width=0.5\textwidth]{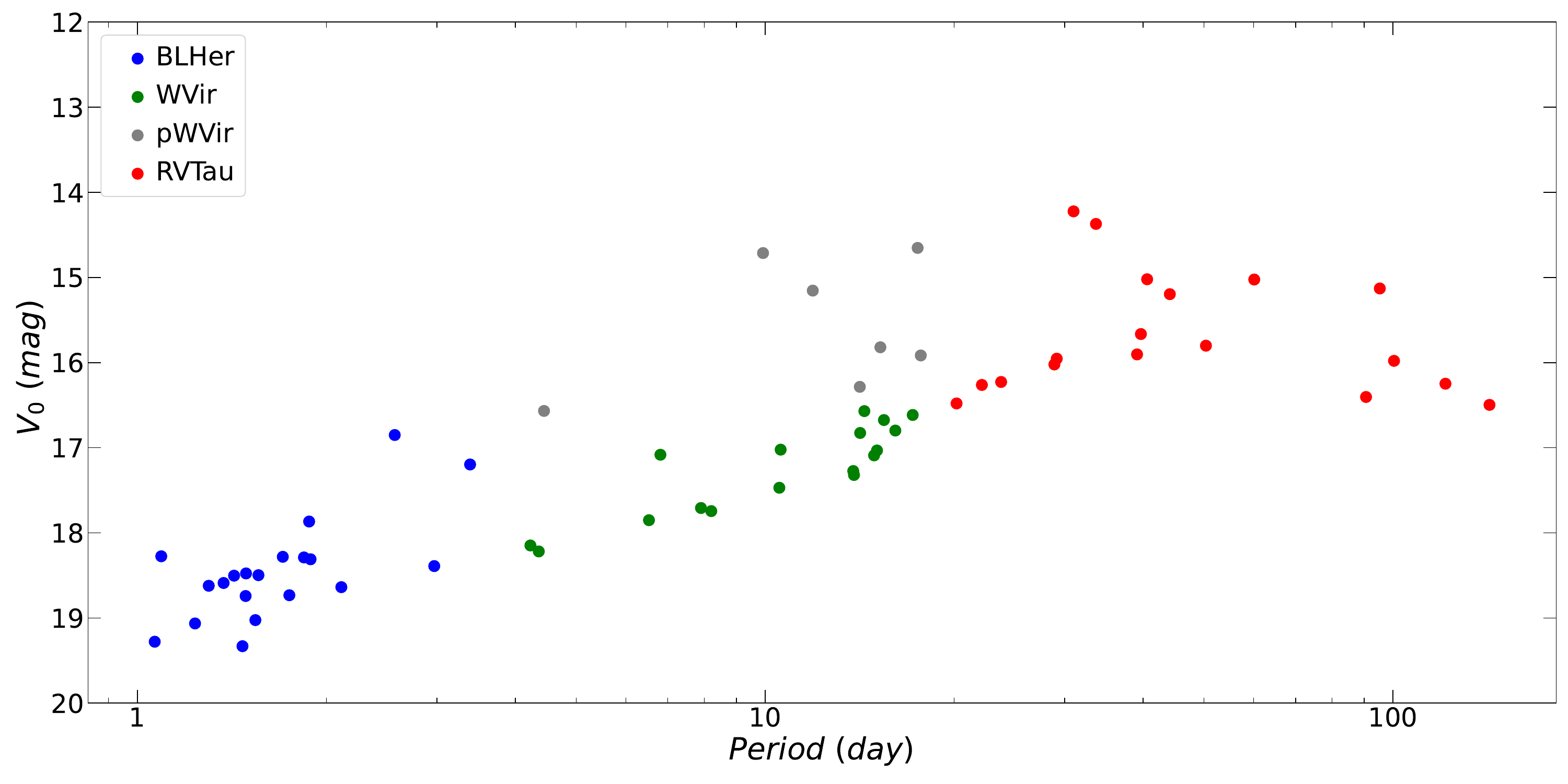}
    }
    \hbox{
    \includegraphics[width=0.5\textwidth]{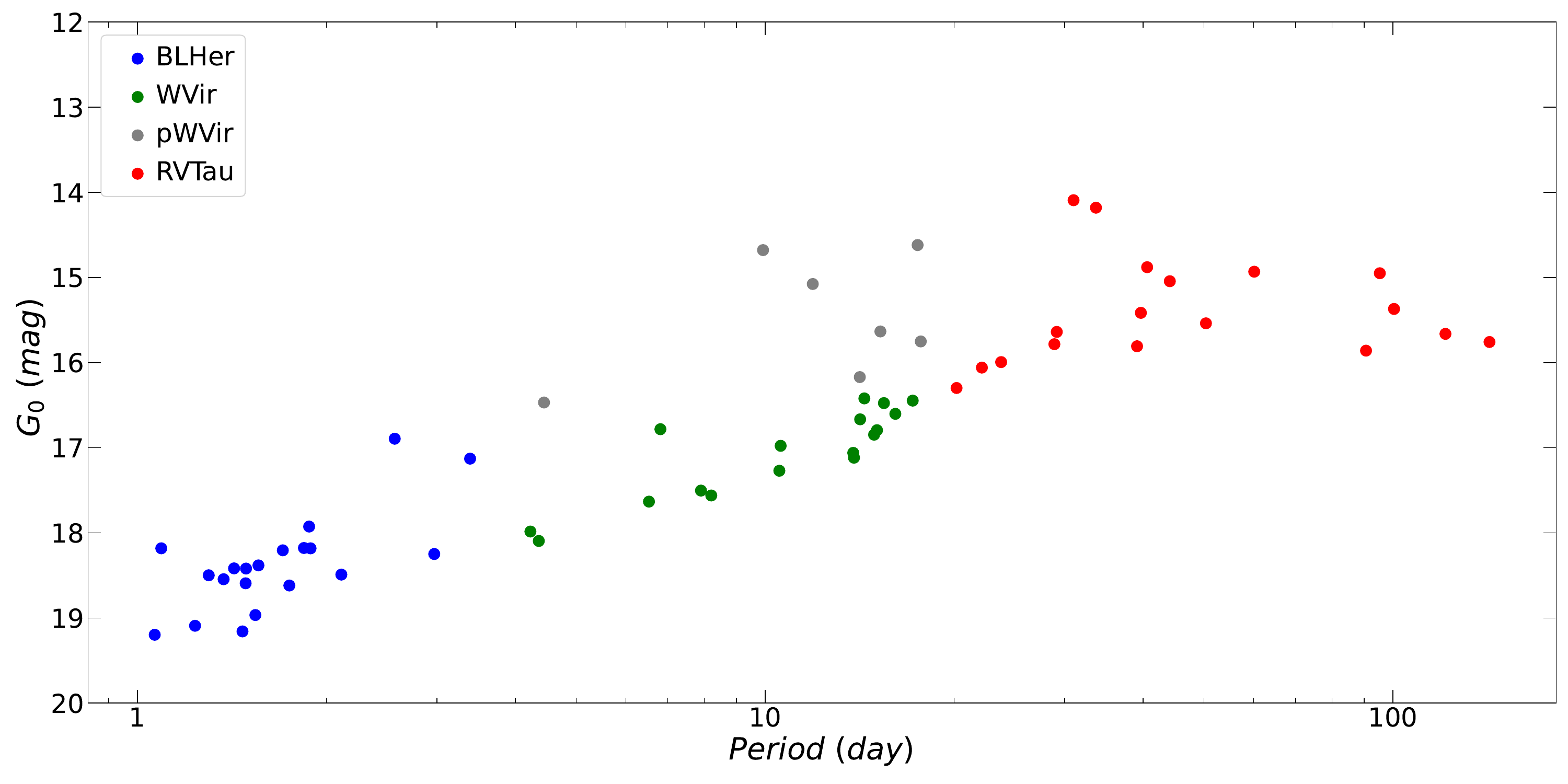}
    \includegraphics[width=0.5\textwidth]{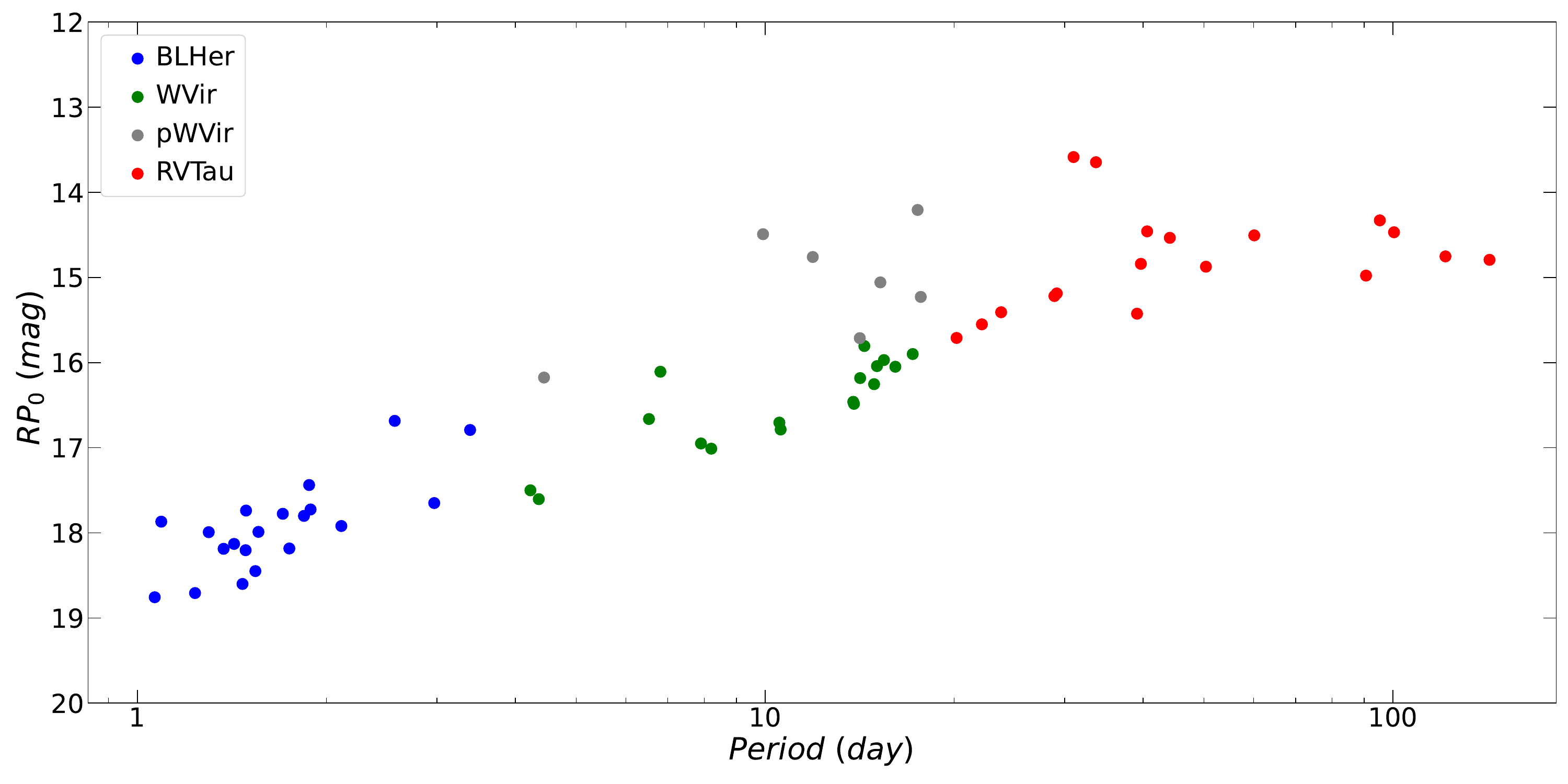}
    }
    \hbox{
    \includegraphics[width=0.5\textwidth]{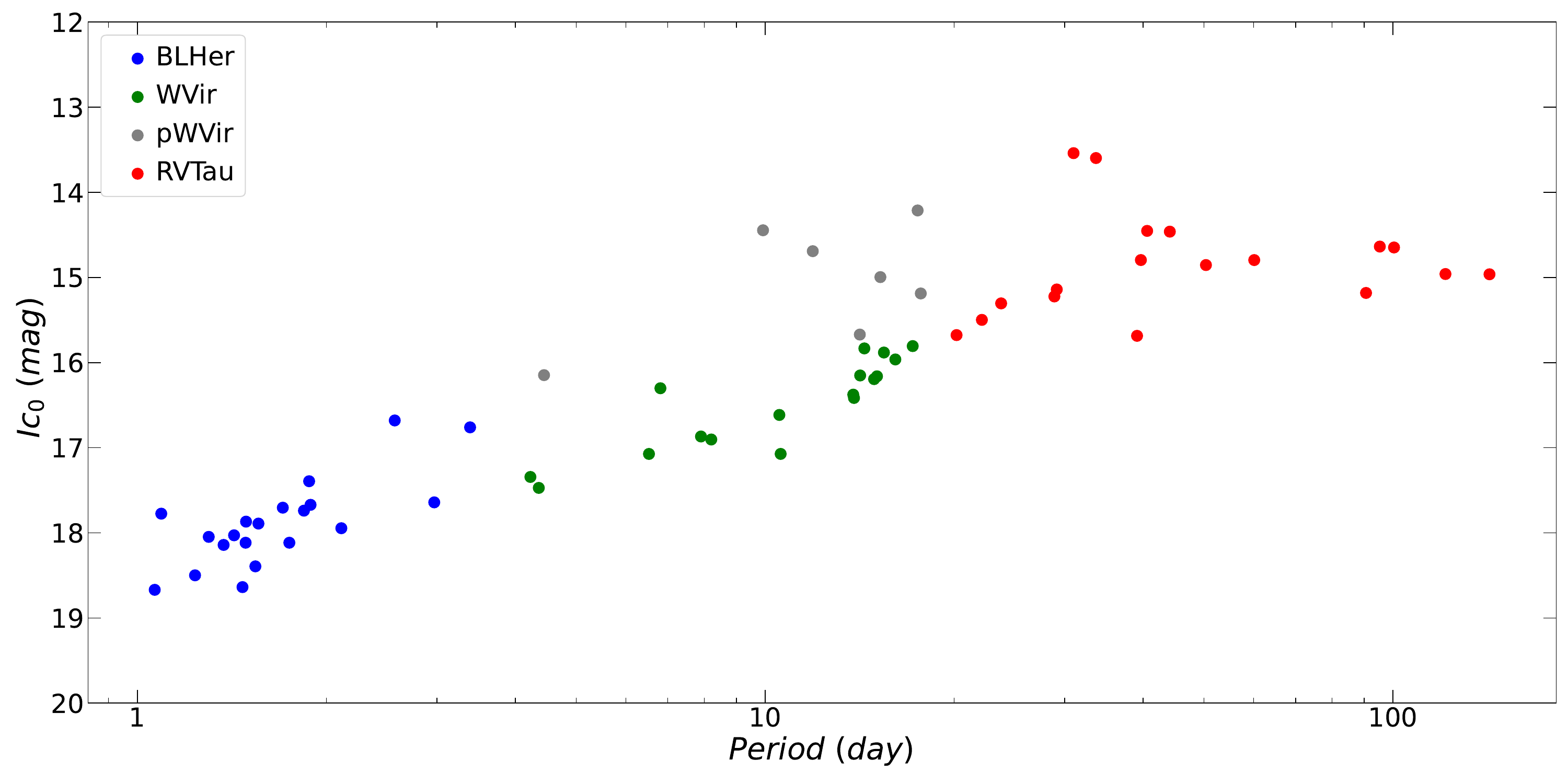}
    \includegraphics[width=0.5\textwidth]{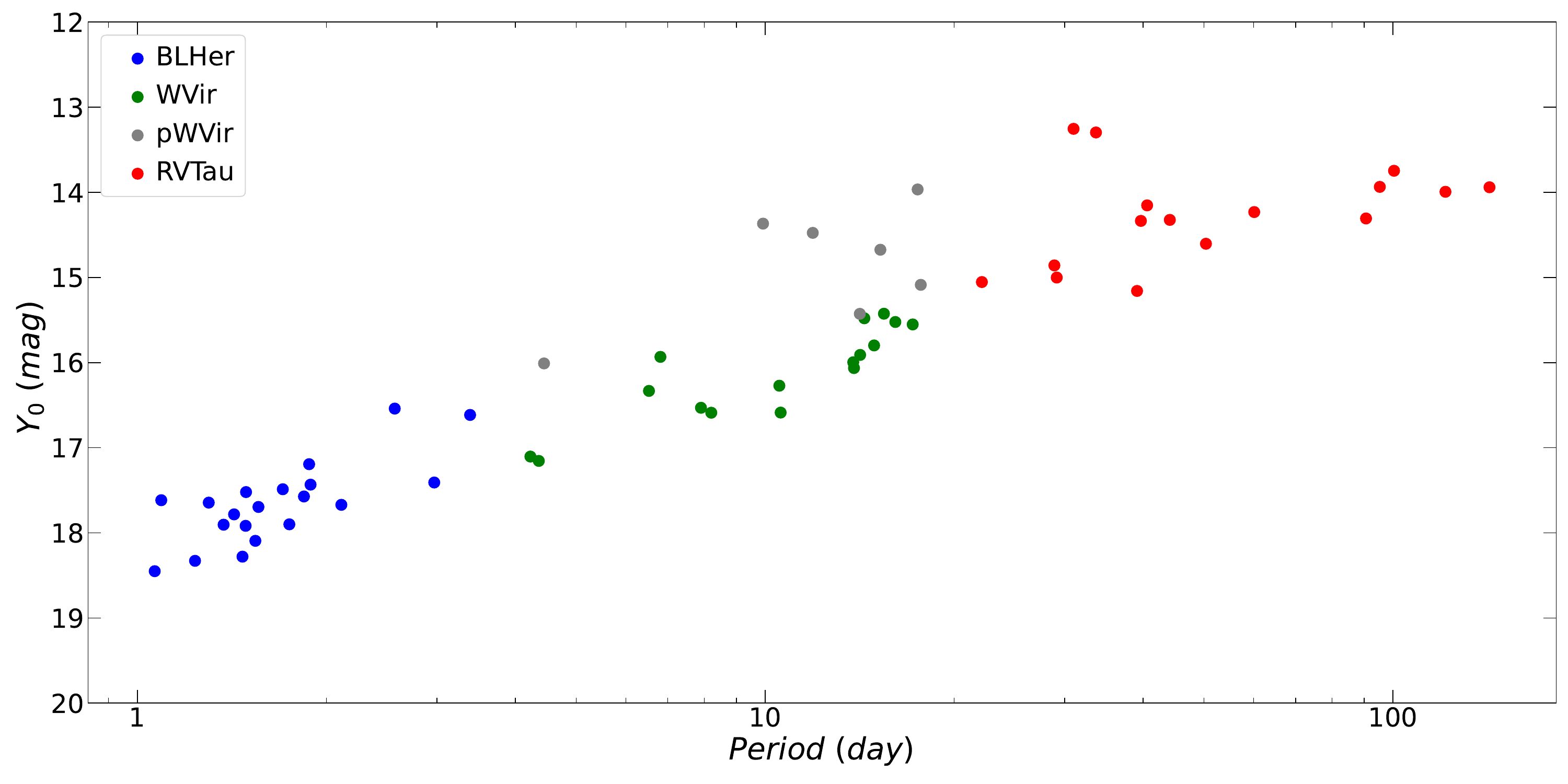}
    }
    \hbox{
    \includegraphics[width=0.5\textwidth]{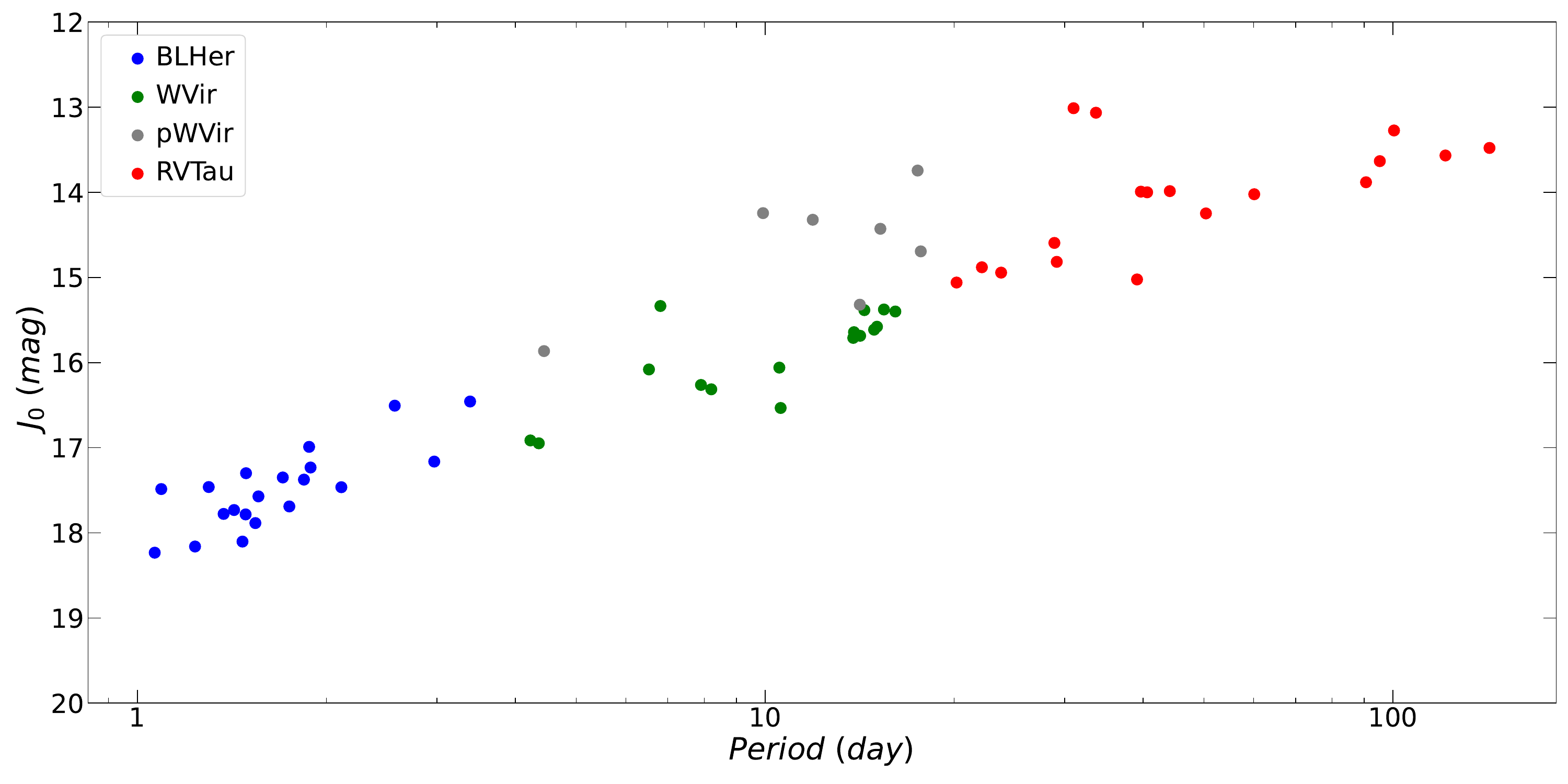}
    \includegraphics[width=0.5\textwidth]{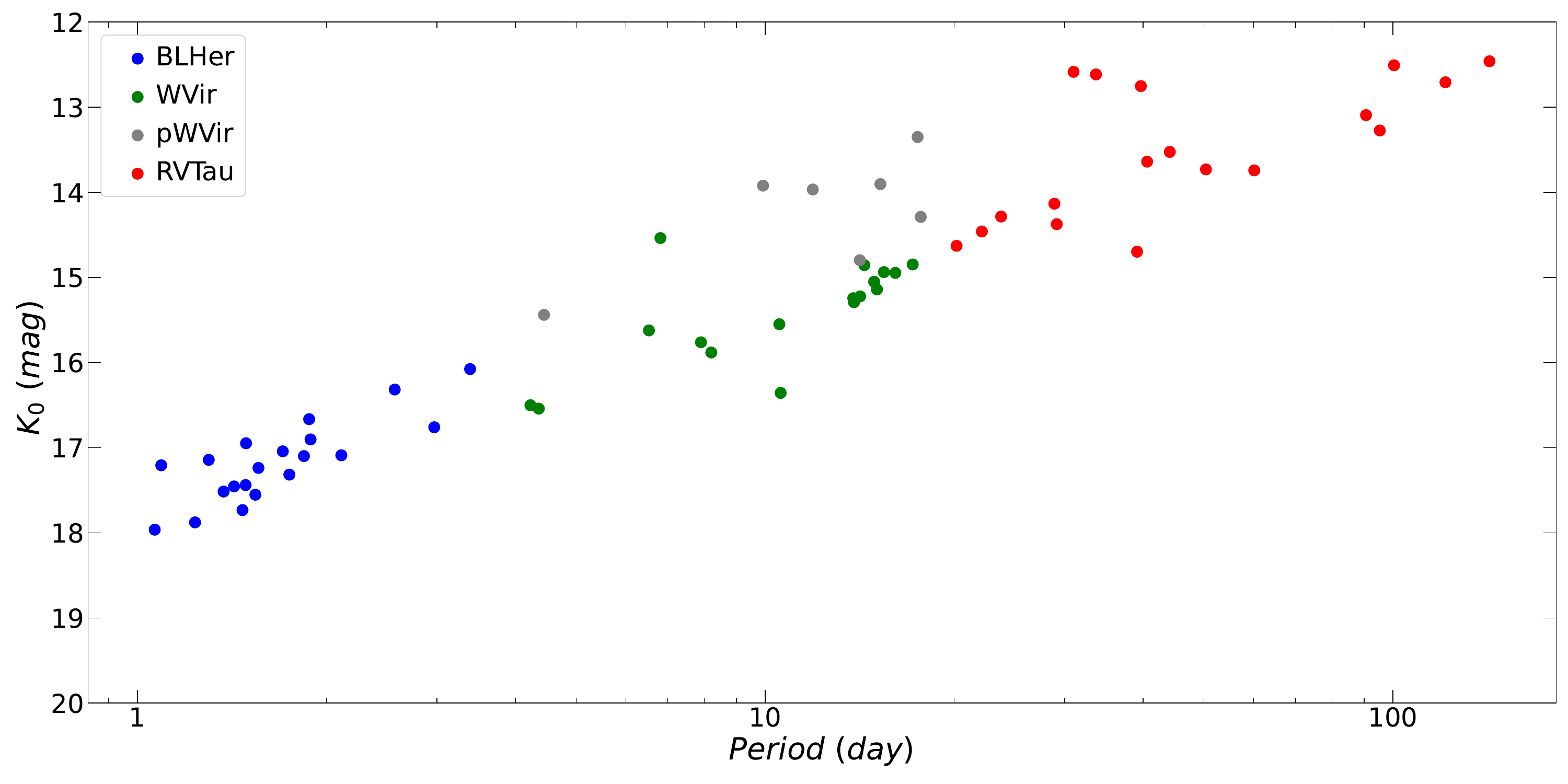}
    }
    
    }
    \caption{\label{plsmc} Same as \ref{pllmc} but for the SMC.}	
	\end{figure*}

\begin{figure*}
    \vbox{
    \hbox{
    \includegraphics[width=0.5\textwidth]{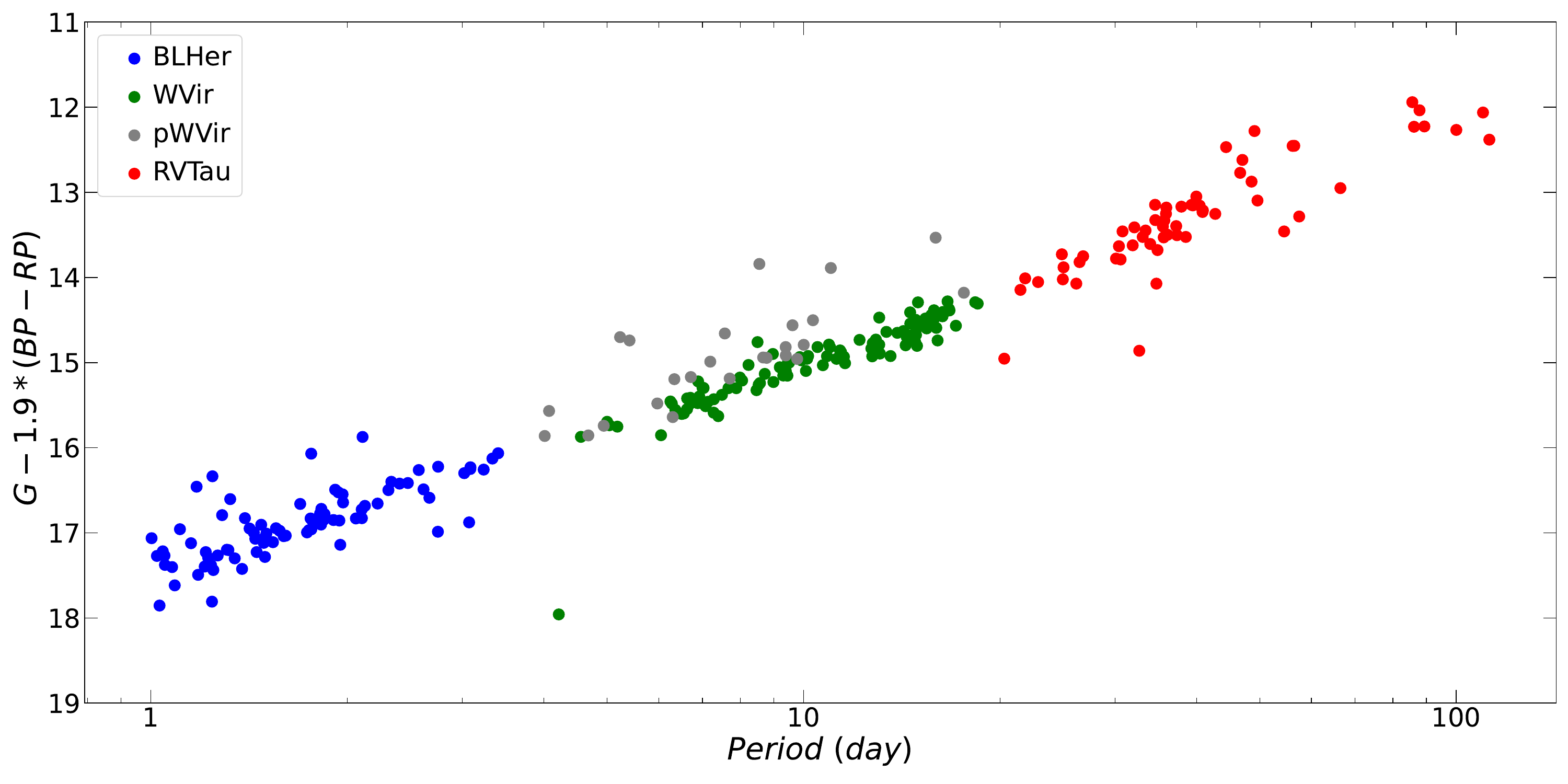}
    \includegraphics[width=0.5\textwidth]{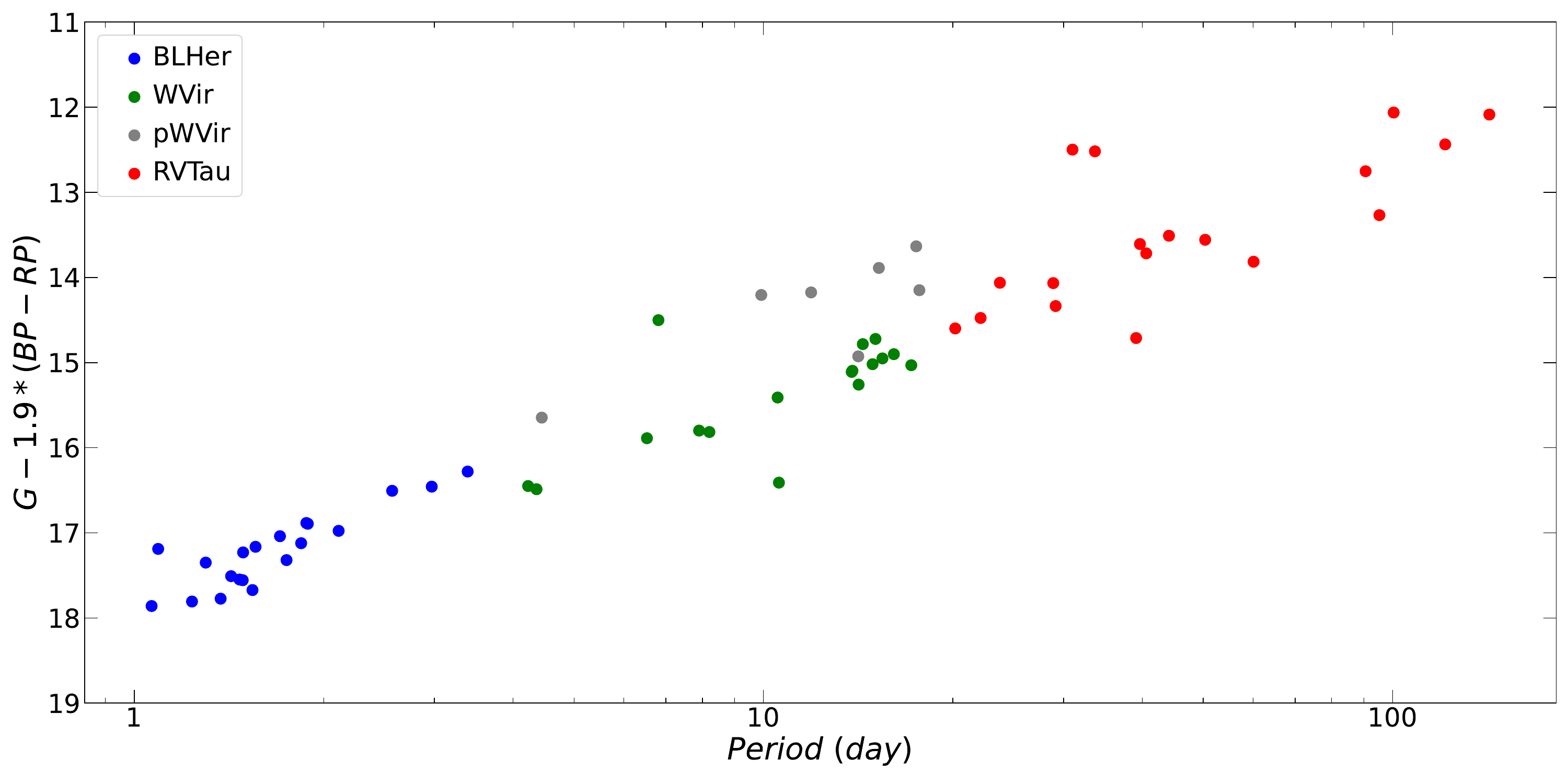}
    }
    \hbox{
    \includegraphics[width=0.5\textwidth]{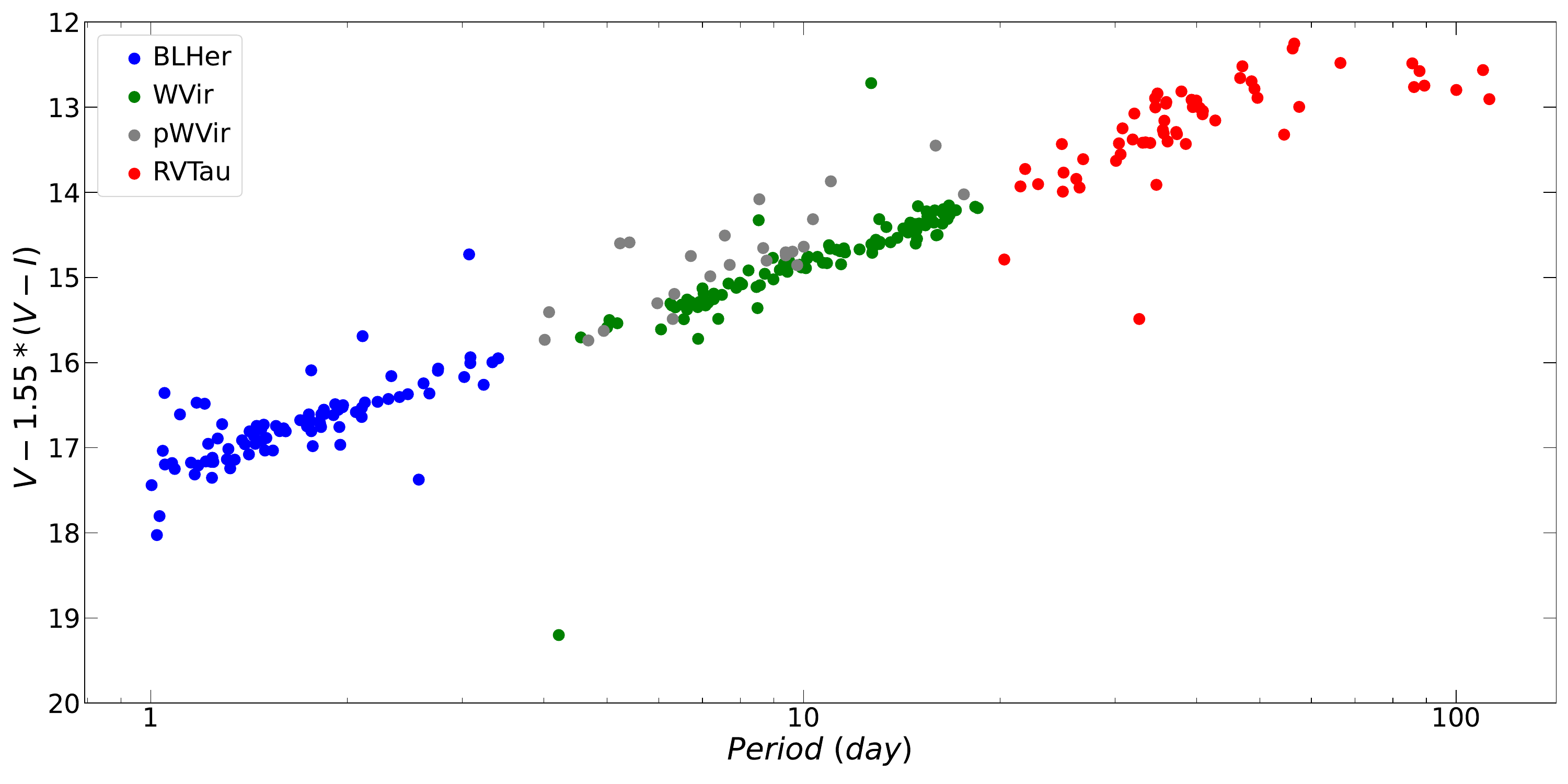}
    \includegraphics[width=0.5\textwidth]{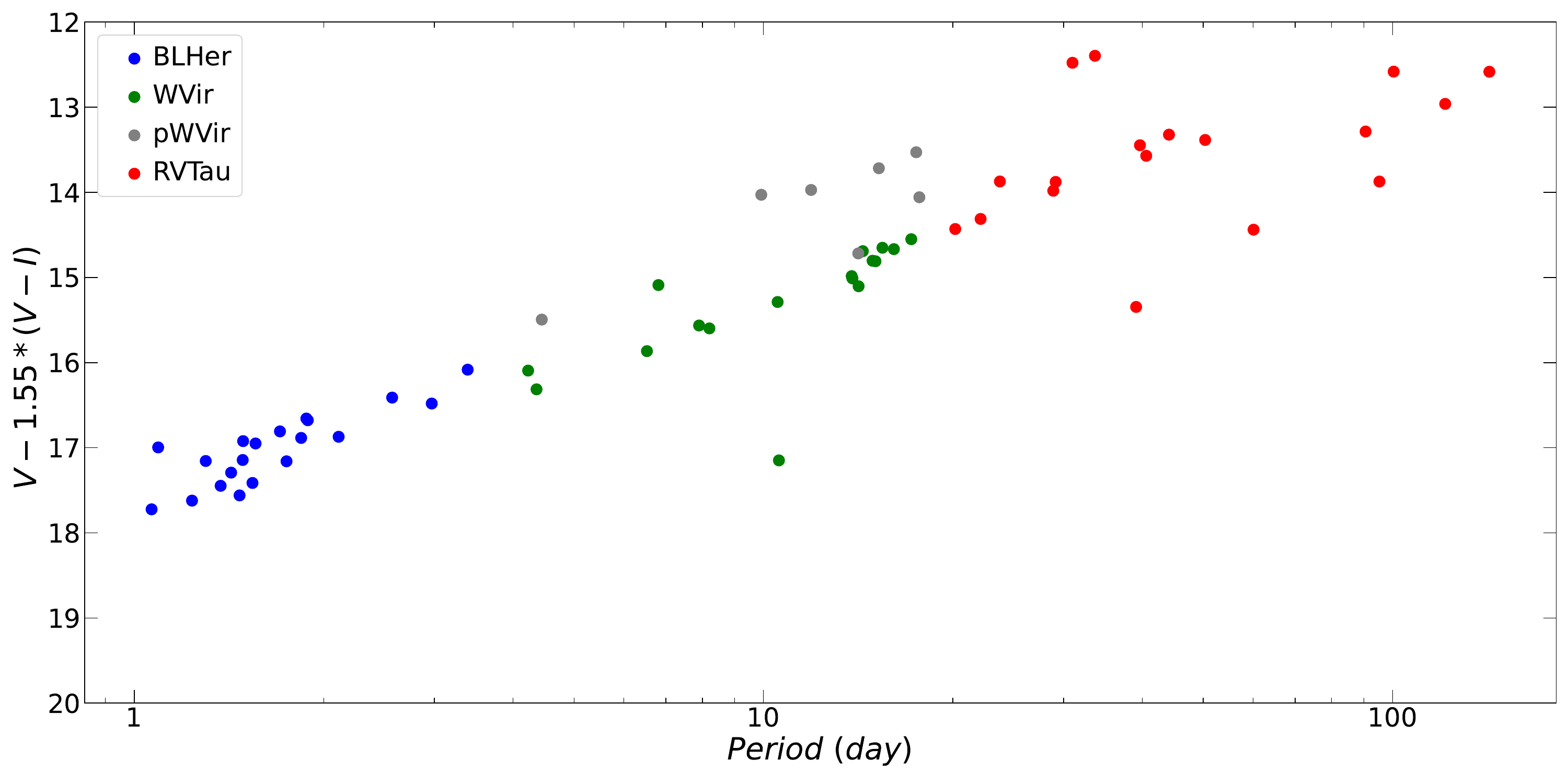}
    }
    \hbox{
    \includegraphics[width=0.5\textwidth]{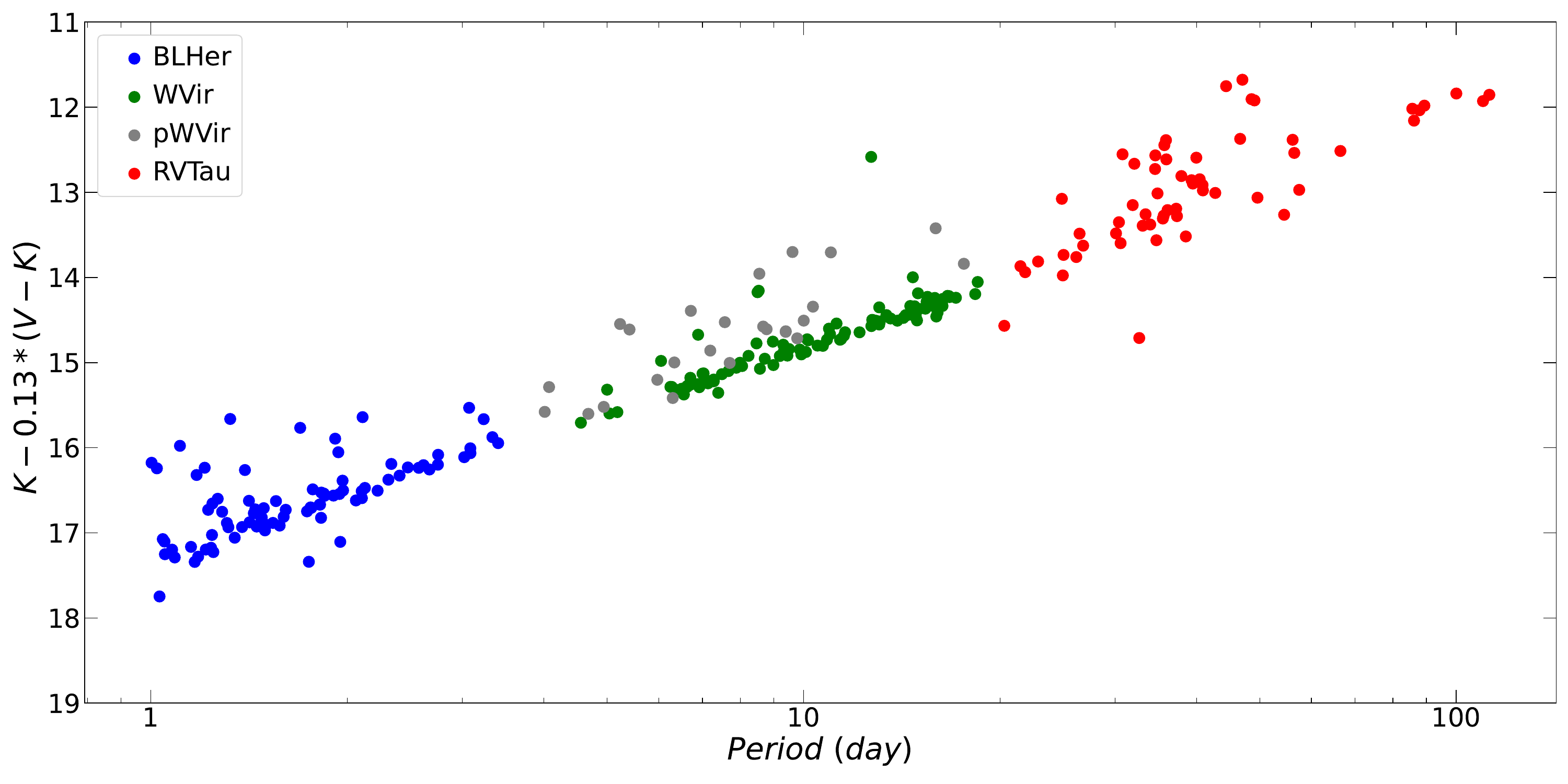}
    \includegraphics[width=0.5\textwidth]{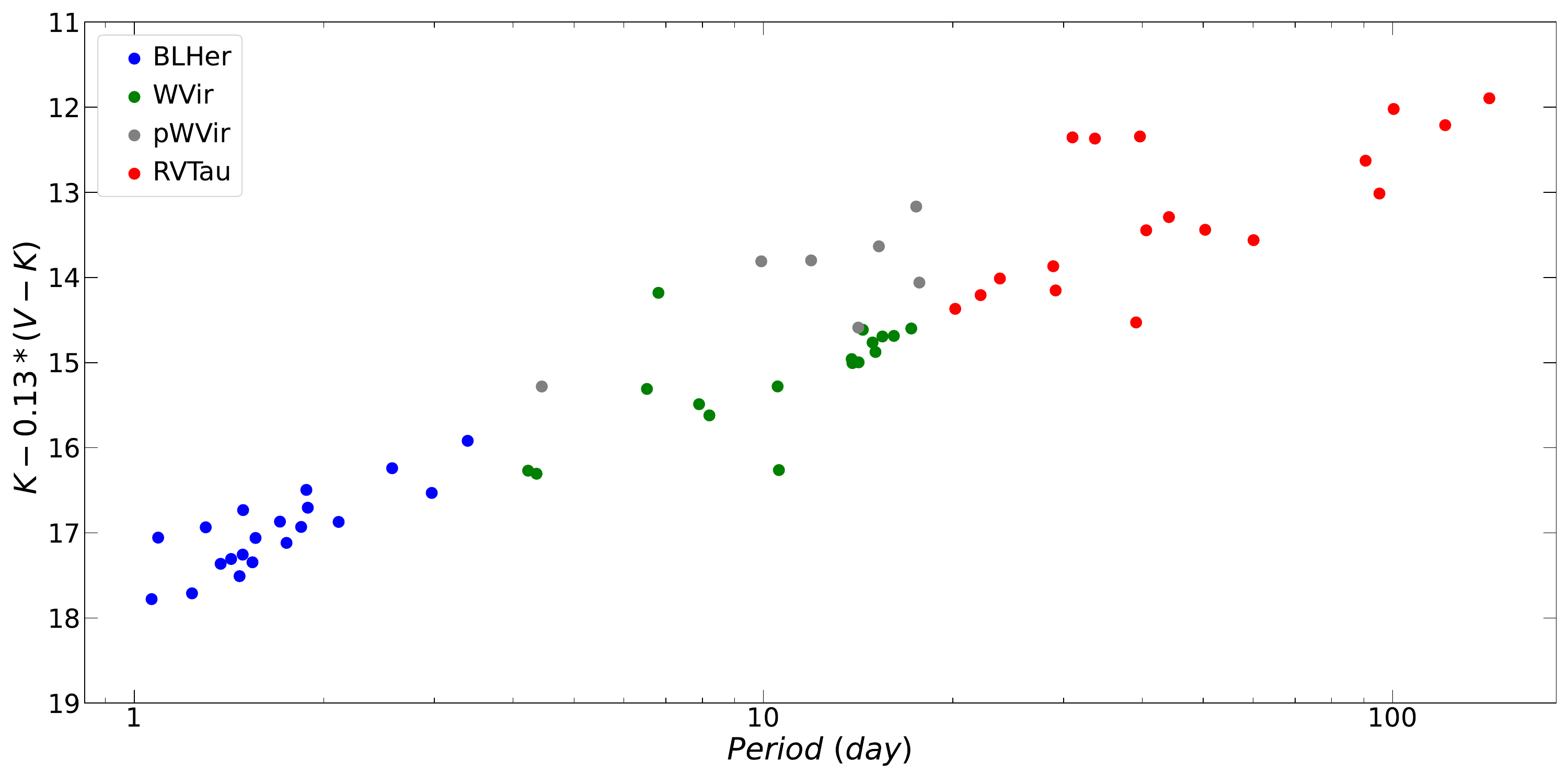}
    }
    \hbox{
    \includegraphics[width=0.5\textwidth]{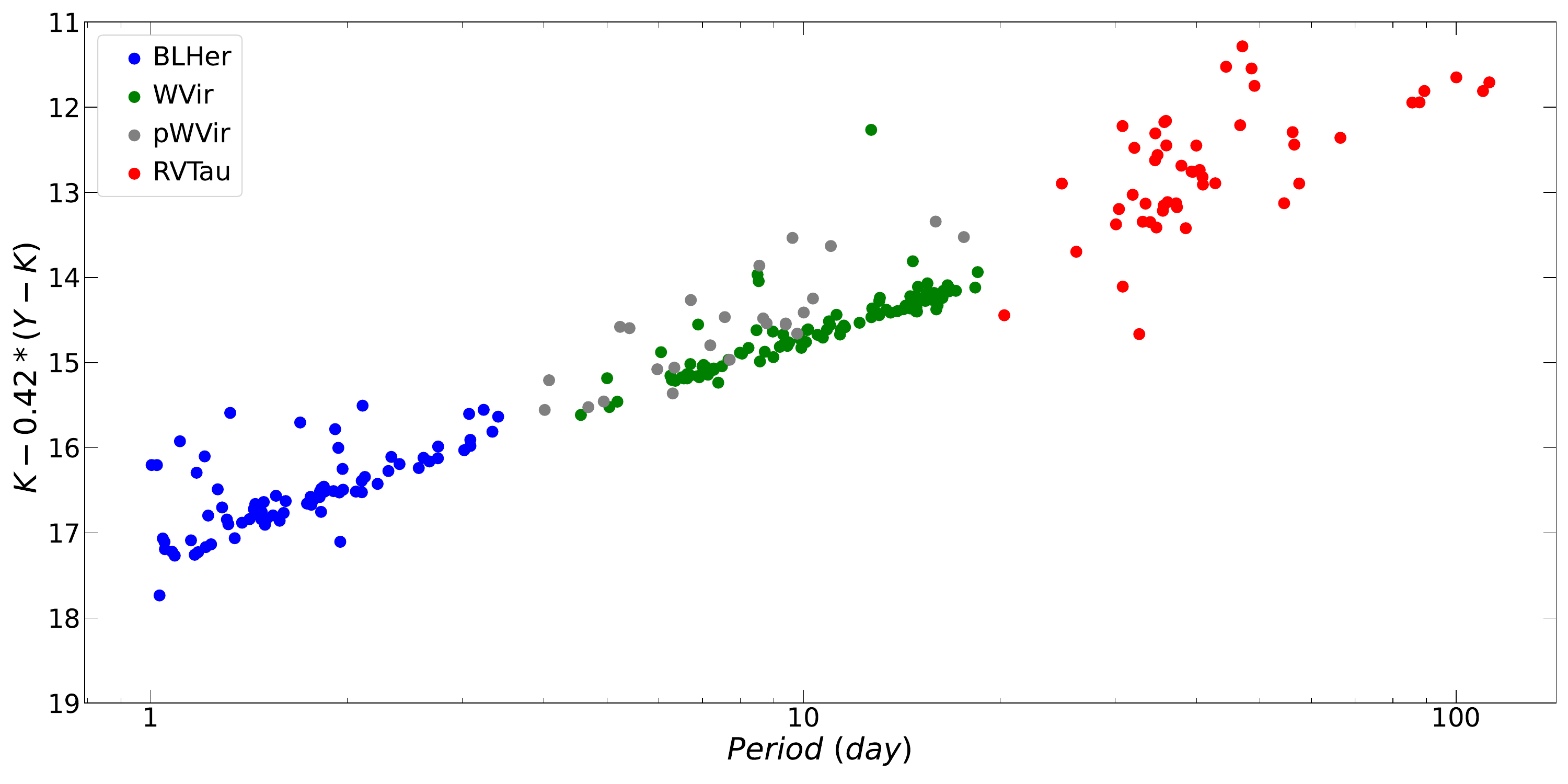}
    \includegraphics[width=0.5\textwidth]{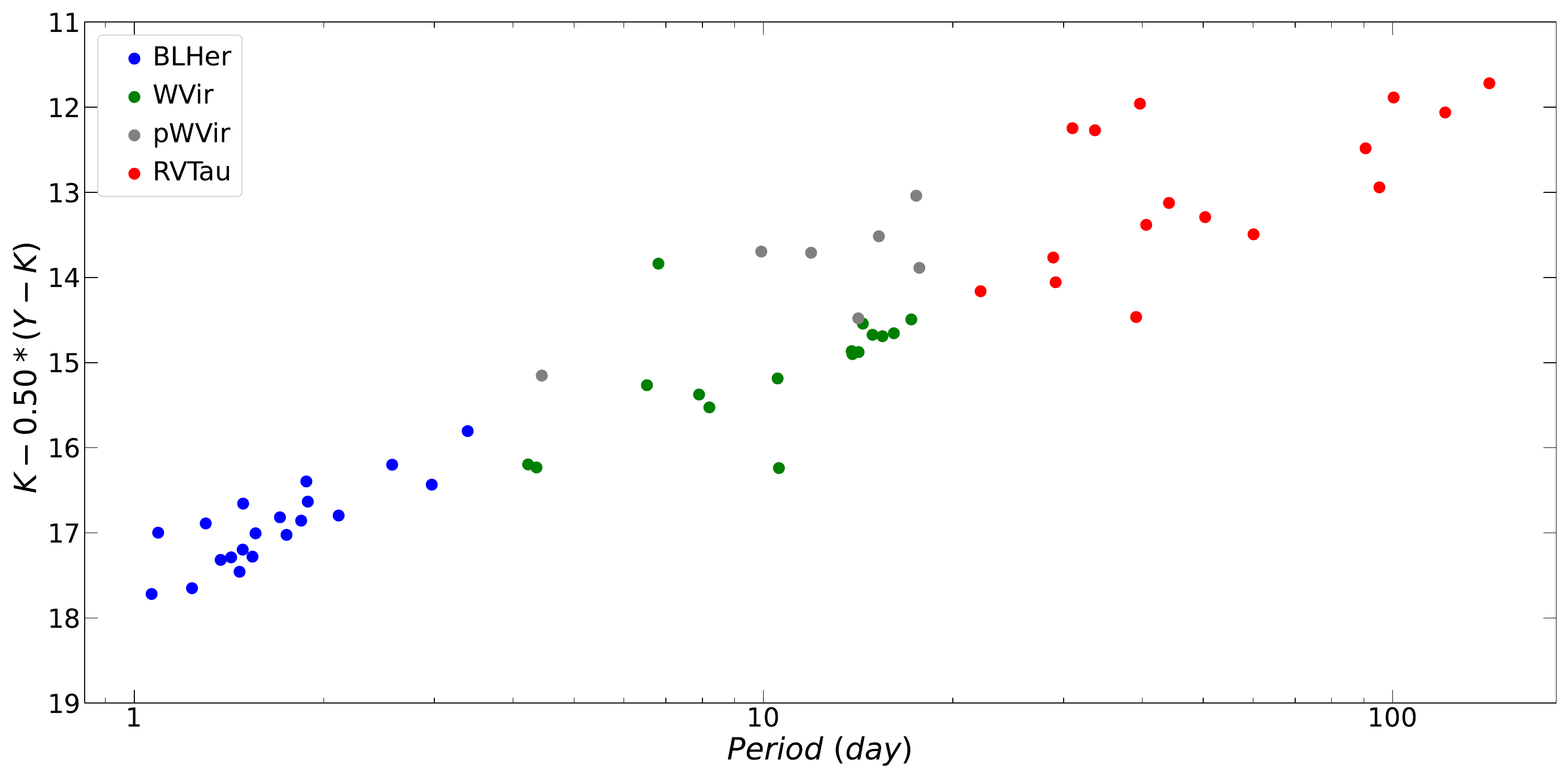}
    }
    \hbox{
    \includegraphics[width=0.5\textwidth]{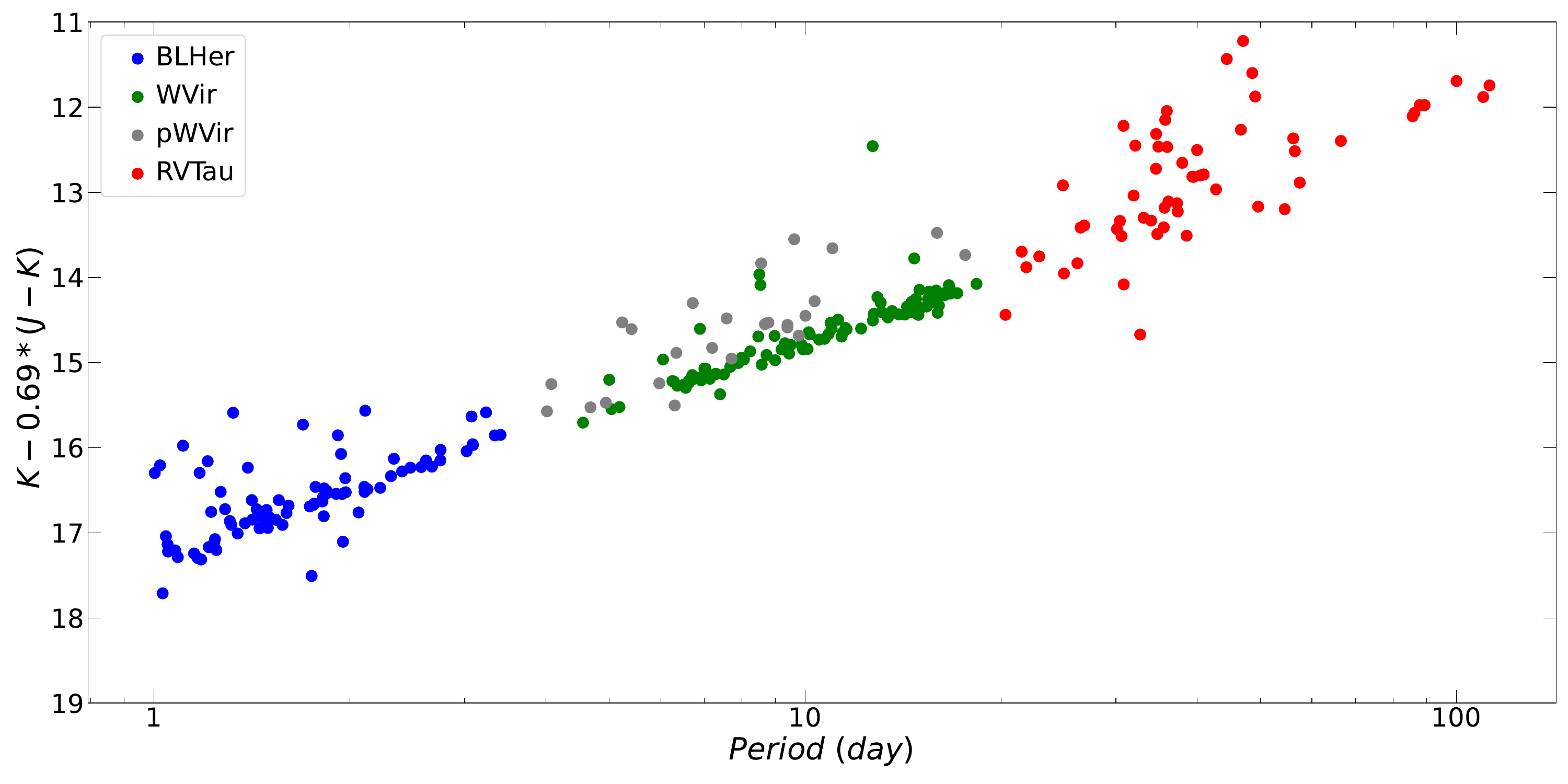}
    \includegraphics[width=0.5\textwidth]{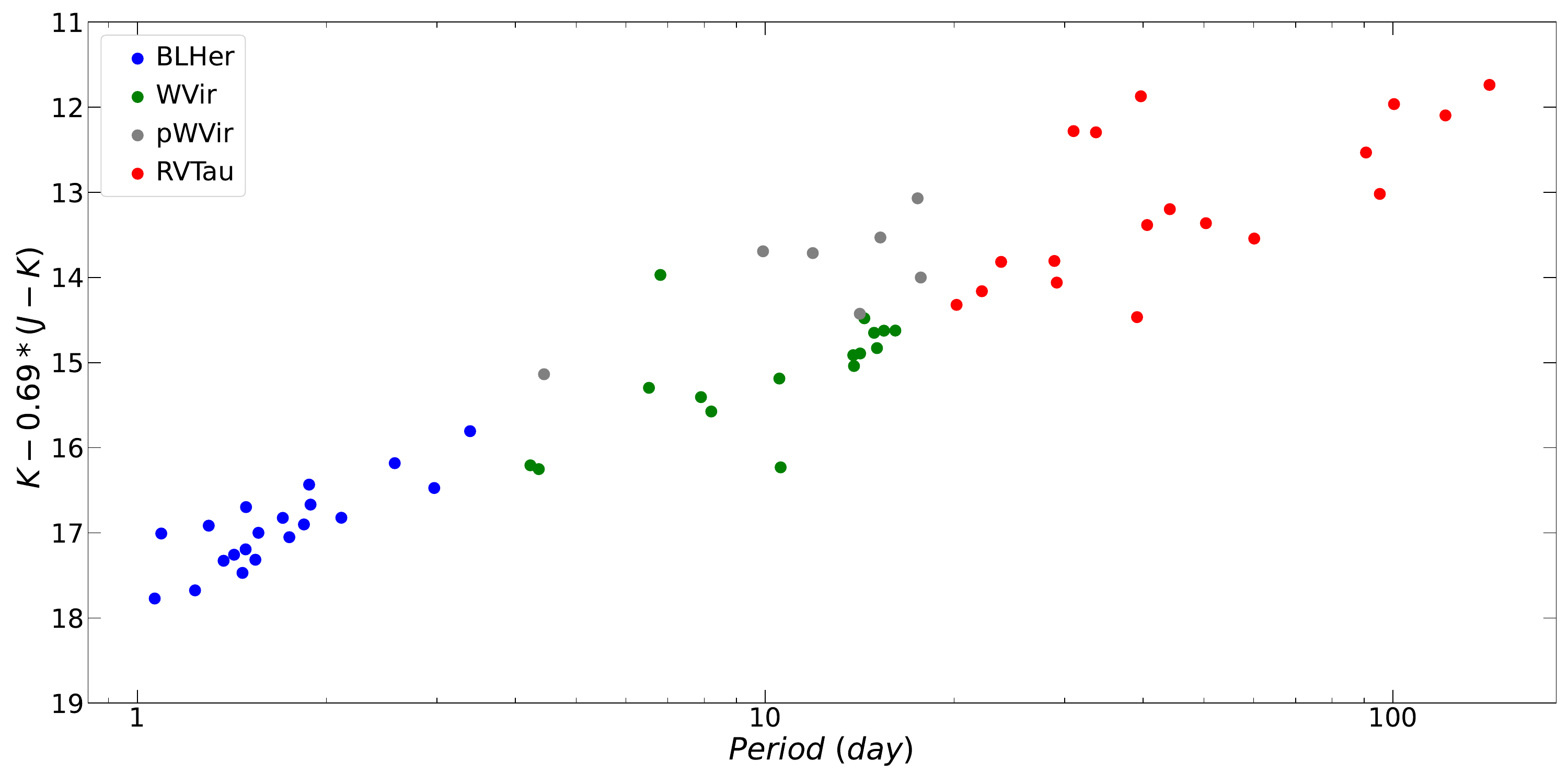}
    }
    }
    \caption{\label{pw} Left panels: $PW$ relations in selected bands for T2Cs in the LMC.
    Right panels: Same as left but for the SMC. The colours are the same as in Fig.~\ref{pllmc}.}	
	\end{figure*}

\subsection{$PL/PW/PLC$ derivation with the Least Trimmed Squares algorithm}

\sectionmark{$PL/PW/PLC$ derivation with LTS}
To determine the coefficients of the $PL/PW/PLC$ relations, we fitted linear relations of the following types: 

\begin{eqnarray}
m_{\lambda_0} & = &\alpha +\beta\times \log P ~~~~~~~~~~~~~~~~~~~~~~~~~~~~~~~~~~~~~PL \\
w(\lambda_1,\lambda_2) &= &\alpha + \beta \times \log P ~~~~~~~~~~~~~~~~~~~~~~~~~~~~~~~~~~~~~PW \\
m_{\lambda_{1,0}} &=& \alpha + \beta \times \log P + \gamma \times (m_{\lambda_1}-m_{\lambda_2})_0~~~~~~~PLC
\end{eqnarray}

\noindent 
where the observed quantities are the periods and the dereddened intensity averaged magnitudes.  
To carry out these linear fits in one or two dimensions, we adopted the {\tt Python} code LTS \citep[Least Trimmed Squares,][]{cappellari2013atlas3d}. This procedure is particularly robust with respect to the outlier removal since the clipping is carried out inside-out, contrary to the standard $\sigma$ clipping.

\subsection{Results for the $PL/PW/PLC$ relations}

The $PL$ relations were obtained in all the bands listed in the first column of Table~\ref{ass}. As for the Wesenheit magnitudes, we calculated the $PW$ relations for the five quantities defined in Table~\ref{tab:coeffwes}, where the coefficient of the colour term was obtained from the wavelength-dependent absorption values listed in the second column of Table \ref{ass}. 
As for the $Gaia$ bands, which have a wider bandwidth compared to Johnson-Cousins bands, we used the empirical Wesenheit function from  \citet{Ripepi2019_reclassificaiton}. 
As for the $PLC$, we considered the same magnitude-colour combinations which are at the base of the Wesenheit magnitudes shown in Table~\ref{tab:coeffwes}, where the colour coefficients are not fixed but free to vary.

An overview of the observational data at the base of all the $PL$ relations to be derived in the LMC and SMC are reported in Fig.~\ref{pllmc} and ~\ref{plsmc}, respectively. Similarly, the left and right panels of Fig.~\ref{pw} show the $PW$ relations for T2Cs in the LMC and SMC, respectively. In all the figures, the different Cepheid types are highlighted with different colours. 
A visual inspection of these figures reveals several clear features, which will be later confirmed from a quantitative point of view:

\begin{itemize}

\item 
In all the considered cases, the apparent dispersion of the data decreases from the optical to NIR bands. This is particularly evident in the LMC where the number of pulsators is much larger than in the SMC. The effect is also more evident among the $PL$ relations compared to the $PW$ ones, as these are in all cases much tighter than the former, with the exception of the $J$ and $K_s$ bands.

\item
In the optical, the BLHer, WVir (and pWVir) and RVTau do not follow a unique linear $PL$ relation. This is true also if we restrict the analysis to BLHer and WVir as is common in the literature. These types start following a unique $PL$ relation for bands redder than $I$. According to \citet{caputo04why}, this is due to the different occupations of the IS in the optical bands for the T2Cs of different types.

\item
In all the cases unique $PW$ relations appear to be followed by all T2C types.  This is due to two factors: i) a better correction for the reddening, in the sense that the individual reddening corrections applied to the magnitudes at the base of the different $PL$ relations are less effective than the use of the Wesenheit magnitudes which are reddening free by construction; ii) the inclusion of the colour term in the construction of the Wesenheit magnitudes, that reduces significantly the difference in occupation of the IS which is likely the cause of the different behaviour of the different T2C subtypes in the $PL$ diagrams.    

\item
In almost all the cases the RVTau stars show a larger dispersion compared to BLHer and WVir. This is due to possibly different origins of these stars, which, as discussed in the Sect. \ref{intro}, could be a mixture of both old and intermediate-age stars. Also, BLHer variables are slightly more dispersed than WVir stars (excluding pWVir). This is likely due to the fainter magnitudes reached by these stars. 

\end{itemize}

\begin{table}[ht]
\centering
\footnotesize\setlength{\tabcolsep}{3pt} 
\caption{Definition of the Wesenheit magnitudes used in this paper (see also Table~\ref{ass}).}
  \label{tab:coeffwes}
\begin{tabular}{ll}
 \hline  
 \noalign{\smallskip} 
       Relation &   \\ 
 \noalign{\smallskip}
 \hline  
 \noalign{\smallskip}  
  W$VI$ & = $I$ $-$ 1.55 ($V$ $-$ $I$) \\ 
  W$VK_s$ & = $K_s$ $-$ 0.14 ($V$ $-$ $K_s$) \\ 
  W$JK_s$ & = $K_s$ $-$ 0.71 ($J$ $-$ $K_s$) \\
  W$YK_s$ & = $K_s$ $-$ 0.50 ($Y$ $-$ $K_s$) \\
  WG & = $G$ $-$ 1.90 ($G_{BP} -G_{RP}$) \\
\noalign{\smallskip}
\hline  
\noalign{\smallskip}
\end{tabular}
\end{table}

\begin{figure*}
    \vbox{
    \hbox{
    \includegraphics[width=0.5\textwidth]{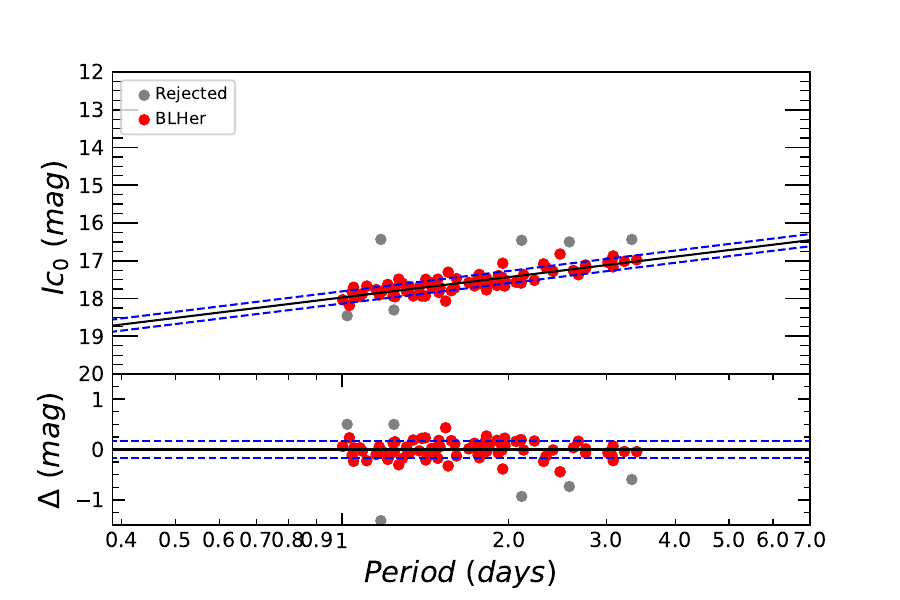}
    \includegraphics[width=0.5\textwidth]{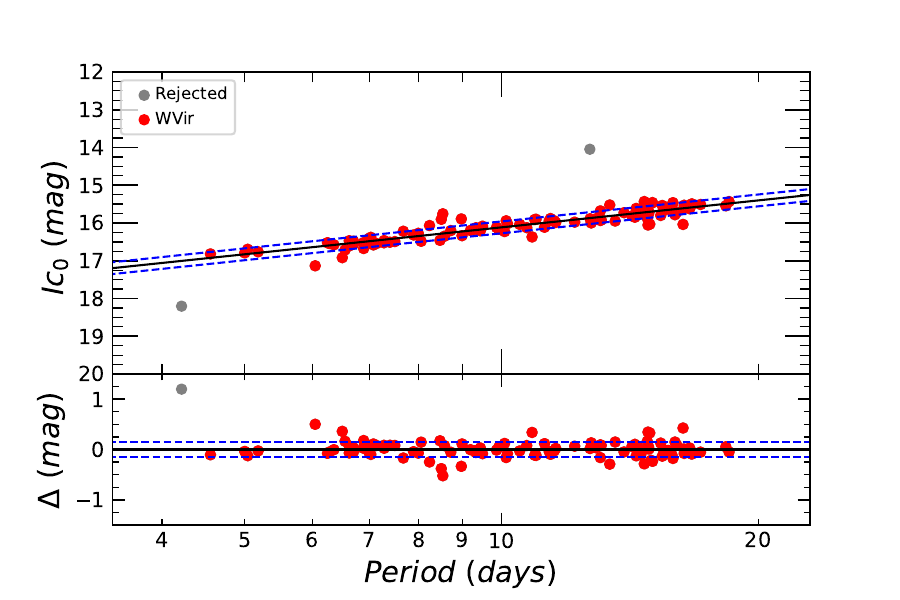}
    }
    \hbox{
    \includegraphics[width=0.5\textwidth]{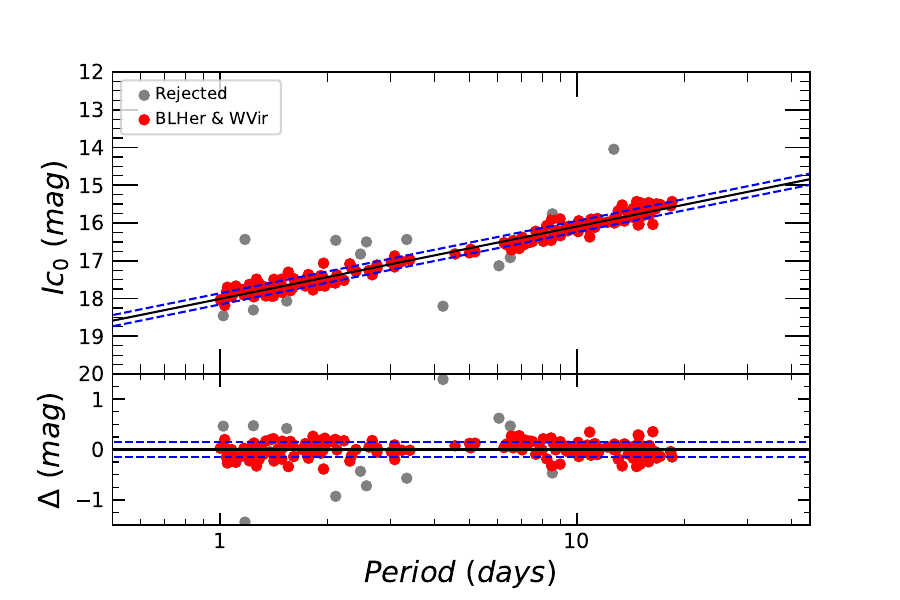}
    \includegraphics[width=0.5\textwidth]{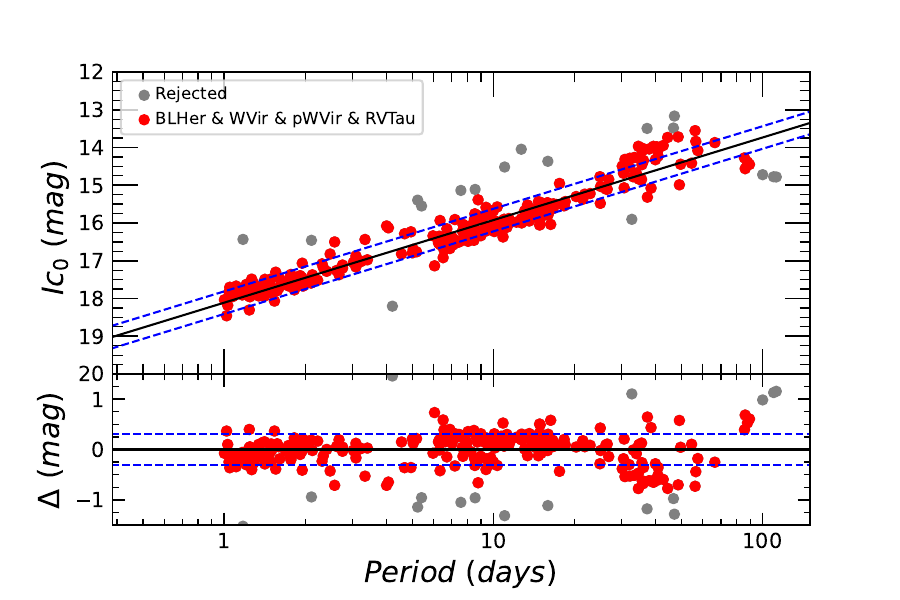}
    }
    }
    \caption{\label{esempifit} Top panels: Examples of $PL$ fitting in the $I$ band for different combinations of T2Cs in the LMC (see labels in the figures). The red and grey filled circles are the data used for the fit and the outliers, respectively. The solid black line is the best fit to the data, while the dashed blue lines show the $\pm 1 \sigma$ levels. Bottom panels: residuals of the fit. The colours are the same as above.
    The fits for the other bands are in the Appendix~\ref{app:additionalPL}. }	
	\end{figure*}

\begin{figure*}
    \vbox{
    \hbox{
    \includegraphics[width=0.5\textwidth]{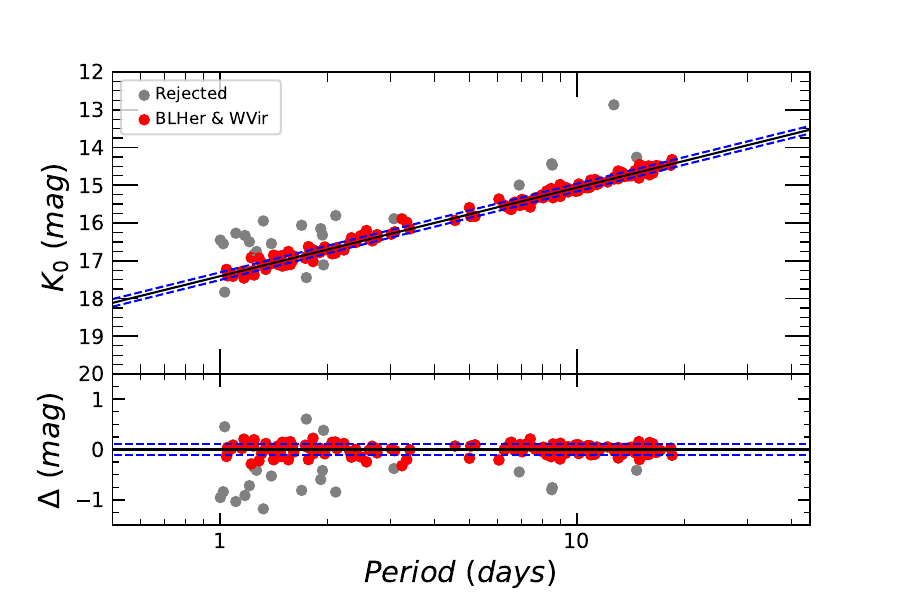}
    \includegraphics[width=0.5\textwidth]{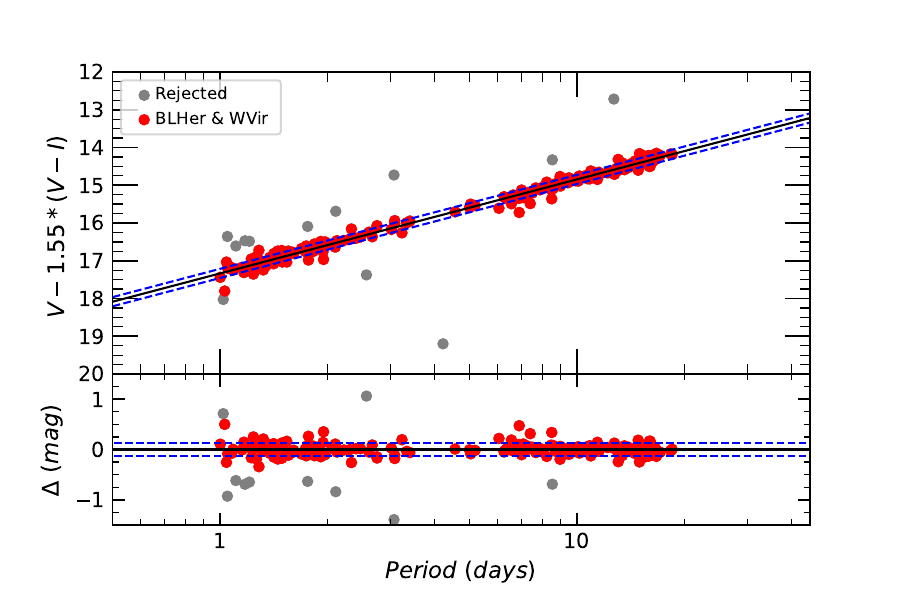}
    }
    \hbox{
    \includegraphics[width=0.5\textwidth]{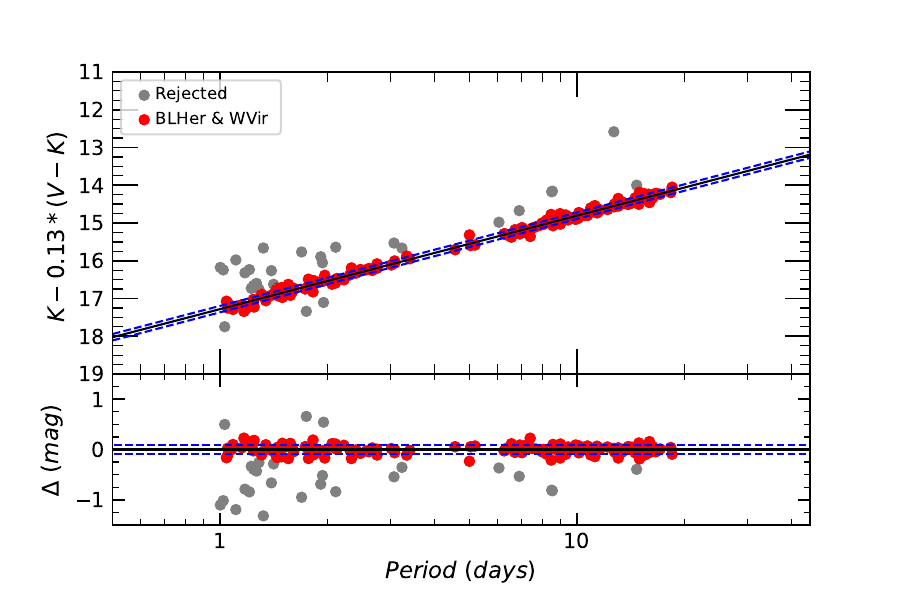}
    \includegraphics[width=0.5\textwidth]{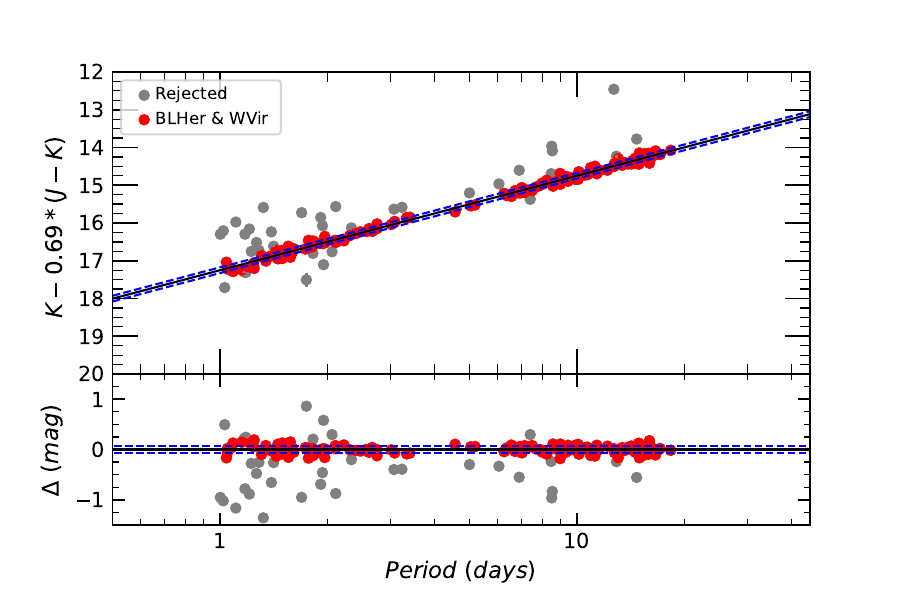}
    }
    \hbox{
    \includegraphics[width=0.5\textwidth]{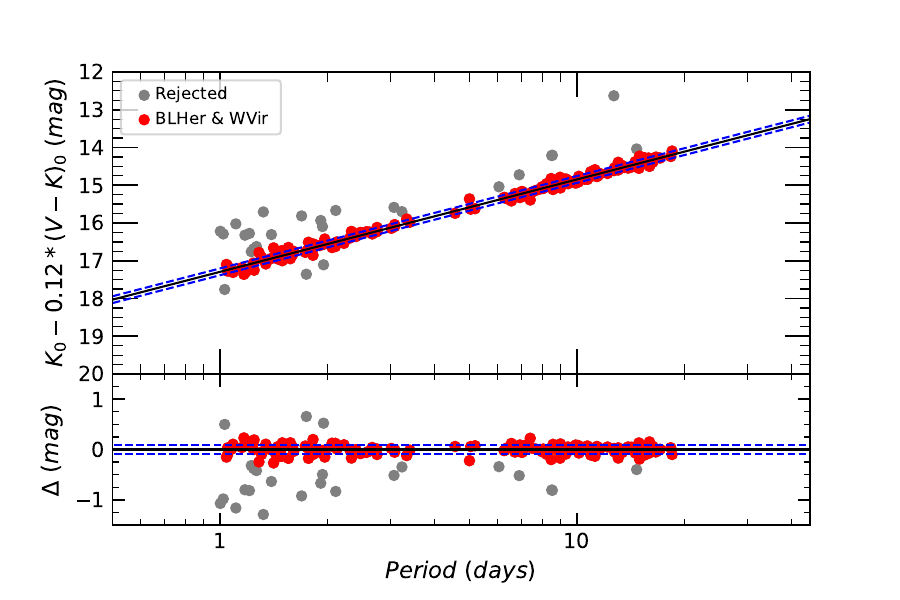}
    \includegraphics[width=0.5\textwidth]{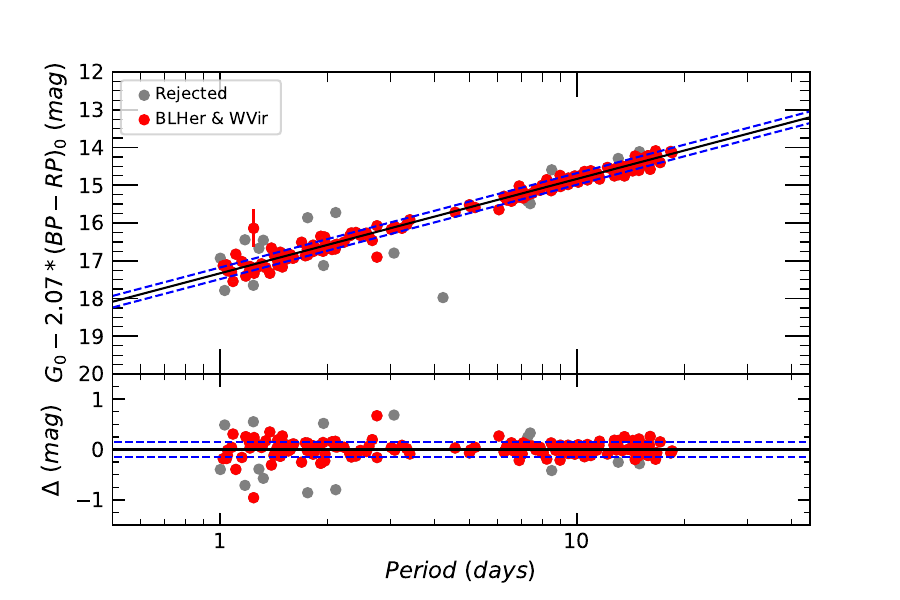}
    }
    }
    \caption{\label{esempifit2} Examples of best fit  for BLHer and WVir in the LMC. From the top we show the $PLK_s$, $PWVI$, $PWVK_s$, $PWJK_s$, $PLCVK_s$, $PLCG$ relations, respectively. The colours of dots and lines are the same as in Fig.~\ref{esempifit}.
    Additional fits are in the Appendix~\ref{app:additionalPL}. }	
	\end{figure*}

Taking into account these considerations, we have carried out the $PL/PW/PLC$ fittings for each T2C type separately and then in different combinations in order to quantify the differences in the slopes between the different types or the possibility of combining the sub-samples.  

Examples of the $PL/PW$ fits are shown in Fig.~\ref{esempifit} and ~\ref{esempifit2} while all the figures showing the remaining fits can be found in Appendix~\ref{app:additionalPL} (Fig.~\ref{fit1lmc} and Fig.~\ref{fit1smc} for the LMC and the SMC, respectively). 

More in detail, Fig.~\ref{esempifit} shows the $PL$ relations in the $I$ band for single and combined T2Cs types, while Fig.~\ref{esempifit2} displays selected tight $PL$ and $PW$ relations we obtain when combining BLHer\& WVir types, only.

The results of the fit for all the considered cases and combination of types are reported in Tables~\ref{tab:lmcT2} and ~\ref{tab:smcT2}, for the T2Cs in the LMC and SMC, respectively. These tables report the most interesting cases for the T2Cs, i.e. those without the more problematic RVTau pulsators. All the remaining cases are in Appendix~\ref{app:additionalPL} (Tables~\ref{tab:lmcT2pt2} and ~\ref{tab:smcT2pt2}). 

Inspecting these tables we find a confirmation of our previous qualitative assessments. 
\begin{itemize}
    \item In the optical, $PL$ relations calculated for BLHer and WVir have slopes which are different at several $\sigma$-levels. As we approach the $J$ and $K_s$ bands the slopes become consistent within less than 1 $\sigma$. 
    \item The dispersion decreases going from optical to NIR bands. We will discuss this trend in more detail in the next Section.
    \item The colour term, obtained from LTS plane fit, for the PLC relations in $K_s, J-K_s$ and $G, G_{BP}-G_{RP}$ is consistent, within 1 $\sigma$, with the Wesenheit coefficient. This evidence demonstrates that the Wesenheit relations are not only reddening-free but also able to significantly account for the width of the IS.
\end{itemize}
For the reasons listed above, in the subsequent analyses, we will use the $PL$ relations in the $J$ and $K_s$ bands and the $PW$ relations in $K_s, J-K_s$, $K_s, V-K_s$ and $G, G_{BP}-G_{RP}$.

\begin{table*}
\footnotesize\setlength{\tabcolsep}{3pt} 
\caption{Coefficients of the $PL$, $PLC$ and $PW$ relations for T2Cs in the LMC. }
  \label{tab:lmcT2}
\begin{center}

\begin{tabular}{lllllllllll}
 \hline  
 \noalign{\smallskip} 
       Relation  &  Group & $\alpha $ &  $\sigma_{\alpha}$ & $\beta$ & $\sigma_{\beta} $ &  $\gamma$ & $\sigma_{\gamma}$ & RMS & Used stars & Total stars \\
   &  & mag & mag & mag/dex & mag/dex & & & mag & \\                     
(1)  & (2)  & (3) & (4) & (5) & (6) & (7) &(8) & (9) & (10) & (11)\\   
 \noalign{\smallskip}
 \hline  
 \noalign{\smallskip}  
  PLBP & BLHer & 18.348 & 0.130 & $-$0.840 & 0.250 &  &  & 0.29 & 74 & 83\\
  PLBP & WVir & 19.266 & 0.039 & $-$2.190 & 0.110 &  &  & 0.18 & 98 & 103\\
  PLBP & BLH\&WVir & 18.505 & 0.020 & $-$1.439 & 0.047 & &  & 0.27 & 175 & 186\\
  PLG & BLHer & 18.382 & 0.082 & $-$1.450 & 0.160 & &  & 0.20 & 78 & 85\\
  PLG & WVir & 19.019 & 0.029 & $-$2.270 & 0.082 &  &  & 0.13 & 93 & 103\\
  PLG & BLH\&WVir & 18.454 & 0.014 & $-$1.722 & 0.034 & &  & 0.19 & 178 & 188\\
  PLRP & BLHer & 17.945 & 0.097 & $-$1.670 & 0.200 & & & 0.25 & 72 & 83\\
  PLRP & WVir & 18.538 & 0.033 & $-$2.363 & 0.092 & &  & 0.14 & 100 & 103\\
  PLRP & BLH\&WVir & 18.092 & 0.012 & $-$1.943 & 0.030 & & & 0.17 & 160 & 186\\
  PLV & BLHer & 18.432 & 0.079 & $-$1.250 & 0.160 & & & 0.18 & 73 & 85\\
  PLV & WVir & 19.066 & 0.039 & $-$2.150 & 0.110 & & & 0.17 & 99 & 104\\
  PLV & BLH\&WVir & 18.520 & 0.016 & $-$1.618 & 0.037 & & & 0.21 & 178 & 189\\
  PLI & BLHer & 17.973 & 0.067 & $-$1.800 & 0.140 & & & 0.16 & 79 & 85\\
  PLI & WVir & 18.483 & 0.036 & $-$2.370 & 0.100 & & & 0.16 & 102 & 104\\
  PLI & BLH\&WVir & 18.028 & 0.012 & $-$1.940 & 0.029 & & & 0.17 & 182 & 189\\
  PLY & BLHer & 17.711 & 0.082 & $-$1.680 & 0.170 & & & 0.19 & 68 & 77\\
  PLY & WVir & 18.266 & 0.028 & $-$2.473 & 0.080 & & & 0.13 & 97 & 100\\
  PLY & BLH\&WVir & 17.823 & 0.012 & $-$2.048 & 0.029 & & & 0.16 & 162 & 177\\
  PLJ & BLHer & 17.657 & 0.069 & $-$2.250 & 0.140 & & & 0.17 & 73 & 83\\
  PLJ & WVir & 17.919 & 0.024 & $-$2.400 & 0.068 & & & 0.10 & 94 & 98\\
  PLJ & BLH\&WVir & 17.664 & 0.010 & $-$2.156 & 0.024 & & & 0.12 & 162 & 181\\
  PLK & BLHer & 17.444 & 0.066 & $-$2.560 & 0.140 & & & 0.16 & 73 & 84\\
  PLK & WVir & 17.508 & 0.019 & $-$2.439 & 0.053 & & & 0.08 & 98 & 103\\
  PLK & BLH\&WVir & 17.410 & 0.009 & $-$2.348 & 0.019 & & & 0.10 & 165 & 187\\
  PWG & BLH\&WVir & 17.445 & 0.009 & $-$2.436 & 0.022 & & & 0.14 & 170 & 186\\
  PWVI & BLH\&WVir & 17.337 & 0.010 & $-$2.491 & 0.022 & & & 0.12 & 177 & 189\\
  PWVK & BLH\&WVir & 17.282 & 0.007 & $-$2.475 & 0.017 &  &  & 0.09 & 160 & 187\\
  PWYK & BLH\&WVir & 17.226 & 0.007 & $-$2.516 & 0.017 & & & 0.09 & 151 & 177\\
  PWJK & BLH\&WVir & 17.251 & 0.006 & $-$2.501 & 0.016 & & & 0.08 & 146 & 181\\
  PLCG & BLH\&WVir & 17.334 & 0.009 & $-$2.501 & 0.033 & 2.070 & 0.062 & 0.15 & 170 & 186\\
  PLCVI & BLH\&WVir & 17.143 & 0.013 & $-$2.604 & 0.036 & 2.912 & 0.092 & 0.10 & 168 & 189\\
  PLCVK & BLH\&WVir & 17.295 & 0.007 & $-$2.447 & 0.029 & 0.118 & 0.030 & 0.09 & 162 & 187\\
  PLCYK & BLH\&WVir & 17.325 & 0.008 & $-$2.421 & 0.026 & 0.223 & 0.059 & 0.09 & 157 & 177\\
  PLCJK & BLH\&WVir & 17.252 & 0.007 & $-$2.493 & 0.025 & 0.691 & 0.099 & 0.075 & 145 & 181\\

\noalign{\smallskip}
\hline  
\noalign{\smallskip}
\end{tabular}      
\end{center}
\tablefoot{The $PL$ relations have the form $y=\alpha +\beta\times x$, where x, and y are the period and magnitude, respectively. For PLCs we have $z= \alpha + \beta \times x + \gamma\times y$, where x,y, and z are the period, colour and magnitude, respectively. Additional fits are in the Appendix \ref{app:pl_Tables} (Table~\ref{tab:lmcT2pt2}). The different columns report: $(1)$ the type of relationship and the band of interest; $(2)$ the pulsating class; $(3)$-$(4)$ the $\alpha$ coefficient (intercept) and relative uncertainty; $(5)$-$(6)$ the $\beta$ coefficient (slope as a function of the period) and relative uncertainty; $(7)$-$(8)$ the $\gamma$ coefficient (colour term of the PLC) and relative uncertainty; (9) the Root Mean Square (RMS) of the relation; $(10)$-$(11)$ the number of stars used in the fit and the total number of stars, respectively.}
\end{table*}

\begin{table*}
\footnotesize\setlength{\tabcolsep}{3pt} 
\caption{As for Table~\ref{tab:lmcT2} but for the T2Cs in the SMC. Additional fits are in Appendix~\ref{app:pl_Tables} (Table \ref{tab:smcT2pt2}).  }
  \label{tab:smcT2}
\begin{center}

\begin{tabular}{llllllllllll}
 \hline  
 \noalign{\smallskip} 
       Relation  &  Group & $\alpha $ &  $\sigma_{\alpha}$ & $\beta$ & $\sigma_{\beta} $ &  $\gamma$ & $\sigma_{\gamma}$ & RMS & Used stars & Total stars \\
   &  & mag & mag & mag/dex & mag/dex & & & mag & \\                     
(1)  & (2)  & (3) & (4) & (5) & (6) & (7) &(8) & (9) & (10) & (11)\\   
 \noalign{\smallskip}
 \hline  
 \noalign{\smallskip}  
  PLBP & BLH\&WVir & 19.059 & 0.061 & $-$1.660 & 0.140 &  &  & 0.35 & 35 & 37\\
  PLG & BLH\&WVir & 18.889 & 0.053 & $-$1.800 & 0.120 & & & 0.30 & 35 & 37\\
  PLRP & BLH\&WVir & 18.439 & 0.055 & $-$1.930 & 0.120 & & & 0.32 & 36 & 37\\
  PLV & BLH\&WVir & 18.962 & 0.055 & $-$1.690 & 0.120 & & & 0.32 & 35 & 37&\\
  PLI & BLH\&WVir & 18.388 & 0.052 & $-$1.890 & 0.120 & & & 0.30 & 36 & 37\\
  PLY & BLH\&WVir & 18.169 & 0.051 & $-$2.030 & 0.110 & & & 0.29 & 35 & 36\\
  PLJ & BLH\&WVir & 18.202 & 0.044 & $-$2.280 & 0.100 & & & 0.12 & 28 & 33\\
  PLK & BLH\&WVir & 17.927 & 0.024 & $-$2.396 & 0.053 & & & 0.12 & 25 & 37\\
  PWG & BLH\&WVir & 17.741 & 0.036 & $-$2.301 & 0.081 & & & 0.21 & 35 & 37\\
  PWVI & BLH\&WVir & 17.592 & 0.033 & $-$2.351 & 0.074 & & & 0.19 & 34 & 37\\
  PWVK & BLH\&WVir & 17.762 & 0.023 & $-$2.504 & 0.052 & & & 0.12 & 26 & 37\\
  PWYK & BLH\&WVir & 17.880 & 0.024 & $-$2.427 & 0.053 & & & 0.11 & 24 & 36\\
  PWJK & BLH\&WVir & 17.712 & 0.025 & $-$2.492 & 0.055 & && 0.12 & 24 & 36\\
  PLCG & BLH\&WVir & 17.368 & 0.035 & $-$2.460 & 0.110 & 2.51 & 0.22 & 0.22 & 35 & 37\\
  PLCVI & BLH\&WVir & 17.157 & 0.051 & $-$2.550 & 0.120 & 3.32 & 0.32 & 0.21 & 36 & 37\\
  PLCVK & BLH\&WVir & 17.239 & 0.041 & $-$2.480 & 0.130 & 0.39 & 0.16 & 0.23 & 35 & 37\\
  PLCYK & BLH\&WVir & 17.257 & 0.043 & $-$2.450 & 0.140 & 0.98 & 0.45 & 0.24 & 34 & 36\\
  PLCJK & BLH\&WVir & 17.380 & 0.046 & $-$2.400 & 0.180 & 1.08 & 0.87 & 0.25 & 34 & 36\\
\noalign{\smallskip}
\hline  
\noalign{\smallskip}
\end{tabular}    
\end{center}
\end{table*}

\section{Results}

In this section, we exploit the $PL/PW/PLC$ relations derived in this work and described in the previous Section. First, we discuss the dependence of the $PL/PW/PLC$ coefficients on the wavelength. Secondly, we compare $PL/PW/PLC$ relations with those in the literature.

\subsection{Wavelength dependence of the $PL/PW/PLC$ coefficients.}

The extensive set of $PL/PW/PLC$ calculations shown in Tables~\ref{tab:lmcT2} and ~\ref{tab:smcT2} (as well as Tables~\ref{tab:lmcT2pt2} and ~\ref{tab:smcT2pt2}) allows us to investigate in detail how the slopes and the dispersions of these relations vary with the wavelength. This exercise permits us to choose the best relationships to use in further analysis and to verify a trend which is already known in the literature for the CCs: the slopes and the dispersions of the $PL$ ($PW$/$PLC$) relations increase (in absolute value) and decrease at longer wavelengths, respectively. This anti-correlation was noted for the CCs by \citet{madore1991cepheid} and explained on a physical basis by \citet{madore2011multi}. Indeed, they demonstrated that this feature is due to the different dependence of the surface brightness on the effective temperature at different wavelengths.

\begin{figure}
\sidecaption
    \includegraphics[width=\hsize]{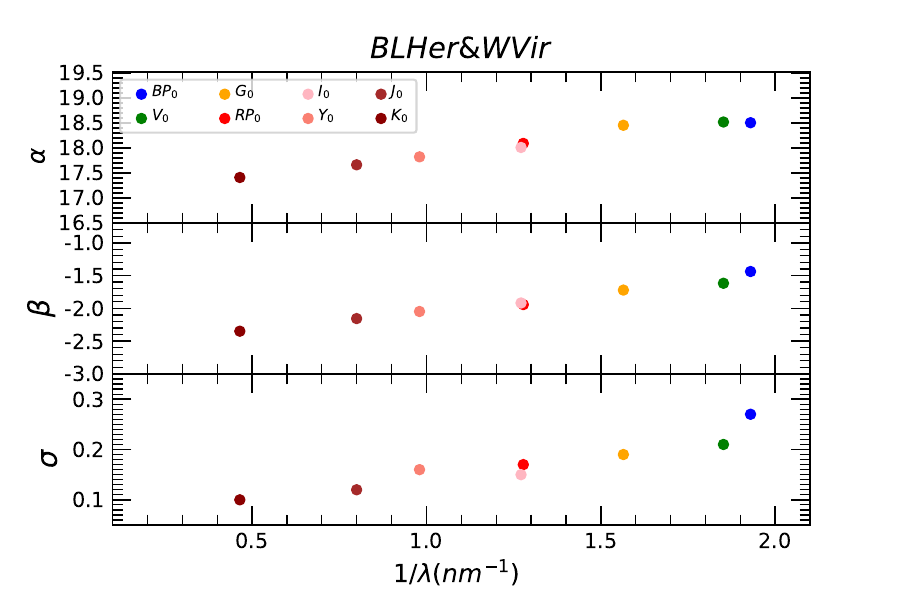}   
    \caption{\label{coeffpllmc} Coefficients for the $PL$ relations in optical and NIR bands for the LMC. $\alpha$ and $\sigma$ are expressed in mag, $\beta$ in mag/dex. Only part of the figures are shown here for illustrative purposes, the remaining figures can be found in Appendix~\ref{app:wave_dependence}. }	
	\end{figure}

\begin{figure}
\sidecaption
        \includegraphics[width=\hsize]{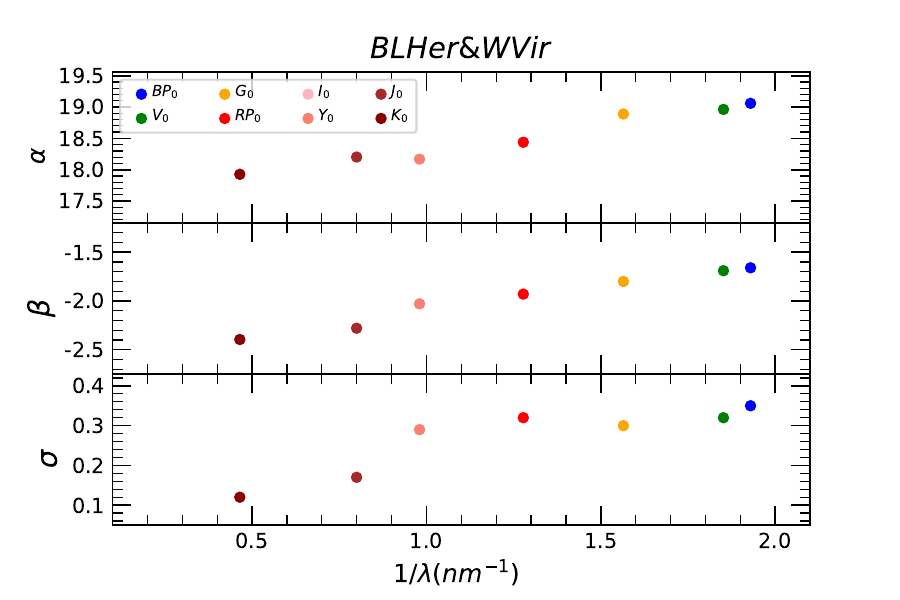}
    \caption{\label{coeffplsmc} As in Fig.~\ref{coeffpllmc} but for the SMC.}	
    \end{figure}

Figures~\ref{coeffpllmc} and ~\ref{coeffplsmc} show the result of this investigation for the $PL$ relations in a few selected cases for the LMC and SMC, respectively. Looking at the panels displaying the $\beta$ (slope) and $\sigma$ (dispersion) coefficients in each figure, the expected trends stand out, especially for the combination BLHer\&WVir, in the LMC, which provides the tightest relations.   
The trends of slope and dispersion are slightly less evident for the SMC, where the significant depth along the line of sight \citep[e.g.,][]{ripepi2017vmc} produces an additional dispersion due to the geometry of the system, which can be larger than the intrinsic width of the IS. 

Similar trends are visible in the other $PL$ relations calculated in this work (see Figs. ~\ref{coeffpllmc2} and ~\ref{coeffplsmc2} ). As for the $PW$ relations (see Figs. ~\ref{coeffpllmc2} and ~\ref{coeffplsmc2} ), also in these cases, the general trend is the same as for the $PL$s but with a less clear behaviour. This is because the inclusion of a colour term in the Wesenheit magnitudes tends to mitigate the effect of the width and the shape of the IS both in the optical and in the NIR bands. These results confirm and expand earlier findings by \citet{Matsunaga2006} and \citet{ngeow2022}.

\subsection{Absolute calibration of PLR/PWR with the geometric distance of LMC}

To compare our $PL/PW/PLC$ with literature as well as to use them to calculate distances, it is mandatory to calibrate their zero points (intercepts) in absolute terms. To carry out this crucial procedure, we decided first to use the geometric distance of the LMC $\mu_{LMC}=18.477\pm0.026$ mag as measured by \citet{pietrzynski2019distance} based on a sample of eclipsing binaries (ECB).

The calibrated $PL/PW/PLC$ relations are listed in  Table~\ref{tab:bestrelationlmcTII}. As expected, the smallest dispersion is obtained for the NIR bands. The $PW$ and $PLC$ provide similar dispersion for the same combination of magnitudes and colours, confirming the goodness of the adopted reddening corrections. Furthermore, the analysis of the colour term coefficients ($\gamma$) of the derived  PLC relations suggests that the optical $PLC$ are less useful than the NIR counterparts. Indeed, the former show values of $\gamma$ which are between ten and twenty times larger, thus small errors in the reddening produce large variations in the derived magnitudes. Particularly interesting is the $PLC$ including $V$ and $K_s$ bands because as already found in our previous investigation \citep[e.g.][]{ripepi2015vmc}, the $\gamma$ coefficient is consistent within the errors with the colour term of the Wesenheit magnitude $W(V,K_s)$. This means that this particular magnitude is not only reddening-free but also correct completely for the width of the IS.

\begin{table*}[h]
  \footnotesize\setlength{\tabcolsep}{4pt} 
  \caption{ Coefficients of the $PL/PLC/PW$ relations for T2Cs in LMC calibrated with $D_{0,LMC}=(18.477\pm 0.027)$ mag according to \citet{pietrzynski2019distance}. }
  \label{tab:bestrelationlmcTII}
 \begin{center}
  \begin{tabular}{lllllllll}
  \hline  
  \noalign{\smallskip}   
       Relation & Group & $\alpha $ &  $\sigma_{\alpha}$ & $\beta$ & $\sigma_{\beta} $ &  $\gamma$ & $\sigma_{\gamma}$ & RMS  \\
    &  & mag & mag & mag/dex & mag/dex & & & mag  \\                                        
  \noalign{\smallskip}
  \hline  
  \noalign{\smallskip}  
  PLJ & BLHer & $-$0.820 &  0.075 & $-$2.250 & 0.140 & & & 0.17\\
  PLJ & WVir & $-$0.558 &  0.038 & $-$2.400 & 0.068 & & & 0.10\\
  PLJ & BLH\&WVir & $-$0.813 &  0.032 & $-$2.156 & 0.024 & & & 0.12\\
  PL$K_s$ & BLHer & $-$1.033 &  0.072 & $-$2.560 & 0.140 & & & 0.16\\
  PL$K_s$ & WVir & $-$0.969 &  0.036 & $-$2.439 & 0.053 & & & 0.08\\
  PL$K_s$ & BLH\&WVir & $-$1.067 &  0.031 & $-$2.348 & 0.019 & & & 0.10\\
  PWG & BLH\&WVir & $-$1.032 &  0.031 & $-$2.436 & 0.022 & & & 0.14\\
  PWVI & BLH\&WVir & $-$1.140 &  0.031 & $-$2.491 & 0.022 & & & 0.12\\
  PW$VK_s$ & BLH\&WVir & $-$1.195 &  0.031 & $-$2.475 & 0.017 && & 0.09\\
  PW$YK_s$ & BLH\&WVir & $-$1.251 &  0.031 & $-$2.516 & 0.017 && & 0.09\\
  PW$JK_s$ & BLH\&WVir & $-$1.226 &  0.031 & $-$2.501 & 0.016 & & & 0.08\\
  PLCG & BLHer & $-$1.103 &  0.058 & $-$2.520 & 0.110 & 2.007 & 0.085 & 0.20\\
  PLCG& WVir & $-$1.149 &  0.040 & $-$2.572 & 0.062 & 2.160 & 0.120 & 0.10\\
  PLCG& BLH\&WVir & $-$1.143 &  0.031 & $-$2.501 & 0.033 & 2.070 & 0.062 & 0.15\\
  PLCVI & BLHer & $-$0.949 &  0.061 & $-$2.460 & 0.110 & 2.150 & 0.120 & 0.12\\
  PLCVI & WVir & $-$1.629 &  0.049 & $-$2.667 & 0.066 & 3.350 & 0.160 & 0.09\\
  PLCVI & BLH\&WVir & $-$1.334 &  0.033 & $-$2.604 & 0.036 & 2.912 & 0.092 & 0.10\\
  PLC$VK_s$ & BLHer & $-$1.161 &  0.052 & $-$2.611 & 0.093 & 0.136 & 0.041 & 0.10\\
  PLC$VK_s$ & WVir & $-$1.133 &  0.036 & $-$2.560 & 0.050 & 0.158 & 0.048 & 0.07\\
  PLC$VK_s$ & BLH\&WVir & $-$1.182 &  0.031 & $-$2.447 & 0.029 & 0.118 & 0.030 & 0.09\\
  PLC$YK_s$ & BLHer & $-$1.165 &  0.057 & $-$2.620 & 0.110 & 0.345 & 0.080 & 0.10\\
  PLC$YK_s$ & WVir & $-$1.013 &  0.037 & $-$2.497 & 0.051 & 0.150 & 0.092 & 0.08\\
  PLC$YK_s$ & BLH\&WVir & $-$1.152 &  0.031 & $-$2.421 & 0.026 & 0.223 & 0.059 & 0.09\\
  \noalign{\smallskip}
  \hline  
  \noalign{\smallskip}
  \end{tabular}
\end{center}
\tablefoot{ The relation are: $M_{\lambda,0}=\alpha +\beta \log P$ for PL($\lambda$); $W_{\lambda_1,\lambda_2}^0=\alpha  +\beta \log P$ for the PW($\lambda_1,\lambda_2 $); $ M_{\lambda_1}^0  =\alpha   +\beta \log P + \gamma (m_{\lambda_1}-m_{\lambda_2})_0 $ for the $PLC$($\lambda_1,\lambda_2 $).}
\end{table*}

\subsubsection{Comparison with the literature}

\begin{figure*}[ht]
    \vbox{
    \hbox{
    \includegraphics[width=0.43\textwidth]{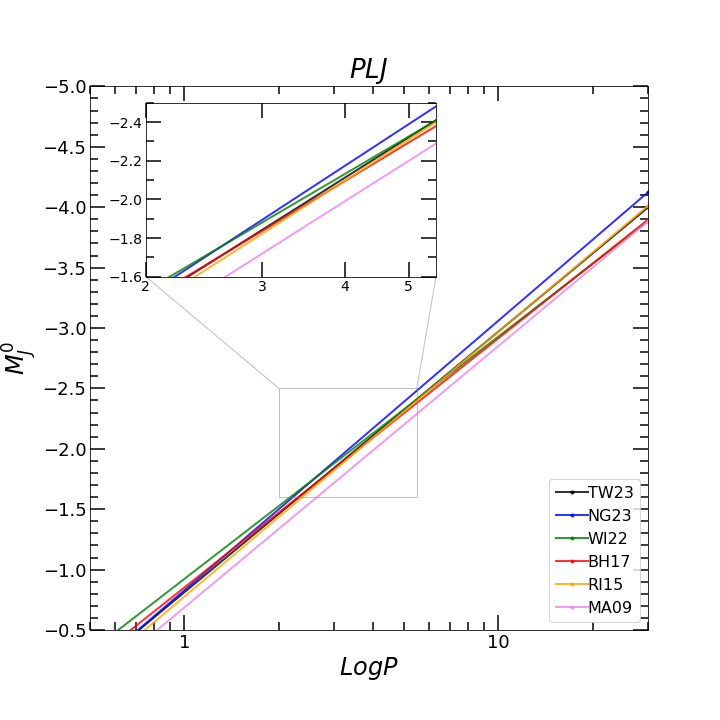}
    \includegraphics[width=0.43\textwidth]{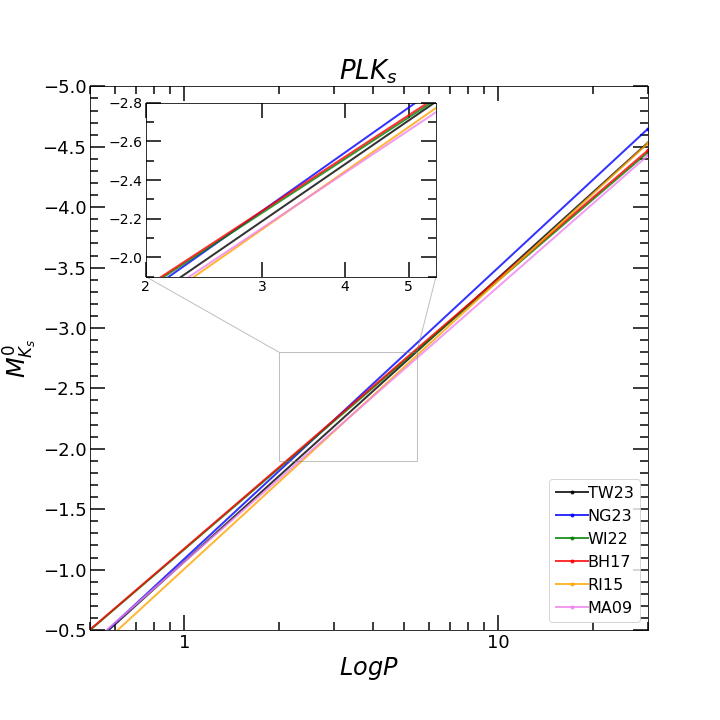}
    }
    \hbox{
    \includegraphics[width=0.43\textwidth]{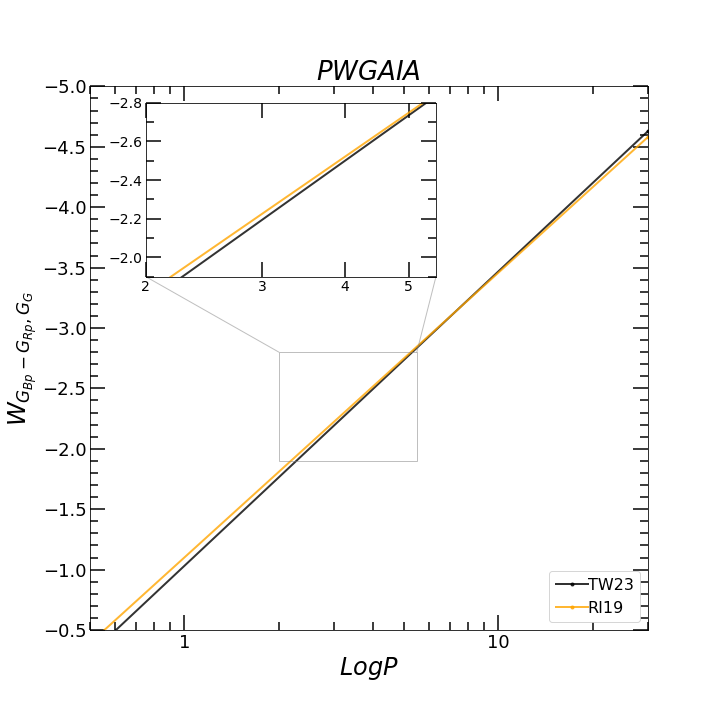}
    \includegraphics[width=0.43\textwidth]{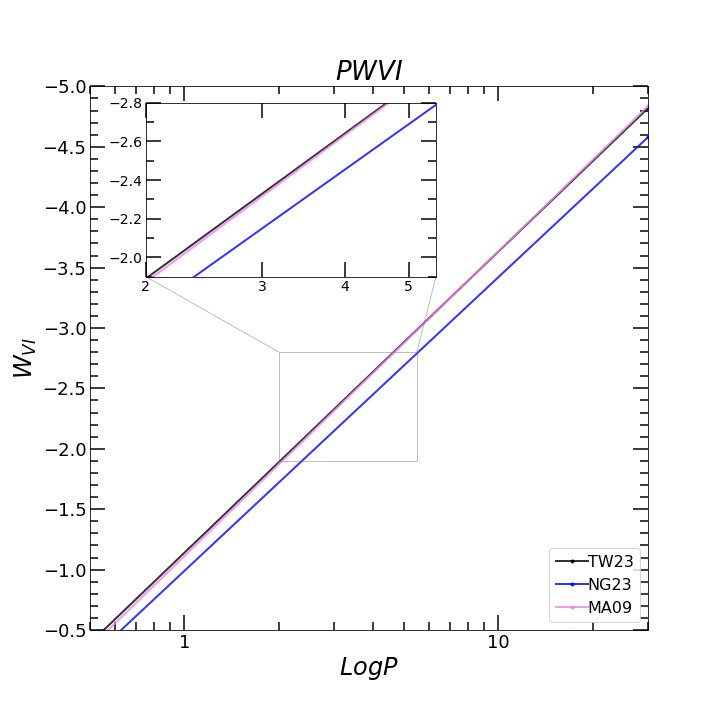}
    }
    \hbox{
    \includegraphics[width=0.43\textwidth]{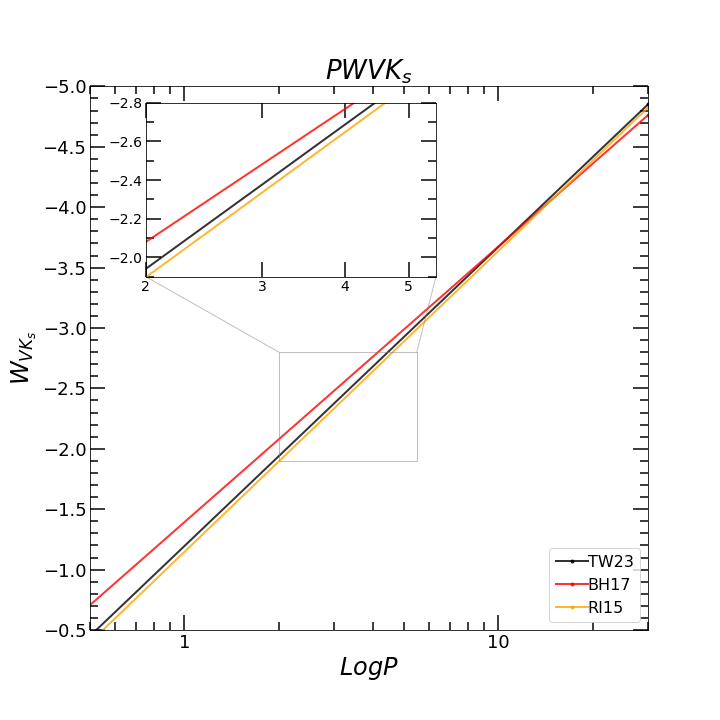}
    \includegraphics[width=0.43\textwidth]{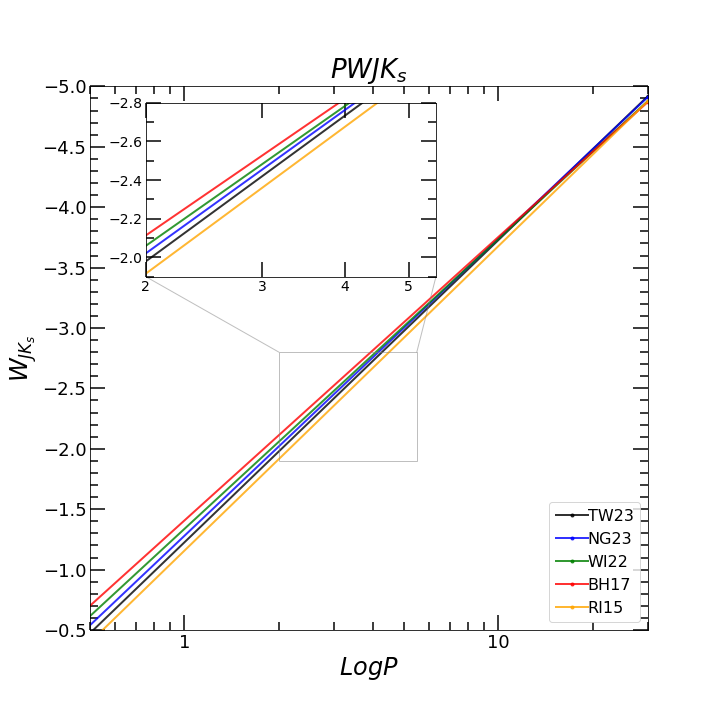}
    }
    }
    \caption{\label{comparison} Comparison of the best $PL/PW$ of this work, labelled as 'TW23' with previous works: \citet['MA09']{Matsunaga2009}, \citet['RI15']{ripepi2015vmc}, \citet['RI19']{Ripepi2019_reclassificaiton}, \citet['c']{Bhardwaj2017_LMC}, \citet['WI22']{Wielgorski2022}, \citet['NG22']{ngeow2022}. The inserts show a zoom-in of regions where there is high overlap between the different relations. }	
	\end{figure*}

The relations in Table~\ref{tab:bestrelationlmcTII} (with absolute intercepts) and Table~\ref{tab:lmcT2} (with relative intercepts) can now be compared to those available in the literature for the LMC and the GGCs.\\
Table~\ref{tab:confrontolmct2} shows the comparison between a variety of $PL/PW/PLC$ relations values from the literature \citep{Matsunaga2006,Matsunaga2011,ripepi2015vmc,Bhardwaj2017_LMC} and our values. Figure~\ref{comparison} shows the comparison of our best relations with \citet{Matsunaga2009, ripepi2015vmc, Ripepi2019_reclassificaiton, Bhardwaj2017_LMC, Wielgorski2022, ngeow2022}. Since the last three literature works were provided in the 2MASS photometric system, we  adopted the conversion equations provided by \citet{Gonzalez-Fernandez2018} to convert from the 2MASS photometric system to the VISTA one: J$^{VISTA}$= J$^{2MASS}$ $-$ 0.0031(J$- K_s$)$^{2MASS}$; $K_s^{VISTA}$= $K_s^{2MASS} -$ 0.0006(J$-K_s$)$^{2MASS}$. 

\begin{table*}[h]
  \footnotesize\setlength{\tabcolsep}{4pt} 
  \caption{Comparison among present $PL/PW$ for T2Cs in LMC and the literature values. } 
  \label{tab:confrontolmct2}
  \begin{center}
  \begin{tabular}{llllllllllll} 
  \hline  
  \noalign{\smallskip}   
        Relation & Galaxy & Sample & $\alpha_{app} $ &  $\sigma_{\alpha}$& $\alpha_{abs} $ &  $\sigma_{\alpha}$ & $\beta$ & $\sigma_{\beta} $ & RMS & $\#$star & Reference \\
    & &  & mag & mag & mag & mag & mag/dex & mag/dex  &mag & &   \\                                        
  \noalign{\smallskip}
  \hline  
  \noalign{\smallskip}
  PLJ & GGC & BLHer & & & -0.666 & 0.041 & $-$2.959 & 0.313 & 0.11 & 7 & M06\\   
 PLJ & LMC & BLHer & 17.768 & 0.038 & & & $-$2.164 & 0.240 & 0.25 & 55 & M11\\ 
  PLJ & LMC & BLHer & 17.669 & 0.113 & & & $-$2.294 & 0.153 & 0.20 & 55 & B17\\
  PLJ & LMC & BLHer & &  & -0.772 & 0.039 & $-$2.356 & 0.259 & 0.16 & 20 & W22\\
  \rowcolor{lightgray}
  PLJ & LMC& BLHer & 17.657 & 0.069 & $-$0.820 &  0.075 & -2.250 & 0.140 &  0.17 & 73 & TW\\
  PLJ & GGC & WVir & & &  -0.911 & 0.027 & $-$2.204 & 0.090 & 0.16 & 39 & M06\\
  PLJ & LMC & WVir & 17.957 & 0.030 & & & $-$2.337 & 0.114 & 0.18 & 82 & M11\\
  PLJ & LMC & WVir & 17.958 & 0.018 & & & $-$2.378 & 0.105 & 0.11 & 72 & B17\\
  \rowcolor{lightgray}
  PLJ & LMC& WVir & 17.919 & 0.024 & $-$0.558 &  0.038 & -2.400 & 0.068 & 0.10 & 94 & TW\\
  PLJ & GGC & BLHer+WVir & & & $-$0.864 & 0.030 & $-$2.230 & 0.070 & 0.16 & 46 & M06\\
  PLJ & LMC & BLHer+WVir & 17.777 & 0.029 & & & $-$2.163 & 0.044 & 0.21 & 137 & M11\\
  PLJ & LMC & BLHer+WVir & 17.700 & 0.035 & & & $-$2.190 & 0.040 & 0.13 &  & R15\\
  PLJ & LMC & BLHer+WVir & 17.634 & 0.039 & & & $-$2.061 & 0.038 & 0.16 & 126 & B17\\
  PLJ & LMC& BLHer+WVir & & & $-$0.789 & 0.015 & -2.177 & 0.040 & 0.12 & 61  & W22\\
  \rowcolor{lightgray}
  PLJ & LMC & BLHer+WVir & 17.664 & 0.010 & $-$0.813 &  0.032& $-$2.156 & 0.020 & 0.12 & 162 & TW\\
  PL$K_s$ & GGC & BLHer & & & $-$1.178 & 0.039 & $-$2.294 & 0.294 & 0.10 & 7 &  M06\\
  PL$K_s$ & LMC & BLHer &  17.329 & 0.040 & & & $-$1.992 & 0.278 & 0.26 & 47 & M11\\
  PL$K_s$ & LMC & BLHer &  17.245 & 0.114 & & & $-$2.083 & 0.154 & 0.26 & 47 & B17\\
  PL$K_s$ & LMC & BLHer & & & $-$0.966 & 0.025 & $-$2.616 & 0.165 & 0.11 & 20 & W22\\
  \rowcolor{lightgray}
  PL$K_s$ & LMC & BLHer & 17.444 & 0.066 & $-$1.033 & 0.072 & $-$2.560 & 0.14  & 0.16 & 73 & TW\\
  PL$K_s$ & GGC & WVir & & & $-$1.0686 & 0.024 & $-$2.442 & 0.082 & 0.15 & 39 & {\bf M06}\\
  PL$K_s$ & LMC & WVir & 17.640 & 0.029 & &  & $-$2.503 & 0.109 & 0.17 & 82 & M11\\
  PL$K_s$ & LMC & WVir & 17.328 & 0.016 & &  & $-$2.250 & 0.097 & 0.12 & 72 & B17\\
  \rowcolor{lightgray}
  PL$K_s$ & LMC & WVir & 17.508 & 0.019 & $-$0.969 &  0.036 & $-$2.439 & 0.053 & 0.08 & 98 & TW\\
  PL$K_s$ & GGC & BLHer+WVir & & & $-$1.110 & 0.020 & $-$2.410 & 0.050 & 0.14 &  & M06\\
  PL$K_s$ & LMC & BLHer+WVir &  17.390 & 0.029 & & & $-$2.278 & 0.047 & 0.21 & 129 & M11\\
  PL$K_s$ & GGC & BLHer+WVir & & & $-$1.116 & 0.021 & $-$2.408 & 0.047 & 0.14 & 46 & M06\\
  PL$K_s$ & LMC & BLHer+WVir &  17.470 & 0.020 & & & $-$2.385 & 0.030 & 0.09 & & R15\\
  PL$K_s$ & LMC & BLHer+WVir &  17.302 & 0.015 & & & $-$2.232 & 0.037 & 0.18 & 119 & B17\\
  PL$K_s$ & LMC & BLHer+WVir &  & & $-$1.013 & 0.011 & $-$2.387 & 0.030 & 0.09 & 62 & W22\\
  \rowcolor{lightgray}
  PL$K_s$ & LMC & BLHer+WVir & 17.410 & 0.008 & $-$1.067 &  0.031 & $-$2.348 & 0.019 & 0.10 & 165 & TW\\
  PWVI & LMC & BLHer+WVir &  17.476 & 0.015 & & & $-$2.521 & 0.022 & 0.11 & 131 & M11\\  
  PWVI & LMC & BLHer& 17.356 & 0.024 & & & $-$2.683& 0.090 & & 79 & I18\\
  PWVI & LMC & WVir& 17.378 & 0.062& & & $-$2.536& 0.060 & & 94 & I18\\ 
  \rowcolor{lightgray}
  PWVI & LMC & BLHer+WVir & 17.337 & 0.009 & $-$1.140 & 0.031 & -2.491 & 0.022 & 0.09 &177 & TW \\
  PWJK & LMC & BLHer+WVir & 17.320 & 0.025 & & & $-$2.520 & 0.030 & 0.22 &  & R15 \\
  PWJK & LMC & BLHer+WVir & 17.070 & 0.021 & & & $-$2.346 & 0.051 & 0.22 & 119 & B17 \\
  PWJK & LMC & BLHer+WVir & & & $-$1.166 & 0.011 & $-$2.544 & 0.029 & 0.09 & 61 &  W22\\
  \rowcolor{lightgray}
  PWJK & LMC & BLHer+WVir & 17.251 & 0.006 & $-$1.226 & 0.031 & -2.501 & 0.016 & 0.08 & 146 & TW\\
  PWVK & LMC & BLHer+WVir & 17.330 & 0.020 & & & $-$2.490 & 0.003 & 0.08 &  & R15 \\
  PWVK & LMC & BLHer+WVir & 17.084 & 0.036 & & & $-$2.281 & 0.036 & 0.25 & 124 & B17 \\
  \rowcolor{lightgray}
  PWVK & LMC & BLHer+WVir & 17.282 & 0.007 & $-$1.195 & 0.031 & $-$2.475 & 0.017  & 0.09 & 160 & TW\\
  \noalign{\smallskip}
  \hline  
  \noalign{\smallskip}
  \end{tabular}
  \end{center}
  \tablefoot{$\alpha_{app}$ and $\alpha_{abs}$ refer to the intercepts of the labelled relations in apparent and absolute magnitudes, respectively. \\
  "M06" refers to \cite{Matsunaga2006}, "M11" refers to \cite{Matsunaga2011}, "R15" refers to \cite{ripepi2015vmc}, "B17" refers to \cite{Bhardwaj2017_LMC}, "I18" refers to \cite{Iwanek2018}, "W22" refers to \cite{Wielgorski2022},  "TW" (\textit{gray line}) refers to this work . "$\#$star" indicates the number of stars used in the fit.}
\end{table*}

An inspection of the tables allows us to reach several considerations:

\begin{itemize}

\item
For each T2C sample and filter combination, our $PL$ relations have been derived with the largest sample of stars and the resulting errors in the relation coefficients are the smallest compared to any literature work, especially in the NIR bands (see Table~\ref{tab:confrontolmct2}). 

\item
The coefficients of the BLHer $PL$ relations (Table~\ref{tab:confrontolmct2}) in the \textit{J} band are in agreement within 1 $\sigma$ with \citet{Bhardwaj2017_LMC}, within 1.5 $\sigma$ with \citet[][LMC]{Matsunaga2011} and within 2 $\sigma$ with \citet[][GGCs]{Matsunaga2006}. In the $K_s$ band, they agree within 1.5 $\sigma$ with \citet[][GGCs]{Matsunaga2006} and within 2 $\sigma$ with the others. 
\item
The coefficients of the WVir $PL$ relations in $J$ agree within 1 $\sigma$ with \citet{Bhardwaj2017_LMC} and \citet[][LMC]{Matsunaga2011}, disagree with \citet[][GGCs]{Matsunaga2006}, while in the $K_s$ band they are in agreement within 1 $\sigma$ only with \citet[][GGCs]{Matsunaga2006}. All these discrepancies can be tentatively explained with the rather large uncertainties introduced in the fitting procedure by the short period range adopted when analysing WVir (and BLHer) separately. Indeed, when the two sub-types are used together, the agreement between the different works improves significantly (see below). 
\item
The coefficients of the $PL$s calculated using both BLHer and WVir in the $J$ band agree within 1 $\sigma$ with \cite{Bhardwaj2017_LMC}, \citet{ripepi2015vmc} and \citet{Matsunaga2006}, within 2 $\sigma$ with \citet{Matsunaga2011}.
In the $K_s$ band, the agreement is generally worse, as in the previous cases, indeed the $PL$ coefficients agree within 2 $\sigma$ with all the literature works, except \cite{Bhardwaj2017_LMC} which is more discrepant.
\item
Concerning the coefficients of the Wesenheit relations, there is a very good agreement with previous VMC results \citep{ripepi2015vmc} while there is a large disagreement with \citet{Bhardwaj2017_LMC}. These differences could be due to $K_s$ band photometry in \citet{Bhardwaj2017_LMC}, which was approaching the faint limit for the shortest-period BL Her stars.
\item
The $PLC$ values can only be compared with the previous VMC work \citep{ripepi2015vmc} as shown in Table~\ref{tab:confrontoplc}. The agreement is very good, especially in $V$ and $K_s$, but with improved uncertainties in this work, owing to the larger sample. 
\item
In the SMC, the coefficients of our T2C $PL/PW$ relations listed in Table~\ref{tab:confrontosmct2} agree all within 1-2 $\sigma$ with \citet{Matsunaga2011} and \citet{Iwanek2018} (for the optical $PWVI$ Wesenheit magnitude).
\item 
The $PL$ slopes ($\beta$ coefficients) for the LMC and the SMC agree within 1 $\sigma$ from the G to the $K_s$ band, with the exception of $G_{BP}$, which agrees within 2 $\sigma$ (see Tables~\ref{tab:lmcT2} and ~\ref{tab:smcT2}). 
\item 
The $PWVK_s$ and the $PWJK_s$ slopes ($\beta$ coefficients) for the LMC and the SMC agree within 1 $\sigma$, while the slopes of the others $PW$ agree within 2 $\sigma$.
\end{itemize}

Overall, the comparison with the literature shows good agreement, as for almost all the relations we find coefficients of $PL/PW/PLC$ in agreement within better than 2 $\sigma$ for both the LMC and SMC. At the same time, in almost all cases, our data improves both the precision and the accuracy with respect to the literature, owing to larger samples, better light curve coverage, and deeper photometry guaranteed by the VMC survey.  

\begin{table*}[h]
  \footnotesize\setlength{\tabcolsep}{4pt} 
  \caption{ Comparison between present results and the literature values for the $PLC$ relations.}
  \label{tab:confrontoplc}
  \begin{center}
  \begin{tabular}{lllllllllll}
  \hline  
  \noalign{\smallskip}   
       Relation & Galaxy & Sample & $\alpha $ & $\sigma_{\alpha}$ & $\beta$ & $\sigma_{\beta} $ &  $\gamma$ & $\sigma_{\gamma}$ & RMS & Reference \\
   & & & mag & mag &  mag/dex & mag/dex  &  & & mag&   \\
  \noalign{\smallskip}
  \hline  
  \noalign{\smallskip}  
PLCVK &LMC&  BLHer+WVir & 17.33 & 0.05 & $-$2.48 & 0.04 & 0.125 & 0.04 & 0.09 & R15\\
  \rowcolor{lightgray}
PLCVK & LMC & BLHer+WVir & 17.30 & 0.01 & $-$2.45 & 0.03 & 0.118 & 0.03 & 0.09 & TW\\
PLCJK &LMC & BLHer+WVir & 17.39 & 0.04 & $-$2.45 & 0.04 & 0.35 & 0.14 & 0.09 & R15\\
  \rowcolor{lightgray}
PLCJK & LMC& BLHer+WVir & 17.25 & 0.01 & $-$2.49 & 0.03 & 0.69 & 0.10 & 0.08 & TW\\
  \noalign{\smallskip}
  \hline  
  \noalign{\smallskip}
  \end{tabular}      
  \end{center}
  \tablefoot{The form of the $PLC$ relations is: $ M_{\lambda_1}^0  =\alpha   +\beta \log P + \gamma (m_{\lambda_1}-m_{\lambda_2})_0 $. "R15" refers to \cite{ripepi2015vmc}, "TW" refers to this work. "$\#$star" indicates the number of stars used in the fit.}
\end{table*}

\begin{table*}
  \footnotesize\setlength{\tabcolsep}{4pt} 
  \caption{As for Table \ref{tab:confrontolmct2} but for SMC. } 
  \label{tab:confrontosmct2}
  \begin{center}
  \begin{tabular}{llllllllll}
  \hline  
  \noalign{\smallskip}   
        Relation & Galaxy & Sample & $\alpha $ &  $\sigma_{\alpha}$& $\beta$ & $\sigma_{\beta} $ & RMS & $\#$star & Reference \\
    & &  & mag & mag & mag/dex & mag/dex  &mag & &   \\                                        
  \noalign{\smallskip}
  \hline  
  \noalign{\smallskip}  
  PLJ & SMC & BLHer+WVir  &  18.070 & 0.116 & $-$2.147 & 0.154 & 0.34 & 25 & M11\\
  PLJ+ & SMC & BLHer+WVir  &  17.993 & 0.092 & $-$2.092 & 0.116 & 0.33 & 47 & M11\\
  \rowcolor{lightgray}
  PLJ & SMC& BLHer+WVir & 18.202 & 0.044 & $-$2.280 & 0.100 & 0.17 & 28 & TW \\
  PL$K_s$ & SMC & BLHer+WVir  &  17.5874 & 0.109 & $-$2.082 & 0.151 & 0.32 & 23 & M11\\
  PL$K_s$+ & SMC & BLHer+WVir  &  17.5966 & 0.082 & $-$2.113 & 0.105& 0.29 & 45 & M11\\
  \rowcolor{lightgray}
  PL$K_s$ & SMC &  BLHer+WVir & 17.927 & 0.024 & $-$2.396 & 0.053 & 0.12 & 25 & TW \\
  PWVI & SMC & BLHer+WVir  & 17.554 & 0.083 & $-$2.304 & 0.107 & 0.23 & 27 & M11\\
  PWVI & SMC & BLHer& 17.630 & 0.104& $-$2.753& 0.403 & &  20 & I18\\
  PWVI & SMC & WVir& 17.976 & 0.164& $-$2.688 & 0.156 & & 15 & I18\\
  \rowcolor{lightgray}
  PWVI & SMC & BLHer+WVir & 17.592 & 0.033 & $-$2.351 & 0.074 & 0.19 & 34 & TW \\
  \noalign{\smallskip}
  \hline  
  \noalign{\smallskip}
  \end{tabular}
  \end{center}
  \tablefoot{"M11" refers to \cite{Matsunaga2011}, "I18" refers to \cite{Iwanek2018}, "TW" (\textit{gray line}) refers to this work. In M11, J+ and $K_s$+ refer to the combination of IRSF and NTT data sets.}
\end{table*}

\section{Applications}

After having calibrated our best relationships by assuming the LMC distance, we derive the distances of selected GGCs hosting T2C variables and of selected field T2C stars having good multi-band photometry and $Gaia$ parallaxes.

\subsection{The distance to the GGCs with T2Cs $PL/PW$ relations calibrated with the LMC geometric distance} \label{calibrlmc}
To test the $PL/PW$ relationships calibrated with the LMC geometric distance, we applied them to GGCs hosting T2Cs. Indeed, GGC distances can be derived from a variety of methods and are usually considered accurate \citep[see][and references therein]{baumgardt2021accurate}. 
To this aim, we collected a sample of 46 T2Cs belonging to 22 GGCs, taking from the literature only stars whose mean magnitudes were calculated as in our work \citep[i.e. as intensity-averaged magnitudes and periods, see][]{Bhardwaj2017_Bulge,bhar2021,braga2020separation}. We adopted distances (to compare with) from \citet{ngeow2022,baumgardt2021accurate} and reddenings from \citet{harris2010new}. The $Gaia$ photometry was missing in the quoted publications so we added it from the $Gaia$ DR3 catalogue \citep[e.g.,][]{ripepi2022gaia} (see Table~\ref{tab:t2csggcs} in Appendix~\ref{tableggcs}). 

\begin{figure*}[h]
\centering
    \vbox{
    \hbox{
    \includegraphics[width=0.35\textwidth]{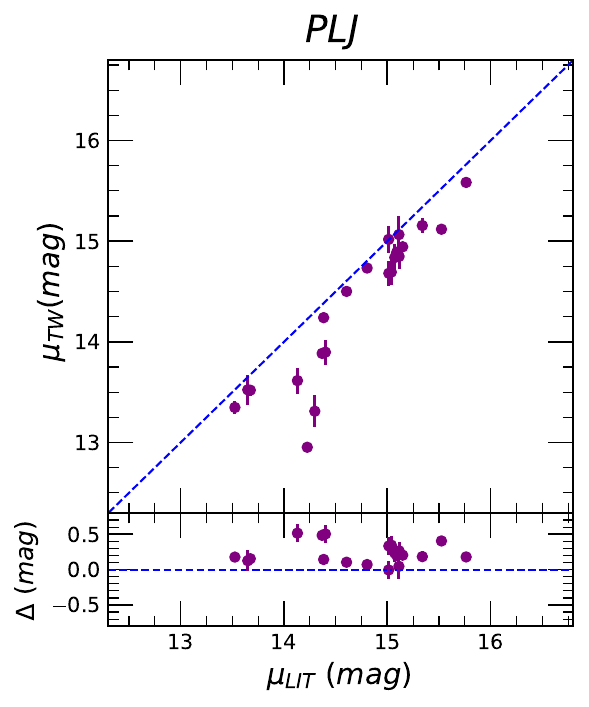}
    \includegraphics[width=0.35\textwidth]{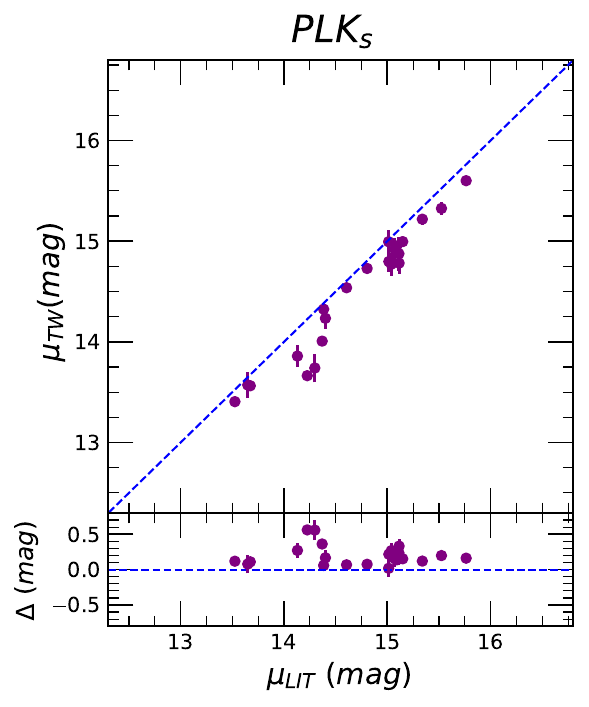}
    }
    \hbox{
    \includegraphics[width=0.35\textwidth]{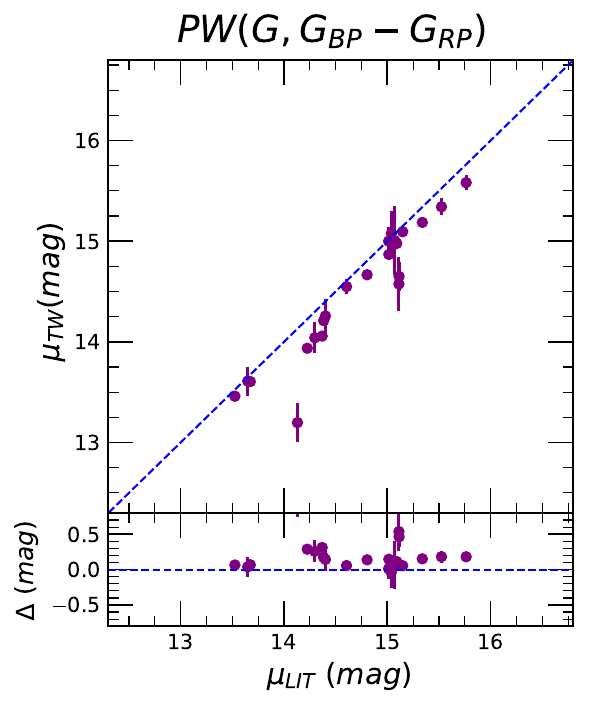}
    \includegraphics[width=0.35\textwidth]{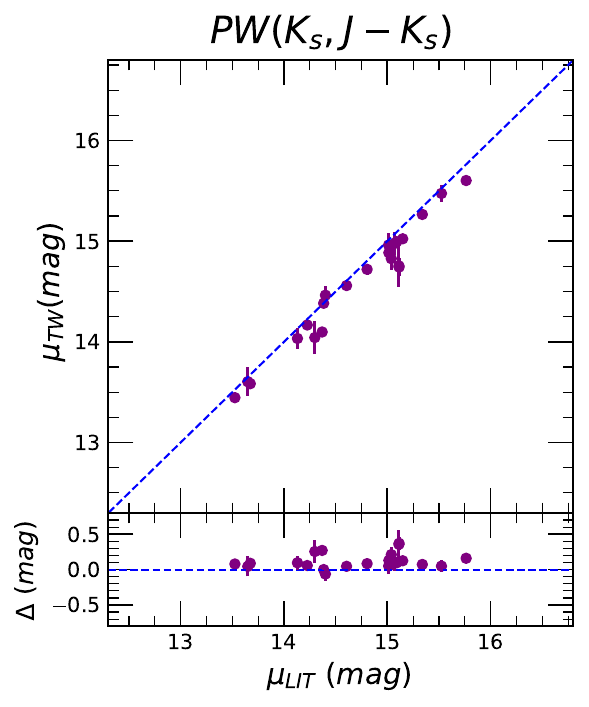}
    }
    \hbox{
    \includegraphics[width=0.35\textwidth]{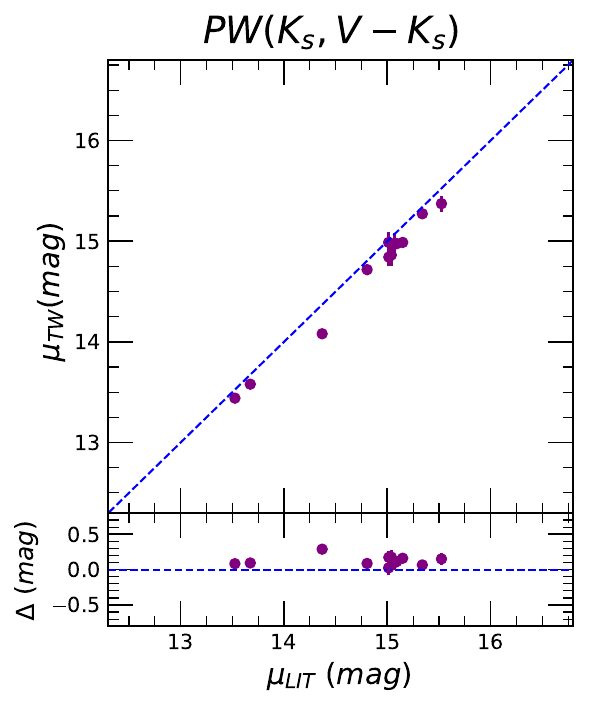}
    }}
    \caption{\label{dmgc} Comparison between our distances and those by \citet{baumgardt2021accurate}. }
\end{figure*}

To obtain the GGC's distance modulus ($\mu$), we inserted the observed periods of the hosted T2Cs in our absolute $PL/PW$ relations, obtaining the star's absolute magnitudes, which, combined with the apparent ones, provide the distances. 
When more than one pulsator was present in one GGC, the individual distances were averaged. The comparison between the $\mu$
calculated with our relations and those compiled by \citet{ngeow2022}, largely based on the work by \citet{baumgardt2021accurate}, are shown in Fig.~\ref{dmgc}. As the sense of the difference in the lower panels is always literature$-$this work, it is clear that the literature distances are systematically larger. The first two columns of Table~\ref{deltadis} quantify these differences. The largest discrepancy is obtained when the $PL$ in the $J$ band is used, while the minimum is found using $PW$ relations, especially in the NIR bands, which also provide the smallest dispersion. The large difference between results based on the NIR $PL$ and the $PW$ relations is not easy to explain. We can hypothesise a not-perfectly homogeneous $J,K_s$ photometry for GGCs data which is mitigated when using Wesenheit magnitudes. In any case, if we take the results for the $PW$ relations as a reference, we have that the distance scale of GGCs is overestimated up to about 0.1 mag, assuming that the distance of the LMC is that provided by \citet{pietrzynski2019distance}.
This supposition can imply a potential error as the spatial distribution of the Late-type ECBs adopted by \citet{pietrzynski2019distance} may differ from that spanned by the T2Cs in our sample. This occurrence could account for at least part of the observed shorter distances of the GGCs found in this section.


\begin{table}[h]
  \footnotesize\setlength{\tabcolsep}{3pt} 
  \caption{ Median and $\sigma$ for each GGC distance moduli difference ($\Delta$) between \citet{baumgardt2021accurate} and this work (see Figs.~\ref{dmgc},~\ref{muplx} and ~\ref{muabl}). }
  \label{deltadis}
  \begin{center}
  \begin{tabular}{lcclclcl}
  \hline  
  \noalign{\smallskip}   
\multicolumn{2}{c}{}  & \multicolumn{2}{c}{Cal. LMC}& \multicolumn{2}{c}{Phot. Par.} & \multicolumn{2}{c}{ABL} \\
\cmidrule(lr){3-4}\cmidrule(lr){5-6} \cmidrule(lr){7-8} 
       & n$_{GGCs}$ & Median $\Delta$ & $\sigma_{med}$ & Median $\Delta$ & $\sigma_{med}$ & Median $\Delta$ & $\sigma_{med}$ \\
 &   & (mag) & (mag) &  (mag) & (mag)  &  (mag) & (mag)\\
  \noalign{\smallskip}
  \hline  
  \noalign{\smallskip}  
  $\Delta_J$       & 22 & 0.203 & 0.037 & 0.157 & 0.037 &  0.140 & 0.037  \\
  $\Delta_{K_s}$   & 22 & 0.158 & 0.026 & 0.101 & 0.026 &  0.086 & 0.026  \\
  $\Delta_{WG}$ & 22 & 0.145 & 0.027 & 0.136 & 0.027 &  0.092 & 0.027  \\
  $\Delta_{WJK_s}$ & 22 & 0.092 & 0.014 & 0.027 & 0.014 & $$-$$0.007 & 0.014  \\
  $\Delta_{WVK_s}$ & 12 & 0.108 & 0.019 & 0.059 & 0.019 &  0.043 & 0.019 \\
\noalign{\smallskip}
  \hline  
  \noalign{\smallskip}
  \end{tabular}
  \tablefoot{"$\#$GGC" indicates the number of GGC used in the calculation. "Cal. LMC" stands from the distance moduli of GGCs obtained through the calibration of PL/PW based on LMC geometric distance, while Phot. Par and ABL show the results obtained using the $Gaia$ parallaxes with two different techniques (see text).}
  \end{center}
\end{table}

\subsection{Absolute calibration of the T2Cs $PL/PW$ relations using $Gaia$ parallaxes}

\begin{figure}
    \includegraphics[width=0.5\textwidth]{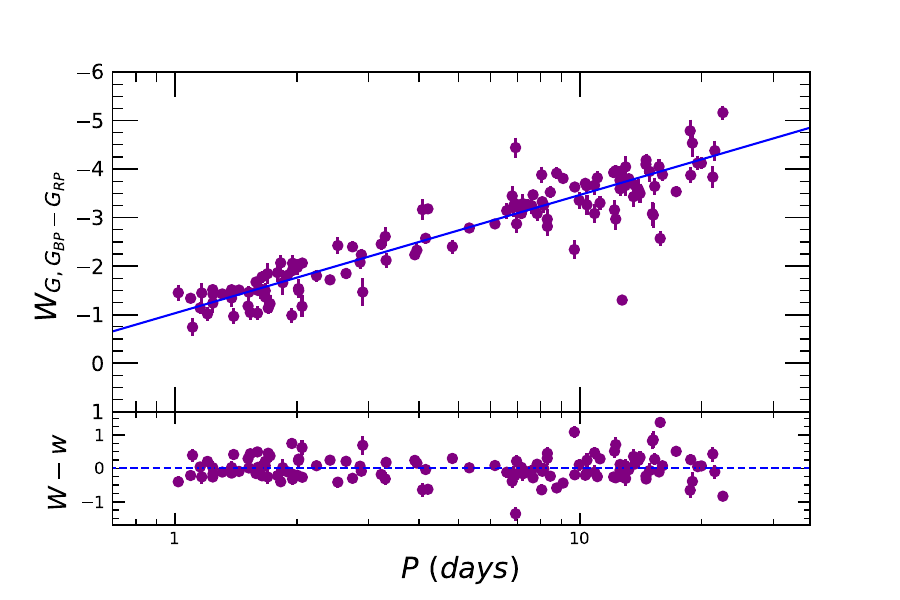}
    \caption{\label{plxgal} Period vs absolute Wesenheit magnitude in the $Gaia$ bands ($W(G,G_{BP}-G_{RP})$ for a sample of Galactic T2Cs selected to have a relative error on the parallax better than 20\% (solid circles). The solid line has a slope calculated for the LMC sample and a zero point calculated using the geometric distance to the same galaxy \citep[][]{pietrzynski2019distance}.}
\end{figure}

Given the unexpected result reported in the previous section, we decided to calibrate differently the zero points of the T2Cs $PL/PW$ relations. Specifically, we adopt a sample of field Galactic T2Cs having optical photometry and parallax from $Gaia$ DR3, while for a small sub-sample, we also used the NIR data published by \citet{Wielgorski2022}.
Given the small size of the latter sample, we decided to fix the slopes of the $PL/PW$ relations to those of the LMC, which has been derived using a much larger number of pulsators. Before proceeding, we verified that the slope of the LMC adapts well to the Galactic data, as shown in Fig~\ref{plxgal}, where we plotted the Period--absolute $W_{G,G_{BP}-G_{RP}}$\footnote{The absolute magnitudes have been derived using $G-1.9 \times (G_{BP}-G_{RP}) + 5 \times \log(\varpi)-10$, where $\varpi$ is the $Gaia$ EDR3 parallax corrected using the \citet{Lindegren2021} relations (see next sections).} relation for a sample of Galactic T2Cs having relative errors on the parallaxes better than 20\% and overlapping the similar relation obtained for the LMC pulsators \citep[with the zero point calculated from the LMC distance by][]{pietrzynski2019distance}. The slope obtained in LMC perfectly describes the distribution of the galactic T2Cs at least in the $Gaia$ Wesenheit.

Having safely fixed the slopes of the different $PL/PW$ relations to those of the LMC, we proceed to calculate the absolute zero points. To this aim, we cannot just invert the $Gaia$ parallaxes to obtain the Galactic T2Cs distances lest to lose the statistical properties of parallaxes' errors \citep[e.g.,][]{Luri2018}. Instead, we adopted two quantities, namely the photometric parallax \citep[][]{feast1997hipparcos} and the astrometric-based luminosity \citep[ABL, e.g.,][]{arenou1998distances} to carry out the calculations, which are described in detail in Appendix~\ref{calplx}. 

The results of this procedure are shown in Table~\ref{tab:0pointsphotpar} which can be compared with Table~\ref{tab:bestrelationlmcTII},  listing the results obtained from the use of the geometric distance to the LMC. To ease the comparison, we used the absolute zero points (i.e. the values of $\alpha$) and the relative ones (Table~\ref{tab:lmcT2}) to directly calculate the distance modulus of the LMC and the SMC for the different $PL/PW$ relations. These values and the relative errors are listed in the last part of Table~\ref{tab:0pointsphotpar}. 

\subsubsection{The LMC distance}
The LMC distances obtained with the photometric parallax are systematically longer by 0.04--0.06 mag than the geometric value of \citet{pietrzynski2019distance}, except the result based on the $PW$ relation in the $Gaia$ bands, but consistent with it within 1$\sigma$.
The LMC distances calculated through the ABL procedure are systematically larger than those obtained from the Photometric Parallax, with the largest deviation (0.04 mag) detected in the $W(J,K_s$) magnitude.   

In general, the $\mu_{LMC}$ derived through the Photometric Parallax technique are in better agreement with the geometric $\mu$ \citep{pietrzynski2019distance} than those derived through the ABL. In any case, the $PL/PW$ relations calibrated using the $Gaia$ parallaxes of Galactic T2Cs provide larger $\mu$ by 0.05-0.06 mag on average than the \citep{pietrzynski2019distance} value. 

This difference may be caused by the adopted  \citet{Lindegren2021} individual zero point offsets which tend to over-correct the parallaxes and that require a counter-correction \citep[see][for discussions and references on this subject]{Groenewegen2021,Riess2021,Ripepi2021}. To test this possibility, we adopted two typical values of the counter-correction among the many available in the literature: $-$14 $\mu$as \citep{Riess2021} and $-$22 $\mu$as \citep{Molinaro2023}. However, the adoption of these offsets on the parallaxes has the effect of increasing the discrepancy, therefore, we retained the $PL/PW$ relations with $\alpha$ values calculated adopting only the individual \citet{Lindegren2021} correction on parallaxes. 

\subsubsection{The SMC distance}

Given the agreement among almost the totality of the slopes for the LMC and the SMC (see previous section), the zero points calculated using Galactic T2Cs can be used to estimate the distance modulus of the SMC (listed in the last columns of Table \ref{tab:0pointsphotpar}.

As for the LMC, the SMC distances calculated through the ABL procedure are systematically larger than those obtained from the Photometric Parallax. 

In general, the $\mu_{SMC}$ derived through the Photometric Parallax technique are in better agreement with the geometric $\mu$ \citep{graczyk2020} than those derived through the ABL. The $PL/PW$ relations calibrated using the $Gaia$ parallaxes of Galactic T2Cs provide larger $\mu$ by 0.05-0.06 mag on average than the \citet{graczyk2020} value (18.977 $\pm$ 0.016 $\pm$ 0.028 mag). 

This result means that the ECB and T2C methods give approximately the same distance offset (0.5 mag) between the two clouds. This occurrence 
 suggests that the method is solid and that the depth effects in the SMC are not a significant issue.

\subsection{The distance of the GGCs based on absolute calibration with $Gaia$ parallaxes}

\begin{figure*}[h]
    \vbox{
    \hbox{
    \includegraphics[width=0.35\textwidth]{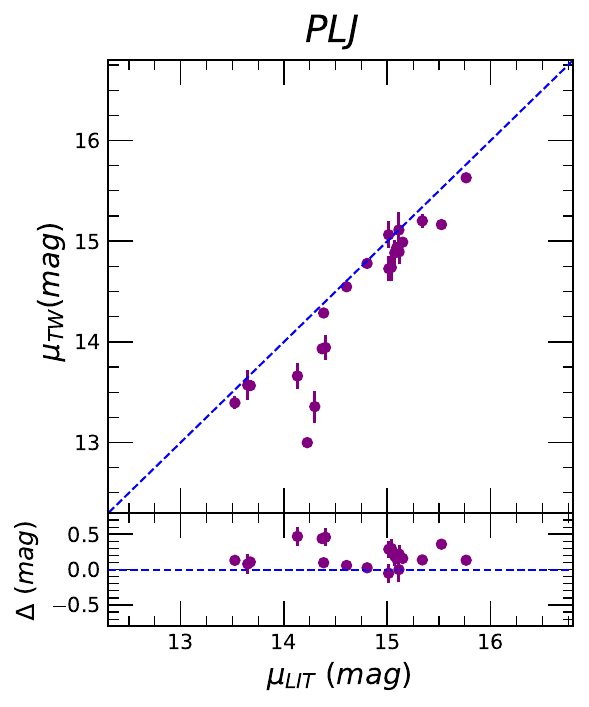}
    \includegraphics[width=0.35\textwidth]{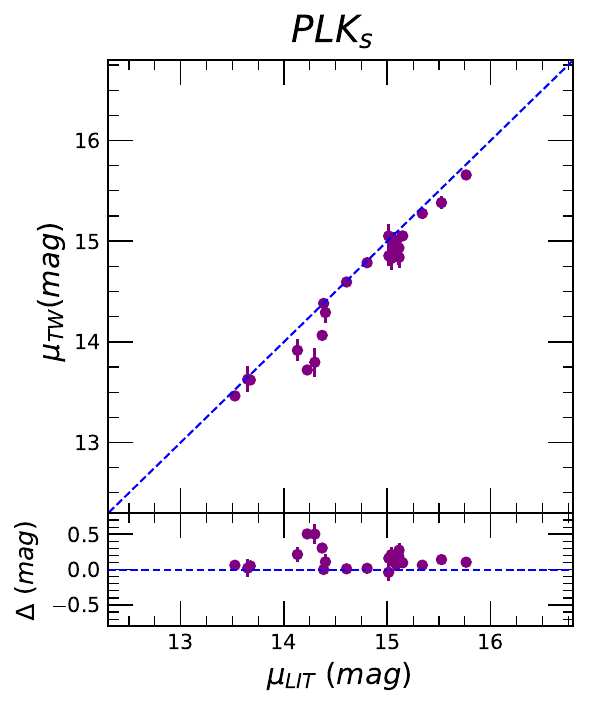}
    }
    \hbox{
    \includegraphics[width=0.35\textwidth]{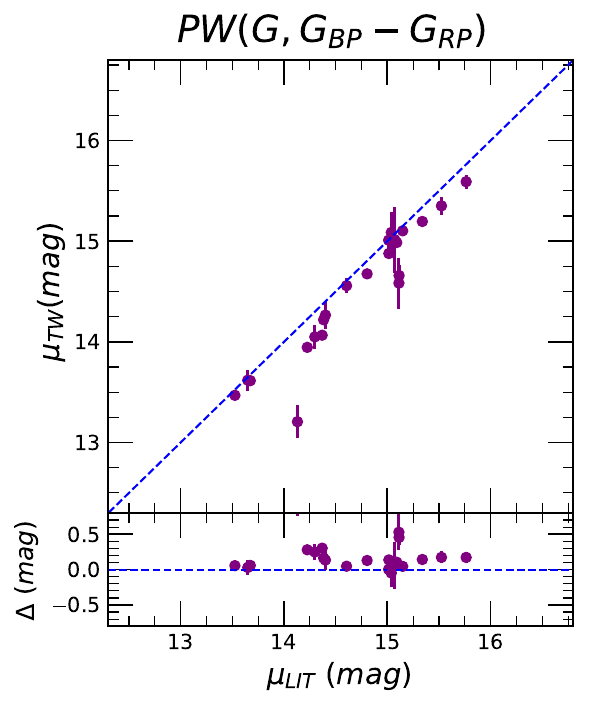}
    \includegraphics[width=0.35\textwidth]{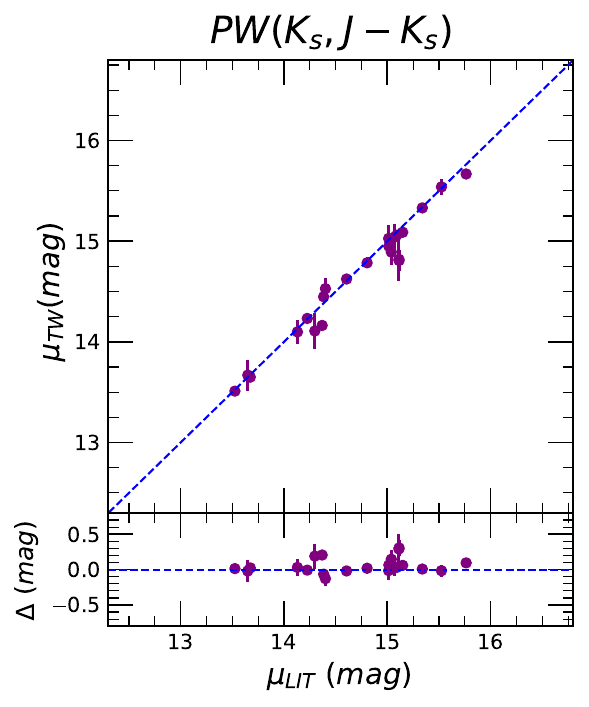}
    }
    \hbox{
    \includegraphics[width=0.35\textwidth]{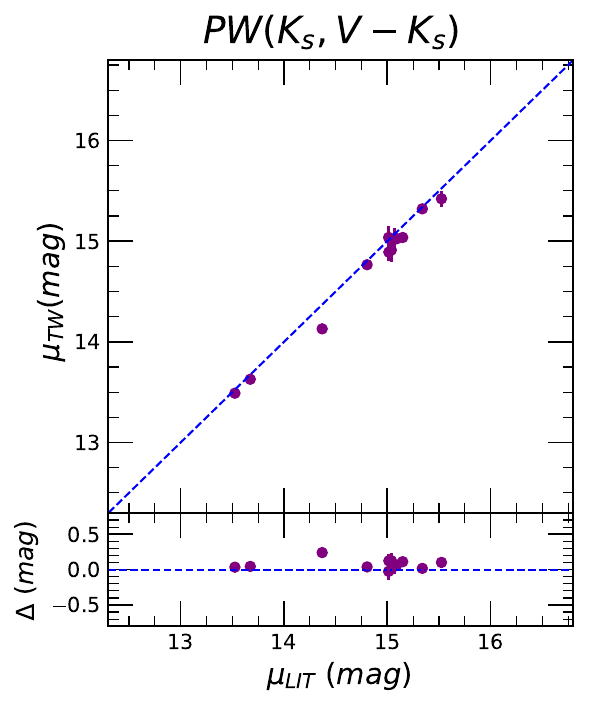}
    }}
    \caption{\label{muplx} Comparison between our distances, based on photometric parallaxes, and those of \citet{baumgardt2021accurate}. }
\end{figure*}

\begin{figure*}[h]
    \vbox{
    \hbox{
    \includegraphics[width=0.35\textwidth]{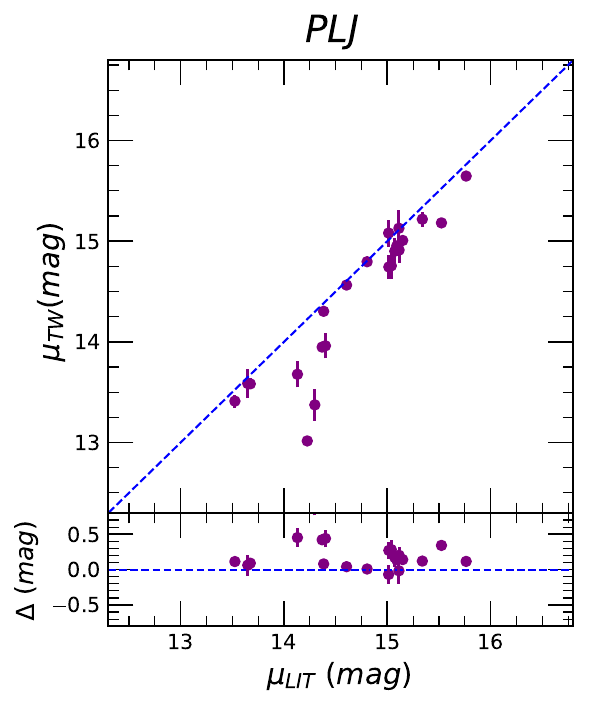}
    \includegraphics[width=0.35\textwidth]{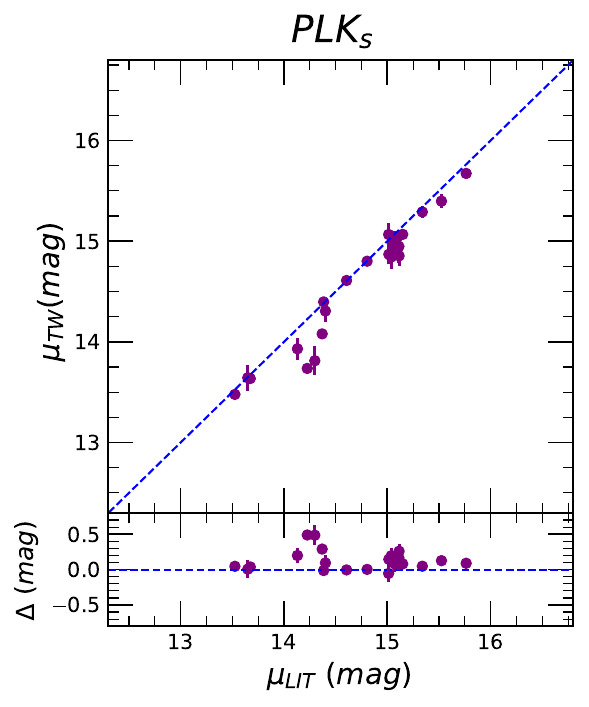}
    }
    \hbox{
    \includegraphics[width=0.35\textwidth]{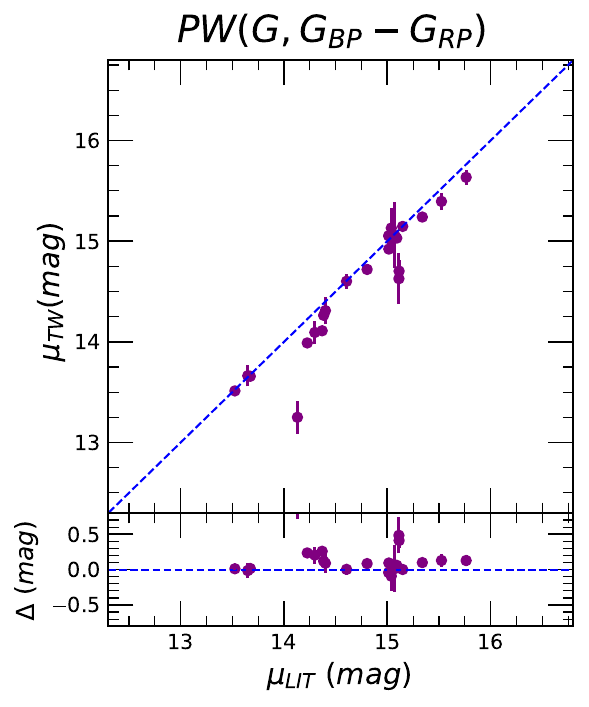}
    \includegraphics[width=0.35\textwidth]{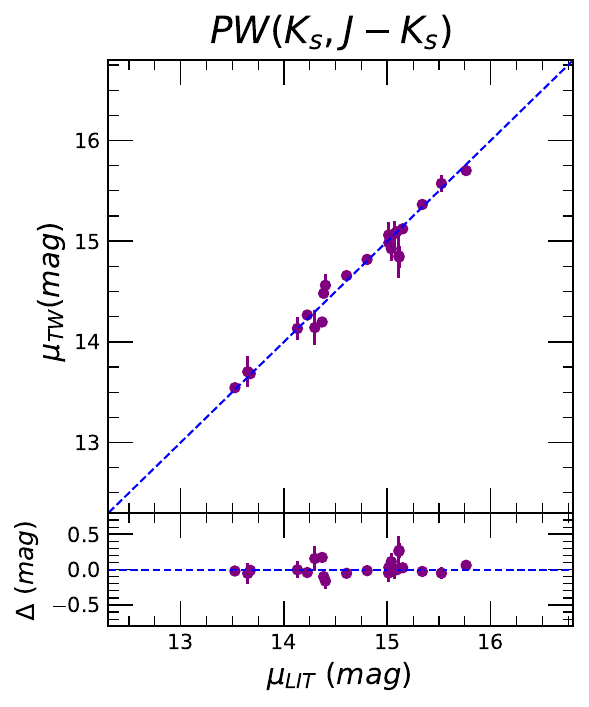}
    }
    \hbox{
    \includegraphics[width=0.35\textwidth]{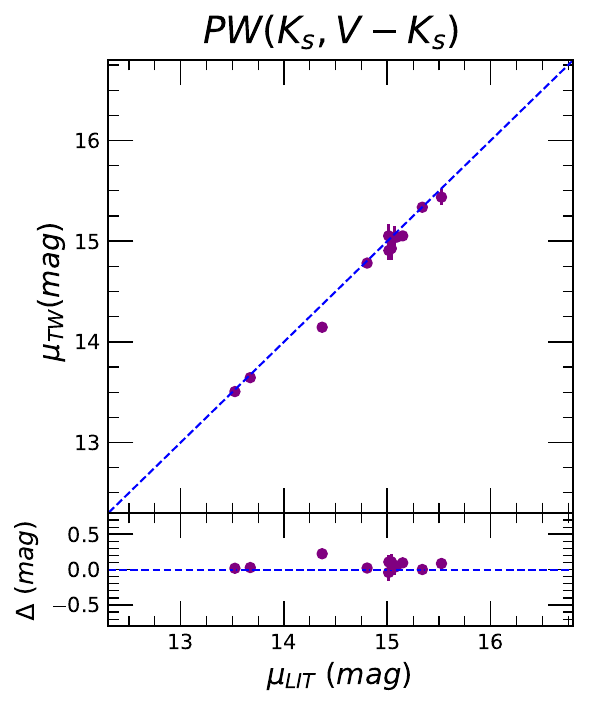}
    }}
    \caption{\label{muabl} Comparison between our distances, based on ABL, and those of \citet{baumgardt2021accurate}. }
\end{figure*}

The relations of Table~\ref{tab:0pointsphotpar} can also be used to derive a new estimate of the distances of the GGCs, working with the same data set as in Sect.~\ref{calibrlmc}. Using the same formalism as in the first column of Table~\ref{deltadis}, we obtain the $\mu$ difference shown in the remaining columns of Table~\ref{deltadis} and in Figs.~\ref{muplx} and ~\ref{muabl}. We find again that the distances by \citet{baumgardt2021accurate} are longer by about 0.03--0.06 mag than ours, even if by a smaller amount, at least when using the $PL$ relations.

\begin{table*}[h]
  \footnotesize\setlength{\tabcolsep}{3pt} 
  \caption{Coefficients of the $PL$ and $PW$ relations for T2Cs with slope calculated in the LMC and zero point calibrated with Galactic T2Cs through Photometric Parallaxes (top) and ABL (bottom). }
  \label{tab:0pointsphotpar}
 \begin{center}
  \begin{tabular}{lllllllllllll}
  \hline  
  \noalign{\smallskip}   
      Relation & $\alpha $ &  $\sigma_{\alpha}$ & $\beta_{LMC}$ & $\sigma_{\beta_{LMC}} $ & n$_{T2Cs}$  & $\mu_{\rm LMC} $ &  $\sigma_{\mu_{\rm LMC}}$ & $\mu_{\rm SMC} $ &  $\sigma_{\mu_{\rm SMC}}$\\
    &  mag & mag &  mag/dex & mag/dex & & mag & mag & mag & mag\\                                        
  \noalign{\smallskip}
  \hline  
  \noalign{\smallskip} 
\multicolumn{9}{c}{Photometric Parallax} \\
  \hline  
  \noalign{\smallskip} 
  PLJ& $-$0.859 &  0.059 & $-$2.156 & 0.024 & 21          & 18.52  & 0.06  & 19.06 & 0.07 \\
  PL$K_s$ & $-$1.124 &  0.037 & $-$2.348 & 0.019 & 21     & 18.53  & 0.04 & 19.05 & 0.04 \\
  PWG& $-$1.041 &  0.019 & $-$2.436 & 0.022 & 1121 & 18.49  & 0.02 & --- & --- \\
  PW$VK_s$ & $-$1.244 &  0.034 & $-$2.475 & 0.017 &21     & 18.53  & 0.04$^a$ & 19.01 & 0.04  \\
  PW$JK_s$ & $-$1.291 &  0.032 & $-$2.501 & 0.016 & 21    & 18.54  & 0.03 & 19.00 & 0.04\\
  \noalign{\smallskip}
  \hline  
  \noalign{\smallskip}
\multicolumn{9}{c}{ABL} \\
  \hline  
  \noalign{\smallskip} 
  PLJ& $-$0.876 &  0.063 & $-$2.156 & 0.024 & 21           & 18.54 &  0.06 & 19.08 & 0.08  \\
  PL$K_s$ & $-$1.139 &  0.036 & $-$2.348 & 0.019 & 21      & 18.55 &  0.04 & 19.07 & 0.04  \\
  PWG & $-$1.085 &  0.017 & $-$2.436 & 0.022 & 1118  &  18.53 &  0.04 & --- & ---\\
  PW$VK_s$ & $-$1.260 &  0.029 & $-$2.475 & 0.017 &21      &  18.54 &  0.03 & 19.02 & 0.04\\
  PW$JK_s$ & $-$1.325 &  0.026 & $-$2.501 & 0.016 & 21     &  18.58 &  0.03 & 19.04 & 0.06\\  
  \noalign{\smallskip}
  \hline  
  \noalign{\smallskip}
  \end{tabular}
  \tablefoot{ Note that the relations are: $M_{\lambda,0}=\alpha +\beta \log P$ for PL($\lambda$); $W_{\lambda_1,\lambda_2}^0=\alpha  +\beta \log P$ for the PW($\lambda_1,\lambda_2 $). \\ a=Same result as in \citet{Wielgorski2022}. \\
  Note that the $\mu_{\rm SMC}$ based on the $PWG$ relation is not listed because in this case, the $\beta_{\rm SMC}$ (see Table \ref{tab:smcT2}) is not compatible with the $\beta_{\rm LMC}$. }
\end{center}
\end{table*}

In analogy with the LMC distance estimation, we can also compare the GGCs distance moduli obtained both from the ABL and the Photometric parallaxes procedures with the literature ones \citep[i.e.][]{baumgardt2021accurate}. As expected, we find smaller differences with the GGCs distances calculated with the ABL than the Photometric Parallax case. In particular, the NIR Wesenheit magnitudes provide agreement within $1~\sigma$.  

\begin{table}[h]
  \footnotesize\setlength{\tabcolsep}{3pt} 
  \caption{As in Table~\ref{deltadis}, but for the comparison with the GGCs distances by \citet{bhardwaj2023}.}
  \label{tab:compbhar}
  \begin{center}
  \begin{tabular}{lcclclcl}
  \hline  
  \noalign{\smallskip}   
\multicolumn{2}{c}{}  & \multicolumn{2}{c}{Cal. LMC}& \multicolumn{2}{c}{Phot. Par.} & \multicolumn{2}{c}{ABL} \\
\cmidrule(lr){3-4}\cmidrule(lr){5-6} \cmidrule(lr){7-8}
       & n$_{GGCs}$ & Median $\Delta$ & $\sigma_{med}$ & Median $\Delta$ & $\sigma_{med}$ & Median $\Delta$ & $\sigma_{med}$ \\
 &   & (mag) & (mag) & (mag) & (mag) &  (mag) & (mag) \\
  \noalign{\smallskip}
  \hline  
  \noalign{\smallskip}  
  $\Delta_J$       & 7  & 0.196 & 0.080 &  0.149 & 0.080 &  0.133 & 0.080  \\
  $\Delta_{K_s}$   & 7  & 0.144 & 0.032 &  0.087 & 0.032 &  0.072 & 0.032  \\
  $\Delta_{WG}$ & 7  & 0.084 & 0.024 &  0.075 & 0.024 &  0.031 & 0.024 \\
  $\Delta_{WJK_s}$ & 7  & 0.085 & 0.034 & 0.020 & 0.034&  $-$0.014 & 0.034  \\
  $\Delta_{WVK_s}$ & 7  & 0.107 & 0.030 &  0.058 & 0.030 &  0.042 & 0.030 \\
\noalign{\smallskip}
  \hline  
  \noalign{\smallskip}
  \end{tabular}
  \end{center}
\end{table}

\subsection{Comparison between GGCs distances based on T2Cs and RRL stars}

Several GGCs hosting T2Cs also host RRL variables (see also the Sect. \ref{intro}). Given that RRL stars are noticeably good distance indicators, especially in the NIR regime, they can be used to obtain GGCs distances which are independent from those by \citet{baumgardt2021accurate}. We can therefore compare the distances of the GGCs hosting both RRL and T2Cs and verify the compatibility of the distance scales associated with these two old population distance indicators.  

To this purpose, we considered the work by \citet{bhardwaj2023}, who calculated the distances for a significant sample of GGCs using NIR $PL/PW$ relations for RRL stars based on homogeneous photometry. In Table~\ref{tab:compbhar} we compare in the usual way the $\mu$ values obtained in different ways in this work (see the previous sections) for 7 GGCs hosting T2Cs and the corresponding values calculated with the RRL variables. 

The resulting $\Delta$ values are similar to those shown in Table~\ref{deltadis} when we compared our results with those of \citet{baumgardt2021accurate}. However, the size of the discrepancy is smaller by a $\sim$0.01-0.02 mag. This is not surprising, as \citet{bhardwaj2023} found that the RRL-based GGCs distances are smaller by about 0.015 mag than those provided by \citet{baumgardt2021accurate}. 
In any case, if we use the $PL/PW$ relations calibrated with the $Gaia$ parallaxes the T2C and RRL distance scales agree better than 1-2$\sigma$ if we consider the $PW(J,K_s)$ and $PW(V,K_s)$ relations, while we have ~3$\sigma$ difference in the other cases. 
The situation is worse if we calibrate our relationships with the geometric distance of the LMC. Indeed, in this case, the discrepancy is generally larger than $3\sigma$ of their quoted uncertainties.

\section{Discussion and conclusions}

In this paper, we exploited $Y$, $J$, and $K_s$ time-series photometry for a sample of more than 330 T2Cs located in the LMC and SMC in the context of the ESO Public Survey "VISTA Magellanic Cloud (VMC) Survey". 

The VMC data was complemented with optical data from the OGLE IV survey and the $Gaia$ mission. From these surveys, we obtained the identification and position of the pulsators, the periods and the photometry in the $V$, $I$, $G$, $G_{BP}$ $G_{RP}$ bands. 

After selecting the best VMC time series, we built a set of light curve templates for each T2C sub-type. These templates have been used to derive accurate intensity-averaged magnitudes (and amplitudes) for all the pulsators at our disposal through a sophisticated pipeline already developed in the context of previous VMC works dealing with CCs. This pipeline was further improved in the course of the present work, by introducing a parameter that takes into account the sampling of the observed light curve, enabling a special treatment of the poorly sampled curves which allows us to keep stars that would otherwise have been rejected. 

The final $Y$, $J$, and $K_s$ photometry obtained in this way complemented with the above-mentioned optical data, was used to build a large set of $PL$, $PW$ and $PLC$ relations for a variety of T2C sub-types (BLHer, WVir, RVTau and combinations). To this aim, we employed a {\tt Python}-based robust algorithm, LTS, that embodies a sound outlier rejection procedure. 

The multi-band $PL$ relations produced during this work were used to investigate systematically for the first time the dependence of T2C relations, namely the slopes and the dispersion, on the adopted wavelength. Similarly to what occurs for CCs, our results show that even for these older pulsators the slope increases and the dispersion decreases as we move towards longer wavelengths. 

Overall, there is good agreement with the literature, as for almost all the relations we find coefficients of the PL/PW/PLC relations in agreement within better than 2$\sigma$ for both the LMC and SMC. At the same time, in almost all cases, our data improves both precision and accuracy compared to previous studies, owing to larger samples, better light curve sampling, and deeper photometry as guaranteed by the VMC survey (see Table~\ref{tab:confrontolmct2}).

The tightest relationships derived in this work, i.e. the $PL$ in J and $K_s$ bands, and all the $PW$ relations in the LMC, all showing low dispersions, were calibrated by anchoring their absolute zero points (intercepts) to the geometric distance of the LMC, as accurately determined from a set of ECBs.

These absolute relationships were used to derive the distances of a sample of 22 GGCs hosting T2Cs, finding that the most up-to-date literature distances \citep[e.g.,][]{baumgardt2021accurate} are larger by some 0.10 mag with a 3$\sigma$ significance. Therefore, the geometric distance of the LMC by \citet{pietrzynski2019distance} appears to be in disagreement with the GGCs distance scale.  

Stimulated by this unexpected result, we calibrated the $PL/PW$ relations independently from the LMC distance, by using a sample of Galactic field T2Cs for which accurate data in optical and NIR bands are available in the literature. 
To this aim, we adopted the slopes of the LMC (which are reasonably compatible with those of the MW and SMC) and calculated the zero points of the $PL/PW$ relations using the $Gaia$ EDR3 parallaxes.

Using these relations we recalculated the distance of the LMC and SMC analysed before. As a result, we obtained a $\mu_{LMC}$ which is $\sim$0.06 mag longer on average than the geometric estimate provided by \citet{pietrzynski2019distance}, albeit in agreement within $1\sigma$. This result is in agreement with previous investigations \citep{Wielgorski2022}. A similar result has been obtained for the SMC in comparison with the geometric distance of this galaxy by \citet{graczyk2020}.
Concerning the GGCs, the $\mu$ obtained with the new calibration is still substantially shorter than those reported by \citet{baumgardt2021accurate} by amounts of the order of 0.03-0.06 mag for the most precise $PW$ relations with a typical significance of 1-2$\sigma$.
A better agreement by 0.01-0.02 mag is obtained when we compare our results with the distances of the GGCs obtained from RRL variables only \citep{bhardwaj2023}.
This occurrence is particularly important, as one of the most important developments of this work is to use both RRL and T2C to calibrate the TRGB in an alternative route to calibrate $H_0$. 

It is not easy to interpret these conflicting results.
One possibility is to invoke a significant effect of the metallicity, which we neglected in this paper, based on both theoretical and empirical results \citep[e.g.,][and references therein]{ripepi2015vmc,ngeow2022}, which do not report a significant metallicity dependence of T2C $PL/PW$ relations. However, even a significant metallicity effect seems not sufficient to reconcile simultaneously the distances of the LMC (and SMC) and GGCs, given the large range of abundances spanned by the latter.   
A more likely possible explanation is that the spatial distributions of the Late-type ECBs adopted by \citet{pietrzynski2019distance} and \citet{graczyk2020} to derive the geometric distances of the LMC and the SMC differ significantly from those spanned by the T2Cs in our samples. This occurrence could account for at least part of the observed discrepancy.   

The advent of large spectroscopic surveys such as those foreseen with the WEAVE (WHT Enhanced Area Velocity Explorer)\footnote{https://www.ing.iac.es/astronomy/instruments/weave/weaveinst.html} and 4MOST (4-metre Multi-Object Spectroscopic Telescope)\footnote{https://www.eso.org/sci/facilities/develop/instruments/4MOST.html}, as well as the availability of more accurate parallaxes thanks to the future $Gaia$ DR4 will certainly provide us with the information needed to draw firmer conclusions on the distance scale of T2C and help us understand the discrepant results we obtained in this work concerning the distances of the LMC and the GGCs system.

\begin{acknowledgements}
{\bf We warmly thank our anonymous referee for their competent and helpful comments.}

This research has made use of the SIMBAD database, operated at CDS, Strasbourg, France.\\

This work has made use of data from the European Space Agency (ESA) mission
{\it Gaia} (\url{https://www.cosmos.esa.int/gaia}), processed by the {\it Gaia}
Data Processing and Analysis Consortium (DPAC,
\url{https://www.cosmos.esa.int/web/gaia/dpac/consortium}). Funding for the DPAC
has been provided by national institutions, in particular, the institutions
participating in the {\it Gaia} Multilateral Agreement.\\      
This research was supported in part by the Australian Research Council Centre of Excellence for All Sky Astrophysics in 3 Dimensions (ASTRO 3D), through project number CE170100013.\\
We thank Jacco Th. van Loon for helpful discussion on the manuscript.
\end{acknowledgements}

%
%

   \bibliographystyle{aa} 
   \bibliography{myBib} 

\begin{appendix}

\section{Templates fitting} 
\label{fittemplates}

The tables in this appendix list the coefficients of the templates adopted to fit the observed light curves in the three $Y,\,J,\,K_s$ bands.

\begin{table*}
\footnotesize\setlength{\tabcolsep}{3pt} 
\caption{Fourier parameters of the light curves templates in $Y$ band. The entire table is available in the online edition.}
  \label{templatesy}
  \begin{tabular}{rlrrrrrrrrr}
 \hline  
 \noalign{\smallskip} 
       &  Type & $A_1$ &  $\Phi_1$ & $A_2$ & $\Phi_2$ & ... & $A_9$ & $\Phi_9$ & $A_{10}$ & $\Phi_{10}$  \\
    &  & mag & rad &mag  & rad  & & mag & rad & mag  & rad    \\    
 \noalign{\smallskip}
 \hline  
 \noalign{\smallskip}
 1& BLHer & 0.4537 & 2.835 & 0.110 & 3.806 & ... & 2.340E$-$4 & 5.323 & 2.167E$-$4 & 2.875\\
 2& BLHer & 0.464 & 2.875 & 0.102 & 3.734& ... & 1.282E$-$4 & 4.284 & 1.917E$-$4 & 1.675\\
 3& BLHer & 0.443 & 3.022 & 0.119 & 3.934& ... & 4.809E$-$4 & 0.004 & 4.697E$-$4 & 4.865\\
 4& WVir & 0.491 & 2.782 & 0.091 & 5.487 & ...& 24.277E$-$4 & 4.850 & 11.325E$-$4 & 3.128\\
 5& WVir & 0.434 & 2.251 & 0.151 & 4.192 & ...& 5.6522E$-$4 & 3.944 & 7.230E$-$4 & 1.075\\
6& WVir & 0.493 & 3.504 & 0.063 & 1.612 & ...& 5.173E$-$4 & 4.741 & 3.377E$-$4 & 0.874\\
7&  WVir & 0.487 & 2.979 & 0.015 & 4.989& ... & 6.949E$-$4 & 4.216 & 9.508E$-$4 & 1.287\\
 8& WVir & 0.499 & 3.157 & 0.057 & 0.249 & ...& 1.193E$-$4 & 5.612 & 5.249E$-$5 & 2.606\\
 9& WVir & 0.498 & 3.080 & 0.051 & 5.627 & ...& 9.093E$-$4 & 4.335 & 7.752E$-$4 & 0.592\\
10&  RVTau & 0.429 & 2.817 & 0.113 & 4.181 & ...& 3.457E$-$4 & 0.170 & 9.107E$-$4 & 3.076\\
\noalign{\smallskip}
\hline  
\noalign{\smallskip}
\end{tabular}
\end{table*}

\begin{table*}
\footnotesize\setlength{\tabcolsep}{3pt} 
\caption{Same as Table \ref{templatefity} but in $J$ band.}
  \label{templatesj}
  \begin{tabular}{rlrrrrrrrrr}
 \hline  
 \noalign{\smallskip} 
       &  Type & $A_1$ &  $\Phi_1$ & $A_2$ & $\Phi_2$ & ... & $A_9$ & $\Phi_9$ & $A_{10}$ & $\Phi_{10}$  \\
    &  & mag & rad &mag  & rad  & & mag & rad & mag  & rad    \\    
 \noalign{\smallskip}
 \hline  
 \noalign{\smallskip}
 1&BLHer & 0.498 & 3.071 & 0.037 & 5.728 & ...& 0.001 & 1.036 & 0.001 & 4.804\\
 2& BLHer & 0.427 & 2.246 & 0.160 & 3.386 & ...& 0.005 & 4.904 & 0.004 & 3.922\\
 3& BLHer & 0.461 & 2.953 & 0.085 & 5.192 & ...& 0.001 & 0.618 & 6.440E$-$4 & 4.653\\
 4& BLHer & 0.449 & 2.670 & 0.135 & 4.292 & ...& 4.344E$-$4 & 4.351 & 4.623E$-$4 & 3.750\\
  5&WVir & 0.478 & 2.918 & 0.059 & 4.274 & ...& 3.782E$-$4 & 1.787 & 1.569E$-$4 & 2.116\\
 6& WVir & 0.520 & 3.289 & 0.027 & 2.074& ...& 3.542E$-$4 & 3.109 & 4.715E$-$4 & 6.199\\
7&  WVir & 0.496 & 3.130 & 0.015 & 1.473 & ...& 2.350E$-$4 & 0.878 & 9.319E$-$5 & 1.108\\
 8& WVir & 0.496 & 3.030 & 0.028 & 6.075 & ...& 5.186E$-$4 & 2.608 & 4.291E$-$4 & 5.851\\
 9& RVTau & 0.500 & 3.132 & 0.013 & 0.223 & ...& 2.664E$-$4 & 3.520 & 1.831E$-$4 & 0.520\\
10&  RVTau & 0.475 & 2.960 & 0.106 & 5.462 & ...& 8.9203E$-$4 & 4.818 & 0.001 & 2.615\\

\noalign{\smallskip}
\hline  
\noalign{\smallskip}
\end{tabular}
\end{table*}

\begin{table*}
\footnotesize\setlength{\tabcolsep}{3pt} 
\caption{Same as Table \ref{templatesy}but in $K_s$ band.}
  \label{templatesk}
  \begin{tabular}{rlrrrrrrrrr}
 \hline  
 \noalign{\smallskip} 
       &  Type & $A_1$ &  $\Phi_1$ & $A_2$ & $\Phi_2$ & ... & $A_9$ & $\Phi_9$ & $A_{10}$ & $\Phi_{10}$  \\
    &  & mag & rad &mag  & rad  & & mag & rad & mag  & rad    \\    
 \noalign{\smallskip}
 \hline  
 \noalign{\smallskip}
1& BLHer & 0.391 & 2.579 & 0.158 & 4.177 & ...& 0.002 & 3.226 & 0.001 & 0.997\\
 2& BLHer & 0.501 & 3.258 & 0.033 & 1.019 & ...& 6.734E$-$4 & 0.915 & 5.902E$-$4 & 5.162\\
3&  BLHer & 0.474& 3.053 & 0.060 & 5.761 & ...& 7.575E$-$4 & 0.327 & 6.965E$-$4 & 3.233\\
4&  BLHer & 0.481 & 2.943 & 0.089 & 5.577& ...& 5.507E$-$4 & 4.853 & 3.445E$-$4 & 5.685\\
 5& BLHer & 0.491 & 3.018 & 0.093 & 5.235 & ...& 0.003 & 1.503 & 0.002 & 5.518\\
6&  BLHer & 0.483 & 2.931 & 0.085 & 5.425& ...& 1.760E$-$4 & 3.488 & 1.188E$-$4 & 0.679\\
7&  WVir & 0.480 & 2.950 & 0.053 & 5.171 & ...& 0.001 & 5.316 & 3.661E$-$4 & 3.852\\
 8& WVir & 0.489 & 2.916 & 0.102 & 5.489 & ...& 0.001 & 0.366 & 0.001 & 4.099\\
 9& WVir & 0.419 & 3.114 & 0.113 & 6.061 & ...& 0.002 & 0.110 & 8.113E$-$4 & 2.422\\
 10& WVir & 0.474 & 3.296 & 0.094 & 0.292 & ...& 0.001 & 0.947 & 5.852E$-$4 & 4.205\\
11&  WVir & 0.463 & 3.221 & 0.075 & 0.556 & ...& 0.003 & 1.598 & 0.003 & 5.395\\
12&  WVir & 0.432 & 2.977 & 0.122 & 3.559 & ...& 0.004 & 1.270 & 0.001 & 2.657\\
13&  WVir & 0.501 & 3.223 & 0.016& 1.295 & ...& 1.757E$-$4 & 0.894 & 1.950E$-$4 & 4.639\\
14&  WVir & 0.502 & 2.802 & 0.117 & 5.742 & ...& 7.765E$-$4 & 2.827 & 7.225E$-$4 & 0.142\\
15&  WVir & 0.486 & 3.158 & 0.003 & 5.508 & ...& 6.150E$-$4 & 4.437 & 4.875E$-$4 & 1.251\\
16&  WVir & 0.485 & 3.220 & 0.019 & 5.880 & ...& 6.872E$-$4 & 0.343 & 6.695E$-$4 & 4.046\\
17&  WVir & 0.495 & 2.865 & 0.038 & 4.772 & ...& 5.463E$-$4 & 4.668 & 3.4157E$-$4 & 0.976\\
18&  WVir & 0.498 & 2.670 & 0.044 & 4.194 & ...& 0.002 & 1.361 & 0.002 & 4.425\\
19&  RVTau & 0.513 & 3.037 & 0.020 & 5.860 & ...& 2.698E$-$4 & 4.269& 1.658E$-$4 & 2.505\\
20&  RVTau & 0.456 & 2.994& 0.106 & 5.726 & ...& 2.472E$-$5 & 5.291 & 2.221E$-$4 & 2.610\\
\noalign{\smallskip}
\hline  
\noalign{\smallskip}
\end{tabular}
\end{table*}

\FloatBarrier


\section{Examples of fitted light curves fitted with templates} \label{app:fittedLCs}

\begin{figure*}
\begin{adjustwidth}{-0.75 cm}{}  
    \includegraphics[width=1.1\textwidth]{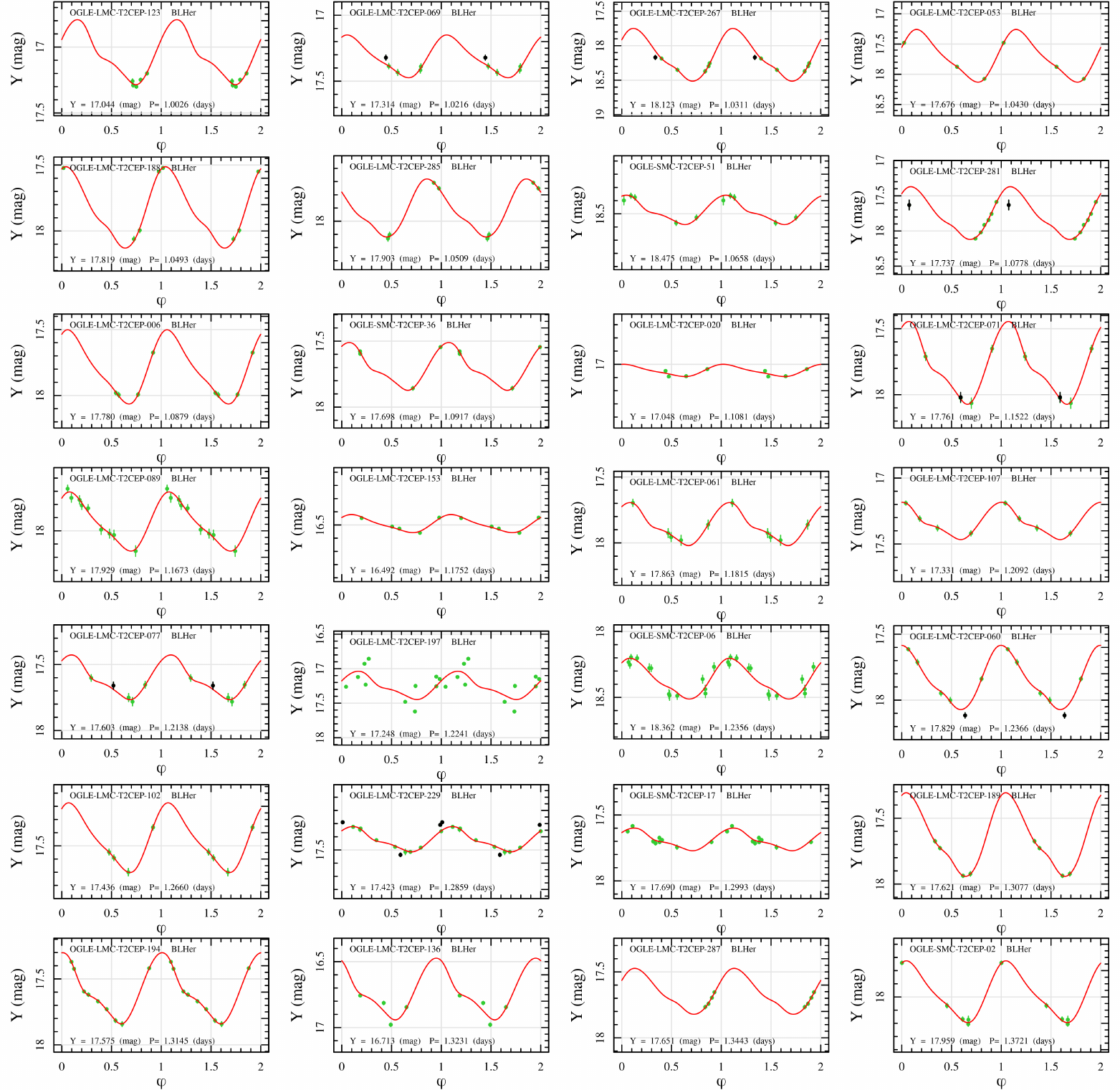}
    \caption{Examples of fitted templates for T2Cs light curves in $Y$.}
    \label{fig:fittemplt2y}
    \end{adjustwidth}
\end{figure*}
\begin{figure*}
    \begin{adjustwidth}{-0.75 cm}{}
    \includegraphics[width=1.1\textwidth]{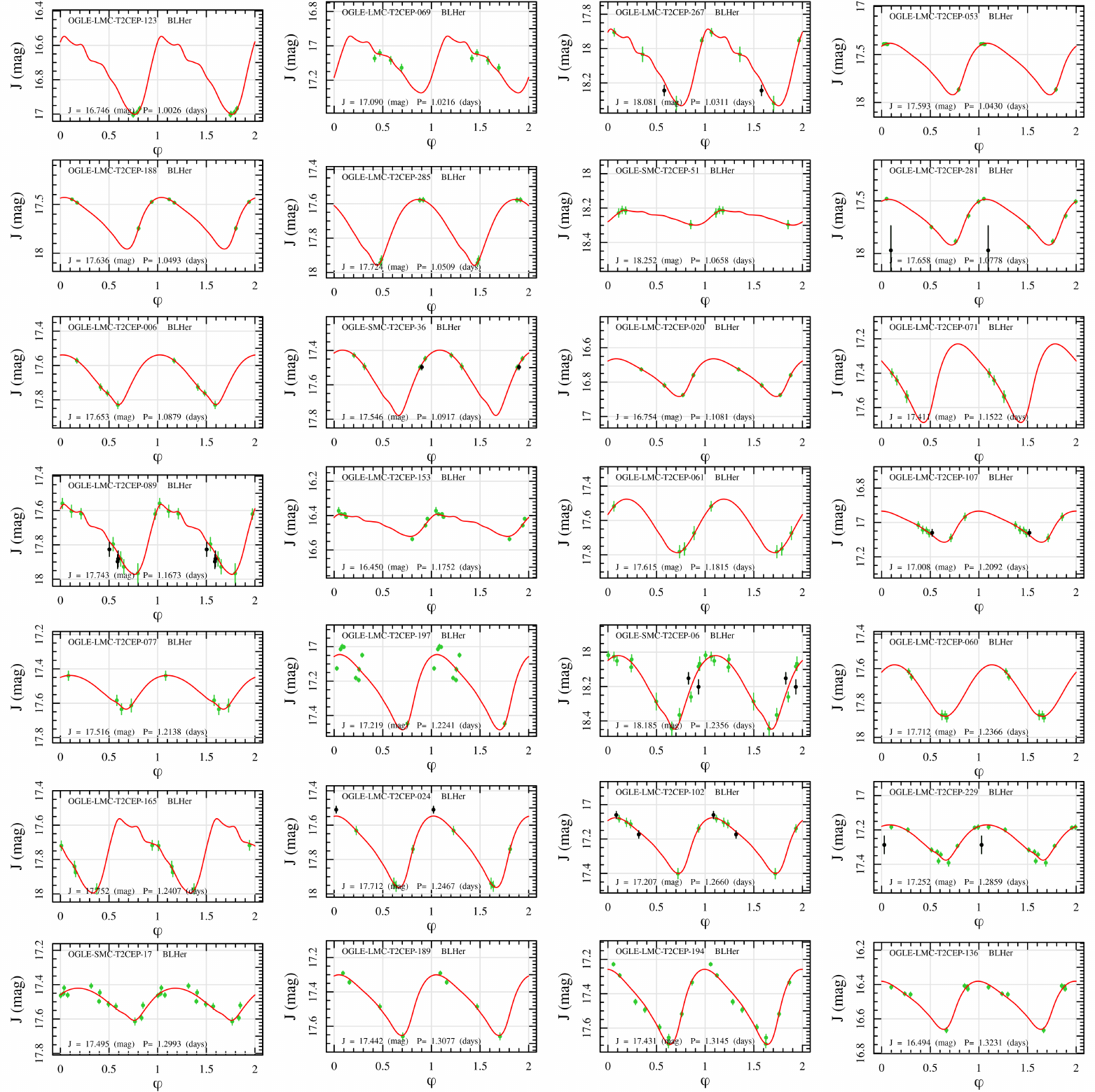}
    \caption{Same as Fig. \ref{fig:fittemplt2y} but in J band.}
    \label{fig:fittemplt2j}  
    \end{adjustwidth}
\end{figure*}
\begin{figure*}
    \begin{adjustwidth}{-0.75 cm}{}
    \includegraphics[width=1.1\textwidth]{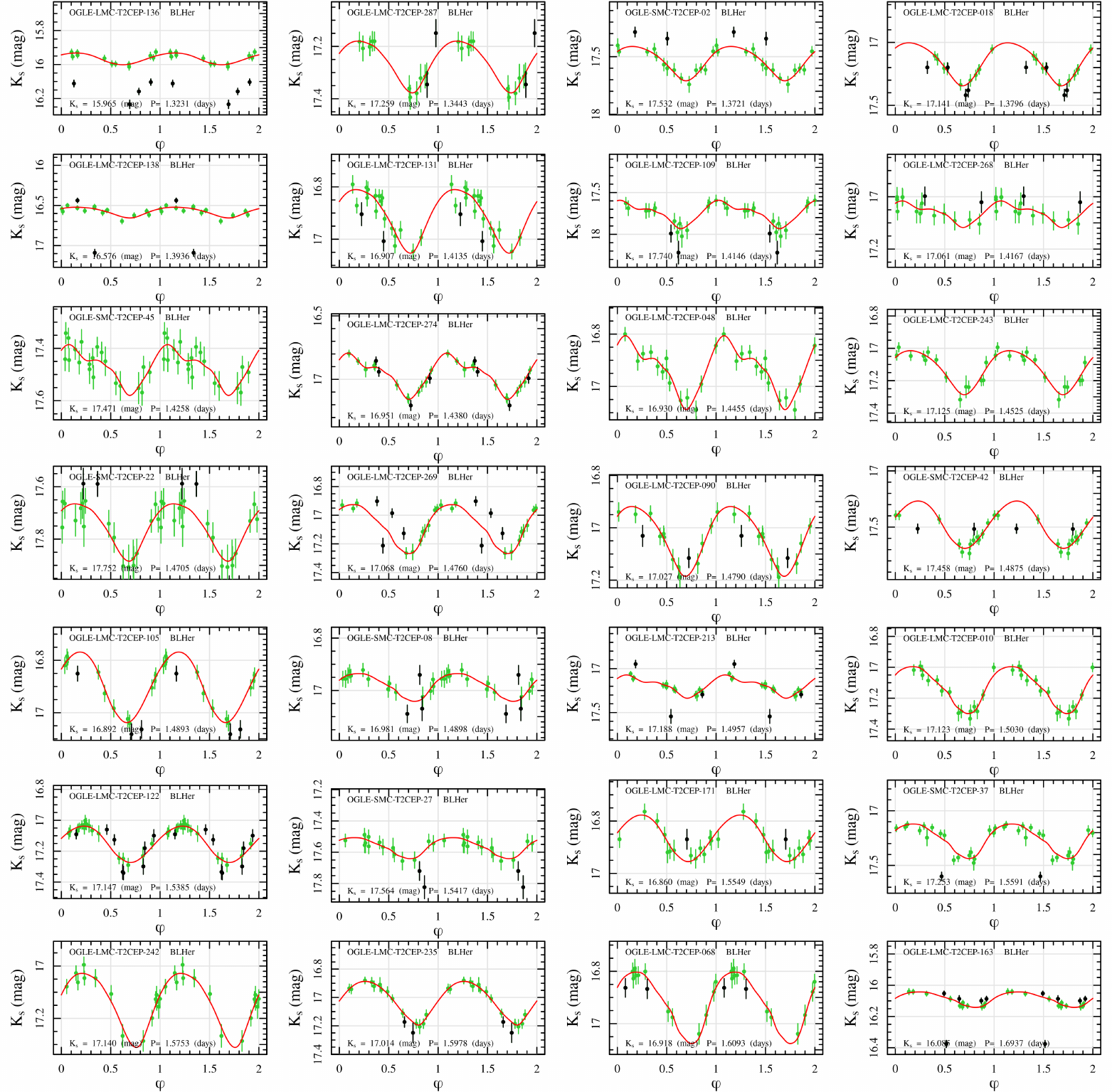}
    \caption{Same as Fig. \ref{fig:fittemplt2y} but in $K_s$ band.}
    \label{fig:fittemplt2k2}
    \end{adjustwidth}
\end{figure*}

\FloatBarrier


\section{Additional $PL/PW/PLC$ relations in the LMC and SMC} \label{app:additionalPL}

In this section, we show the $PL/PW/PLC$ fitting to the data for a variety of 
magnitudes, colours and Wesenheit magnitudes, as well as the parameters from the fitting.

\subsection{Figures} \label{app:figuresPL_fitting}

\begin{figure*}
    \vbox{
    \hbox{
    \includegraphics[width=0.34\textwidth]{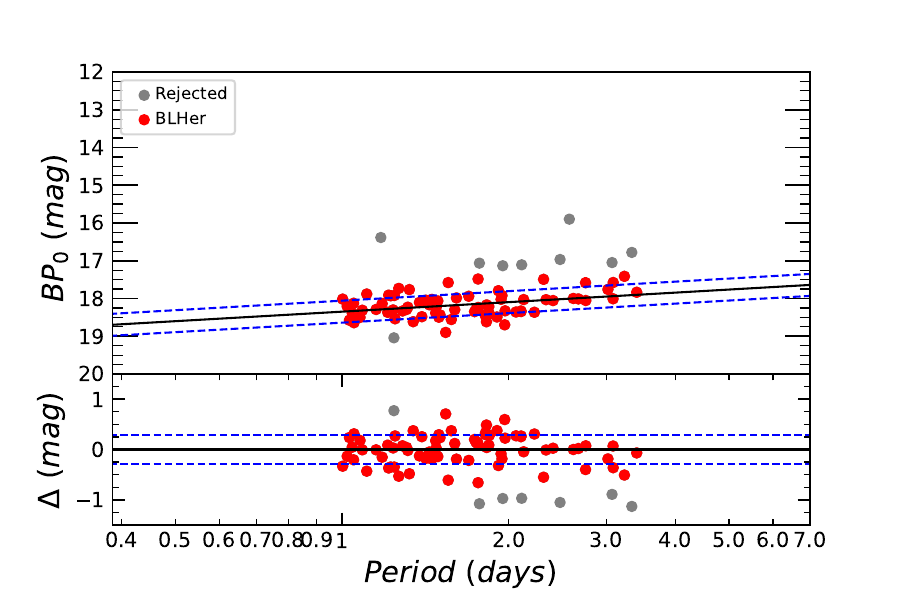}
    \includegraphics[width=0.34\textwidth]{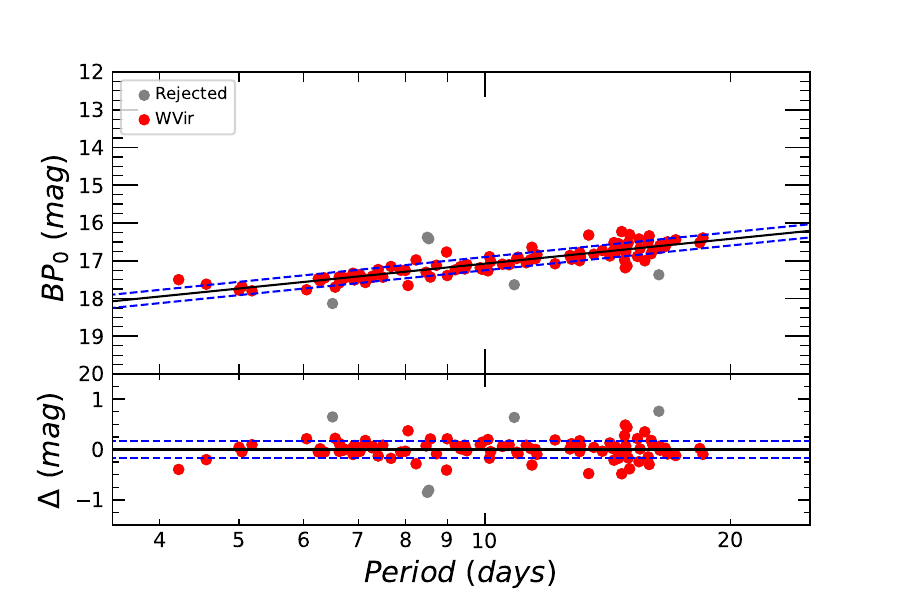}
    \includegraphics[width=0.34\textwidth]{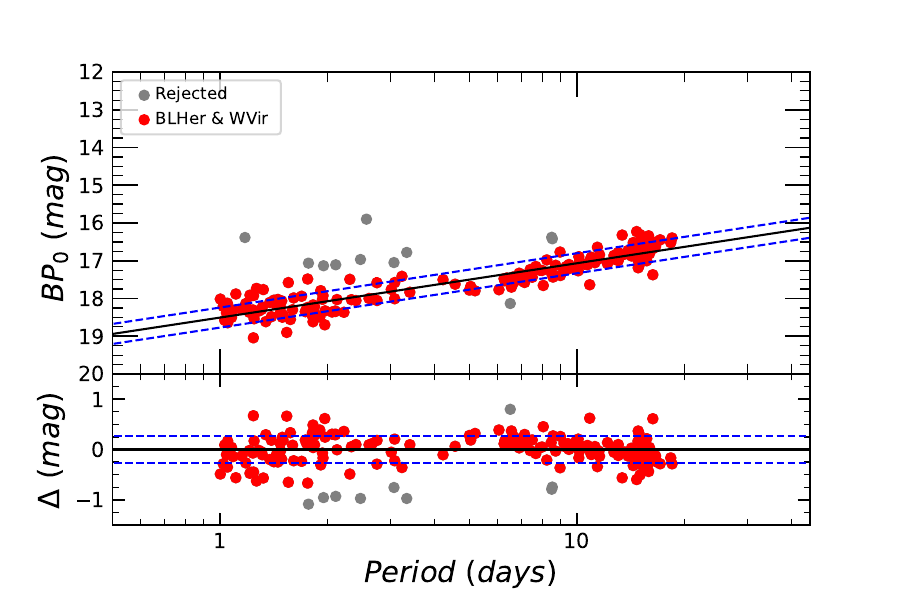}
    }
    \hbox{
    \includegraphics[width=0.34\textwidth]{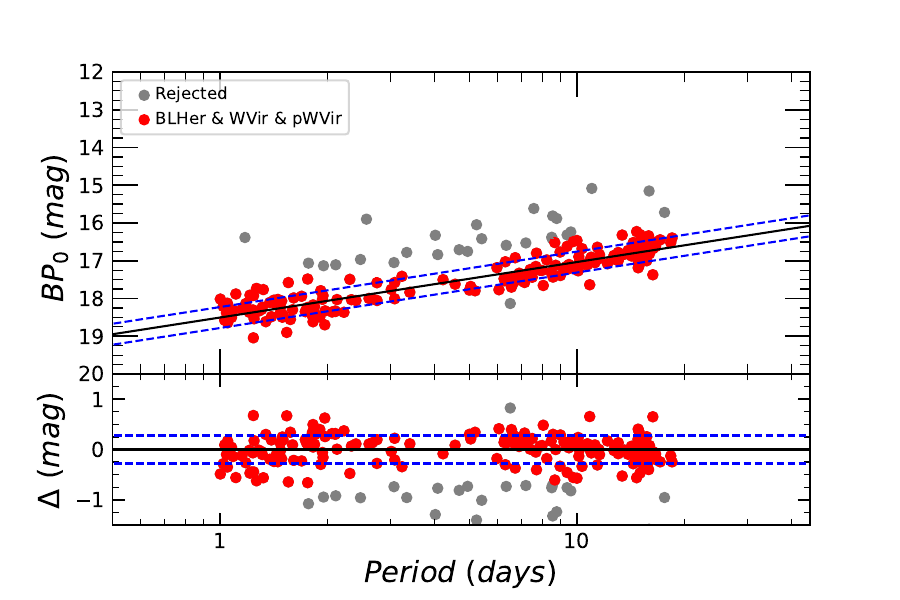}
    \includegraphics[width=0.34\textwidth]{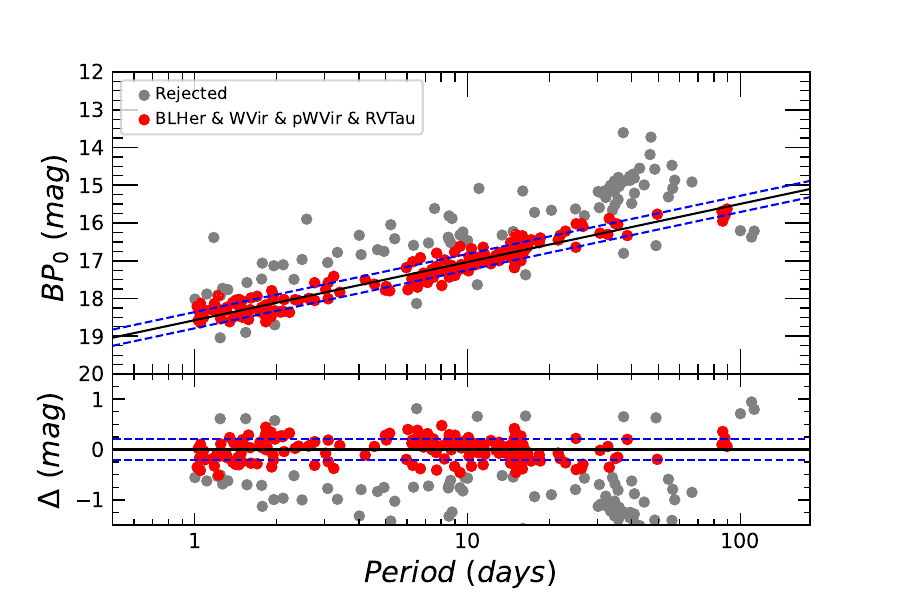}
    \includegraphics[width=0.34\textwidth]{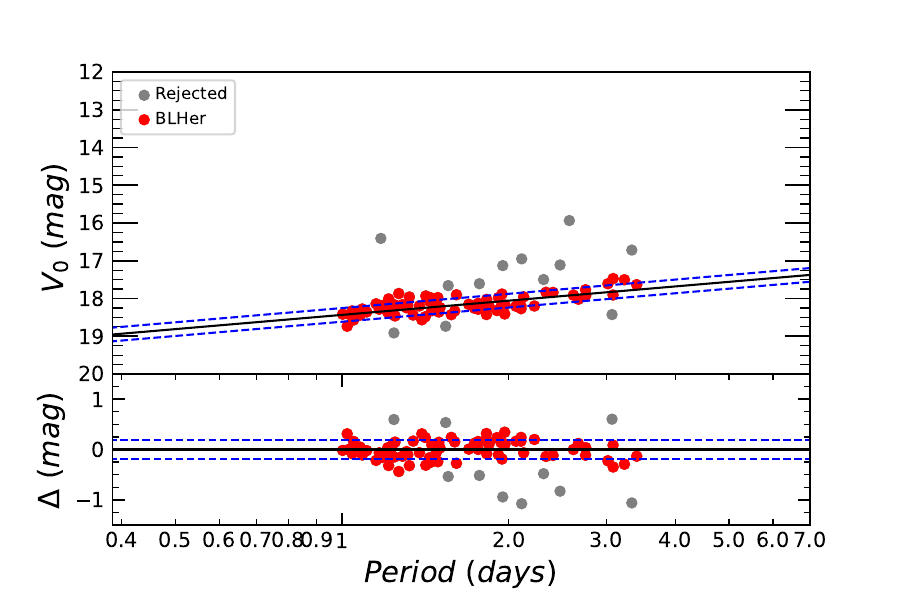}
    }
    \hbox{
    \includegraphics[width=0.34\textwidth]{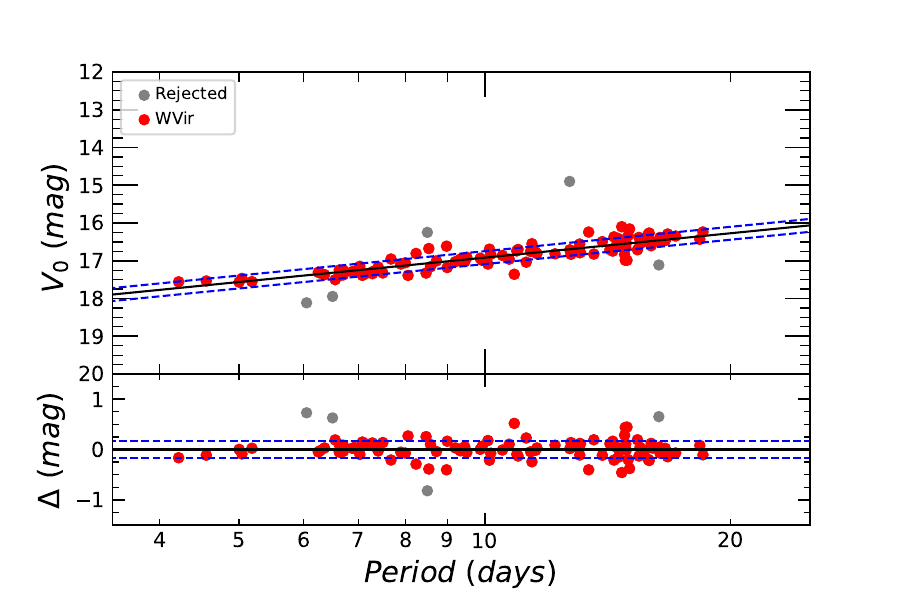}
    \includegraphics[width=0.34\textwidth]{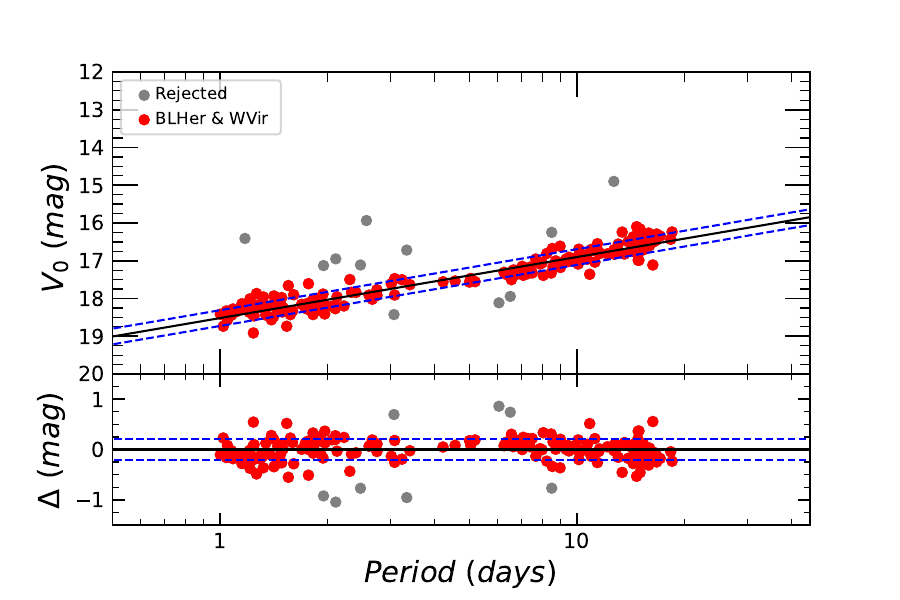}
    \includegraphics[width=0.34\textwidth]{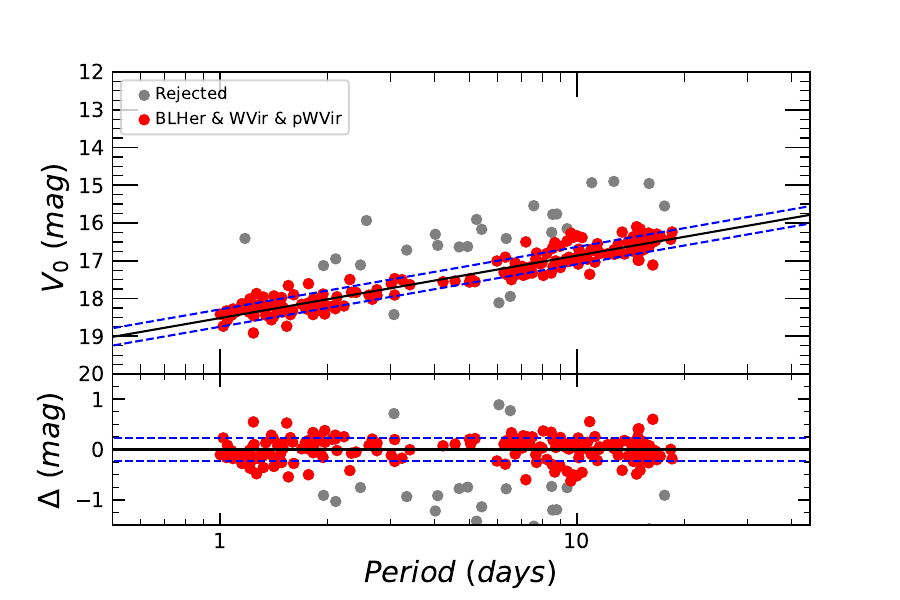}
    }
    \hbox{
    \includegraphics[width=0.34\textwidth]{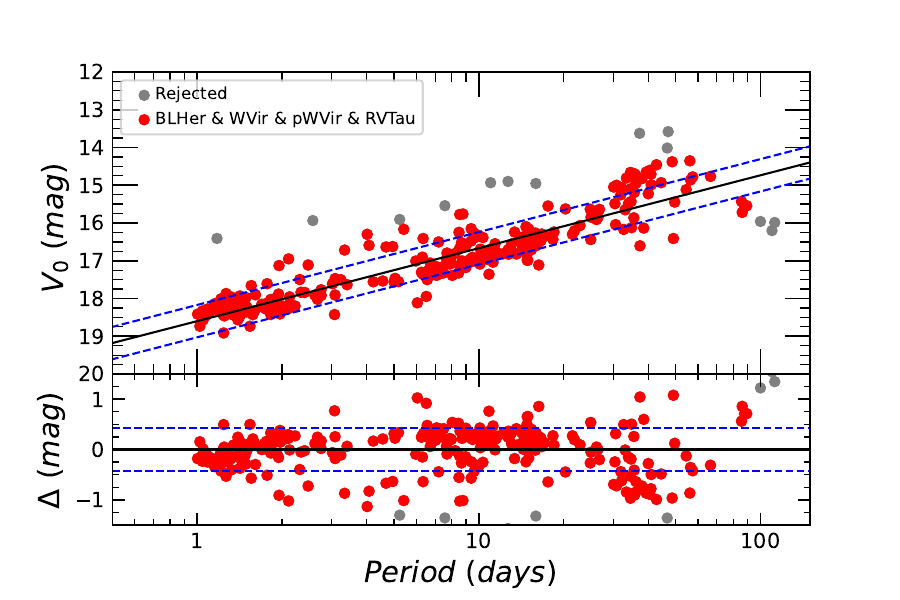}
    \includegraphics[width=0.34\textwidth]{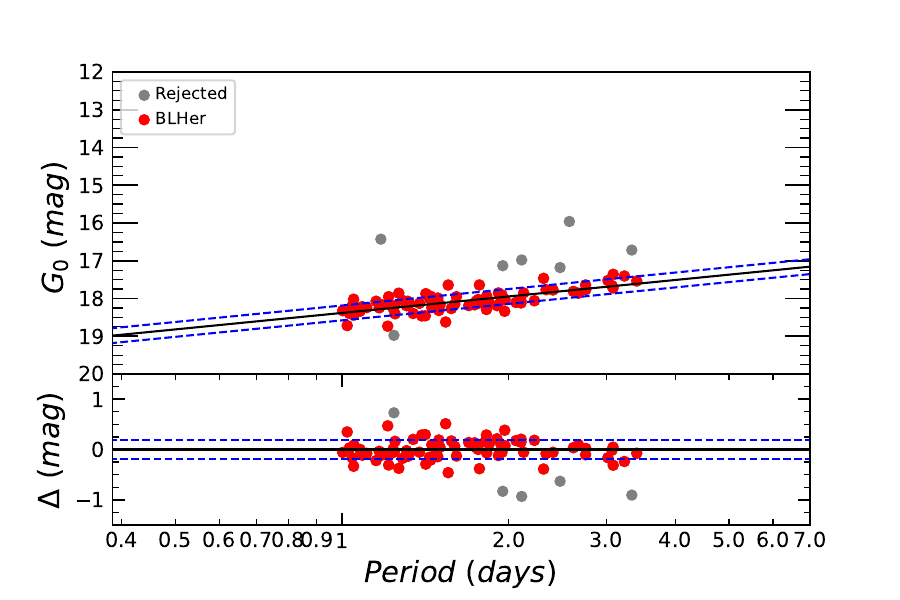}
    \includegraphics[width=0.34\textwidth]{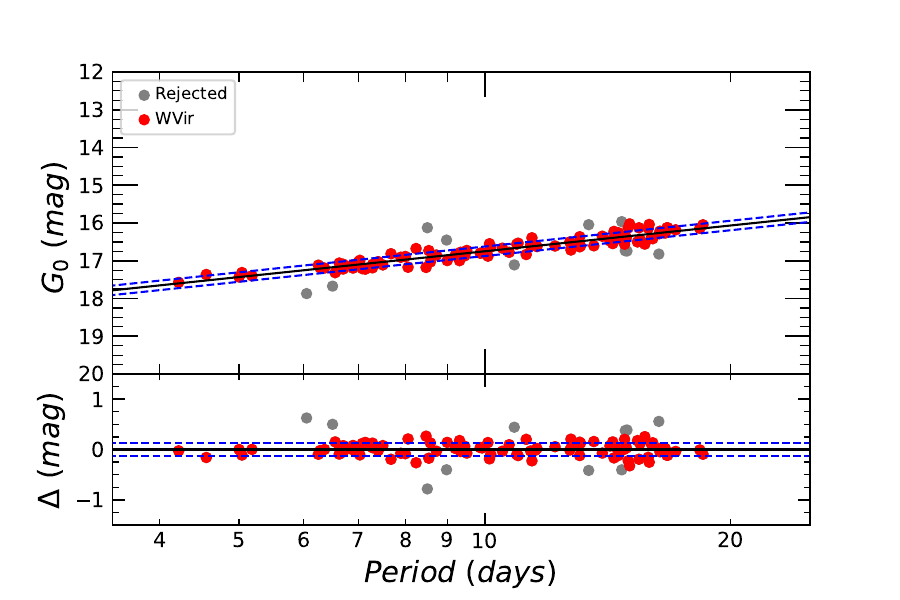}
    }
    \hbox
    {
    \includegraphics[width=0.34\textwidth]{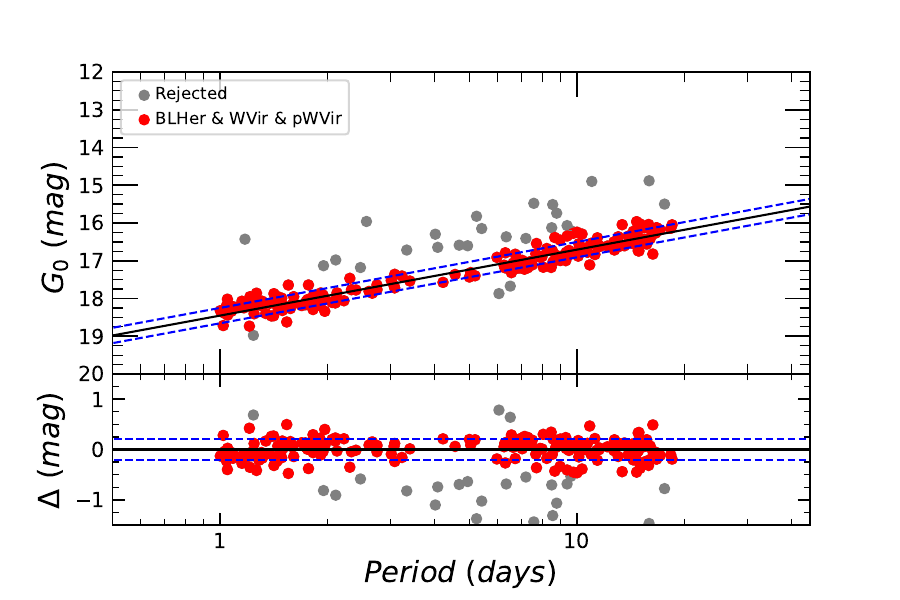}
    \includegraphics[width=0.34\textwidth]{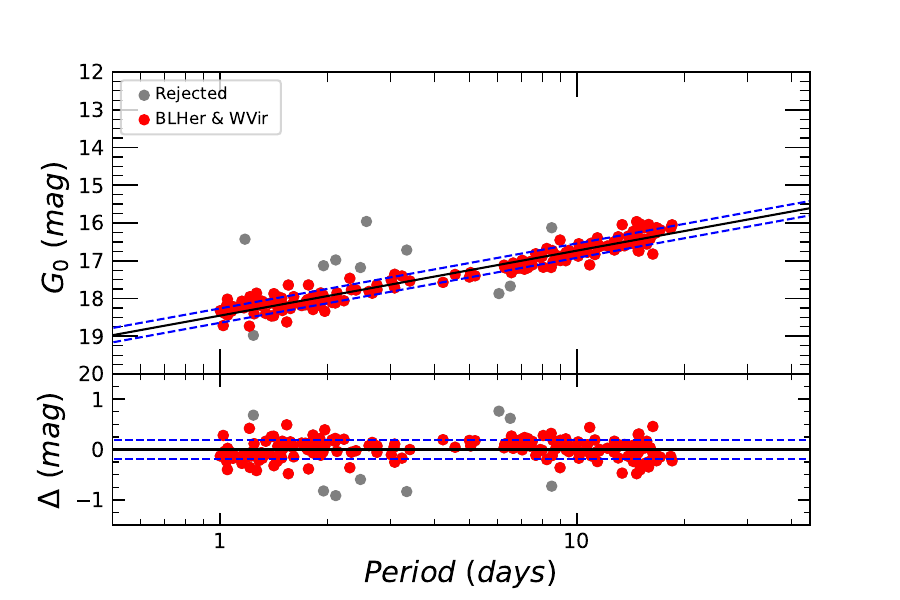}
    \includegraphics[width=0.34\textwidth]{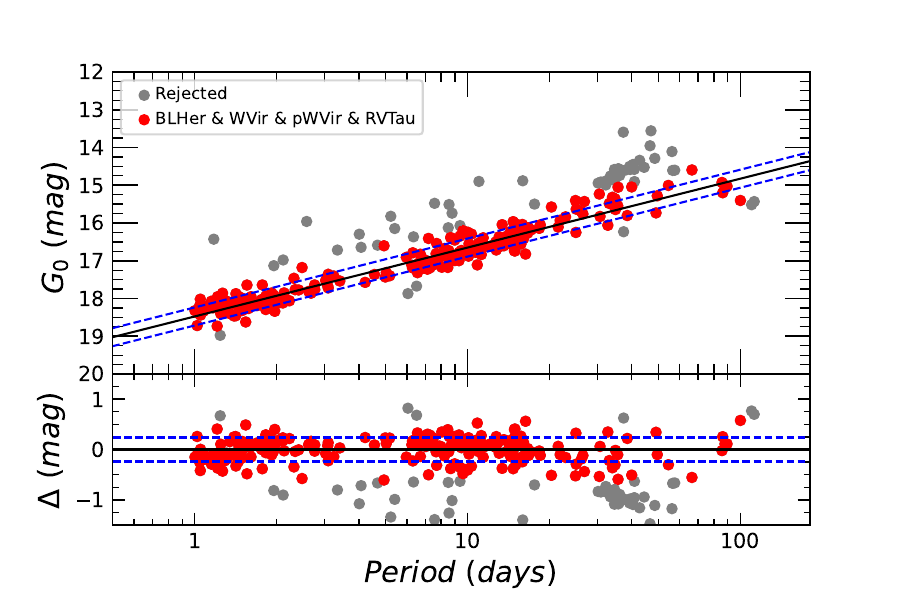}
    }
    \hbox{
    \includegraphics[width=0.34\textwidth]{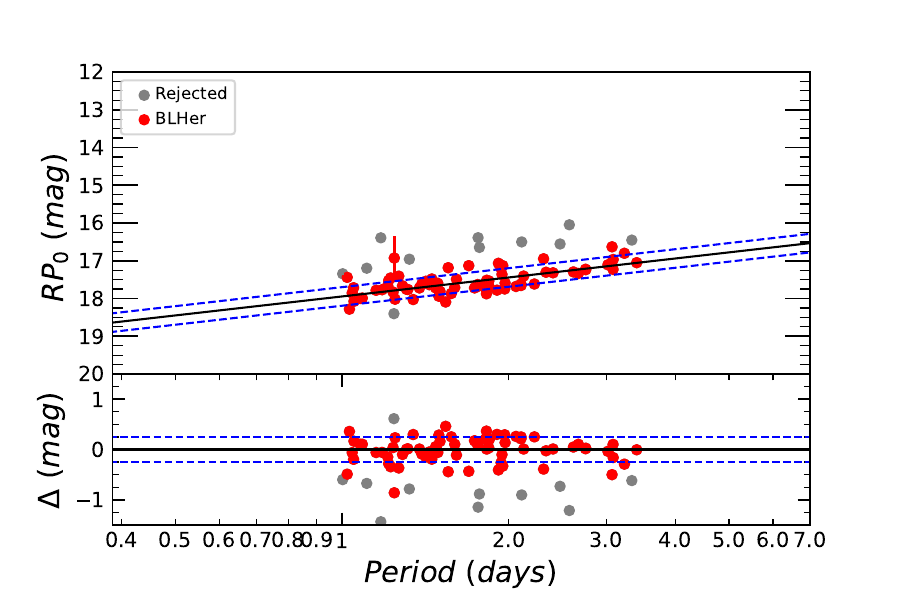}
    \includegraphics[width=0.34\textwidth]{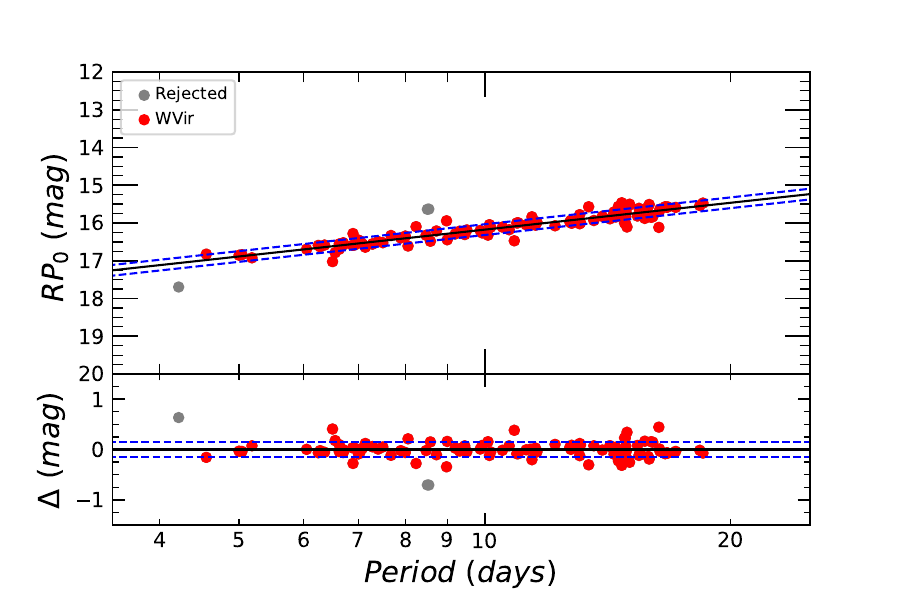}
    \includegraphics[width=0.34\textwidth]{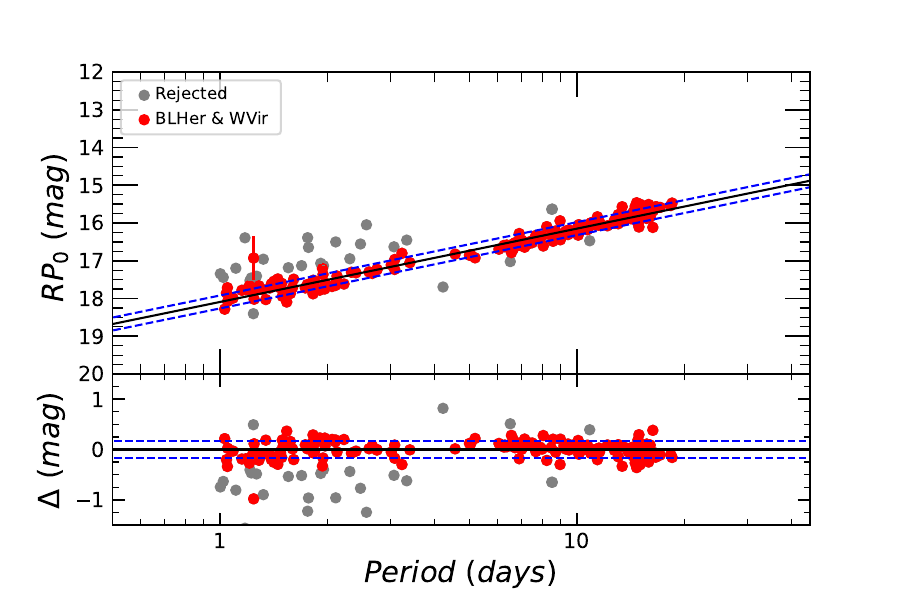}
    }
    }
   \caption{\label{fit1lmc} Relation fitting in different bands for a variety of T2C class combinations in the LMC. }
\end{figure*}
 
\begin{figure*}
    \vbox{
    \hbox{
    \includegraphics[width=0.34\textwidth]{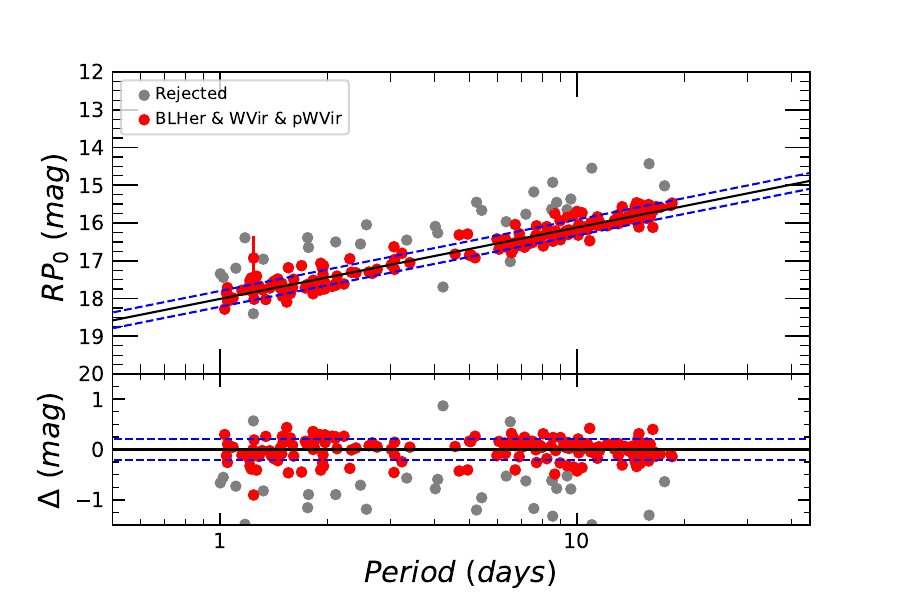}
    \includegraphics[width=0.34\textwidth]{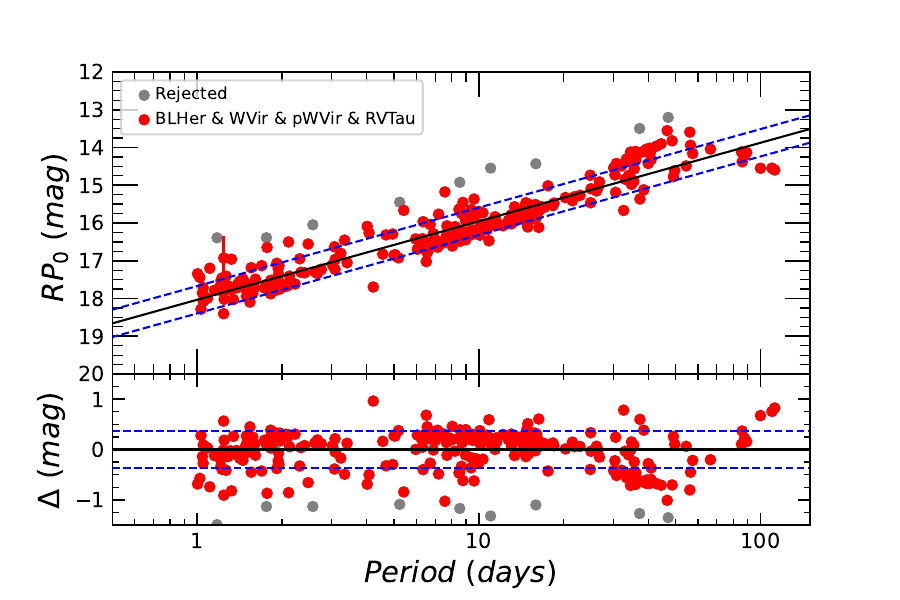}
    \includegraphics[width=0.34\textwidth]{PLIBLHerlmc.pdf}
    }
    \hbox{
    \includegraphics[width=0.34\textwidth]{PLIWVirlmc.pdf}
    \includegraphics[width=0.34\textwidth]{PLI1lmc.pdf}
    \includegraphics[width=0.34\textwidth]{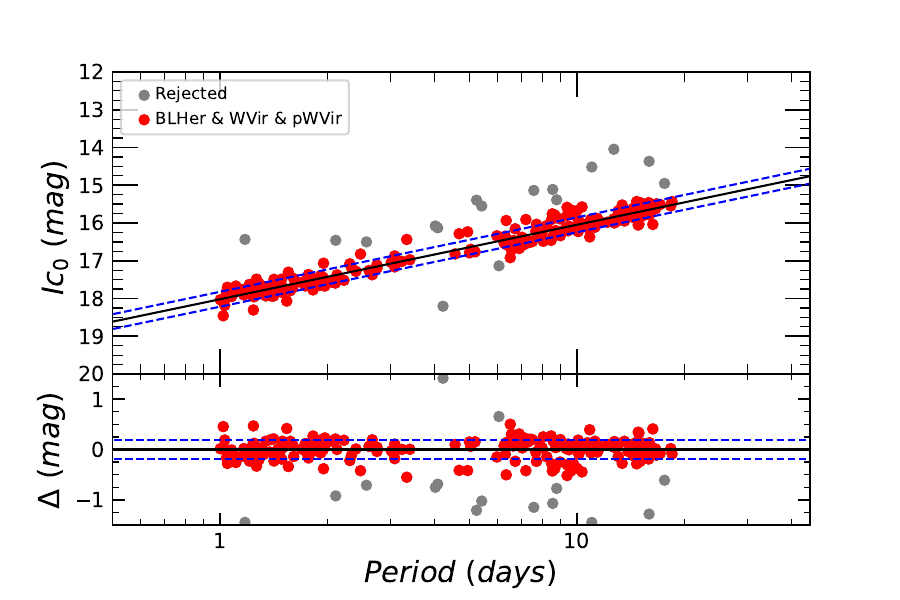}
    }
    \hbox{
    \includegraphics[width=0.34\textwidth]{PLI3lmc.pdf}
    \includegraphics[width=0.34\textwidth]{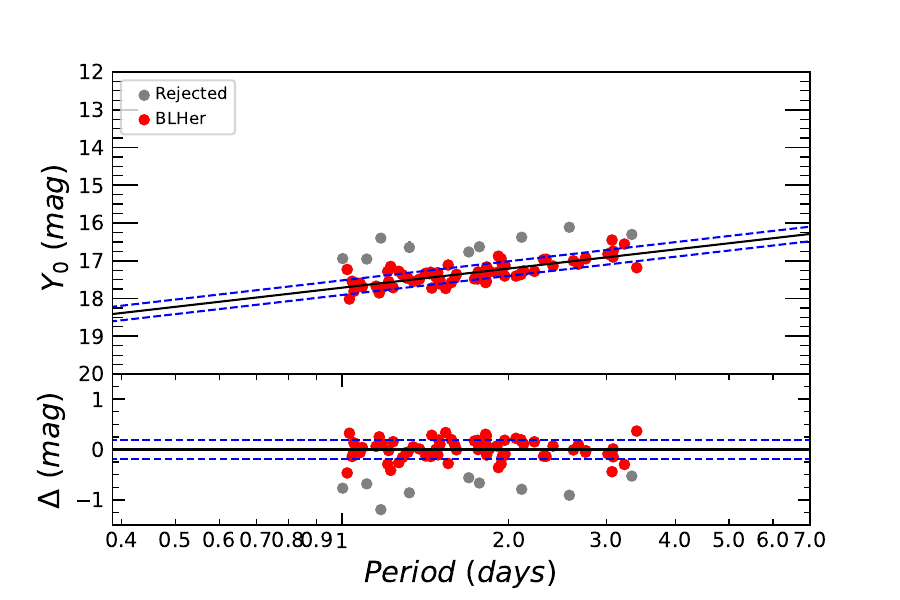}
    \includegraphics[width=0.34\textwidth]{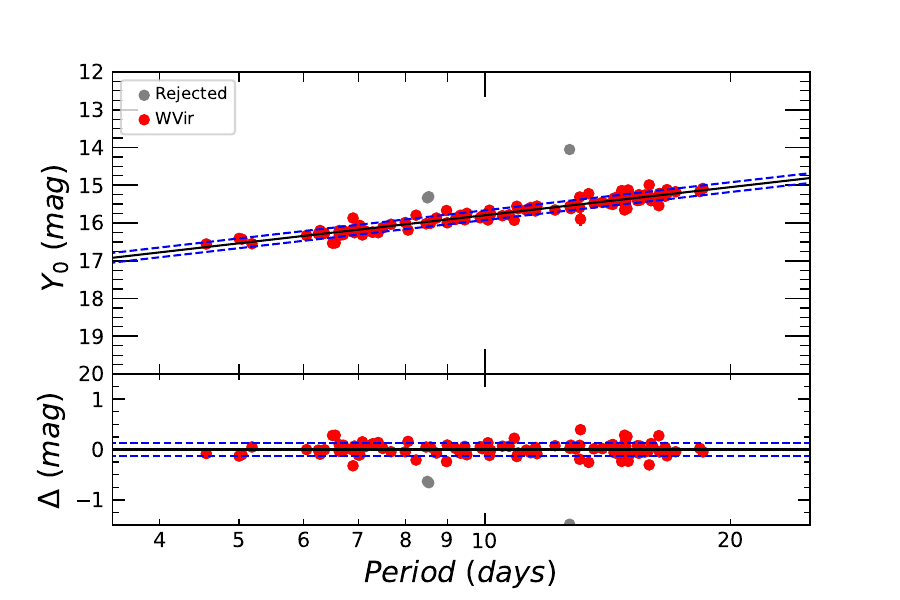} 
    }
    \hbox{
    \includegraphics[width=0.34\textwidth]{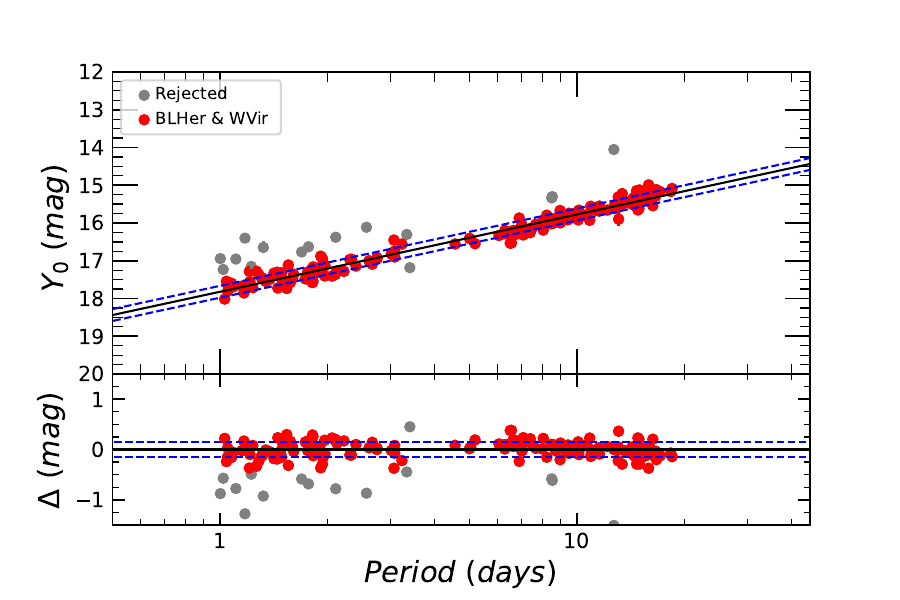}
    \includegraphics[width=0.34\textwidth]{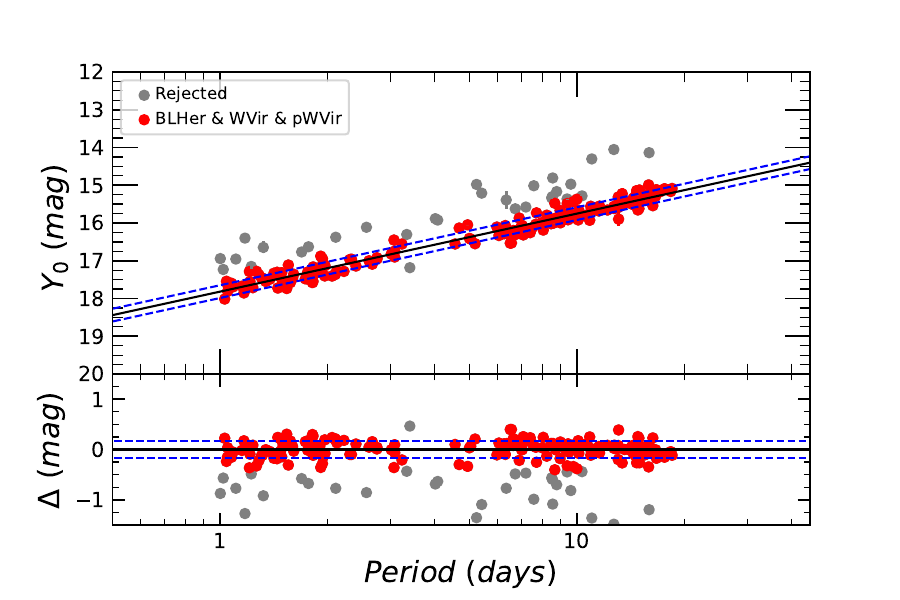}
    \includegraphics[width=0.34\textwidth]{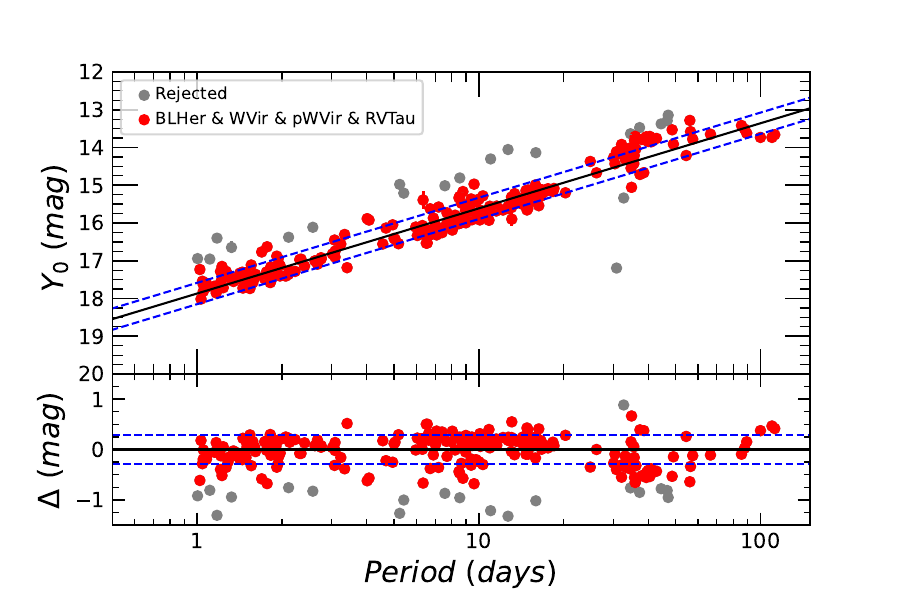}
    }
    \hbox{
    \includegraphics[width=0.34\textwidth]{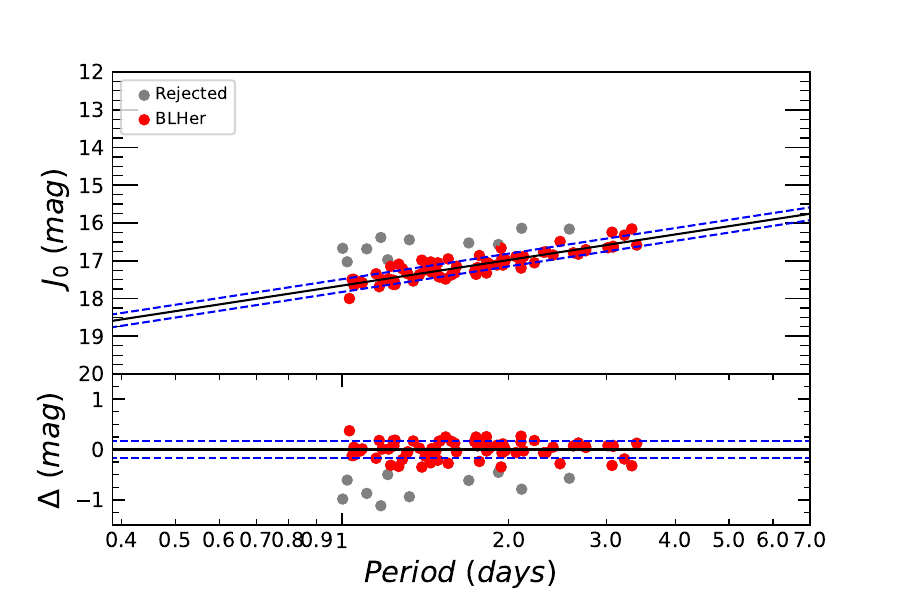}
    \includegraphics[width=0.34\textwidth]{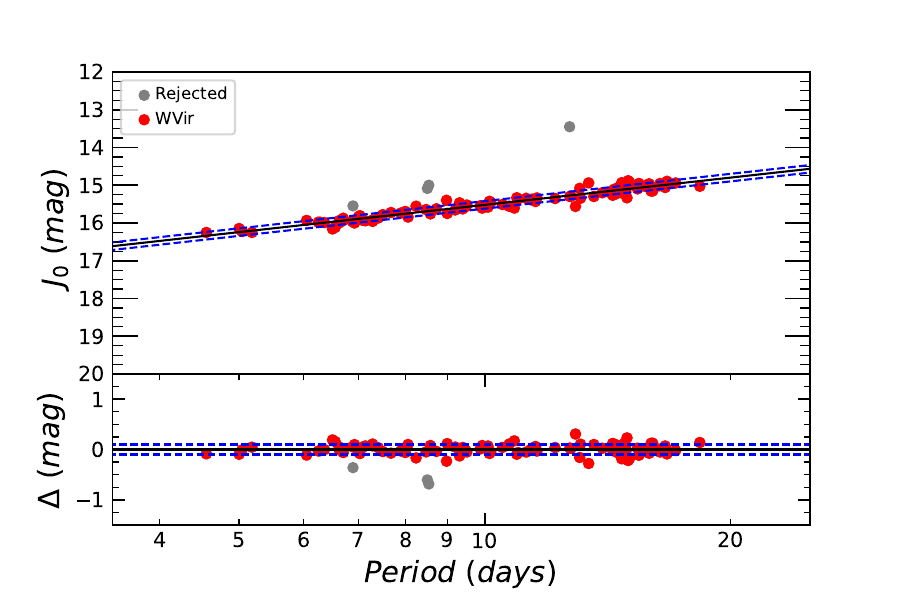}
    \includegraphics[width=0.34\textwidth]{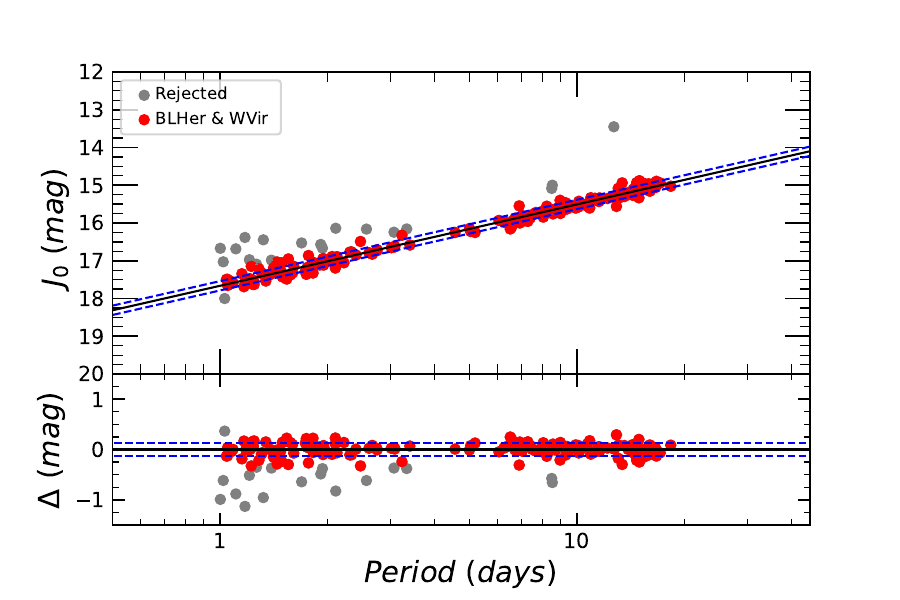}
    }
    \hbox{
    \includegraphics[width=0.34\textwidth]{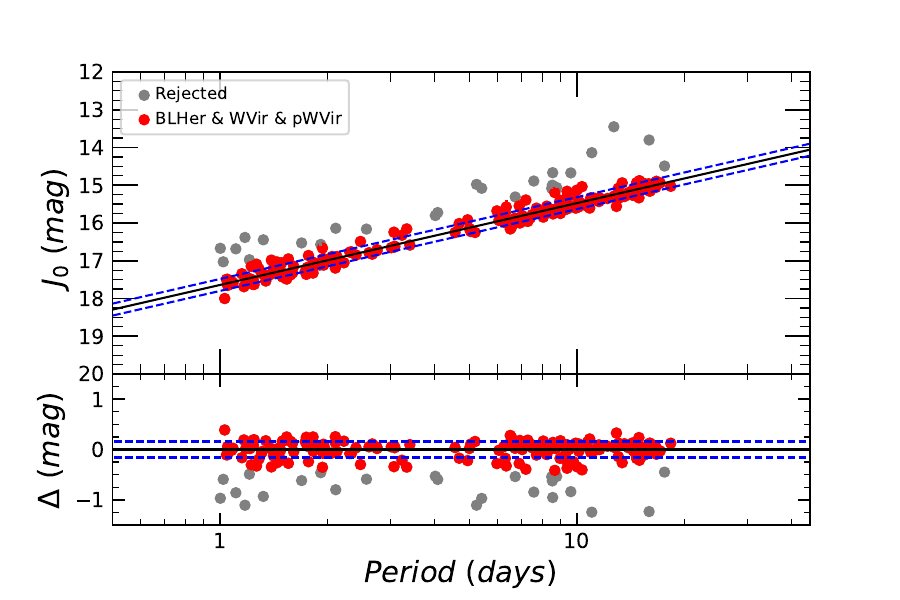}
    \includegraphics[width=0.34\textwidth]{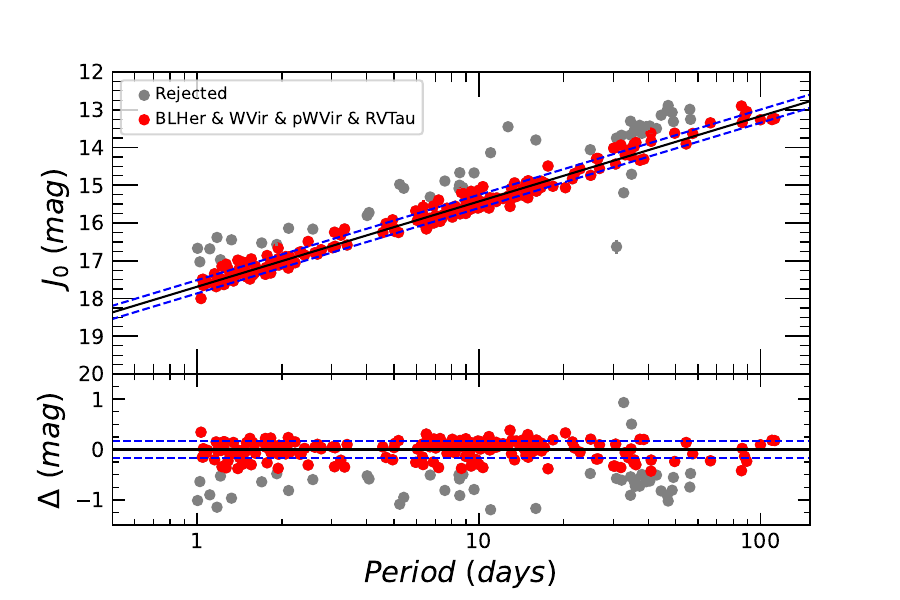}
    \includegraphics[width=0.34\textwidth]{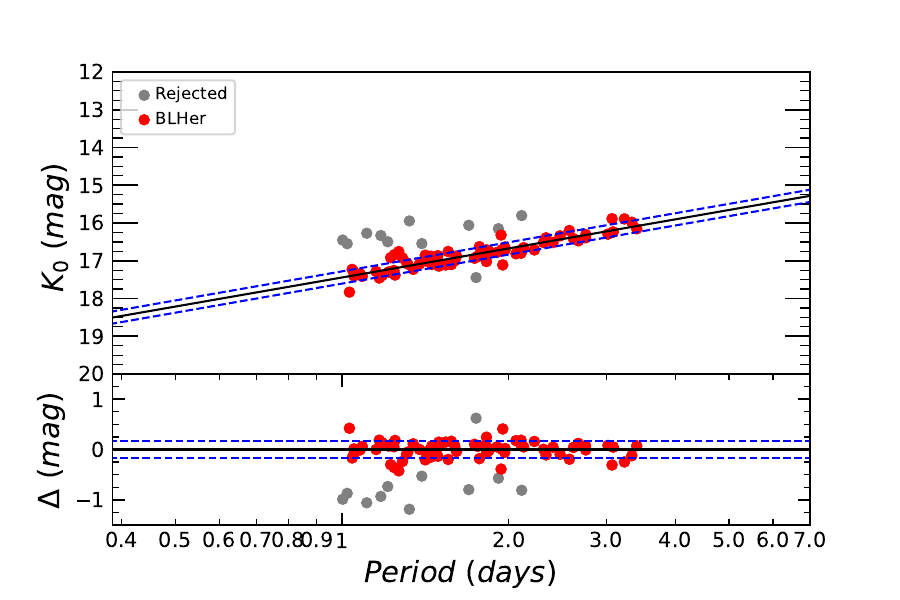}
    }
    }
    \ContinuedFloat
    \caption{continued. }	
	\end{figure*}

\begin{figure*}[h]
    \vbox{
    \hbox{
    \includegraphics[width=0.34\textwidth]{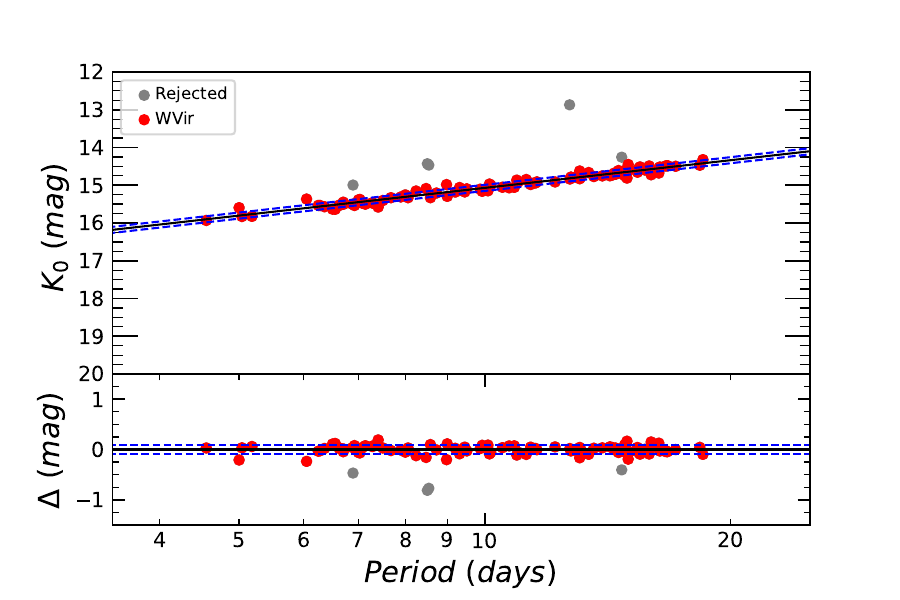}
    \includegraphics[width=0.34\textwidth]{PLK1lmc.pdf}
    \includegraphics[width=0.34\textwidth]{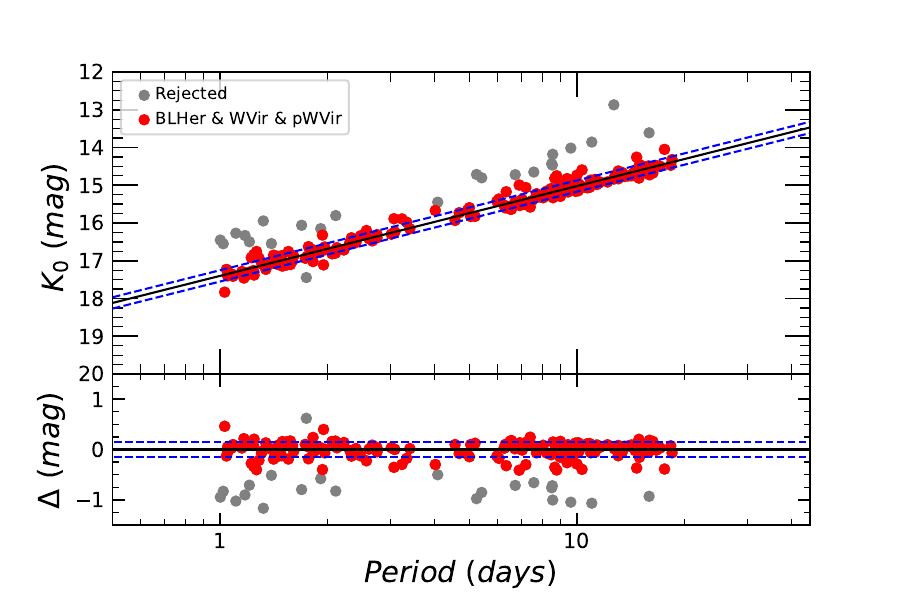}
    }
    \hbox{
    \includegraphics[width=0.34\textwidth]{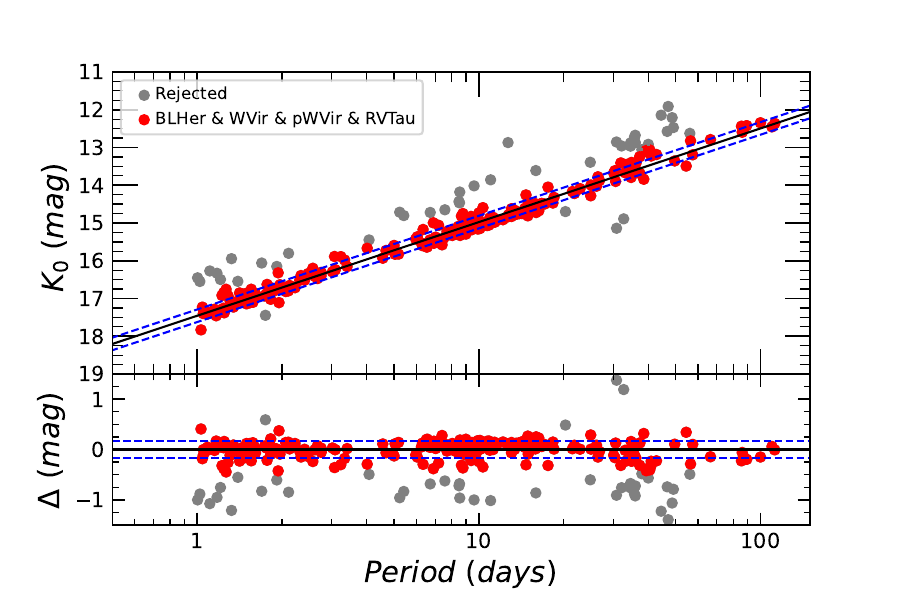}
    \includegraphics[width=0.34\textwidth]{PWVI1lmc.pdf}
    \includegraphics[width=0.34\textwidth]{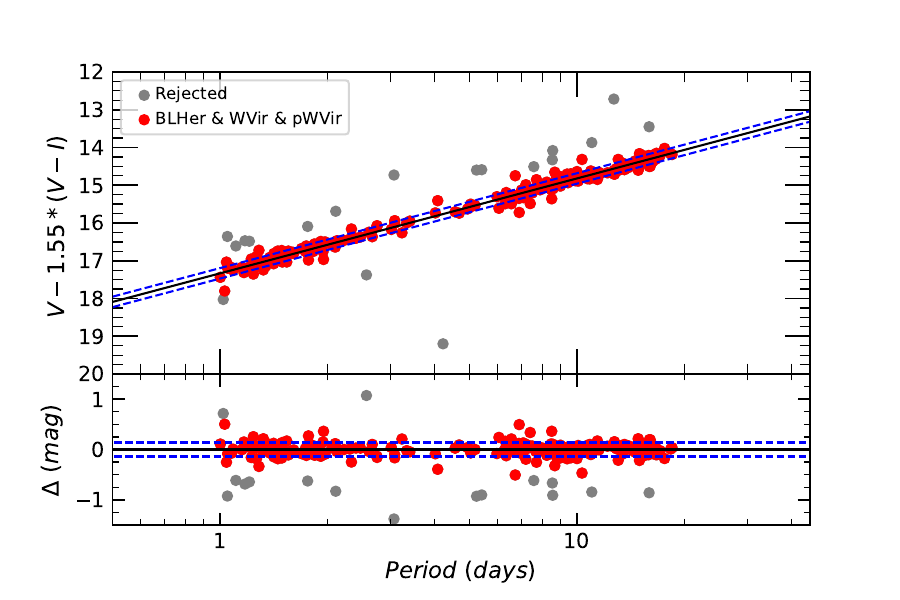}
    }
    \hbox{
    \includegraphics[width=0.34\textwidth]{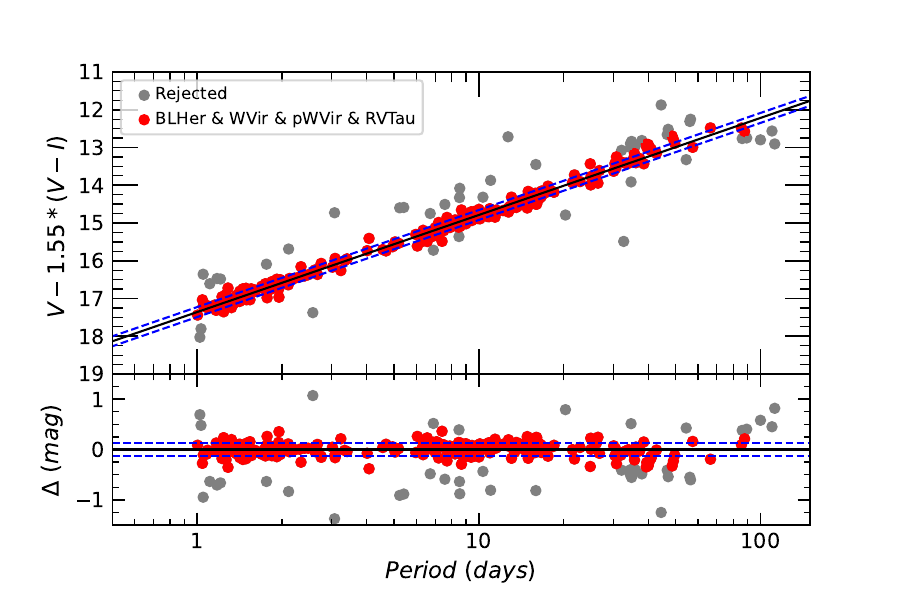}
    \includegraphics[width=0.34\textwidth]{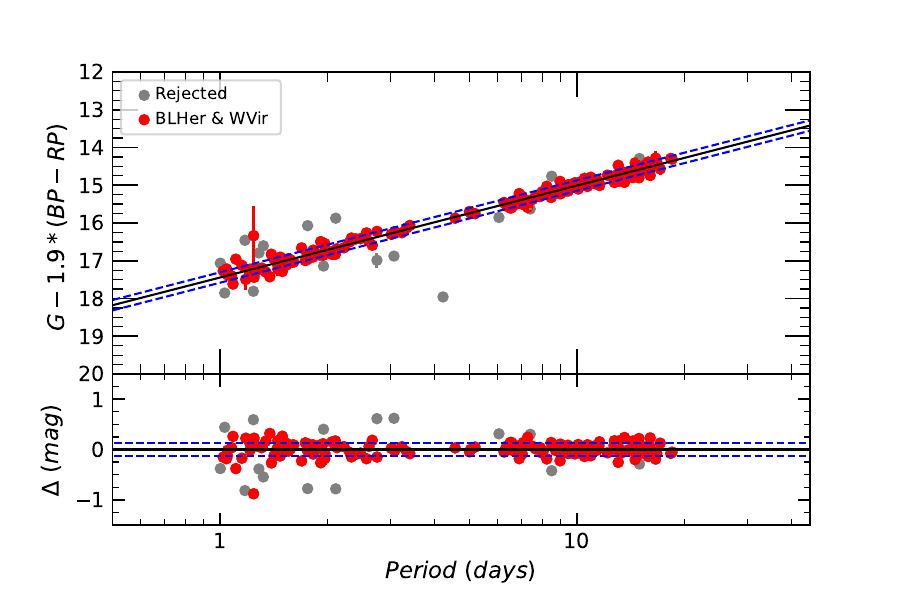}
    \includegraphics[width=0.34\textwidth]{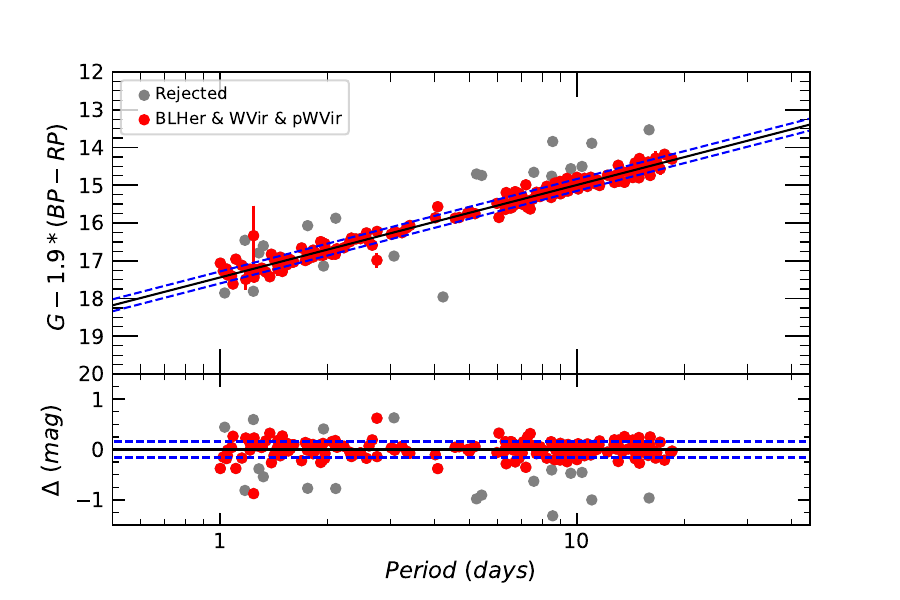}
    }
    \hbox{
    \includegraphics[width=0.34\textwidth]{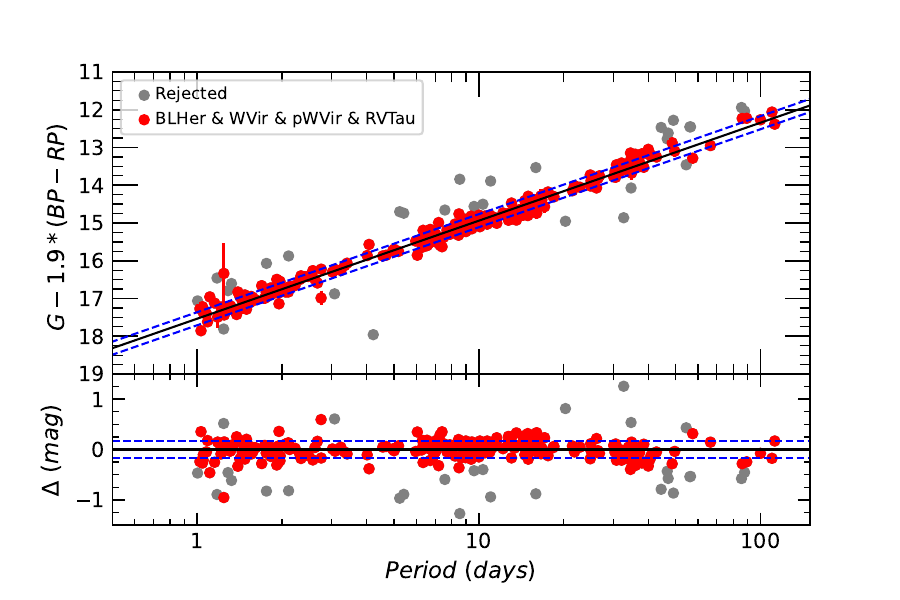}
    \includegraphics[width=0.34\textwidth]{PWVK1lmc.pdf}
    \includegraphics[width=0.34\textwidth]{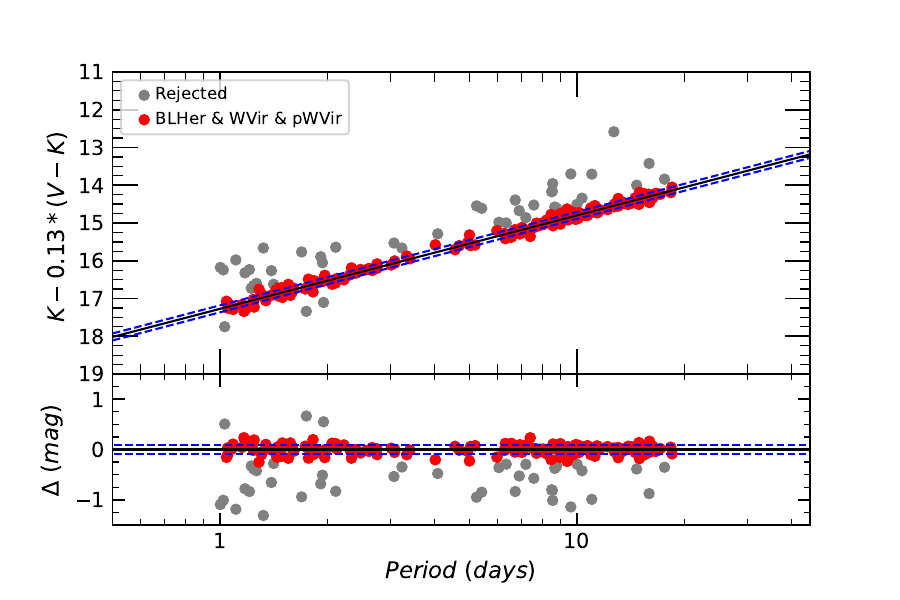}
    }
    \hbox{
    \includegraphics[width=0.34\textwidth]{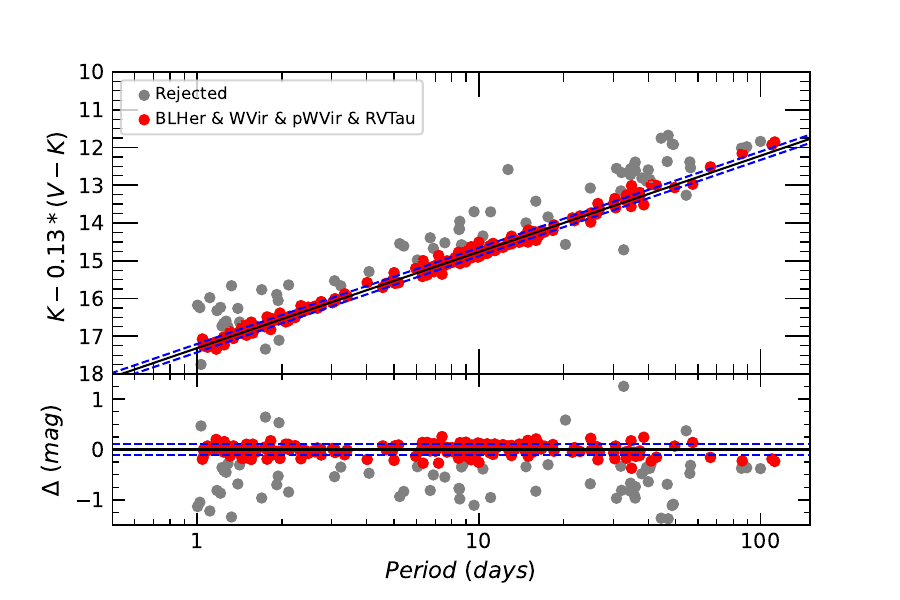}
    \includegraphics[width=0.34\textwidth]{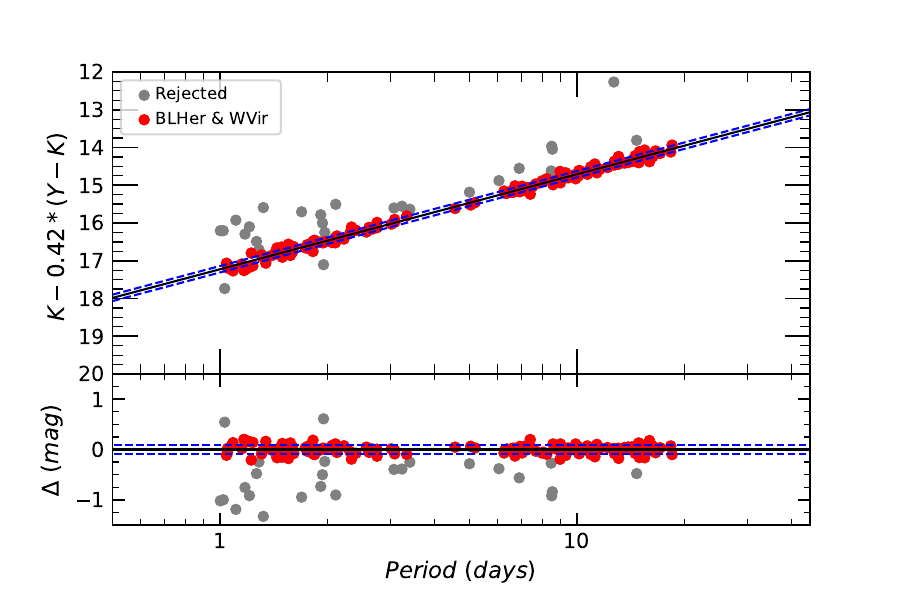}
    \includegraphics[width=0.34\textwidth]{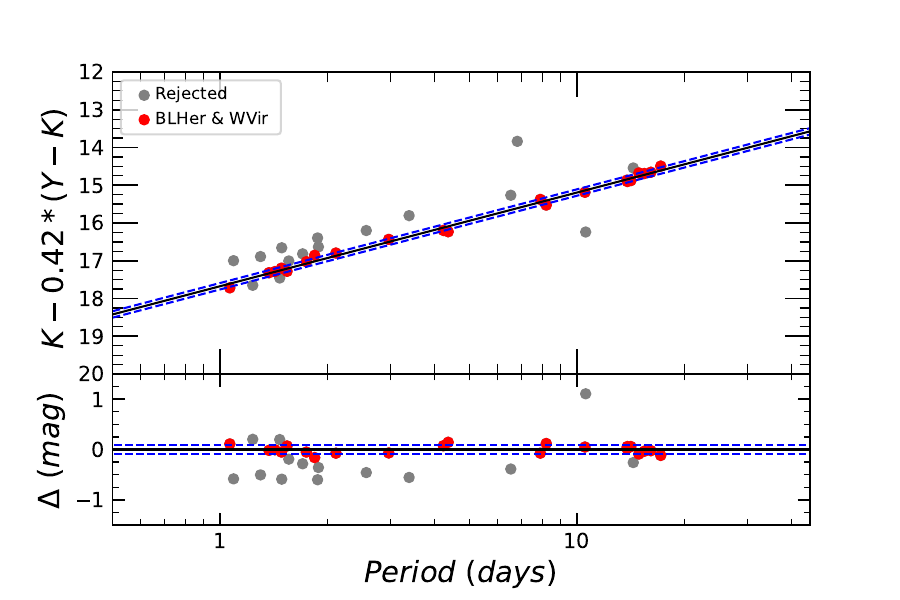}
    }
    \hbox{
    \includegraphics[width=0.34\textwidth]{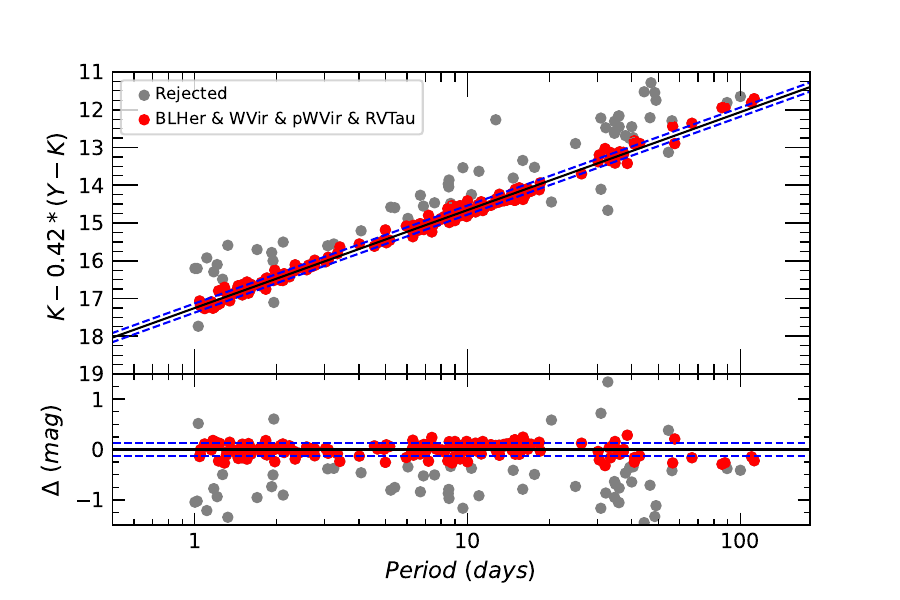}
    \includegraphics[width=0.34\textwidth]{PWJK1lmc.pdf}
    \includegraphics[width=0.34\textwidth]{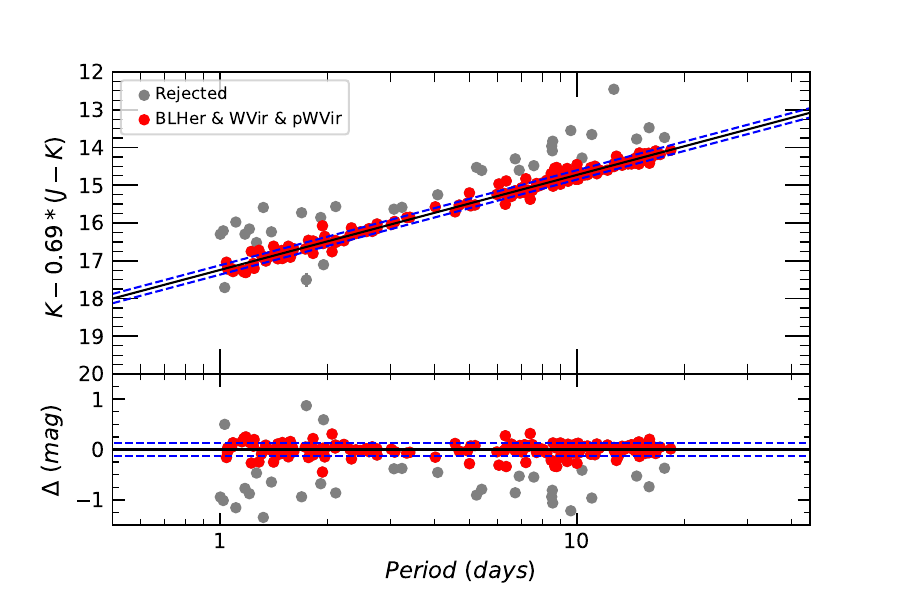}
    }
    }
    \ContinuedFloat
    \caption{continued. }	
    \end{figure*}

\begin{figure*}[h]
    \vbox{
    \hbox{
    \includegraphics[width=0.34\textwidth]{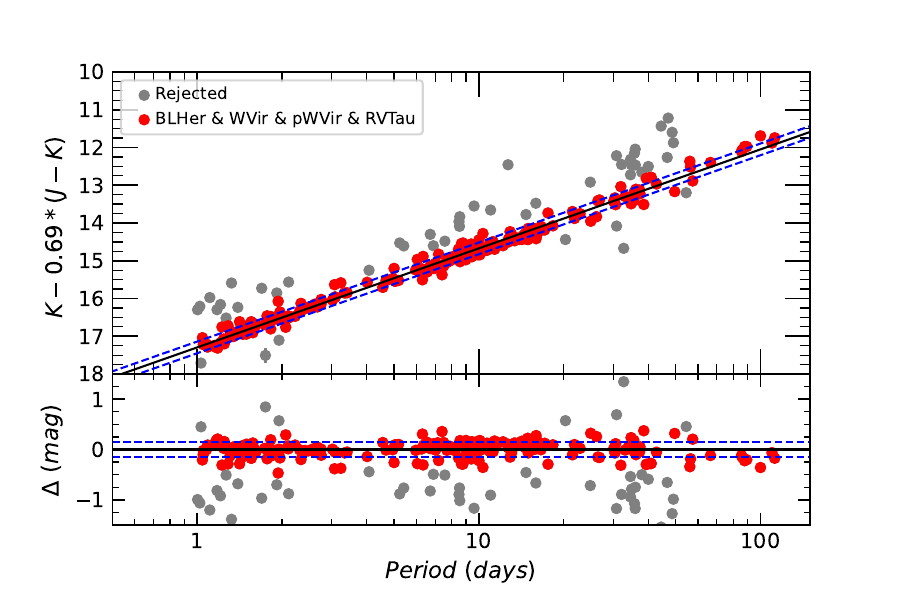}
    \includegraphics[width=0.34\textwidth]{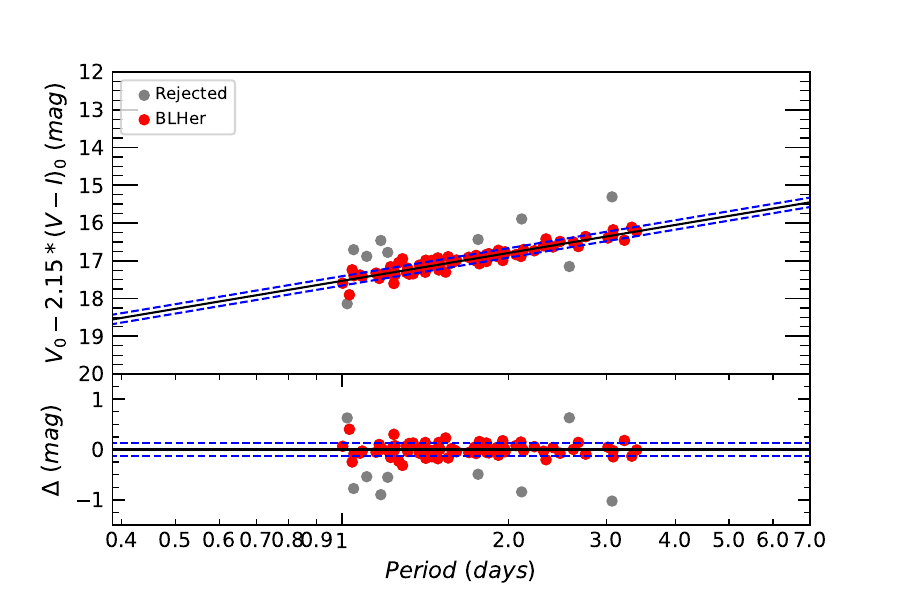}
    \includegraphics[width=0.34\textwidth]{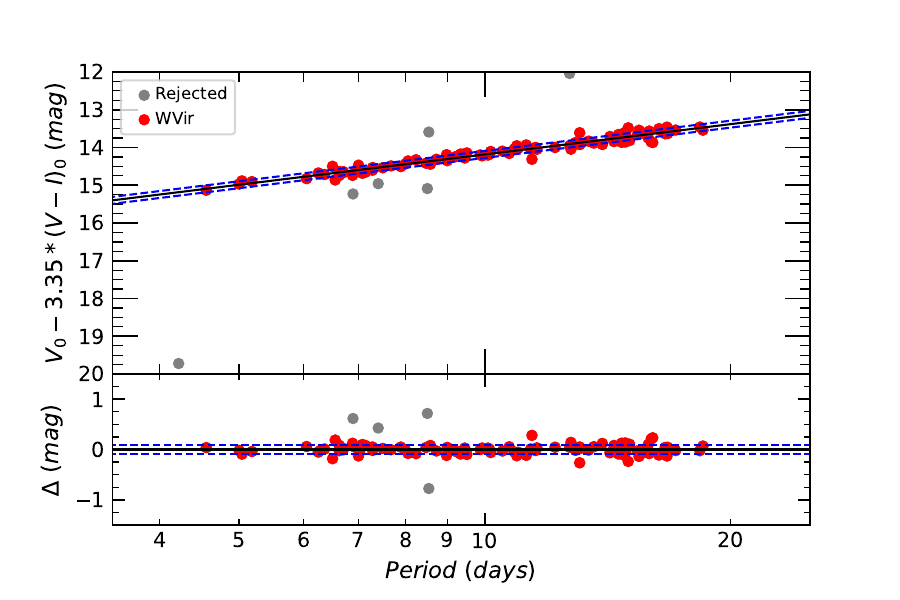}
    }
    \hbox{
    \includegraphics[width=0.34\textwidth]{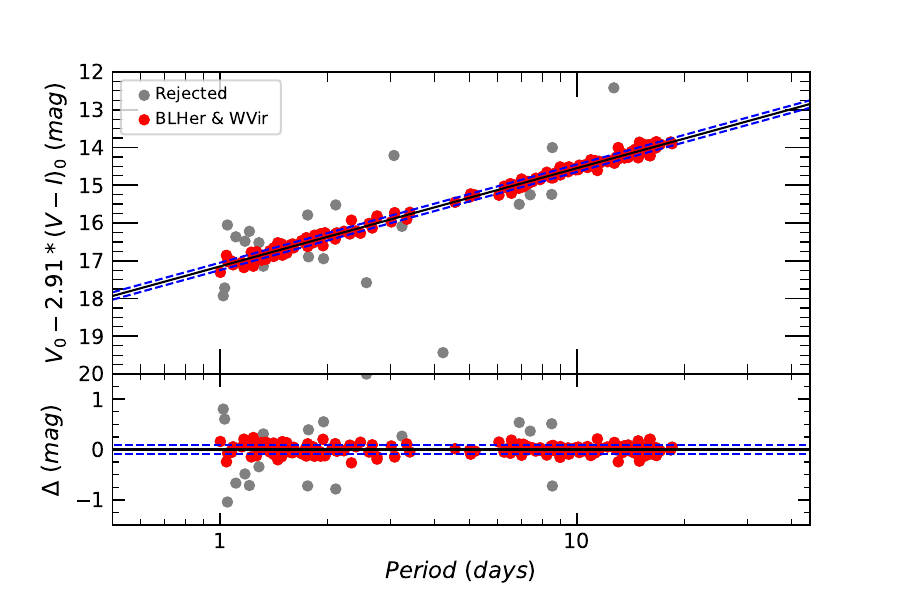}
    \includegraphics[width=0.34\textwidth]{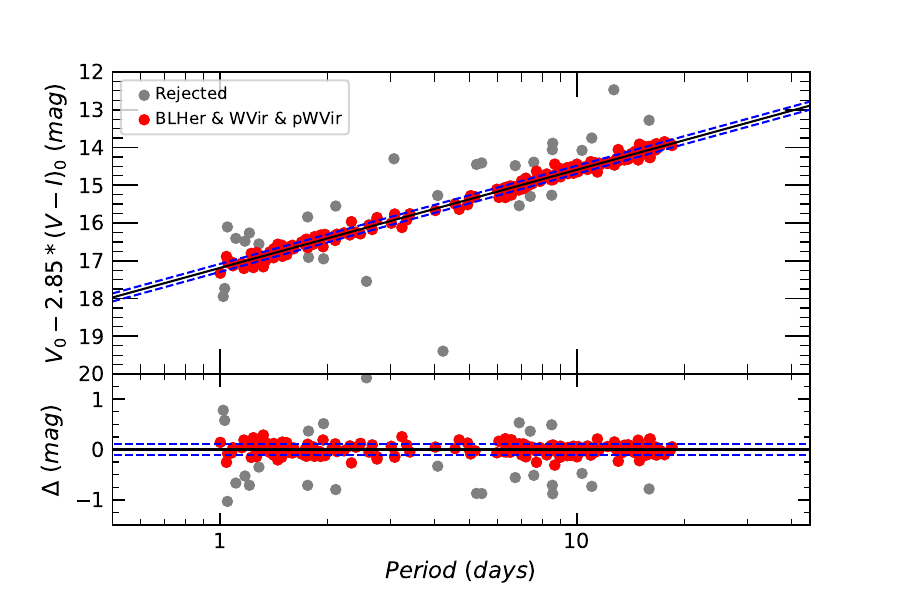}
    \includegraphics[width=0.34\textwidth]{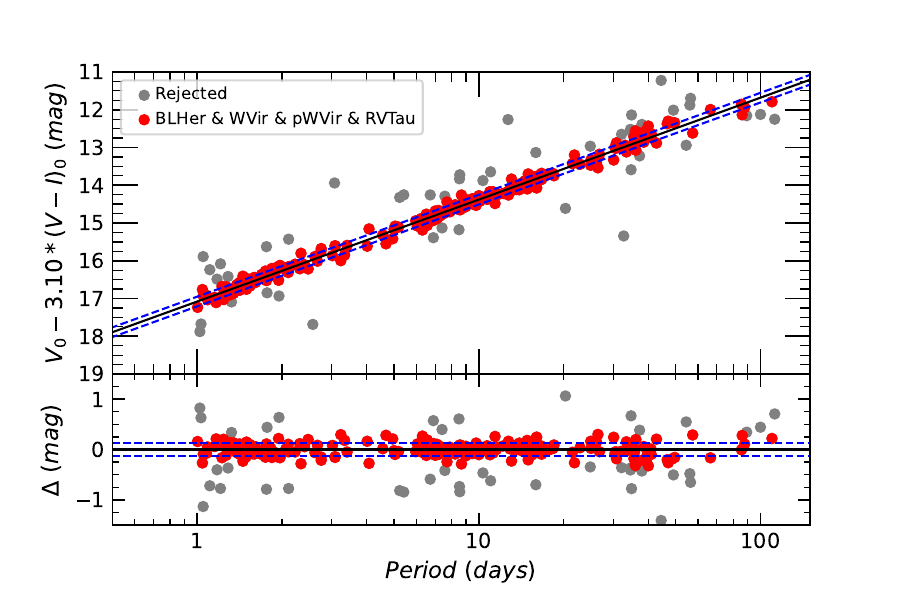}
    }
    \hbox{
    \includegraphics[width=0.34\textwidth]{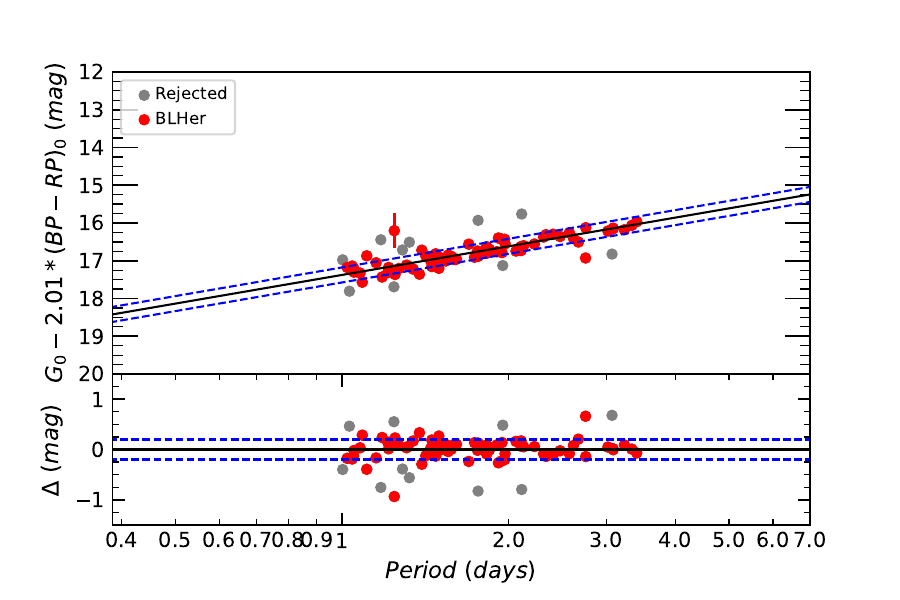} 
    \includegraphics[width=0.34\textwidth]{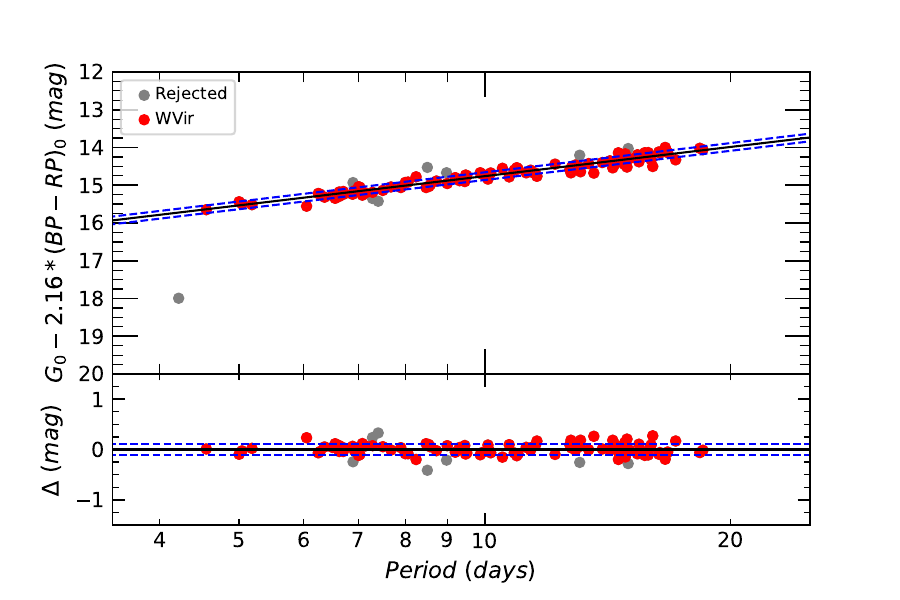}
    \includegraphics[width=0.34\textwidth]{PLCBPRP1lmc.pdf}
    }
    \hbox{
    \includegraphics[width=0.34\textwidth]{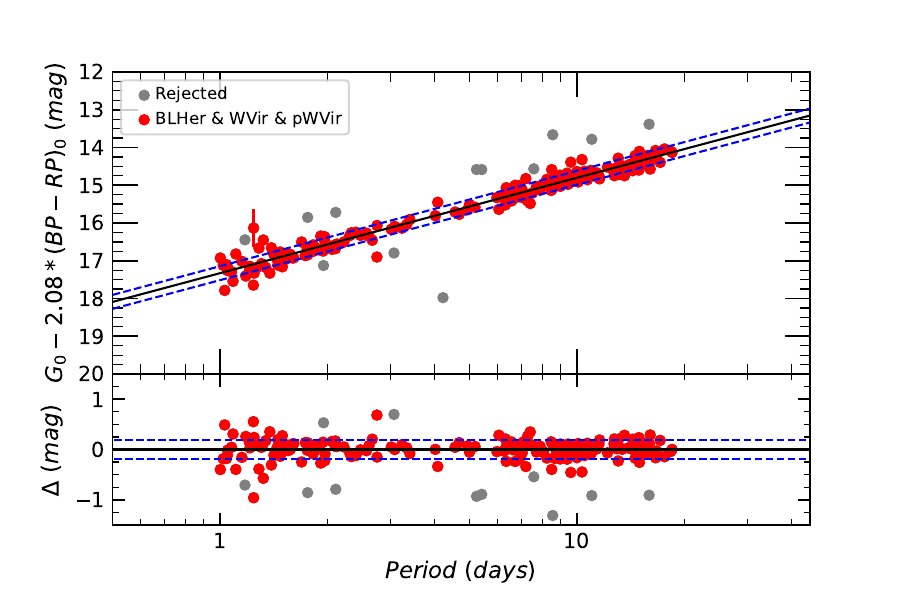}
    \includegraphics[width=0.34\textwidth]{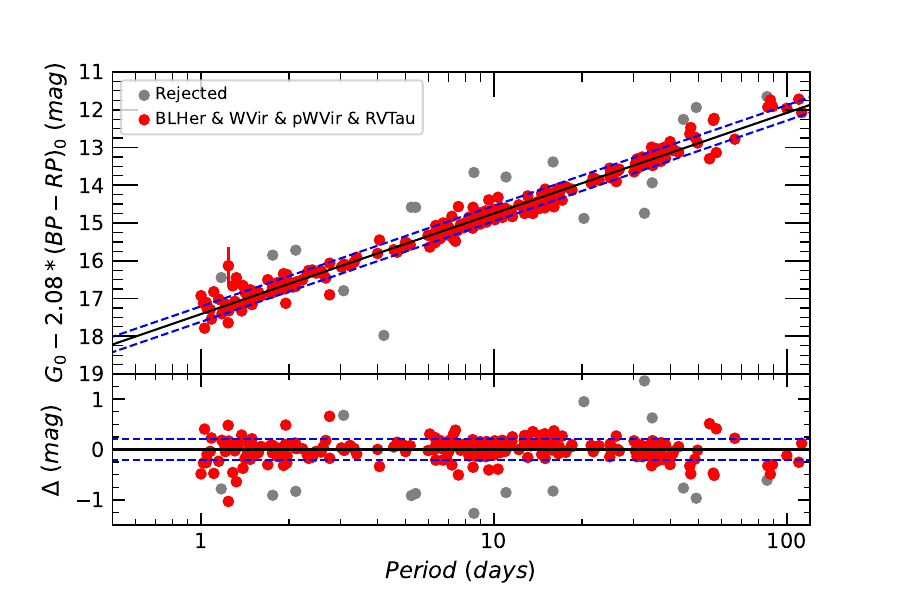}
    \includegraphics[width=0.34\textwidth]{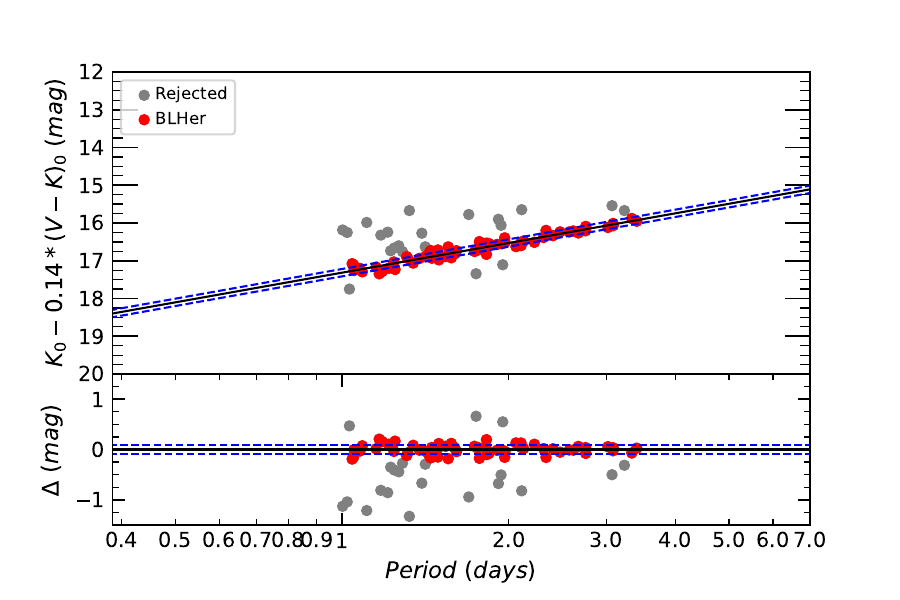}
    }
    \hbox{
    \includegraphics[width=0.34\textwidth]{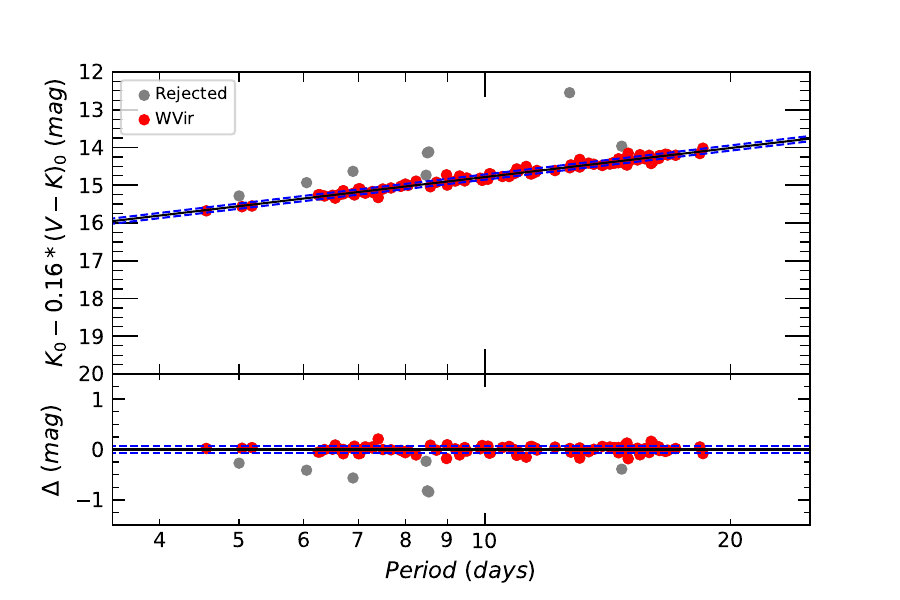}
    \includegraphics[width=0.34\textwidth]{PLCVK1lmc.pdf}
    \includegraphics[width=0.34\textwidth]{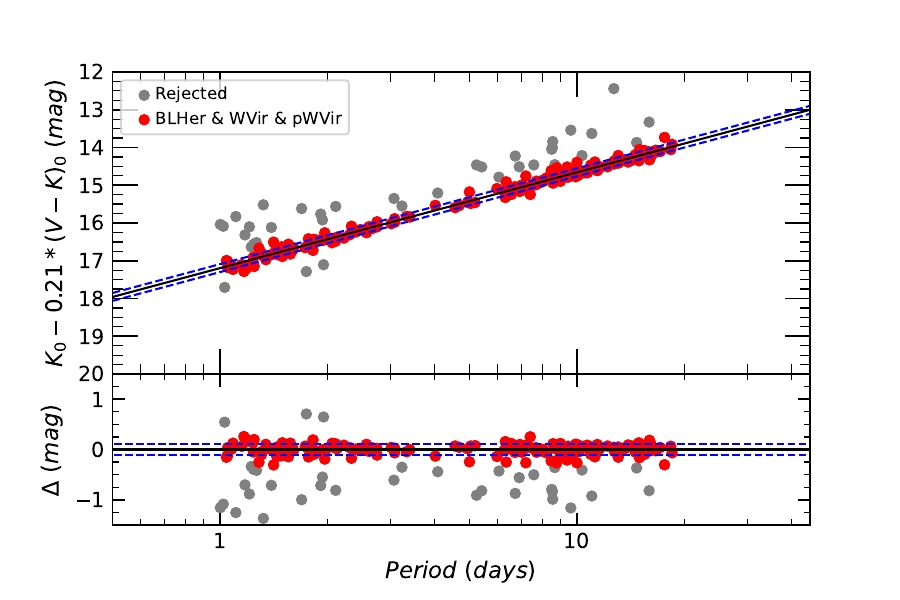}
    }
    \hbox{
    \includegraphics[width=0.34\textwidth]{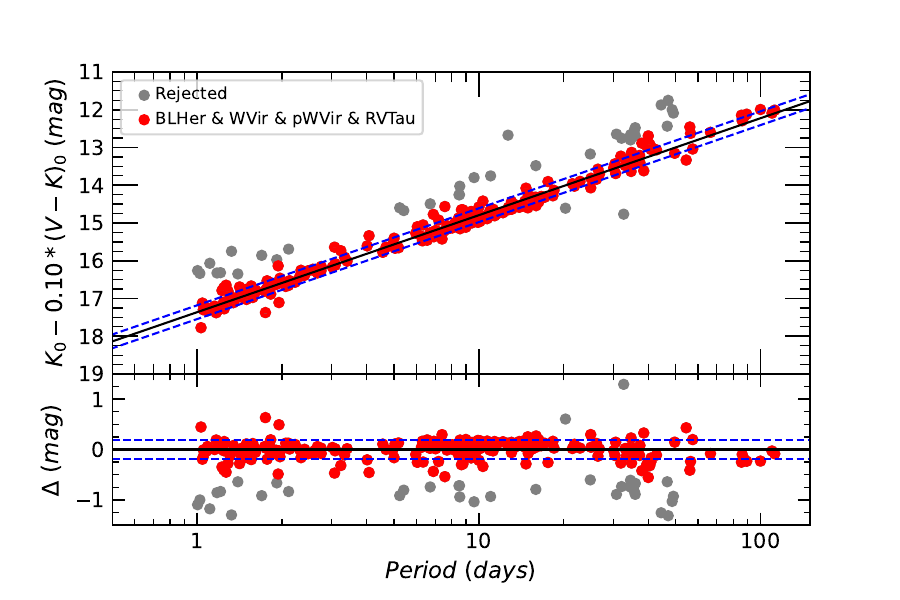}
    \includegraphics[width=0.34\textwidth]{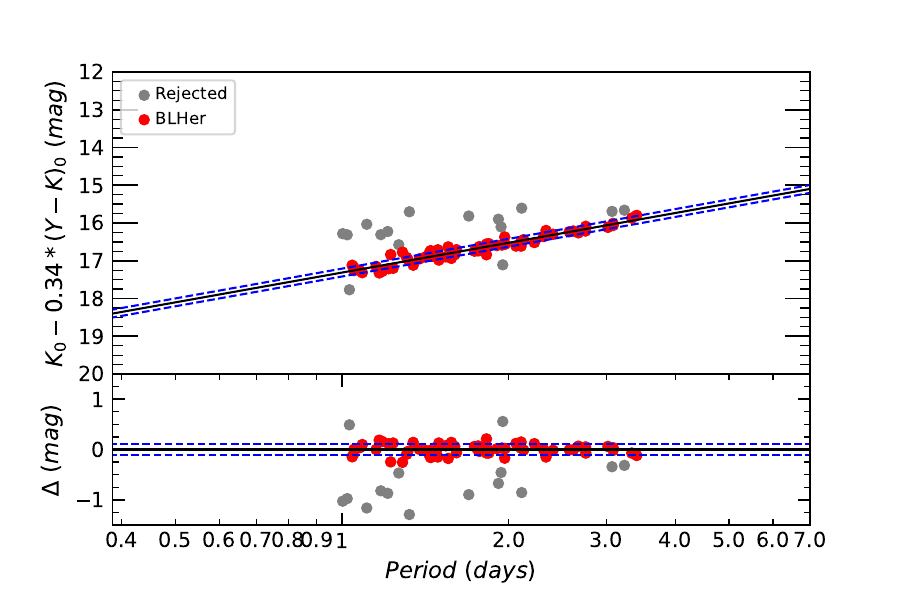}
    \includegraphics[width=0.34\textwidth]{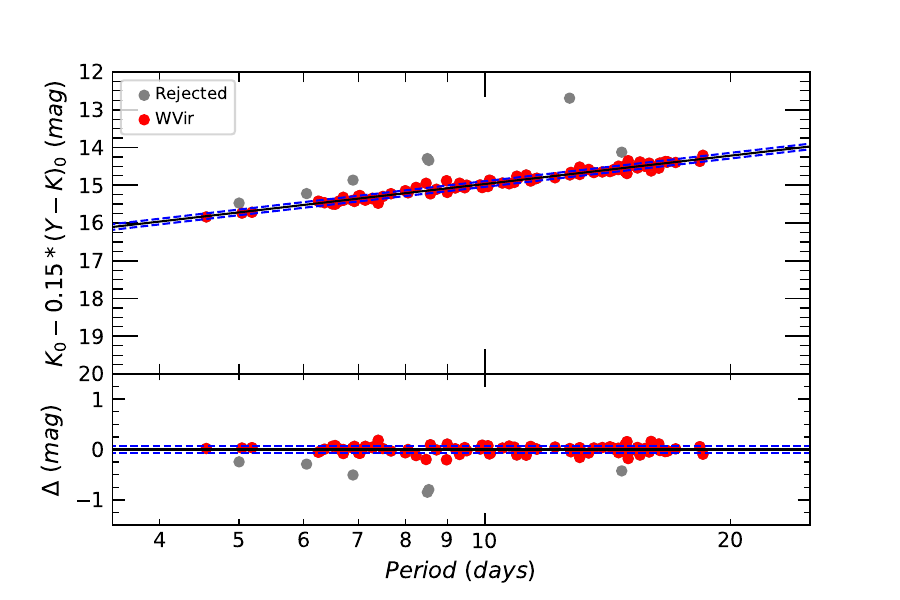}
    }
    }
    \ContinuedFloat
    \caption{continued. }	
	\end{figure*}

\begin{figure*}[h]
    \captionsetup{labelformat=simple}
    \ContinuedFloat
    \vbox{
    \hbox{ 
    \includegraphics[width=0.34\textwidth]{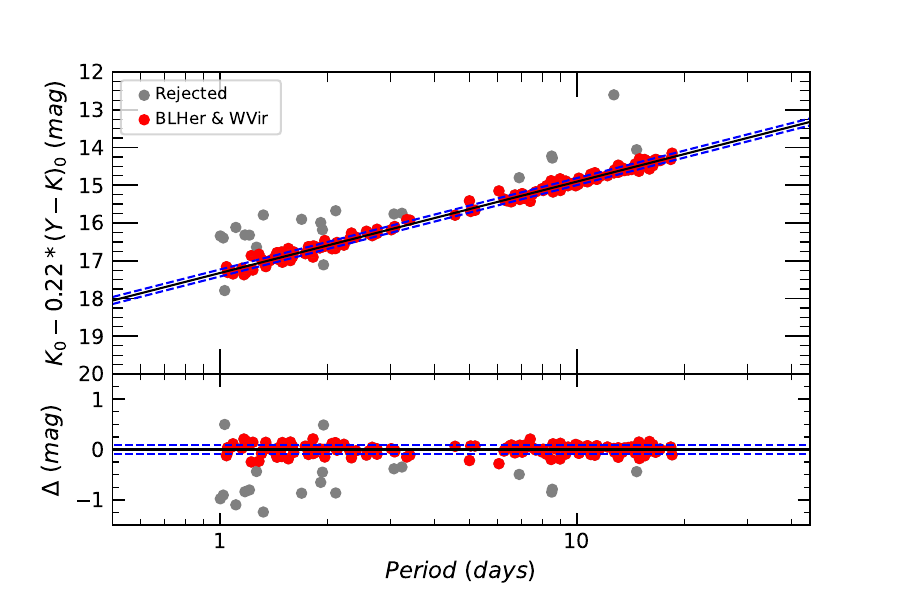}
    \includegraphics[width=0.34\textwidth]{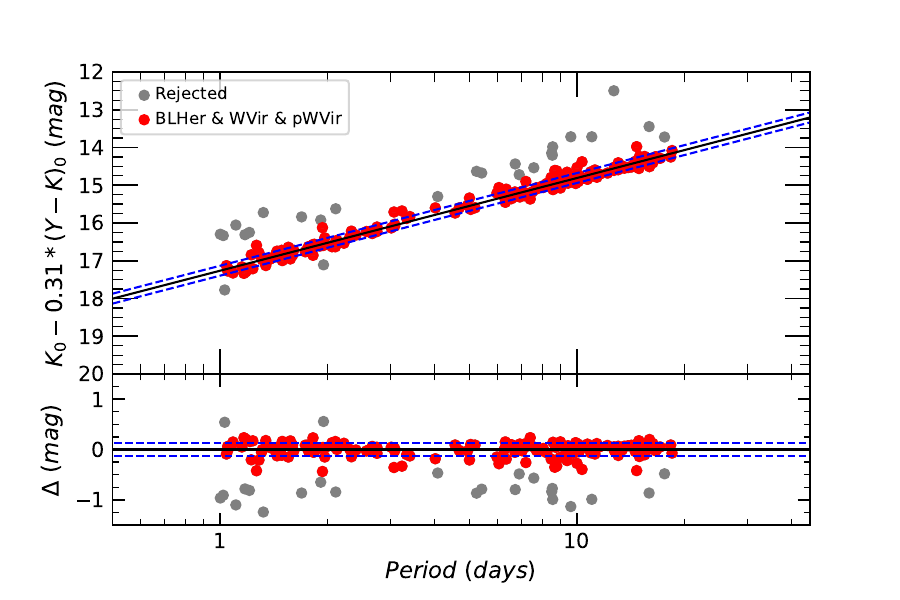}
    \includegraphics[width=0.34\textwidth]{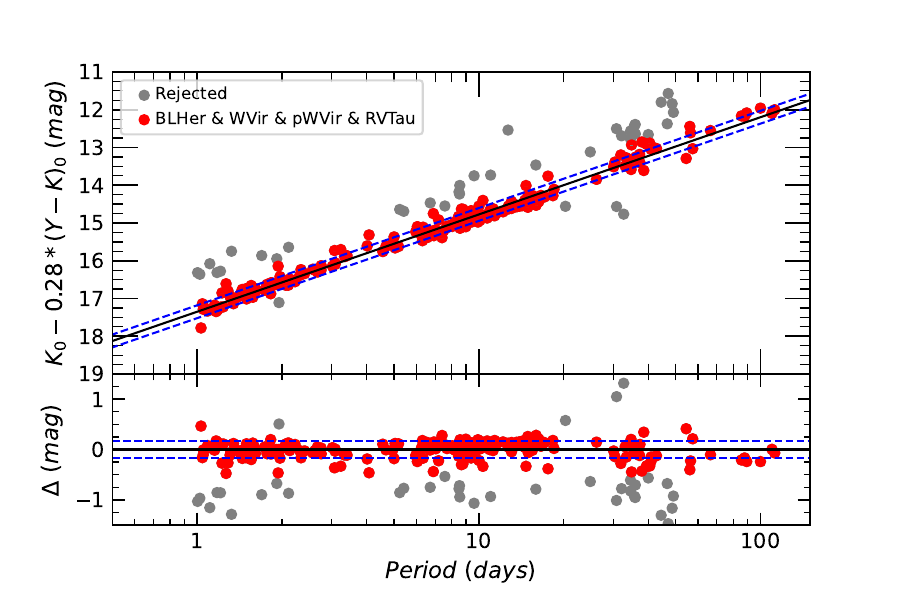}
    } 
    }
    \ContinuedFloat
    \caption{continued. }	
	\end{figure*}

\begin{figure*}[h]
    \vbox{
    \hbox{
    \includegraphics[width=0.34\textwidth]{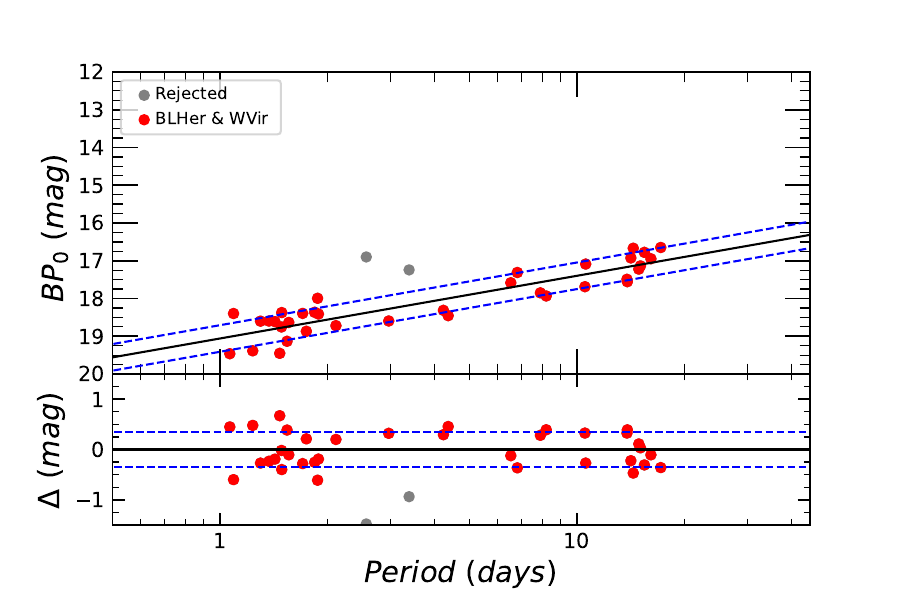}
    \includegraphics[width=0.34\textwidth]{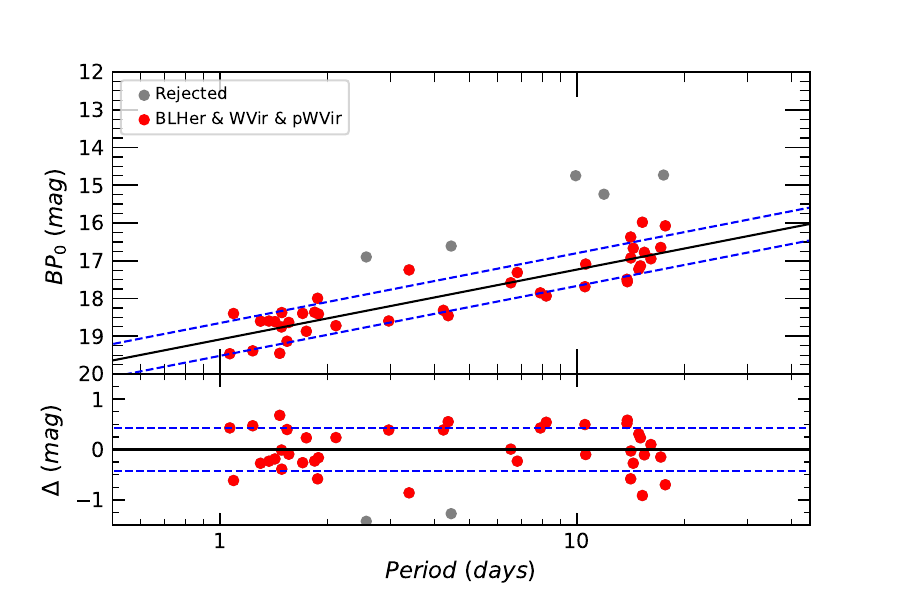}
    \includegraphics[width=0.34\textwidth]{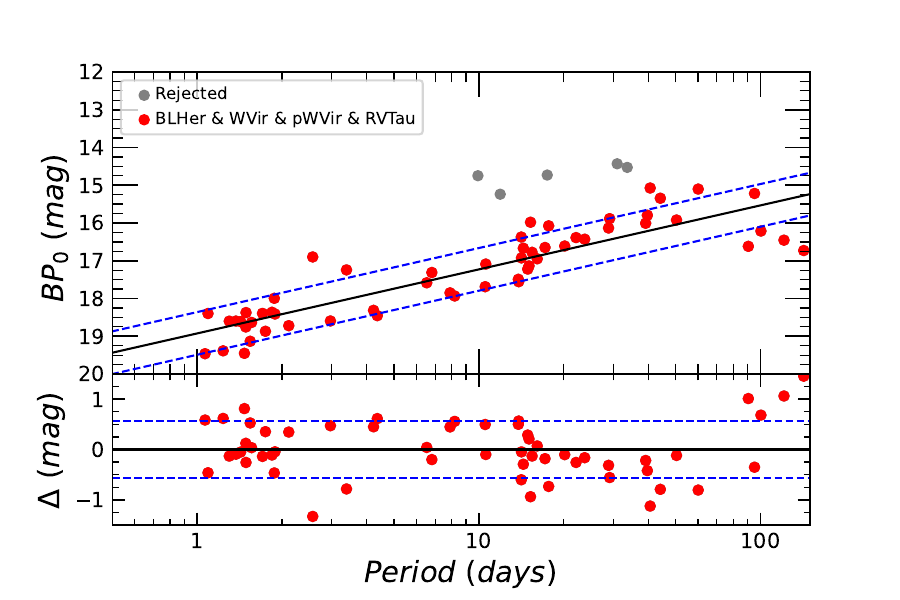}
    }
    \hbox{
    \includegraphics[width=0.34\textwidth]{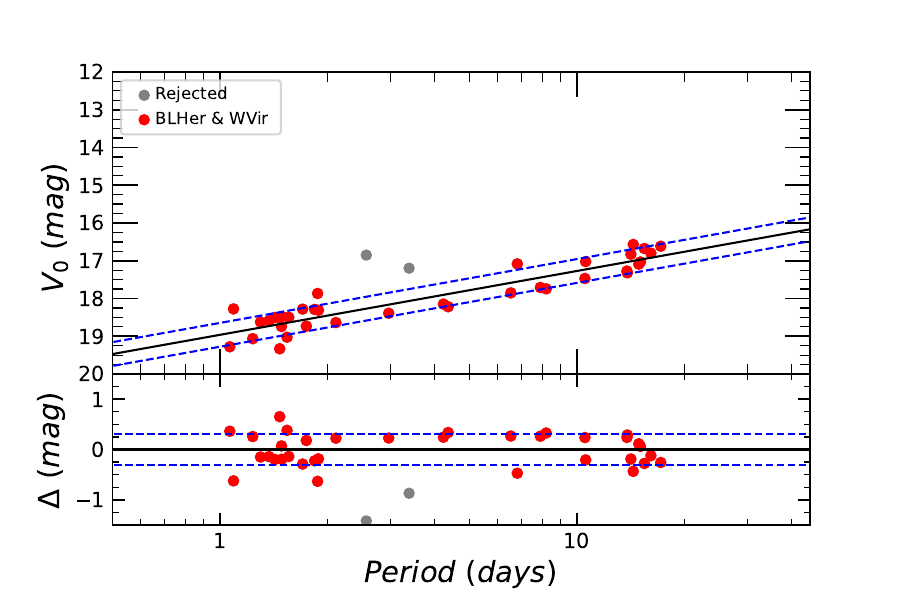}
    \includegraphics[width=0.34\textwidth]{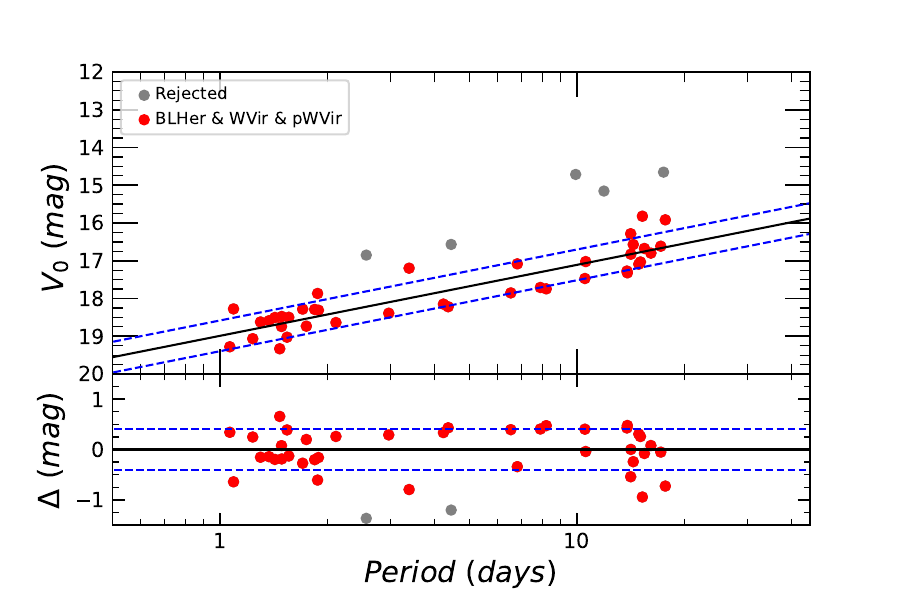}
    \includegraphics[width=0.34\textwidth]{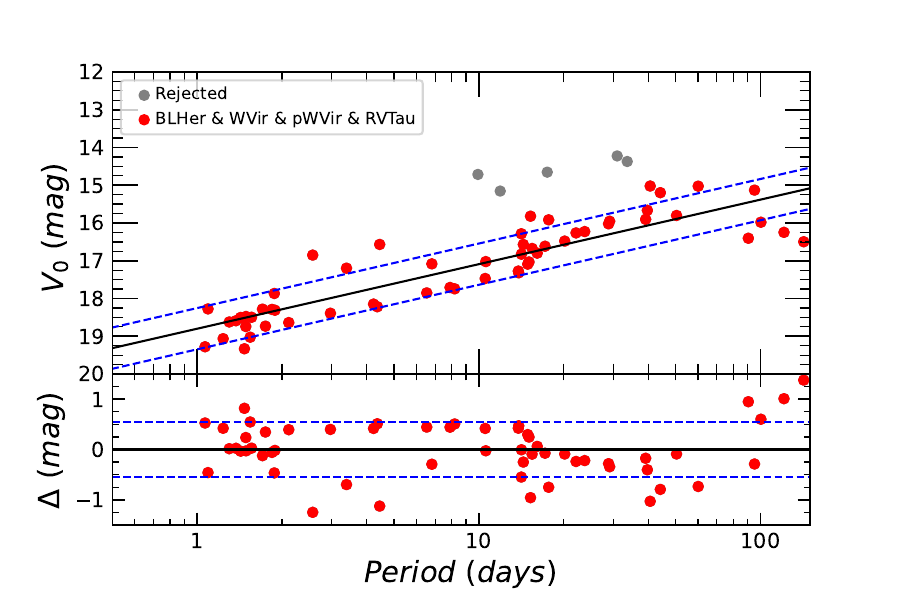}
    }
    \hbox{
    }
    \hbox{
    \includegraphics[width=0.34\textwidth]{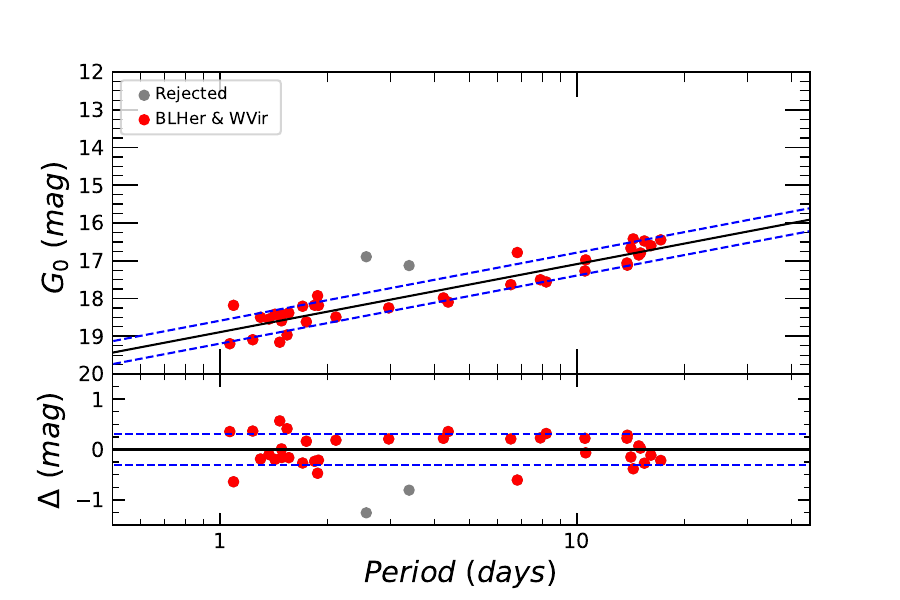}
    \includegraphics[width=0.34\textwidth]{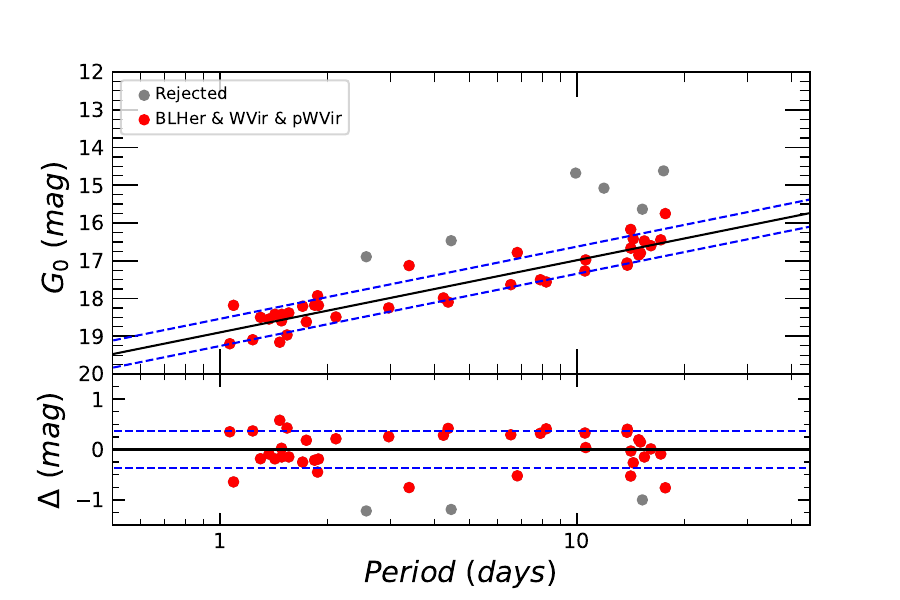}
    \includegraphics[width=0.34\textwidth]{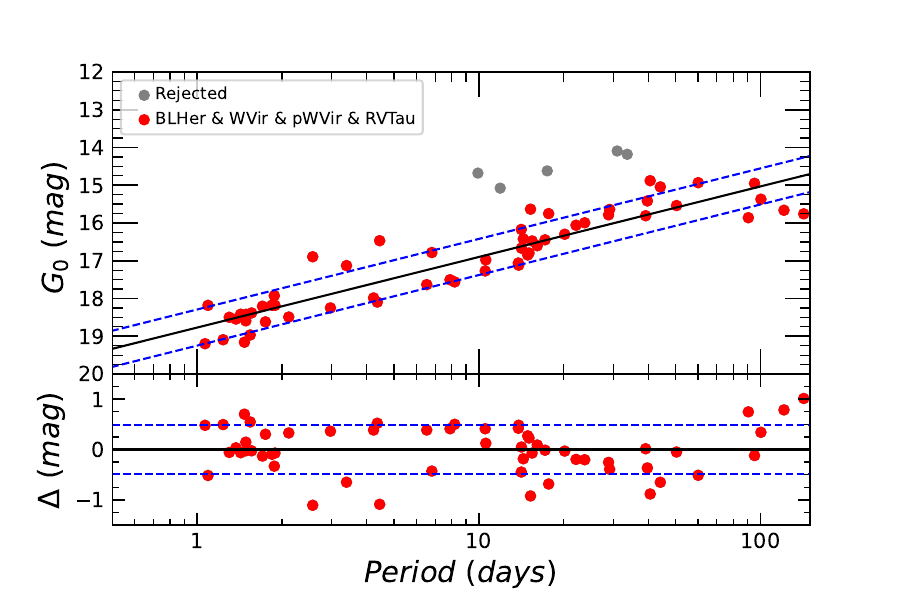}
    }
    \hbox{
    \includegraphics[width=0.34\textwidth]{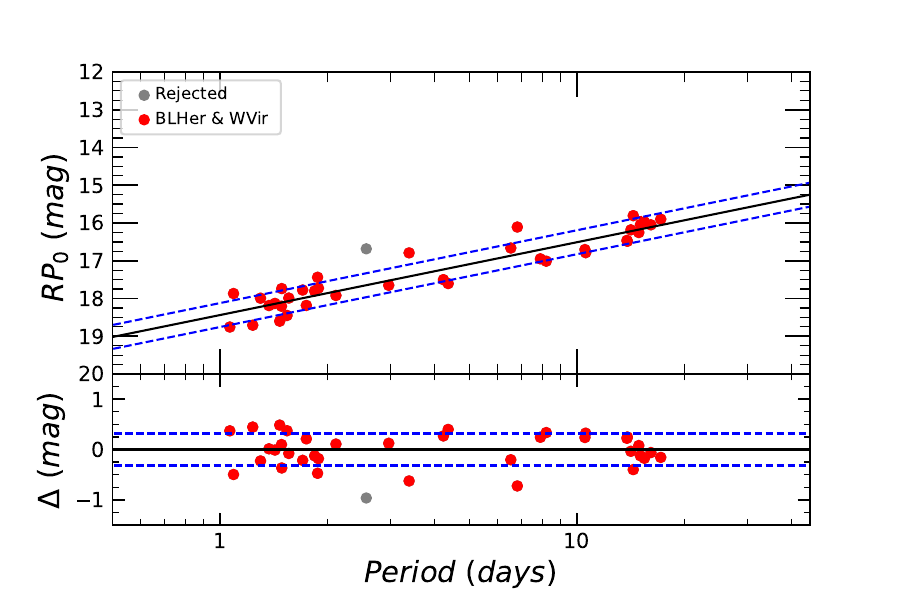}
    \includegraphics[width=0.34\textwidth]{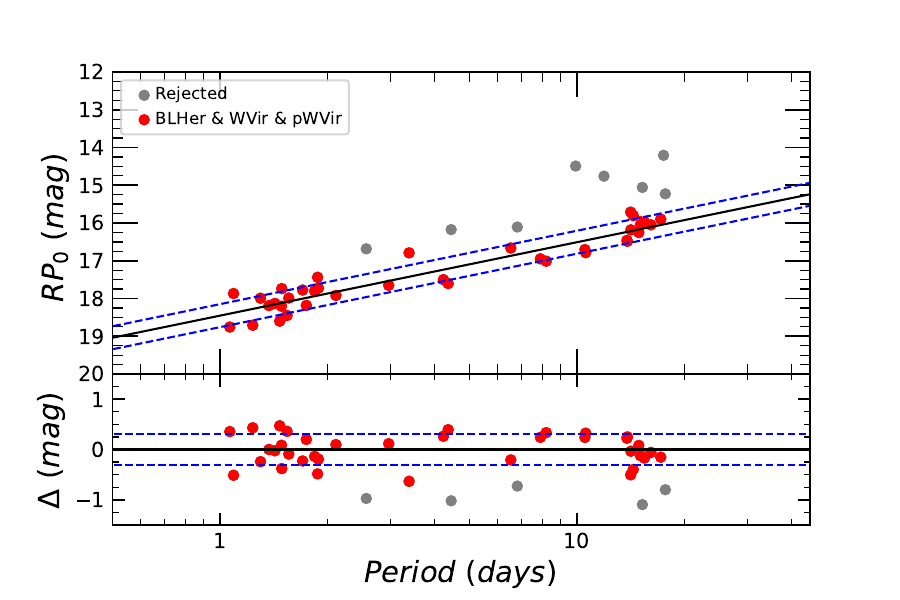}
    \includegraphics[width=0.34\textwidth]{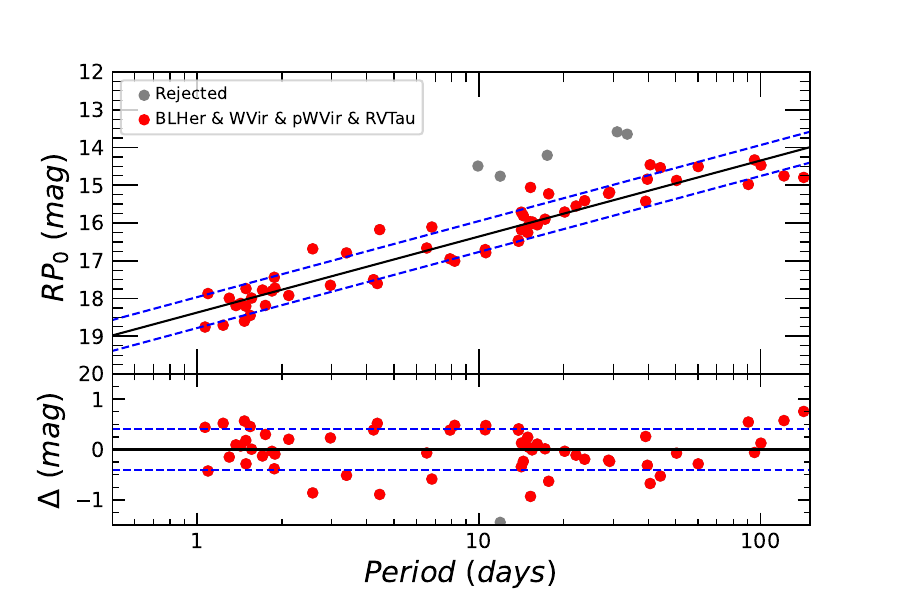}
    }
    }
    \caption{\label{fit1smc} Same as Fig.\ref{fit1lmc} but for SMC. }
	\end{figure*}

\begin{figure*}[h]
    \vbox{
    \hbox{
    \includegraphics[width=0.34\textwidth]{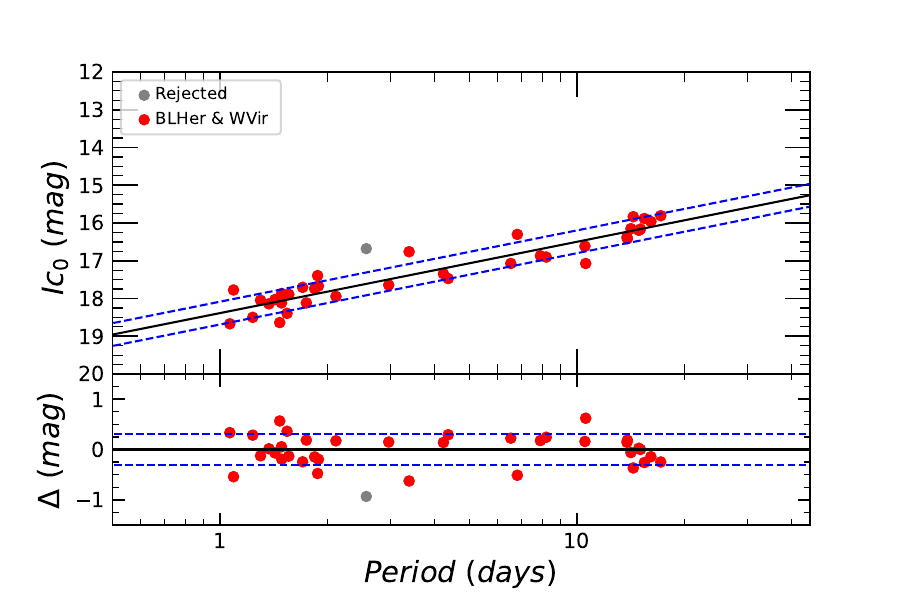}
    \includegraphics[width=0.34\textwidth]{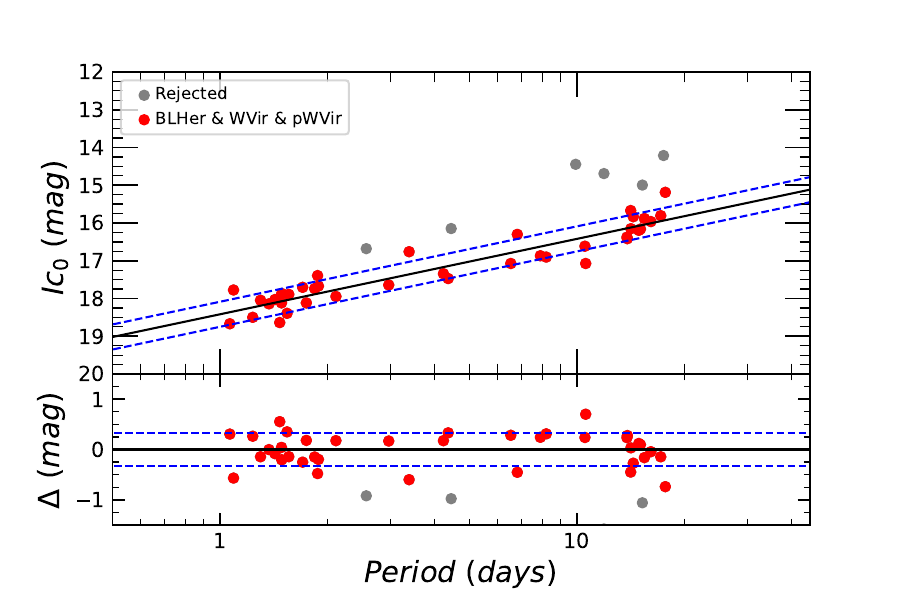}
    \includegraphics[width=0.34\textwidth]{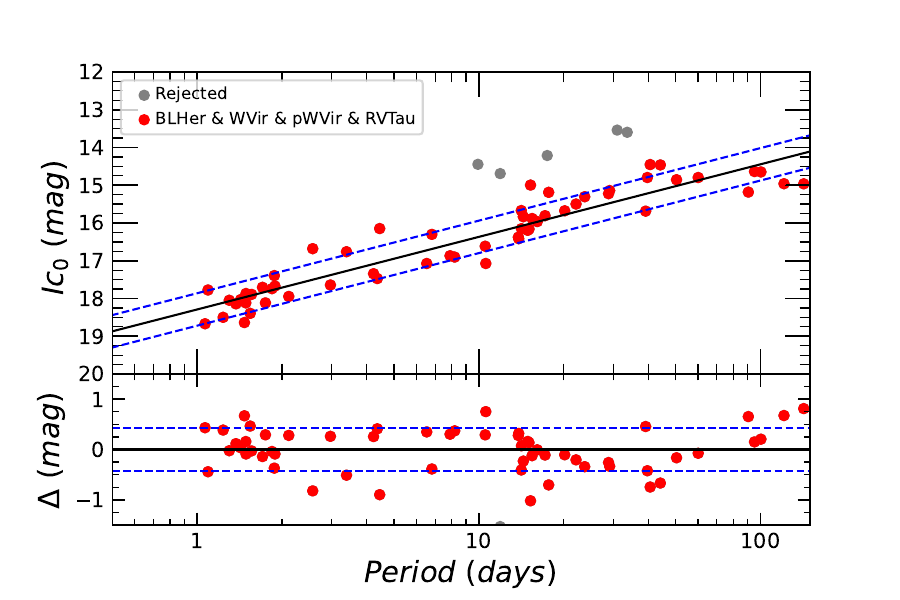}
    }
    \hbox{
    \includegraphics[width=0.34\textwidth]{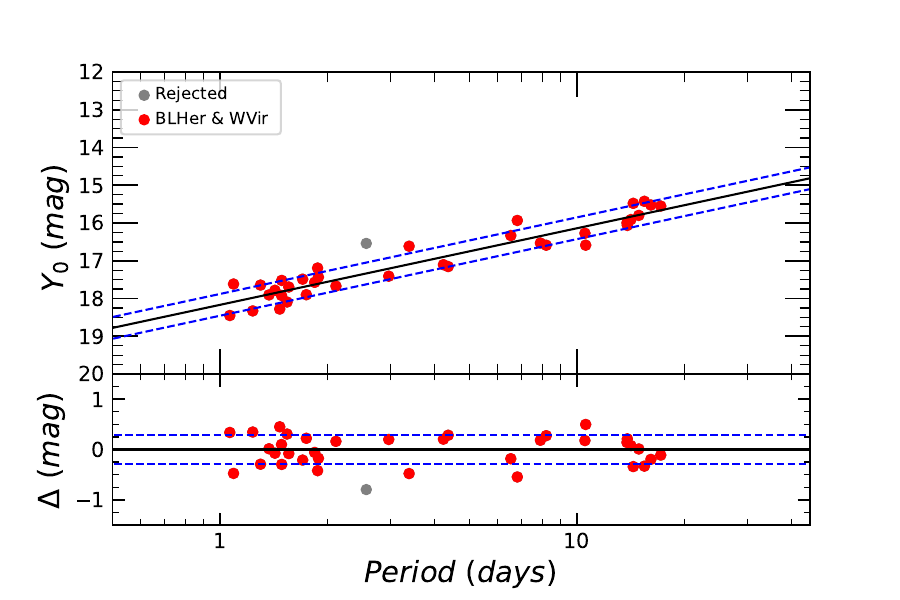}
    \includegraphics[width=0.34\textwidth]{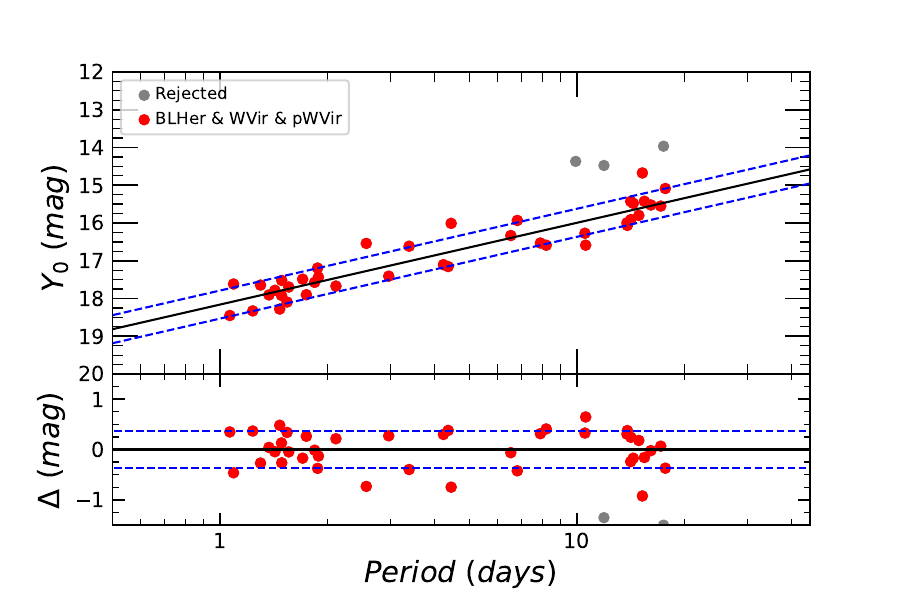}
    \includegraphics[width=0.34\textwidth]{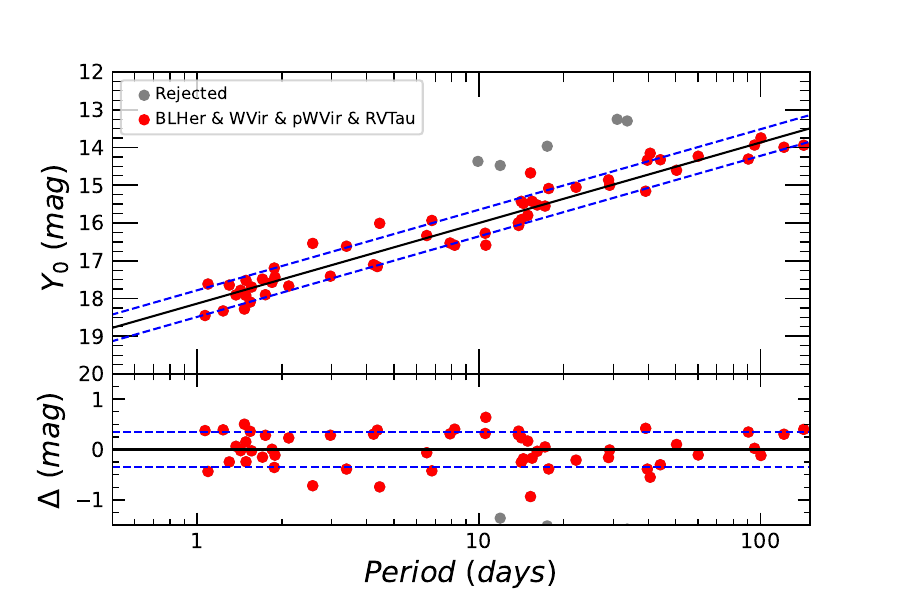}
    }
    \hbox{
    \includegraphics[width=0.34\textwidth]{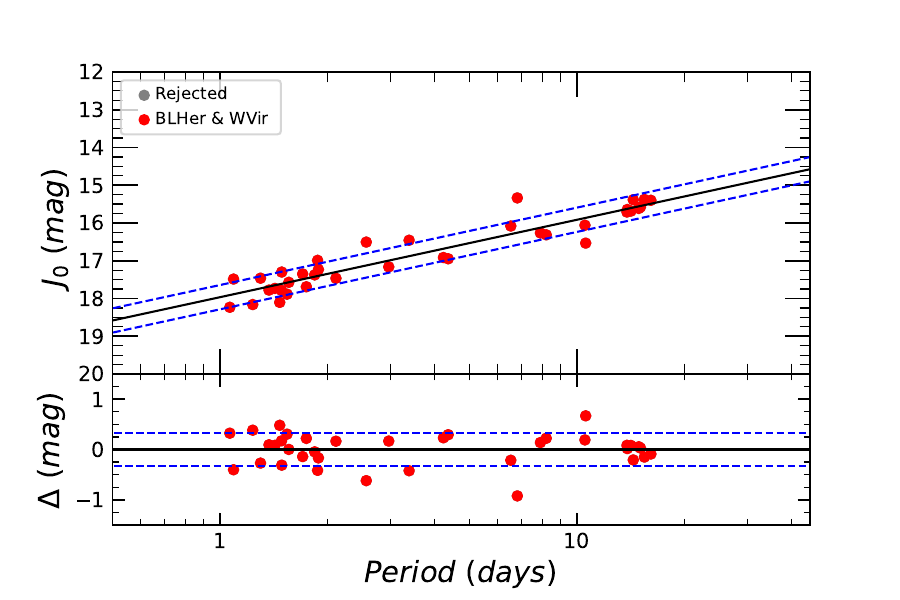}
    \includegraphics[width=0.34\textwidth]{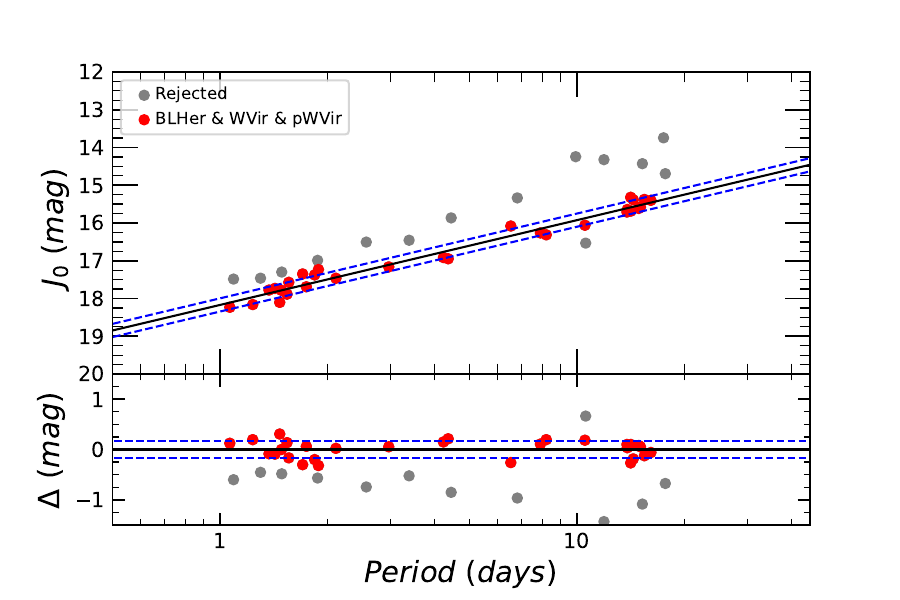}
    \includegraphics[width=0.34\textwidth]{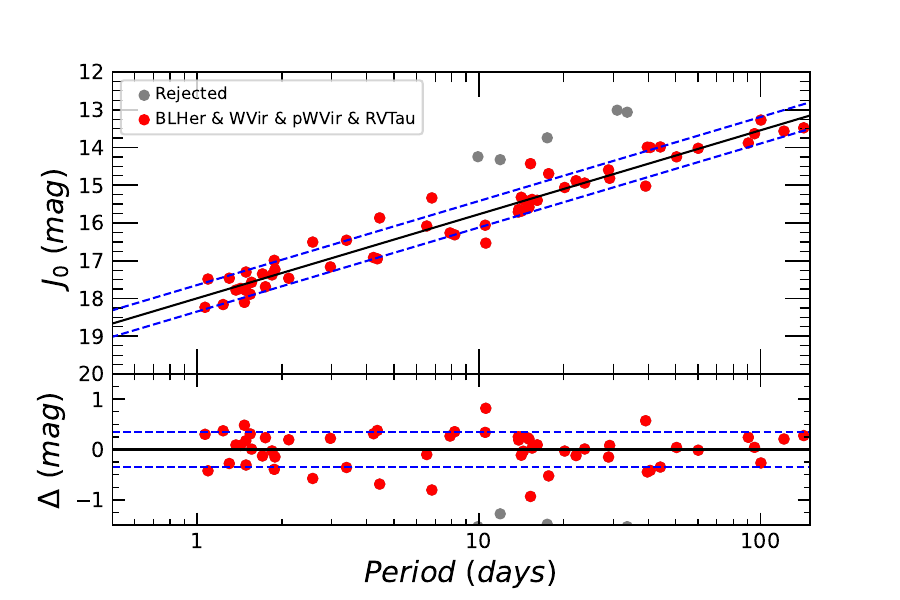}
    }
    \hbox{
    \includegraphics[width=0.34\textwidth]{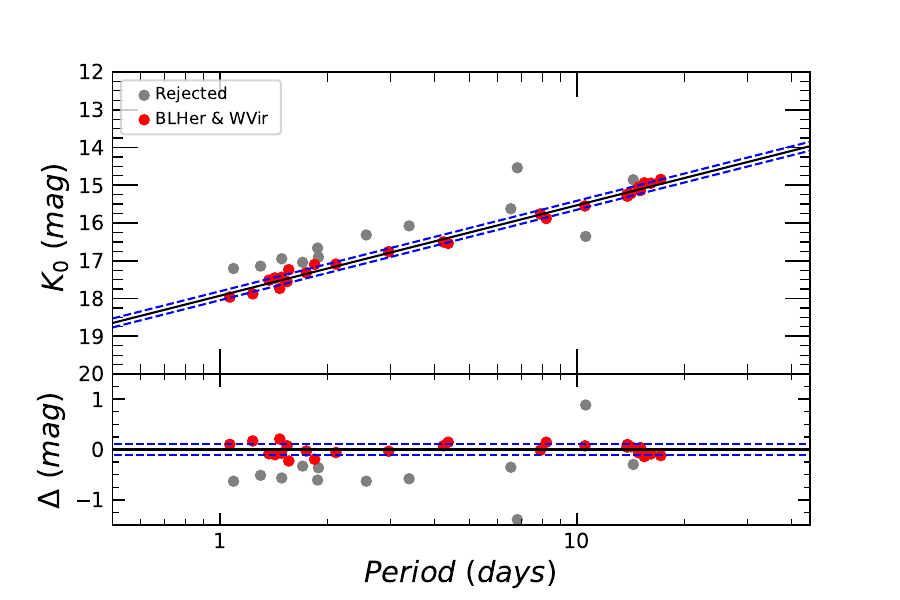}
    \includegraphics[width=0.34\textwidth]{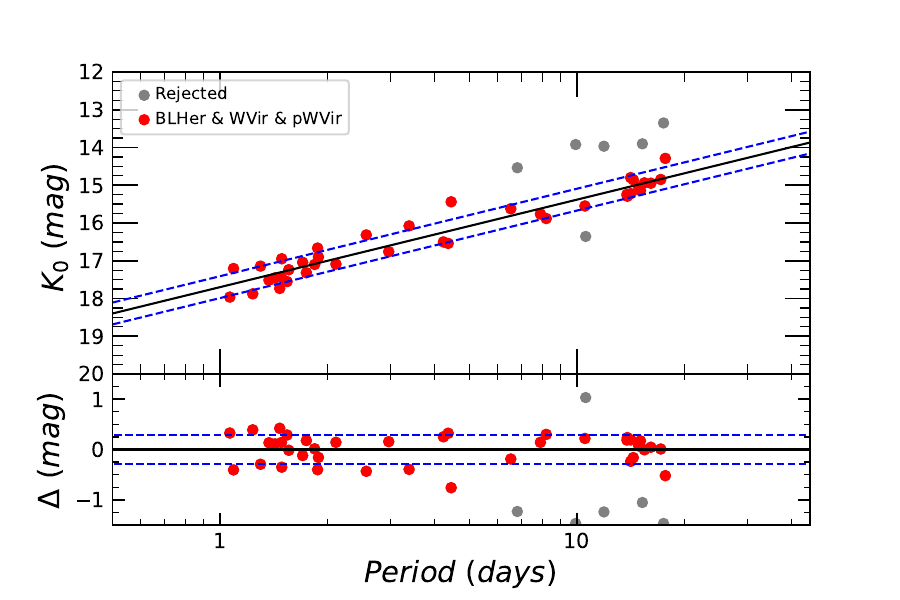}
    \includegraphics[width=0.34\textwidth]{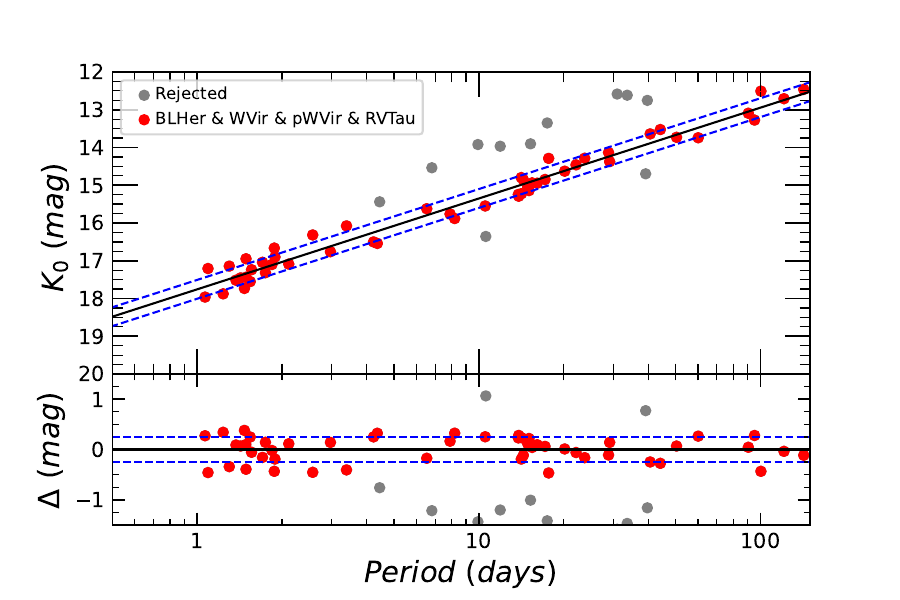}
    }
    \hbox{
    \includegraphics[width=0.34\textwidth]{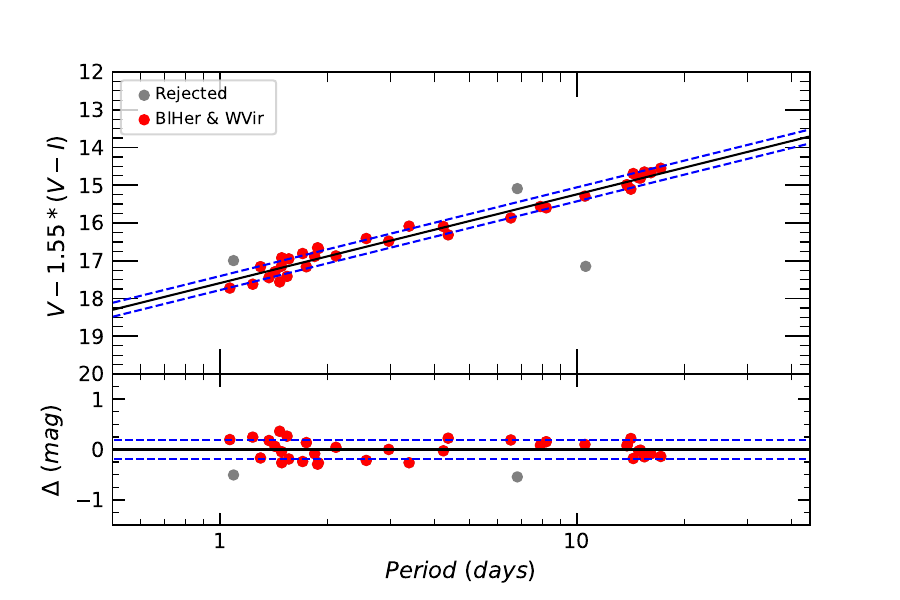}
    \includegraphics[width=0.34\textwidth]{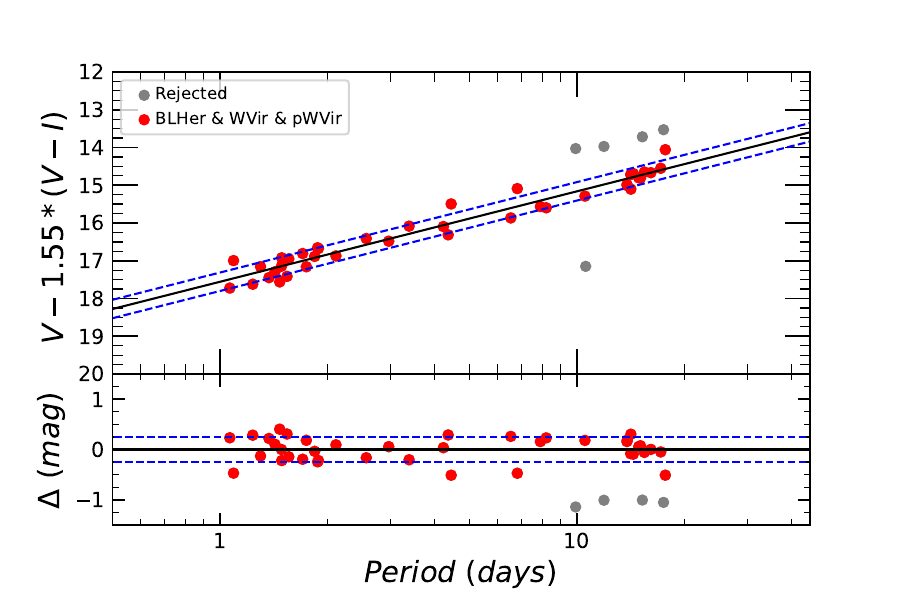}
    \includegraphics[width=0.34\textwidth]{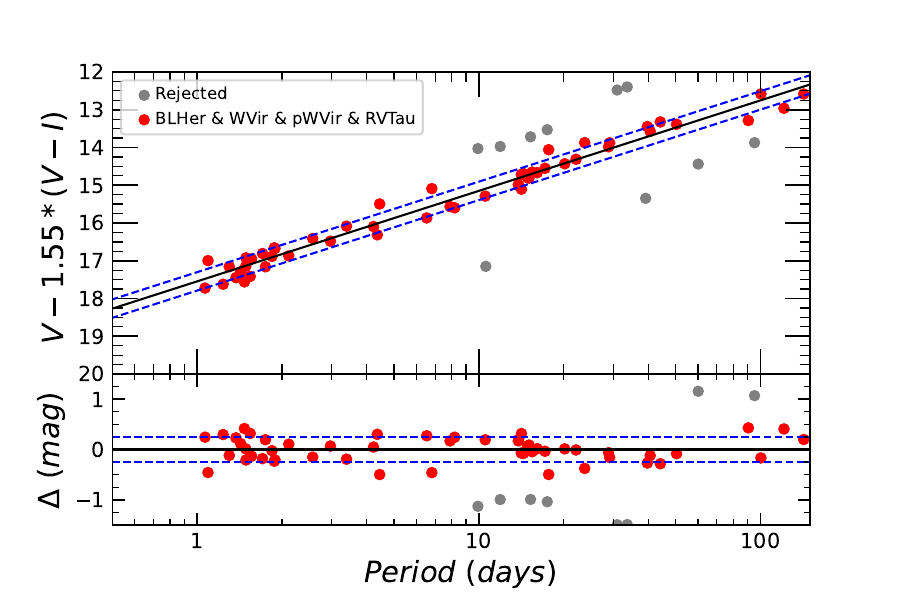}
    }
    \hbox{
    \includegraphics[width=0.34\textwidth]{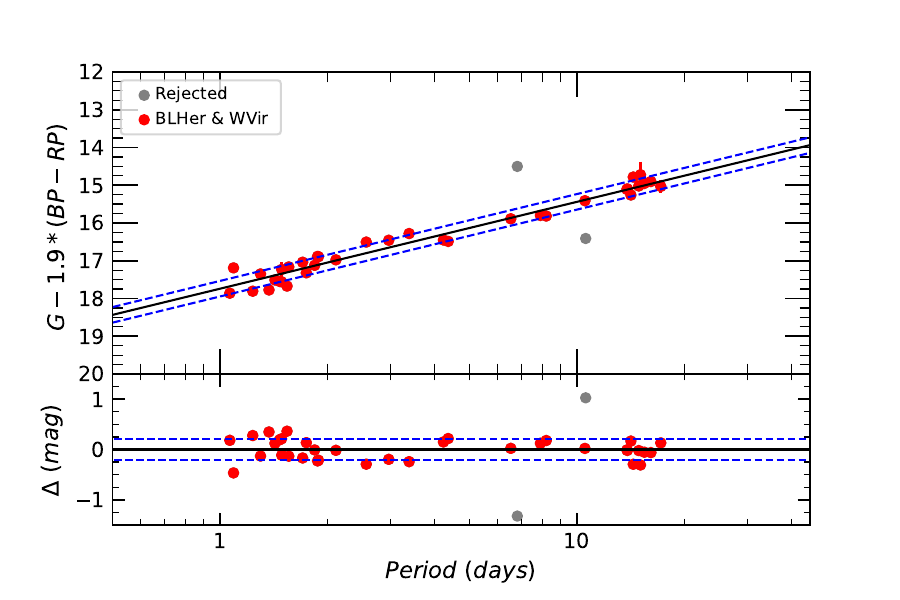}
    \includegraphics[width=0.34\textwidth]{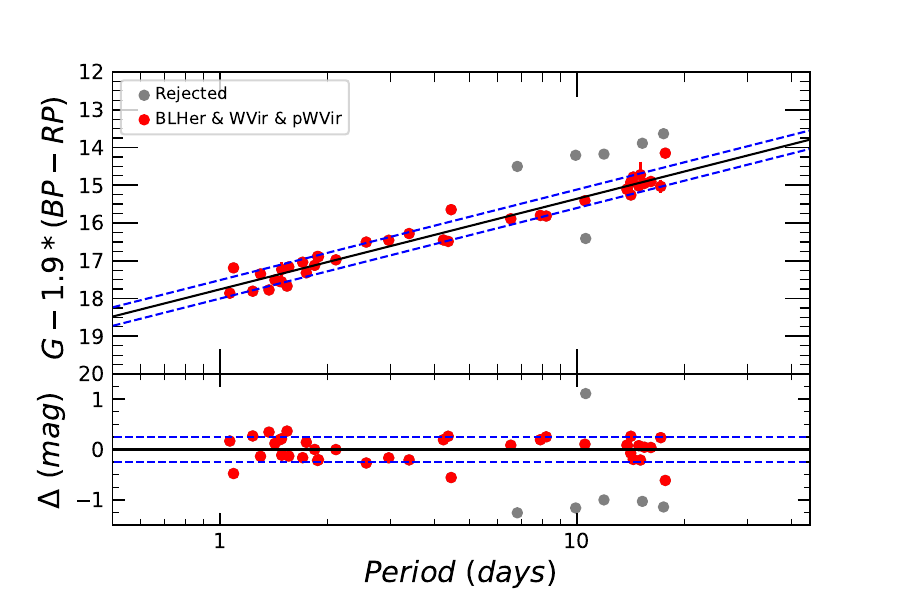}
    \includegraphics[width=0.34\textwidth]{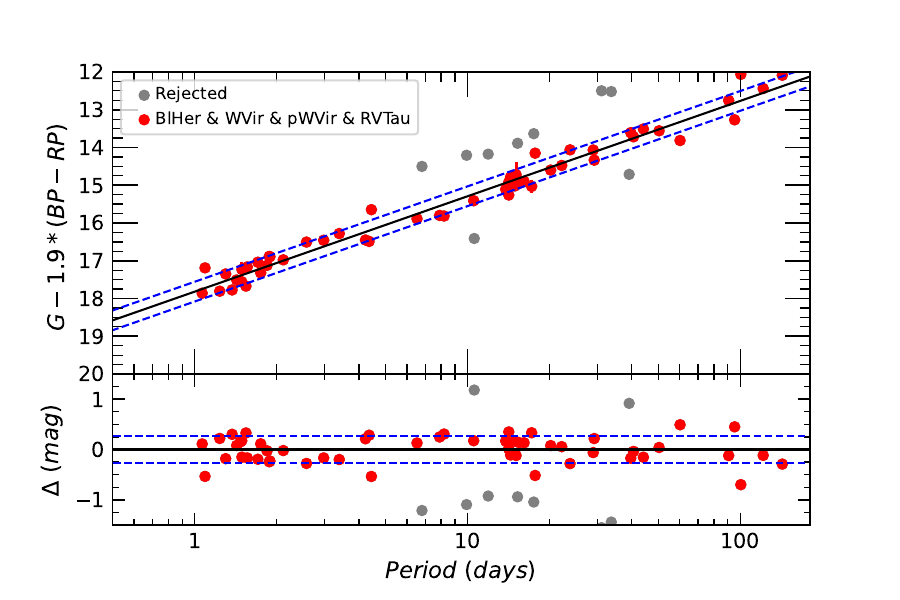}
    }
    }
    \ContinuedFloat
    \caption{continued. }	
	\end{figure*}

\begin{figure*}[h]
    \vbox{
    \hbox{
    \includegraphics[width=0.34\textwidth]{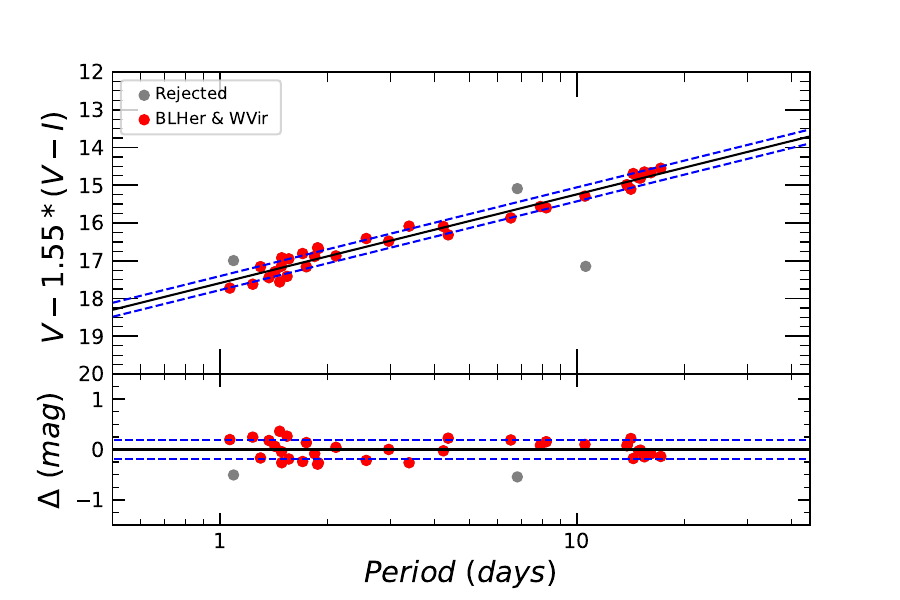}
    \includegraphics[width=0.34\textwidth]{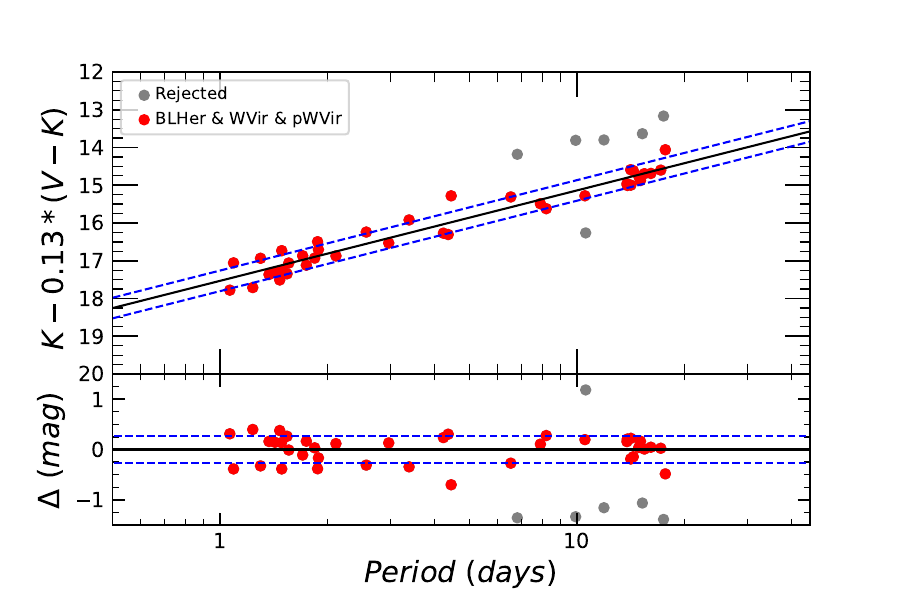}
    \includegraphics[width=0.34\textwidth]{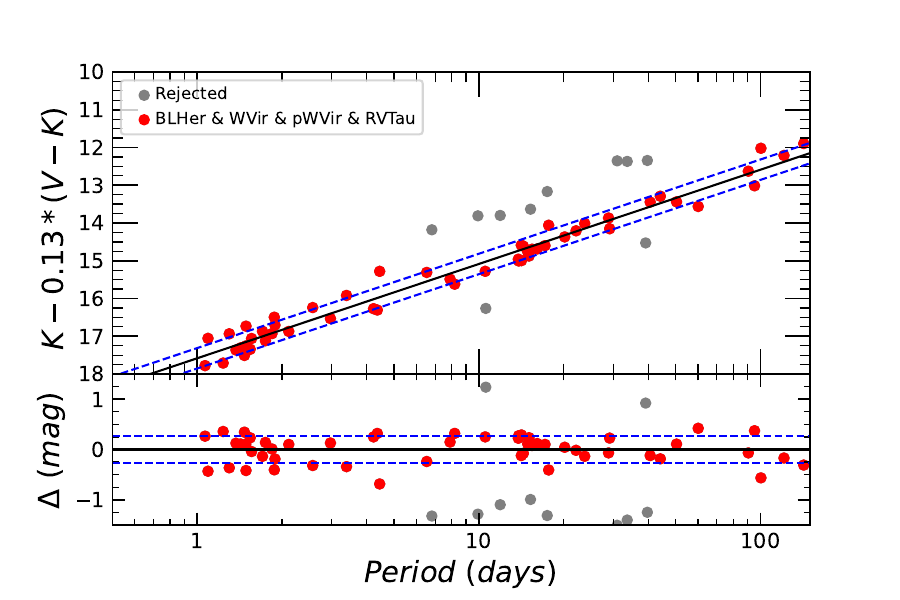}
    }
    \hbox{
    \includegraphics[width=0.34\textwidth]{PWYK1smc.pdf}
    \includegraphics[width=0.34\textwidth]{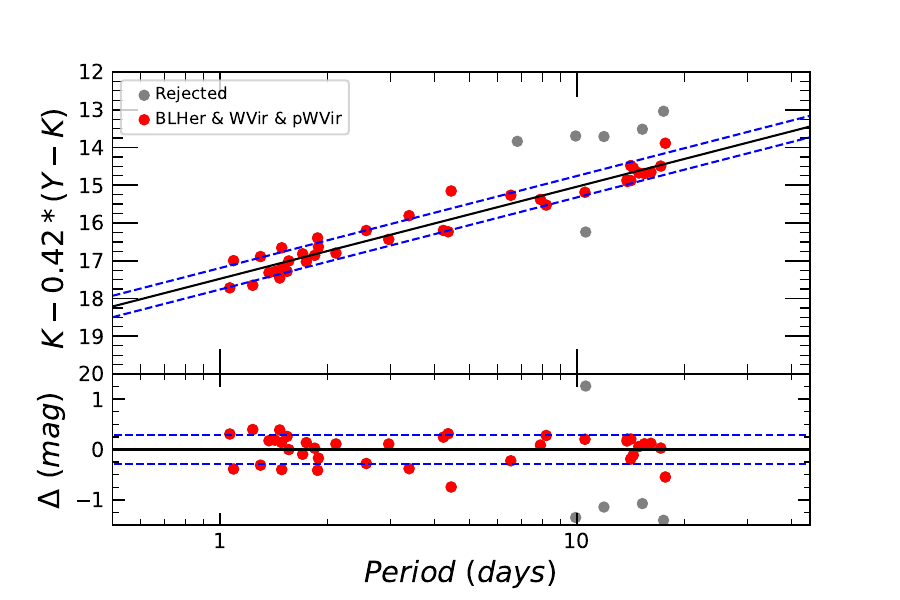}
    \includegraphics[width=0.34\textwidth]{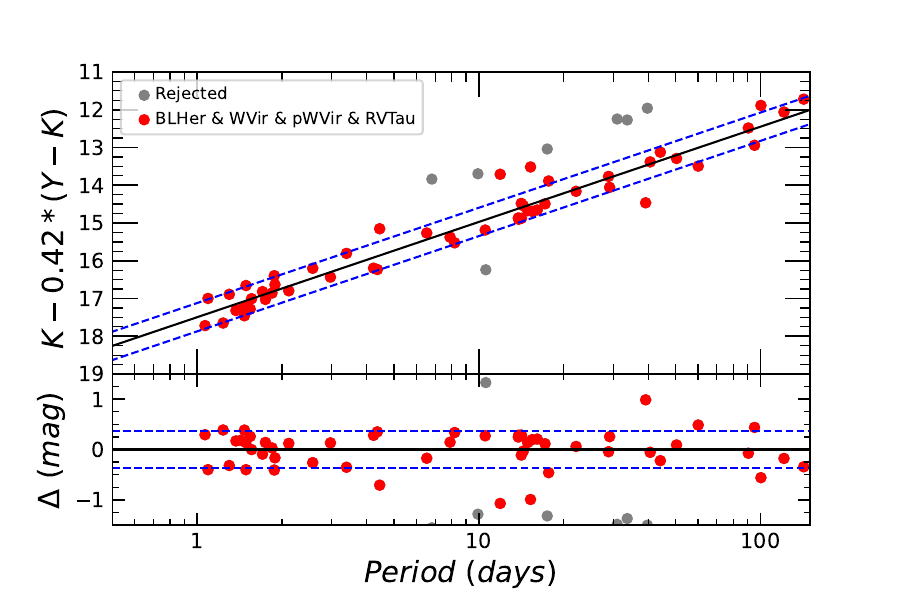}
    }
    \hbox{
    \includegraphics[width=0.34\textwidth]{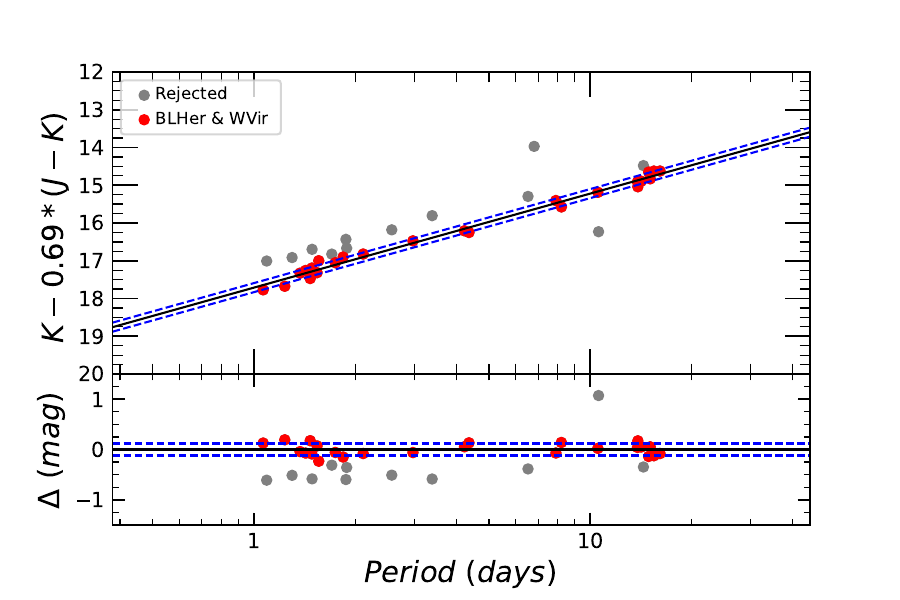}
    \includegraphics[width=0.34\textwidth]{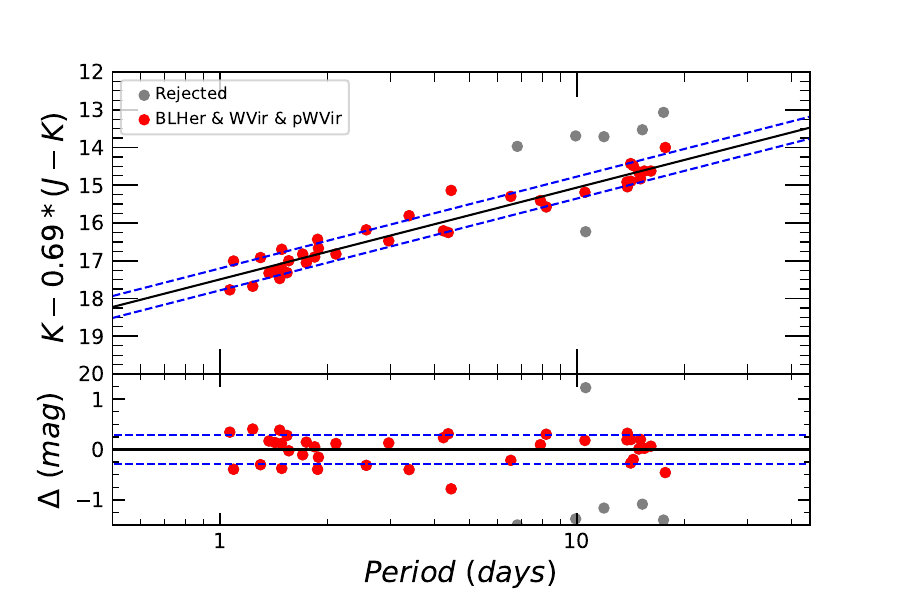}
    \includegraphics[width=0.34\textwidth]{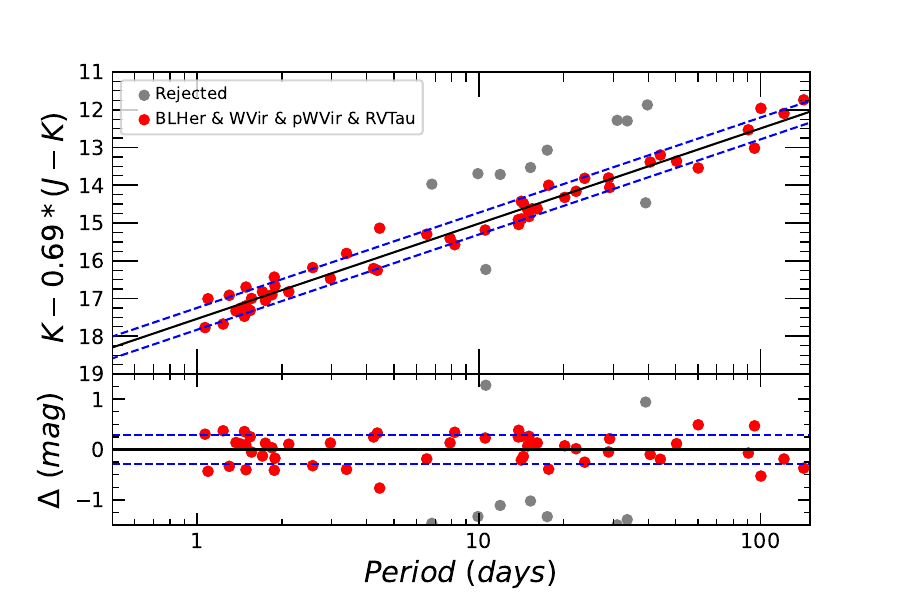}
    }
    \hbox{
    \includegraphics[width=0.34\textwidth]{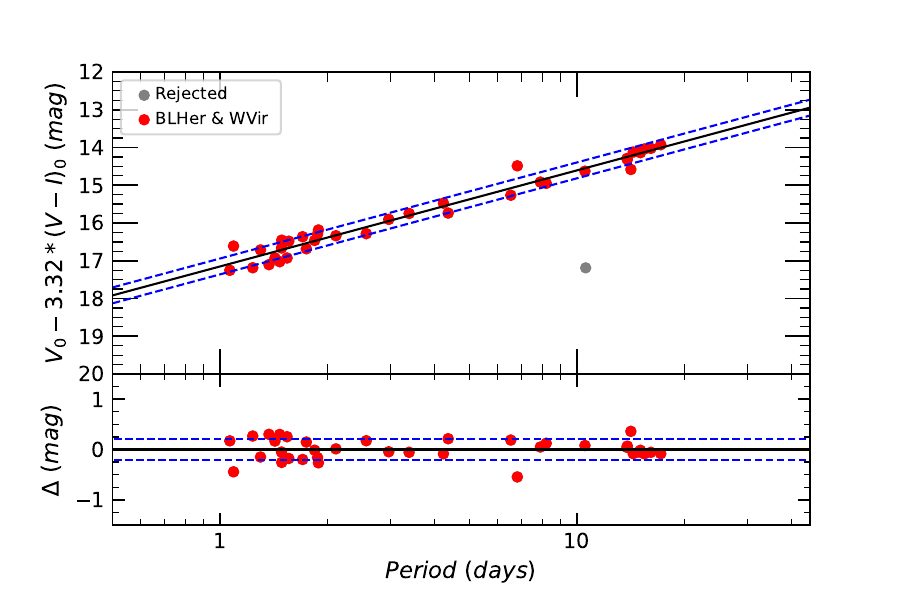}
    \includegraphics[width=0.34\textwidth]{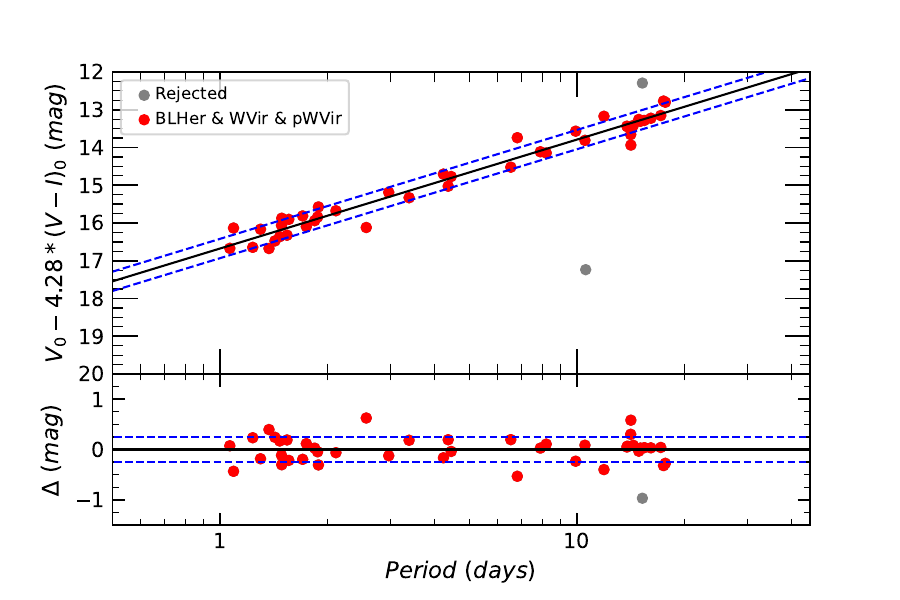}
    \includegraphics[width=0.34\textwidth]{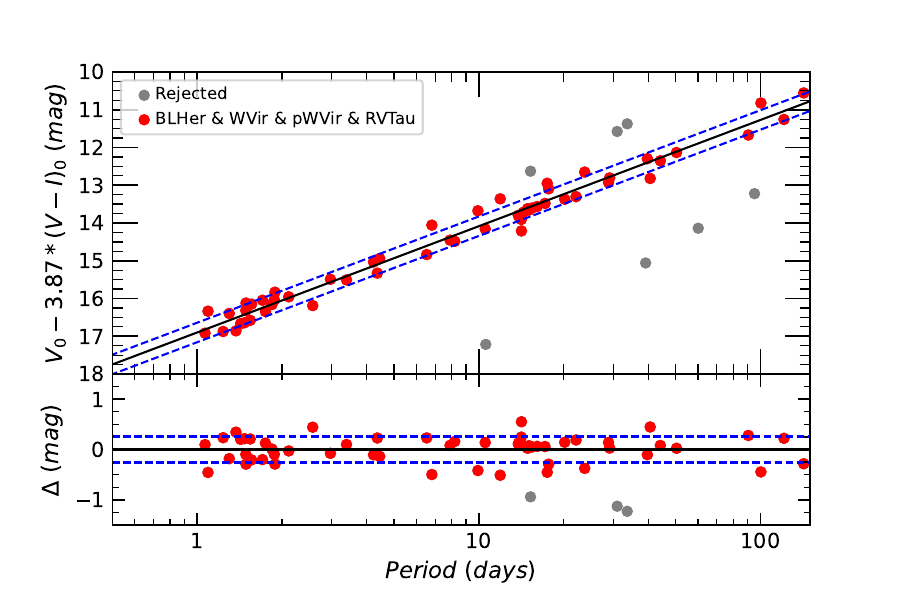}
    }
    \hbox{
    \includegraphics[width=0.34\textwidth]{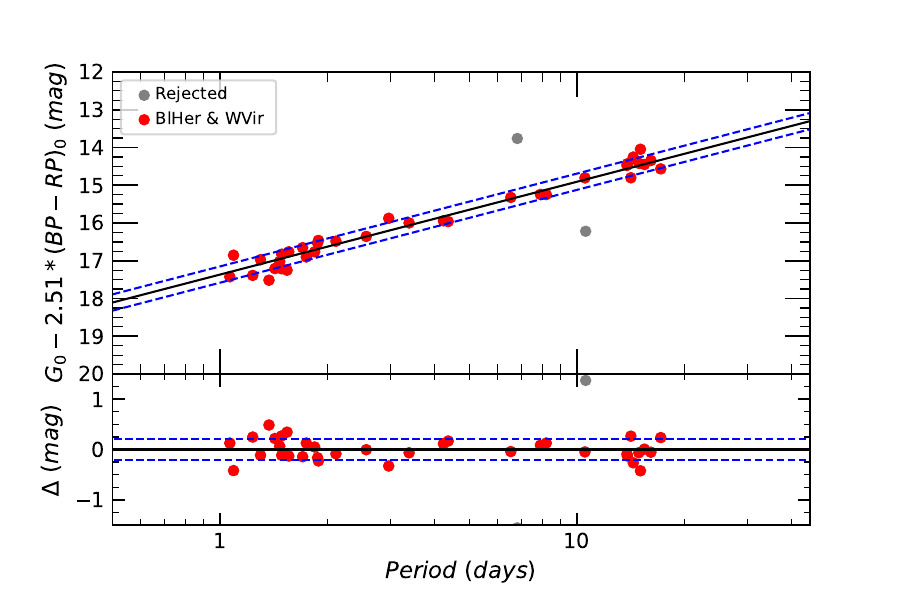}
    \includegraphics[width=0.34\textwidth]{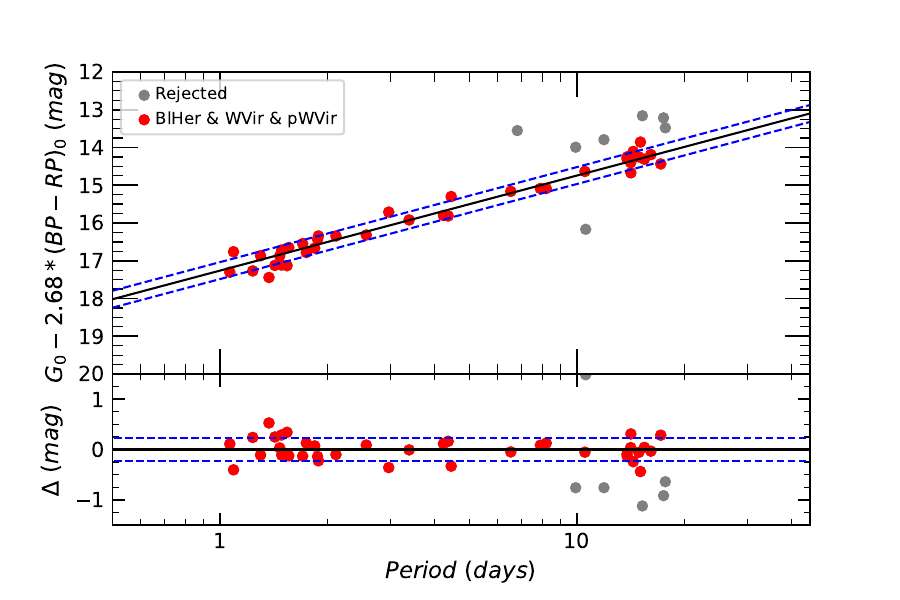}
    \includegraphics[width=0.34\textwidth]{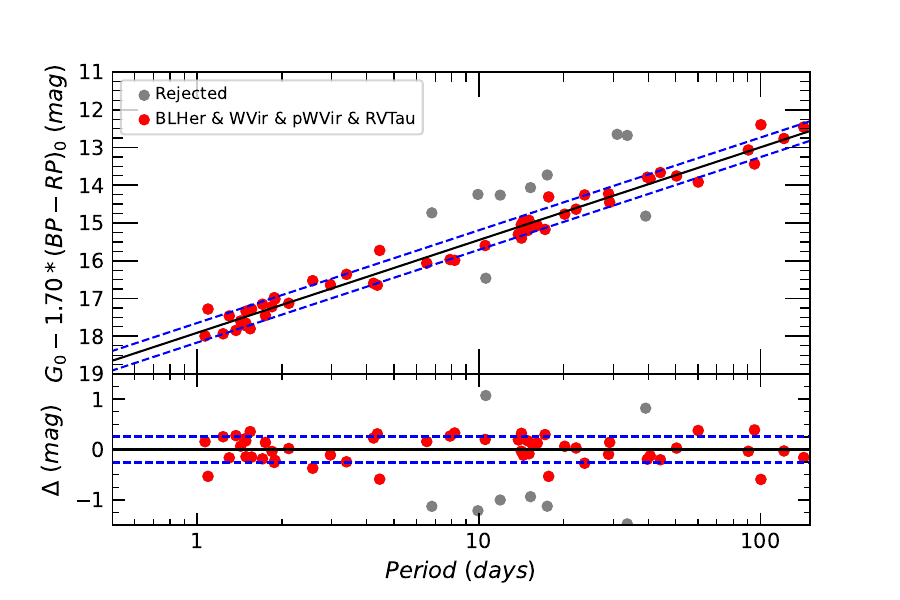}
    }
    \hbox{
    \includegraphics[width=0.34\textwidth]{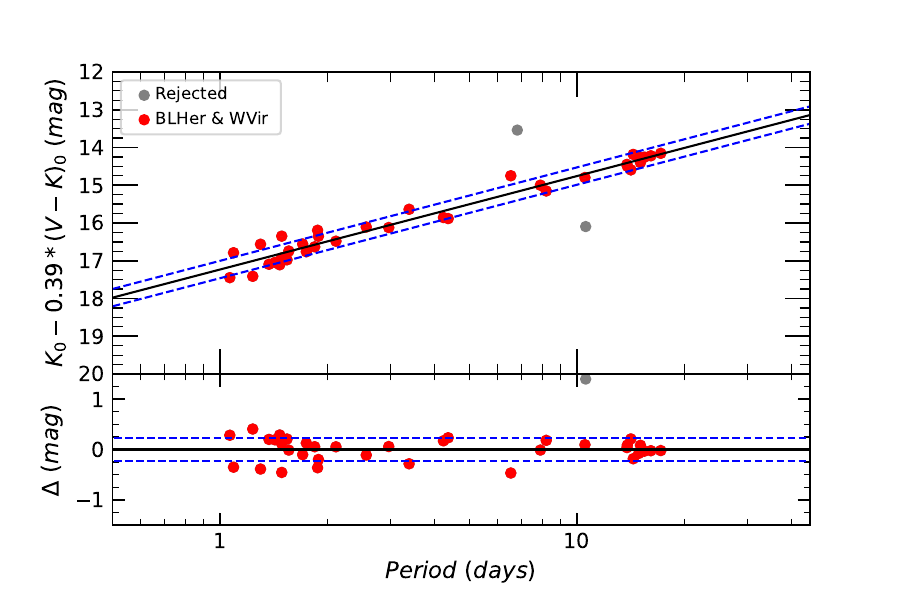}
    \includegraphics[width=0.34\textwidth]{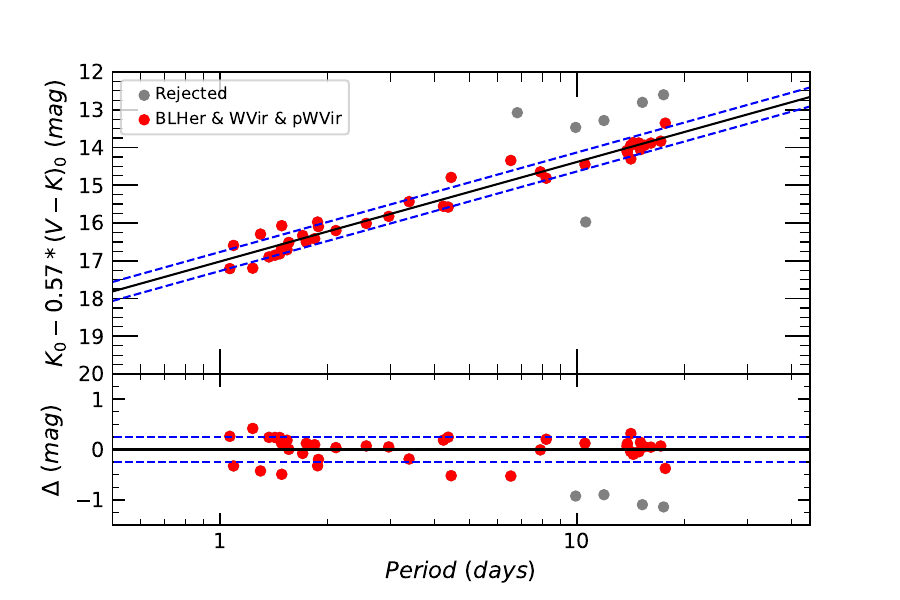}
    \includegraphics[width=0.34\textwidth]{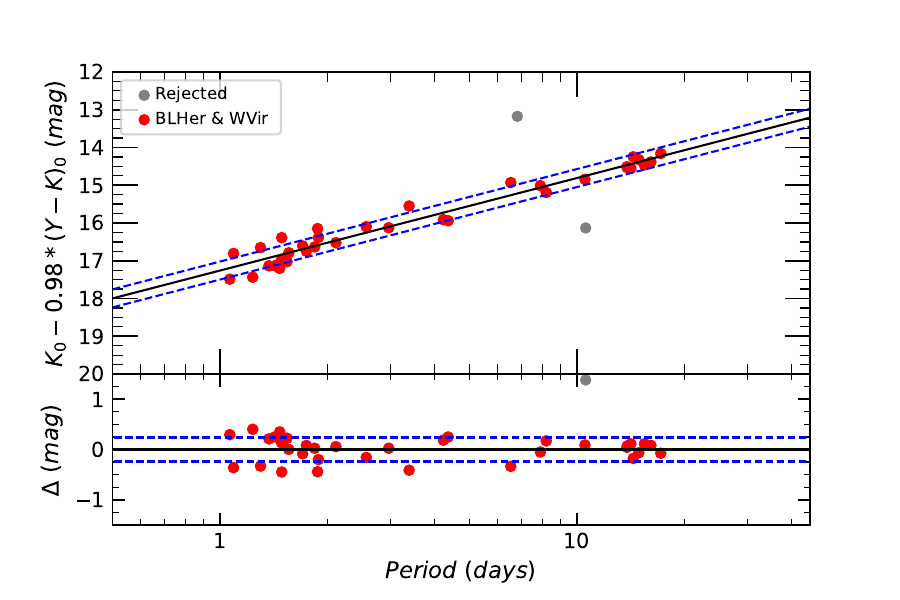}
    }
    }
    \ContinuedFloat
    \caption{continued. }	
	\end{figure*}

\begin{figure*}[h]
    \vbox{
    \hbox{
    \includegraphics[width=0.34\textwidth]{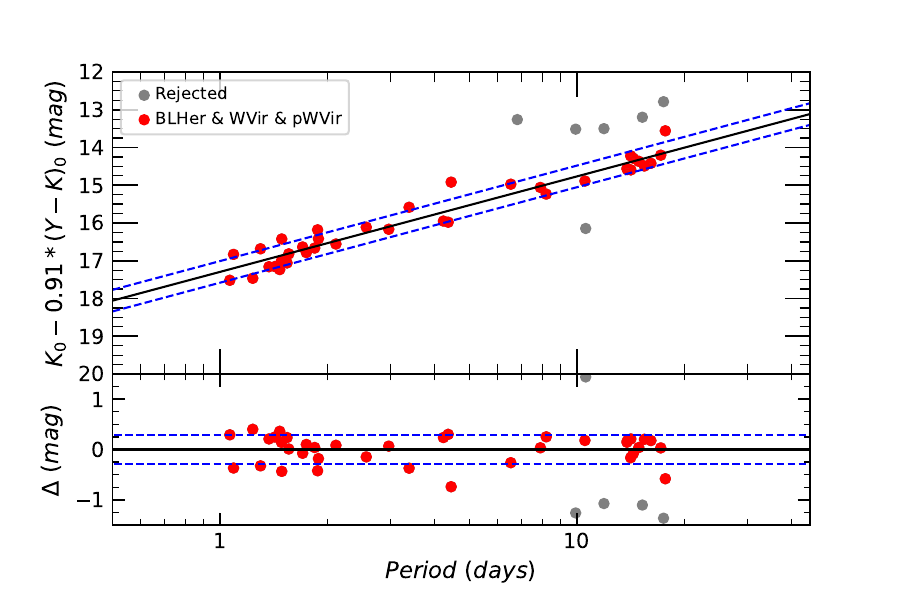}
    \includegraphics[width=0.34\textwidth]{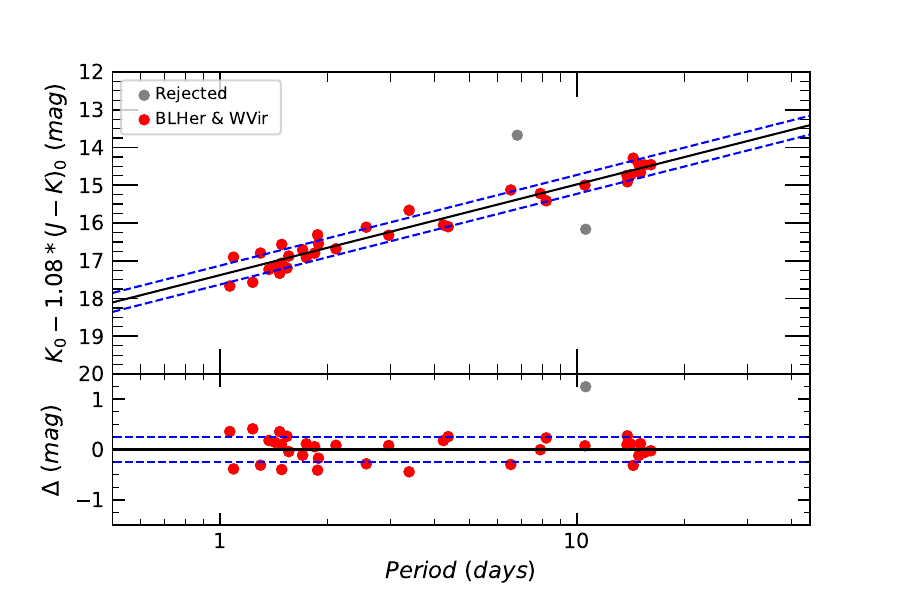}
    \includegraphics[width=0.34\textwidth]{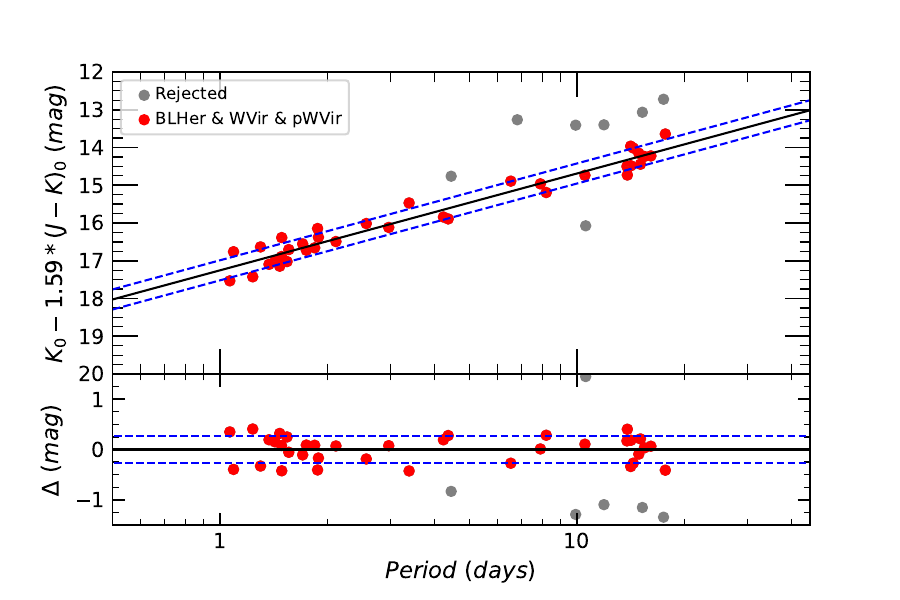}
    }
    }
    \ContinuedFloat
    \caption{continued. }	
	\end{figure*}

\FloatBarrier

\subsection{Tables}\label{app:pl_Tables}

\begin{table*}
\footnotesize\setlength{\tabcolsep}{3pt} 
\caption{Continued from Table~\ref{tab:lmcT2}}
  \label{tab:lmcT2pt2}
\begin{tabular}{llllllllllll}
 \hline  
 \noalign{\smallskip} 
       Relation  &  Group & $\alpha $ &  $\sigma_{\alpha}$ & $\beta$ & $\sigma_{\beta} $ &  $\gamma$ & $\sigma_{\gamma}$ & RMS & Used stars & Total stars & Notes  \\
   &  & mag & mag &  &  & & &  & &      \\                     
(1)  & (2)  & (3) & (4) & (5) & (6) & (7) &(8) & (9) & (10) & (11) & (12)\\   
 \noalign{\smallskip}
 \hline  
 \noalign{\smallskip}
  PLBP & BLH\&WVir\&pWVir & 18.504 & 0.020 & $-$1.470 & 0.049 &  &  & 0.28 & 184 & 211\\
  PLBP & BLH\&WVir\&pWVir\&RVTau & 18.507 & 0.037 & $-$1.761 & 0.066 &  &  & 0.57 & 270 & 270\\
  PLG & BLH\&WVir\&pWVir & 18.454 & 0.015 & $-$1.748 & 0.036 & &  & 0.20 & 187 & 213\\
  PLG & BLH\&WVir\&pWVir\&RVTau & 18.477 & 0.016 & $-$1.825 & 0.032 &  &  & 0.24 & 221 & 272\\
  PLRP & BLH\&WVir\&pWVir & 18.011 & 0.015 & $-$1.891 & 0.037 & & & 0.21 & 181 & 211\\
  PLRP & BLH\&WVir\&pWVir\&RVTau & 18.033 & 0.024 & $-$2.078 & 0.042 &  &  & 0.36 & 261 & 270\\
  PLV & BLH\&WVir\&pWVir & 18.520 & 0.017 & $-$1.655 & 0.040 & & & 0.23 & 189 & 214\\
  PLV & BLH\&WVir\&pWVir\&RVTau & 18.599 & 0.028 & $-$1.931 & 0.052 & & & 0.43 & 260 & 273\\
  PLI & BLH\&WVir\&pWVir & 18.019 & 0.014 & $-$1.972 & 0.034 & & & 0.20 & 198 & 214\\
  PLI & BLH\&WVir\&pWVir\&RVTau & 18.112 & 0.020 & $-$2.188 & 0.036 & & & 0.30 & 256 & 273\\
  PLY & BLH\&WVir\&pWVir & 17.819 & 0.013 & $-$2.067 & 0.031 & & & 0.17 & 172 & 202\\
  PLY & BLH\&WVir\&pWVir\&RVTau & 17.867 & 0.019 & $-$2.254 & 0.036 & & & 0.28 & 232 & 252\\
  PLJ & BLH\&WVir\&pWVir & 17.642 & 0.012 & $-$2.168 & 0.029 & & & 0.16 & 181 & 206\\
  PLJ & BLH\&WVir\&pWVir\&RVTau & 17.688 & 0.012 & $-$2.258 & 0.024 & & & 0.18 & 215 & 264\\
  PLK & BLH\&WVir\&pWVir & 17.401 & 0.011 & $-$2.379 & 0.027 & & & 0.15 & 189 & 212\\
  PLK & BLH\&WVir\&pWVir\&RVTau & 17.457 & 0.012 & $-$2.481 & 0.021 & & & 0.17 & 230 & 272\\
  PWG & BLH\&WVir\&pWVir & 17.443 & 0.010 & $-$2.450 & 0.026 & & & 0.16 & 192 & 211\\
  PWG & BLH\&WVir\&pWVir\&RVTau & 17.535 & 0.011 & $-$2.595 & 0.021 & & & 0.17 & 241 & 270\\
  PWVI & BLH\&WVir\&pWVir & 17.334 & 0.010 & $-$2.514 & 0.025 & & & 0.14 & 196 & 214\\
  PWVI & BLH\&WVir\&pWVir\&RVTau & 17.358 & 0.009 & $-$2.570 & 0.018 &  & & 0.13 & 229 & 273\\
  PWVK & BLH\&WVir\&pWVir & 17.274 & 0.007 & $-$2.475 & 0.018 & & & 0.09 & 170 & 212\\
  PWVK & BLH\&WVir\&pWVir\&RVTau & 17.308 & 0.008 & $-$2.540 & 0.016 & & & 0.11 & 200 & 271\\
  PWYK & BLH\&WVir\&pWVir & 17.209 & 0.008 & $-$2.519 & 0.020 & & & 0.10 & 169 & 202\\
  PWYK & BLH\&WVir\&pWVir\&RVTau & 17.253 & 0.009 & $-$2.595 & 0.018 & & & 0.12 & 193 & 252\\
  PWJK & BLH\&WVir\&pWVir & 17.245 & 0.009 & $-$2.519 & 0.023 & & & 0.12 & 174 & 206\\
  PWJK & BLH\&WVir\&pWVir\&RVTau & 17.294 & 0.011 & $-$2.622 & 0.020 & & & 0.15 & 216 & 264\\
  PLCG & BLHer & 17.374 & 0.050 & $-$2.520 & 0.110 & 2.007 & 0.085 & 0.20 & 73 & 83\\
  PLCG & WVir & 17.328 & 0.027 & $-$2.572 & 0.062 & 2.160 & 0.120 & 0.10 & 95 & 103\\
  PLCG & BLH\&WVir\&pWVir & 17.329 & 0.011 & $-$2.527 & 0.041 & 2.078 & 0.077 & 0.18 & 199 & 211\\
  PLCG & BLH\&WVir\&pWVir\&RVTau & 17.415 & 0.013 & $-$2.666 & 0.028 & 2.078 & 0.056 & 0.2 & 254 & 270\\
  PLCVI & BLHer & 17.528 & 0.053 & $-$2.460 & 0.110 & 2.150 & 0.120 & 0.12 & 76 & 85\\
  PLCVI & WVir & 16.848 & 0.039 & $-$2.667 & 0.066 & 3.350 & 0.160 & 0.09 & 98 & 104\\
  PLCVI & BLH\&WVir\&pWVir & 17.164 & 0.010 & $-$2.612 & 0.031 & 2.886 & 0.073 & 0.11 & 187 & 214\\
  PLCVI & BLH\&WVir\&pWVir\&RVTau & 17.072 & 0.012 & $-$2.717 & 0.026 & 3.106 & 0.067 & 0.16 & 242 & 273\\
  PLCVK & BLHer & 17.316 & 0.043 & $-$2.611 & 0.093 & 0.136 & 0.041 & 0.10 & 63 & 84\\
  PLCVK & WVir & 17.344 & 0.019 & $-$2.560 & 0.050 & 0.158 & 0.048 & 0.07 & 95 & 103\\
  PLCVK & BLH\&WVir\&pWVir & 17.197 & 0.008 & $-$2.537 & 0.030 & 0.211 & 0.030 & 0.11 & 176 & 212\\
  PLCVK & BLHer\&WVir\&pWVir\&RVTau & 17.361 & 0.013 & $-$2.568 & 0.031 & 0.097 & 0.031 & 0.18 & 236 & 272\\
  PLCYK & BLHer & 17.312 & 0.048 & $-$2.620 & 0.110 & 0.345 & 0.080 & 0.10 & 62 & 77\\
  PLCYK & WVir & 17.464 & 0.022 & $-$2.497 & 0.051 & 0.150 & 0.092 & 0.08 & 93 & 100\\
  PLCYK & BLH\&WVir\&pWVir & 17.265 & 0.010 & $-$2.456 & 0.033 & 0.315 & 0.073 & 0.13 & 177 & 202\\
  PLCYK & BLH\&WVir\&pWVir\&RVTau & 17.349 & 0.012 & $-$2.575 & 0.03 & 0.280 & 0.072 & 0.17 & 214 & 252\\
  PLCJK & BLHer & 17.330 & 0.043 & $-$2.704 & 0.097 & 0.610 & 0.150 & 0.10 & 62 & 83\\
  PLCJK & WVir & 17.499 & 0.017 & $-$2.536 & 0.047 & 0.250 & 0.120 & 0.07 & 88 & 98\\
  PLCJK & BLH\&WVir\&pWVir & 17.381 & 0.012 & $-$2.413 & 0.041 & 0.110 & 0.140 & 0.15 & 184 & 206\\
  PLCJK & BLH\&WVir\&pWVir\&RVTau & 17.421 & 0.013 & $-$2.530 & 0.030 & 0.180 & 0.110 & 0.17 & 225 & 264\\

\noalign{\smallskip}
\hline  
\noalign{\smallskip}
\end{tabular}
\end{table*}

\begin{table*}
\footnotesize\setlength{\tabcolsep}{3pt} 
\caption{Continued from Table~\ref{tab:smcT2}.}
  \label{tab:smcT2pt2}
\begin{tabular}{llllllllllll}
 \hline  
 \noalign{\smallskip} 
       Relation  &  Group & $\alpha $ &  $\sigma_{\alpha}$ & $\beta$ & $\sigma_{\beta} $ &  $\gamma$ & $\sigma_{\gamma}$ & RMS & Used stars & Total stars & Notes  \\
   &  & mag & mag &  &  & & &  & &      \\                     
(1)  & (2)  & (3) & (4) & (5) & (6) & (7) &(8) & (9) & (10) & (11) & (12)\\   
 \noalign{\smallskip}
 \hline  
 \noalign{\smallskip}
  PLBP & BLHer\&WVir\&pWVir & 19.084 & 0.070 & $-$1.850 & 0.160 &  & & 0.43 & 39 & 44&\\
  PLBP & BLHer\&WVir\&pWVir\&RVTau & 18.928 & 0.081 & $-$1.700 & 0.120 & & & 0.57 & 56 & 62&\\
  PLG & BLHer\&WVir\&pWVir & 18.897 & 0.059 & $-$1.910 & 0.130 & &  & 0.36 & 38 & 44&\\
  PLG & BLHer\&WVir\&pWVir\&RVTau & 18.770 & 0.068 & $-$1.870 & 0.100 & & & 0.48 & 57 & 62&\\
  PLRP & BLHer\&WVir\&pWVir & 18.458 & 0.052 & $-$1.950 & 0.120 & & & 0.30 & 36 & 44&\\
  PLRP & BLHer\&WVir\&pWVir\&RVTau & 18.372 & 0.058 & $-$2.013 & 0.088 &  &  & 0.41 & 57 & 62&\\
  PLV & BLHer\&WVir\&pWVir & 18.989 & 0.066 & $-$1.880 & 0.150 & &  & 0.41 & 39 & 44&\\
  PLV & BLHer\&WVir\&pWVir\&RVTau & 18.800 & 0.077 & $-$1.710 & 0.120 &  &  & 0.55 & 57 & 62&\\
  PLI & BLHer\&WVir\&pWVir & 18.420 & 0.055 & $-$2.000 & 0.120 & & & 0.33 & 38 & 44&\\
  PLI & BLHer\&WVir\&pWVir\&RVTau & 18.289 & 0.061 & $-$1.922 & 0.092 & & & 0.43 & 57 & 62&\\
  PLY & BLHer\&WVir\&pWVir & 18.164 & 0.060 & $-$2.170 & 0.140 & & & 0.37 & 40 & 43&\\
  PLY & BLHer\&WVir\&pWVir\&RVTau & 18.134 & 0.051 & $-$2.133 & 0.076 & & & 0.35 & 54 & 59&\\
  PLJ & BLHer\&WVir\&pWVir & 18.171 & 0.033 & $-$2.246 & 0.074 & & & 0.17 & 29 & 43&\\
  PLJ & BLHer\&WVir\&pWVir\&RVTau & 17.993 & 0.050 & $-$2.224 & 0.075 & & & 0.35 & 56 & 61&\\
  PLK & BLHer\&WVir\&pWVir & 17.701 & 0.048 & $-$2.320 & 0.110 & & & 0.29 & 38 & 44&\\
  PLK & BLHer\&WVir\&pWVir\&RVTau & 17.756 & 0.037 & $-$2.406 & 0.055 & & & 0.25 & 51 & 62&\\
  PWG & BLHer\&WVir\&pWVir & 17.748 & 0.041 & $-$2.399 & 0.093 & & & 0.24 & 38 & 44\\
  PWG & BLHer\&WVir\&pWVir\&RVTau & 17.819 & 0.039 & $-$2.527 & 0.057 &  &  & 0.26 & 53 & 62&\\
  PWVI & BLHer\&WVir\&pWVir & 17.557 & 0.040 & $-$2.396 & 0.090 & & & 0.24 & 39 & 44&\\
  PWVI & BLHer\&WVir\&pWVir\&RVTau & 17.545 & 0.035 & $-$2.396 & 0.056 & & & 0.24 & 52 & 62&\\
  PWVK & BLHer\&WVir\&pWVir & 17.536 & 0.045 & $-$2.400 & 0.100 & & & 0.27 & 38 & 44&\\
  PWVK & BLHer\&WVir\&pWVir\&RVTau & 17.583 & 0.039 & $-$2.497 & 0.059 & & & 0.27 & 52 & 62&\\
  PWYK & BLHer\&WVir\&pWVir & 17.515 & 0.048 & $-$2.420 & 0.110 & & & 0.28 & 37 & 43&\\
  PWYK & BLHer\&WVir\&pWVir\&RVTau & 17.753 & 0.028 & $-$2.631 & 0.044 & & & 0.17 & 38 & 59&\\
  PWJK & BLHer\&WVir\&pWVir & 17.493 & 0.049 & $-$2.430 & 0.110 & & & 0.29 & 37 & 43&\\
  PWJK & BLHer\&WVir\&pWVir\&RVTau & 17.536 & 0.042 & $-$2.520 & 0.064 & & & 0.29 & 51 & 61&\\
  PLCG & BLHer\&WVir\&pWVir & 17.265 & 0.034 & $-$2.520 & 0.100 & 2.680 & 0.210 & 0.22 & 37 & 44&\\
  PLCG & BLHer\&WVir\&pWVir\&RVTau & 17.9102 & 0.038 & $-$2.456 & 0.078 & 1.700 & 0.150 & 0.26 & 53 & 62&\\
  PLCVI & BLHer\&WVir\&pWVir & 16.678 & 0.043 & $-$2.890 & 0.100 & 4.280 & 0.260 & 0.26 & 42 & 44&\\
  PLCVI & BLHer\&WVir\&pWVir\&RVTau & 16.901 & 0.040 & $-$2.814 & 0.083 & 3.870 & 0.200 & 0.26 & 55 & 62&\\
  PLCVK & BLHer\&WVir\&pWVir & 17.028 & 0.042 & $-$2.640 & 0.130 & 0.570 & 0.160 & 0.25 & 38 & 44&\\
  PLCYK & BLHer\&WVir\&pWVir & 17.299 & 0.049 & $-$2.530 & 0.160 & 0.910 & 0.530 & 0.29 & 37 & 43&\\
  PLCJK & BLHer\&WVir\&pWVir & 17.252 & 0.046 & $-$2.560 & 0.170 & 1.590 & 0.860 & 0.27 & 36 & 43&\\
\noalign{\smallskip}
\hline  
\noalign{\smallskip}
\end{tabular}
\end{table*}

\FloatBarrier


\section{Additional wavelength dependence of $PL/PW/PLC$ relations} \label{app:wave_dependence}

In this section, we show the wavelength dependence of the coefficients of several $PL/PW/PLC$ relations. 
These figures complete the results shown in Fig.~\ref{coeffpllmc} and ~\ref{coeffplsmc}.

\begin{figure*}
\sidecaption
    \vbox{
    \hbox{
    \includegraphics[width=0.48\textwidth]{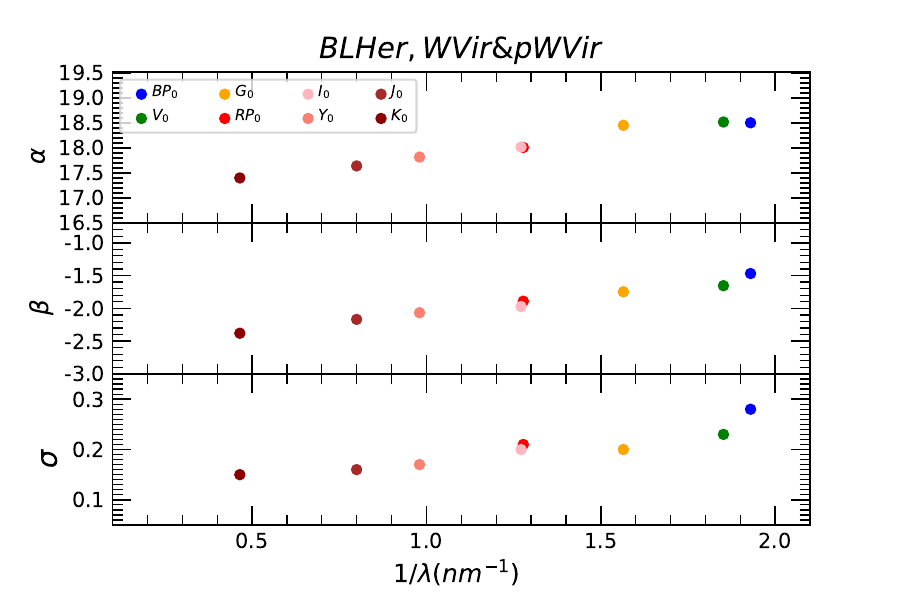}   
    \includegraphics[width=0.48\textwidth]{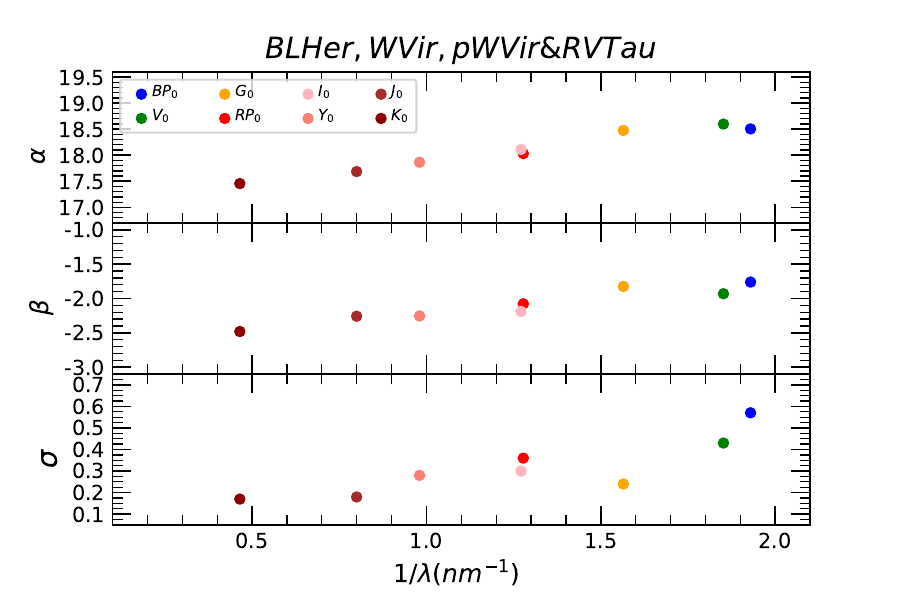}}
    \hbox{
    \includegraphics[width=0.48\textwidth]{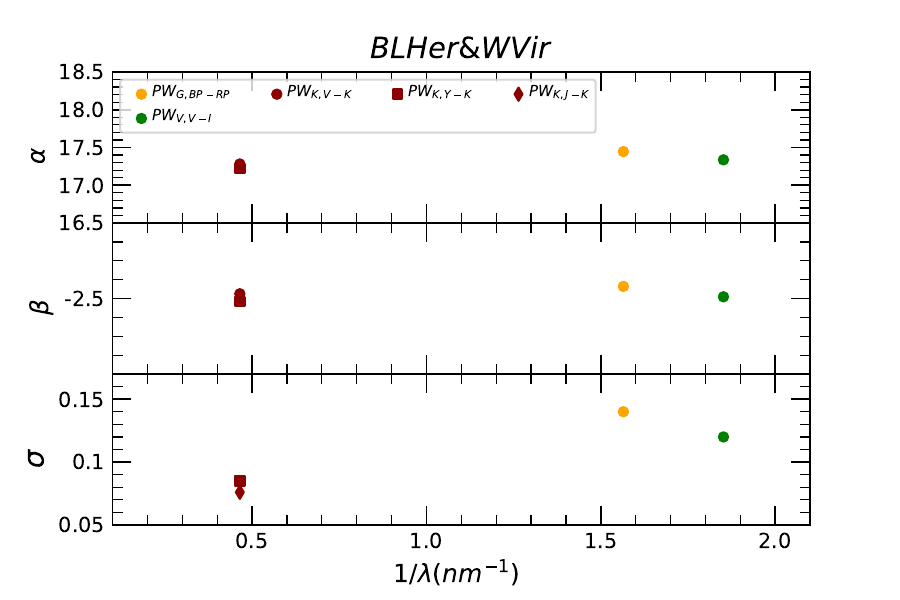}
    \includegraphics[width=0.48\textwidth]{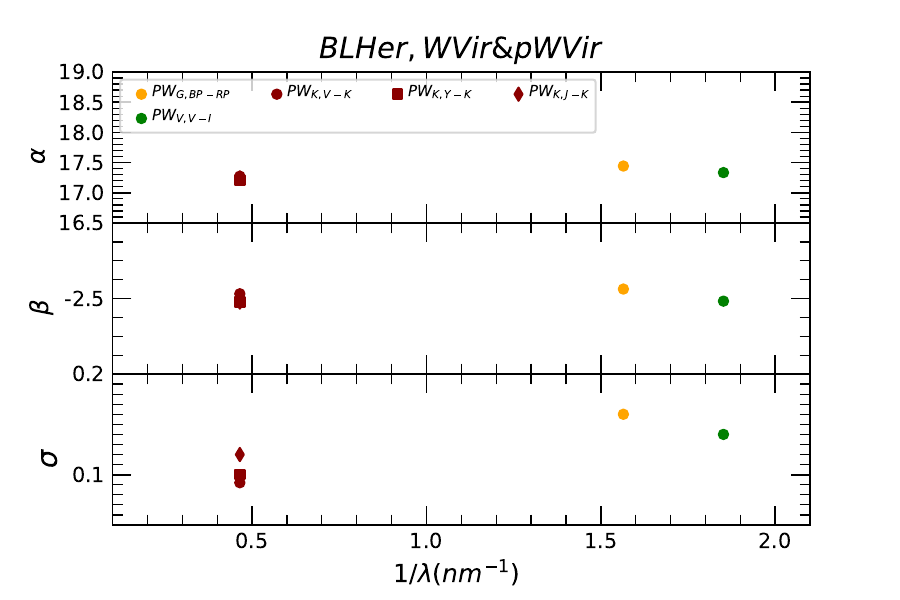}
    }
    \hbox{
    \includegraphics[width=0.48\textwidth]{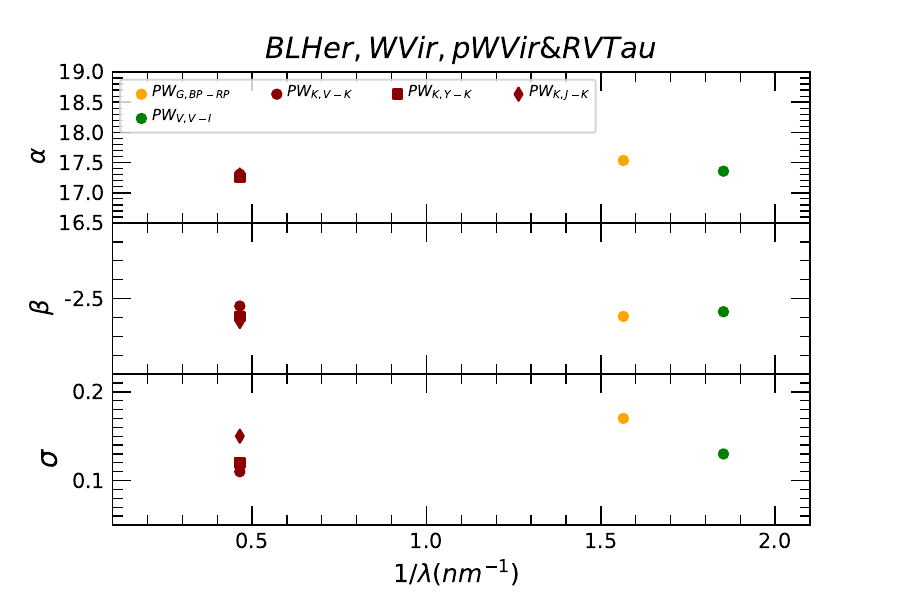}
    }
    }
    \caption{\label{coeffpllmc2} Coefficients for the $PL/PW$ relations in optical and NIR bands for LMC. $\alpha$ and $\sigma$ are expressed in mag, $\beta$ in mag/dex. }	
	\end{figure*}
 
\begin{figure*}
    \vbox{
    \hbox{
    \includegraphics[width=0.48\textwidth]{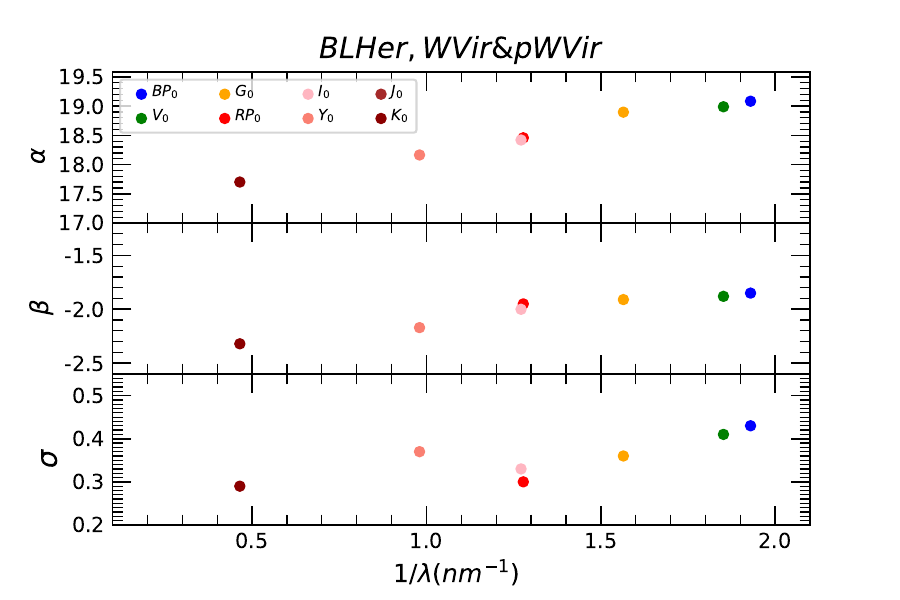}
    \includegraphics[width=0.48\textwidth]{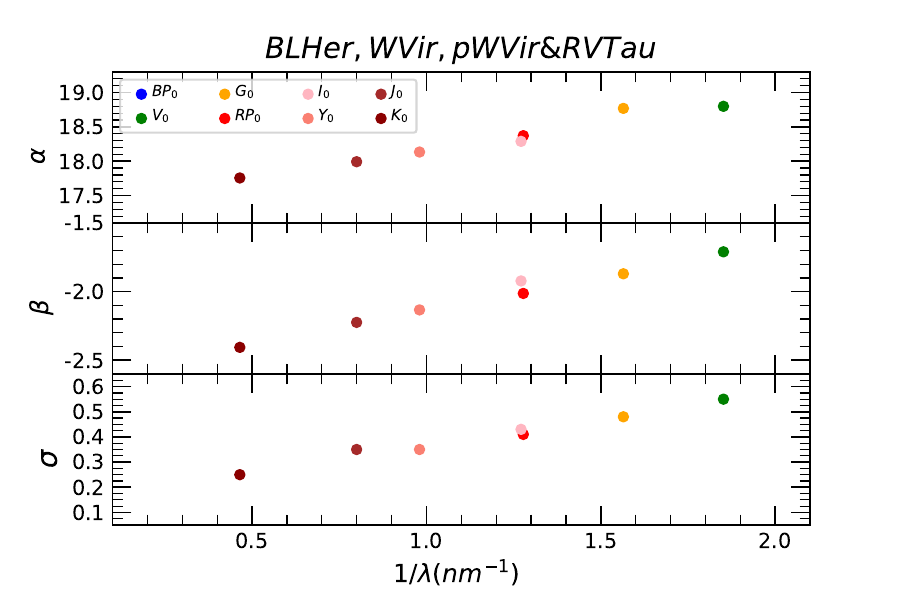}
    }
    \hbox{
    \includegraphics[width=0.48\textwidth]{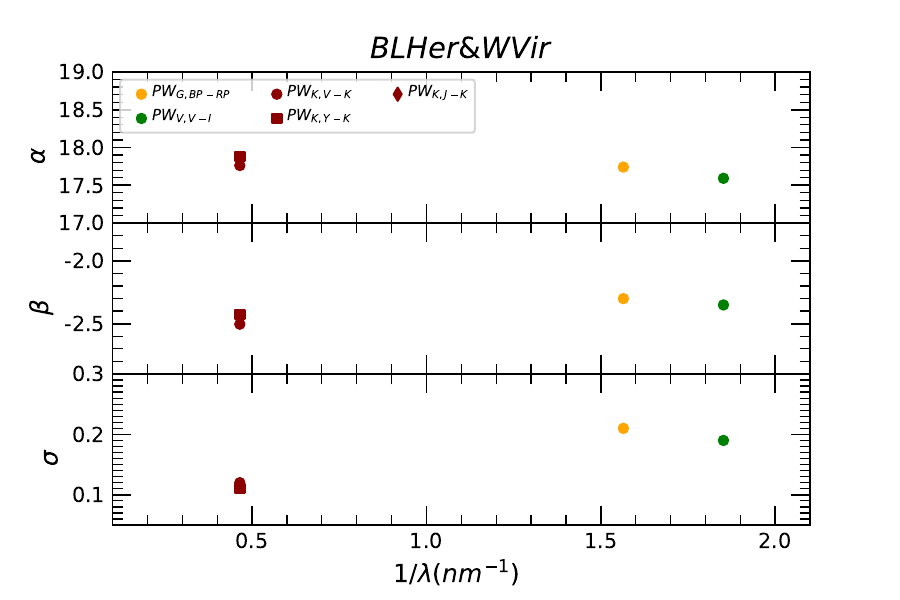}
    \includegraphics[width=0.48\textwidth]{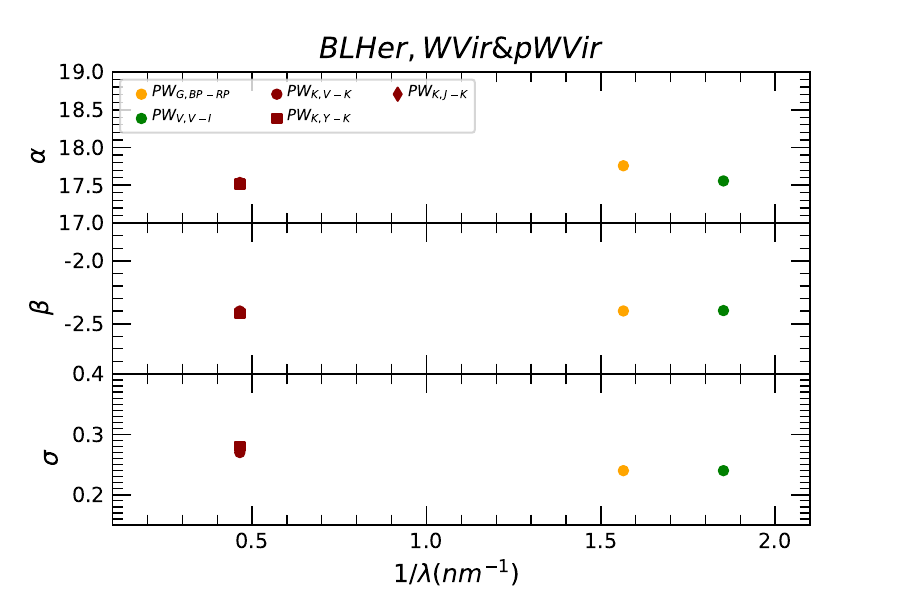}
    }
    \hbox{
    \includegraphics[width=0.48\textwidth]{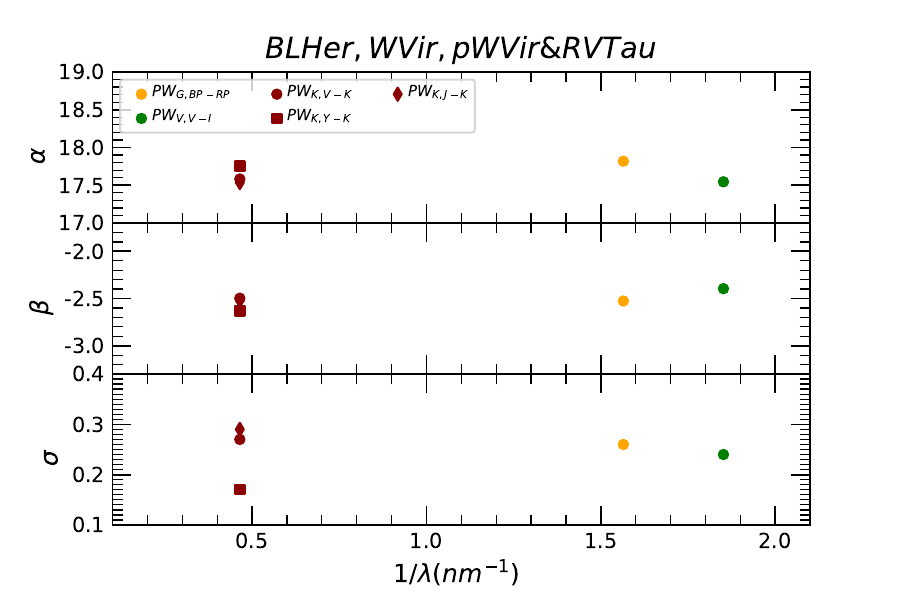}
    }
    }
    \caption{\label{coeffplsmc2} Same as Fig.~\ref{coeffpllmc2} but for the SMC.}	
	\end{figure*}

\FloatBarrier


\section{Sample of T2Cs hosted in the GGs} \label{tableggcs}
\begin{sidewaystable*}
\footnotesize\setlength{\tabcolsep}{3pt} \caption{Optical and NIR photometric parameters for the 46 GGCs T2Cs analysed in this work. }
\label{tab:t2csggcs}
\begin{tabular}{lcccccccccccccccccccccccc}
\hline  
\noalign{\smallskip}   
  GGC & VAR & RA & DEC & D & $\sigma_D$ & P  & V & I  & $\langle G \rangle$ & $\sigma_{\langle G \rangle}$& $\langle G_{BP} \rangle$ & $\sigma_{\langle G_{BP} \rangle}$ & $\langle G_{RP} \rangle$ & $\sigma_{\langle G_{RP} \rangle}$  & $\langle J \rangle$ & $\sigma_{\langle J \rangle}$ & $\langle K_s \rangle$ & $\sigma_{\langle K_s \rangle}$ & E(B-V) & $\sigma_{\langle E(B-V) \rangle}$ & Notes \\
  & & dec & dec & kpc & kpc & days & mag & mag &mag  &   mag & mag & mag &  mag & mag & mag  & mag & mag& mag & mag &    \\                     
(1)  & (2)  & (3) & (4) & (5) & (6) & (7) &(8) & (9) & (10) & (11) & (12)  & (13) & (14) & (15) &(16)  & (17) &(18) & (19)  & (20) & (21) & (22)\\                     
\noalign{\smallskip}
\hline  
\noalign{\smallskip}  
HP1 & V16 & 262.78641 & $-$30.00606 & 7.0 & 0.14 & 16.399 & --- & --- & 14.877 & 0.005 & 16.290 & 0.049 & 13.681 & 0.005 & 11.768 & 0.022 & 10.675 & 0.024 & 2.39 & 0.05 & a,c\\
  HP1 & V17 & 262.77393 & $-$29.99057 & 7.0 & 0.14 & 14.466 & --- & --- & 15.055 & 0.009 & 16.435 & 0.015 & 13.829 & 0.007 & 11.872 & 0.044 & 10.783 & 0.04 & 2.32 & 0.05 & a,c\\
  NGC2808 & V10 & 137.98691 & $-$64.88978 & 10.06 & 0.11 & 1.76544 & 15.28 & 14.47 & 15.088 & 0.012 & 15.392 & 0.012 & 14.487 & 0.009 & 13.893 & 0.039 & 13.436 & 0.044 & 0.22 & 0.002 & a,c\\
  NGC5139 & V61 & 201.80820 & $-$47.45864 & 5.43 & 0.05 & 2.273 & 13.661 & 12.821 & 13.443 & 0.008 & 13.740 & 0.010 & 12.833 & 0.009 & 12.19 & 0.007 & 11.771 & 0.008 & 0.14 & 0.001 & a,d\\
  NGC5139 & V60 & 201.64868 & $-$47.54681 & 5.43 & 0.05 & 1.350 & 13.624 & 13.001 & 13.615 & 0.024 & 13.832 & 0.075 & 13.149 & 0.052 & 12.584 & 0.005 & 12.281 & 0.008 & 0.14 & 0.001 & b,d\\
  NGC5139 & V48 & 201.65750 & $-$47.50709 & 5.43 & 0.05 & 4.476 & 12.924 & 12.092 & 12.779 & 0.005 & 13.085 & 0.026 & 12.138 & 0.018 & 11.47 & 0.013 & 11.034 & 0.011 & 0.14 & 0.001 & a,d\\
  NGC5139 & V43 & 201.64376 & $-$47.44938 & 5.43 & 0.05 & 1.157 & 13.759 & 13.149 & 13.707 & 0.030 & 13.864 & 0.087 & 13.139 & 0.054 & 12.73 & 0.013 & 12.426 & 0.013 & 0.14 & 0.001 & b,d\\
  NGC5139 & V29 & 201.61343 & $-$47.47989 & 5.43 & 0.05 & 14.700 & 12.015 & 11.049 & 11.802 & 0.033 & 12.241 & 0.018 & 11.151 & 0.014 & 10.379 & 0.013 & 9.854 & 0.026 & 0.14 & 0.001 & a,d\\
  NGC5139 & V92 & 201.56174 & $-$47.35413 & 5.43 & 0.05 & 1.346 & 13.946 & 13.199 & 13.807 & 0.005 & 14.105 & 0.004 & 13.285 & 0.007 & 12.7 & 0.004 & 12.313 & 0.008 & 0.13 & 0.001 & a,d\\
  NGC5139 & V1 & 201.52153 & $-$47.39518 & 5.43 & 0.05 & 29.348 & 10.829 & 10.058 & 10.886 & 0.022 & 11.201 & 0.067 & 10.222 & 0.051 & 9.334 & 0.022 & 8.879 & 0.023 & 0.13 & 0.001 & b,d\\
  NGC5272 & V154 & 205.54851 & 28.37045 & 10.18 & 0.08 & 15.299 & 12.33 & 11.68 & 12.427 & 0.004 & 12.475 & 0.086 & 11.808 & 0.035 & 11.348 & 0.044 & 10.929 & 0.034 & 0.01 & 0.00 & a,c\\
  NGC5904 & V84 & 229.65062 & 2.07117 & 7.48 & 0.06 & 26.465 & 11.287 & 10.451 & 11.213 & 0.034 & 11.569 & 0.067 & 10.687 & 0.059 & 10.097 & 0.107 & 9.625 & 0.093 & 0.04 & 0.00 & a,c\\
  NGC5904 & V42 & 229.60327 & 2.04808 & 7.48 & 0.06 & 25.678 & 11.659 & 10.74 & 11.088 & 0.018 & 11.368 & 0.027 & 10.591 & 0.015 & 10.043 & 0.062 & 9.642 & 0.077 & 0.04 & 0.00 & a,c\\
  NGC5986 & V13 & 236.50102 & $-$37.80646 & 10.54 & 0.13 & 40.620 & --- & --- & 13.098 & 0.004 & 13.953 & 0.013 & 12.164 & 0.010 & 10.912 & 0.024 & 10.075 & 0.017 & 0.34 & 0.003 & b,c\\
  NGC6093 & V1 & 244.21914 & $-$22.95444 & 10.34 & 0.12 & 16.310 & 13.365 & --- & 13.154 & 0.007 & 13.497 & 0.100 & 12.374 & 0.129 & 11.624 & 0.039 & 11.09 & 0.038 & 0.21 & 0.002 & a,c\\
  NGC6254 & V2 & 254.29894 & $-$4.06660 & 5.07 & 0.06 & 18.678 & 12.05 & 10.934 & 11.553 & 0.022 & 12.048 & 0.024 & 10.867 & 0.010 & 9.991 & 0.055 & 9.416 & 0.059 & 0.29 & 0.003 & a,c\\
  NGC6254 & V3 & 254.23316 & $-$4.07125 & 5.07 & 0.06 & 7.835 & 12.75 & 11.721 & 12.450 & 0.007 & 12.938 & 0.004 & 11.792 & 0.004 & 10.971 & 0.045 & 10.402 & 0.074 & 0.27 & 0.003 & a,c\\
  NGC6256 & V1 & 254.89584 & $-$37.12304 & 7.24 & 0.29 & 12.504 & --- & 13.402 & 14.560 & 0.015 & 15.697 & 0.024 & 13.474 & 0.021 & 11.766 & 0.081 & 10.767 & 0.061 & 1.71 & 0.05 & a,c\\
  NGC6266 & V2 & 255.29568 & $-$30.13317 & 6.41 & 0.1 & 10.604 & 13.418 & 12.065 & 12.955 & 0.006 & 13.577 & 0.008 & 12.131 & 0.008 & 11.068 & 0.054 & 10.409 & 0.065 & 0.47 & 0.005 & a,c\\
  NGC6273 & V2 & 255.66211 & $-$26.23261 & 8.34 & 0.16 & 14.142 & --- & 12.242 & 13.054 & 0.004 & 13.573 & 0.004 & 12.347 & 0.003 & 11.466 & 0.038 & 10.879 & 0.041 & 0.32 & 0.003 & a,c\\
  NGC6273 & V1 & 255.65890 & $-$26.25326 & 8.34 & 0.16 & 16.920 & --- & 12.26 & 13.343 & 0.017 & 13.796 & 0.051 & 12.392 & 0.039 & 11.357 & 0.03 & 10.736 & 0.029 & 0.32 & 0.003 & b,c\\
  NGC6273 & V4 & 255.65651 & $-$26.27543 & 8.34 & 0.16 & 2.433 & --- & 13.947 & 14.630 & 0.015 & 14.760 & 0.042 & 13.595 & 0.025 & 13.225 & 0.026 & 12.684 & 0.043 & 0.31 & 0.003 & b,c\\
  NGC6284 & V4 & 256.12640 & $-$24.77058 & 14.21 & 0.42 & 2.818 & --- & 14.786 & 15.505 & 0.015 & 15.833 & 0.015 & 14.700 & 0.032 & 14.111 & 0.037 & 13.605 & 0.042 & 0.31 & 0.003 & a,c\\
  NGC6284 & V1 & 256.11222 & $-$24.75607 & 14.21 & 0.42 & 4.481 & --- & 14.504 & 15.309 & 0.010 & 15.624 & 0.0262 & 14.442 & 0.020 & 13.66 & 0.041 & 13.12 & 0.034 & 0.3 & 0.003 & b,c\\
  NGC6325 & V2 & 259.49082 & $-$23.77680 & 7.53 & 0.32 & 10.744 & --- & 13.632 & 14.849 & 0.007 & 15.959 & 0.020 & 13.750 & 0.020 & 12.131 & 0.014 & 11.221 & 0.015 & 0.96 & 0.010 & b,c\\
  NGC6325 & V1 & 259.51034 & $-$23.76247 & 7.53 & 0.32 & 12.468 & --- & 13.436 & 14.646 & 0.006 & 15.748 & 0.009 & 13.579 & 0.013 & 11.985 & 0.037 & 11.053 & 0.038 & 0.95 & 0.010 & a,c\\
  NGC6402 & V76 & 264.37111 & $-$3.24557 & 9.14 & 0.25 & 1.890 & 15.978 & 14.75 & 16.901 & 0.004 & 99.99 & 99.99 & 99.99 & 99.99 & 13.82 & 0.016 & 13.124 & 0.022 & 0.48 & 0.005 & b,c\\
  NGC6402 & V7 & 264.41842 & $-$3.27237 & 9.14 & 0.25 & 13.595 & 14.745 & 13.224 & 14.156 & 0.013 & 14.947 & 0.016 & 13.254 & 0.012 & 12.035 & 0.033 & 11.303 & 0.026 & 0.48 & 0.005 & a,c\\
  NGC6402 & V2 & 264.36823 & $-$3.27902 & 9.14 & 0.25 & 2.795 & 15.629 & 14.337 & 20.905 & 0.035 & 99.99 & 99.99 & 99.99 & 99.99 & 13.405 & 0.018 & 12.802 & 0.021 & 0.48 & 0.005 & b,c\\
  NGC6402 & V1 & 264.40577 & $-$3.23319 & 9.14 & 0.25 & 18.758 & 14.21 & 12.633 & 13.640 & 0.018 & 14.398 & 0.019 & 12.728 & 0.017 & 11.558 & 0.026 & 10.834 & 0.048 & 0.48 & 0.005 & a,c\\
  NGC6441 & V129 & 267.55350 & $-$37.05501 & 12.73 & 0.16 & 17.832 & 15.128 & 13.61 & 14.508 & 0.019 & 13.799 & 0.028 & 12.026 & 0.024 & 12.146 & 0.022 & 11.471 & 0.068 & 0.62 & 0.006 & b,c\\
  NGC6441 & V6 & 267.56511 & $-$37.03784 & 12.73 & 0.16 & 22.521 & 14.885 & 13.231 & 14.240 & 0.011 & 14.770 & 0.022 & 13.137 & 0.014 & 12.045 & 0.041 & 11.418 & 0.089 & 0.61 & 0.006 & a,c\\
  NGC6453 & V2 & 267.72093 & $-$34.58600 & 10.07 & 0.22 & 27.195 & 14.231 & 12.375 & 13.566 & 0.018 & 14.340 & 0.060 & 12.623 & 0.048 & 11.245 & 0.04 & 10.482 & 0.019 & 0.66 & 0.007 & b,c\\
  NGC6453 & V1 & 267.71712 & $-$34.60133 & 10.07 & 0.22 & 30.983 & 14.601 & 12.789 & 13.904 & 0.054 & 14.763 & 0.027 & 12.836 & 0.020 & 11.47 & 0.037 & 10.632 & 0.026 & 0.66 & 0.007 & a,c\\
  NGC6569 & V16 & 273.40684 & $-$31.82040 & 10.53 & 0.26 & 87.5 & --- & --- & 13.537 & 0.026 & 14.777 & 0.106 & 12.291 & 0.060 & 10.502 & 0.105 & 9.422 & 0.085 & 0.43 & 0.004 & b,c\\
  NGC6626 & V4 & 276.12551 & $-$24.86016 & 5.37 & 0.1 & 13.470 & --- & 11.734 & 12.636 & 0.011 & 13.281 & 0.007 & 11.802 & 0.011 & 10.757 & 0.055 & 10.047 & 0.062 & 0.49 & 0.005 & a,c\\
  NGC6749 & V1 & 286.33323 & 1.93256 & 7.59 & 0.21 & 4.481 & --- & --- & 16.504 & 0.010 & 17.901 & 0.040 & 15.340 & 0.025 & 13.352 & 0.027 & 12.323 & 0.019 & 1.75 & 0.05 & b,c\\
  NGC6779 & V6 & 289.14905 & 30.19413 & 10.43 & 0.14 & 44.966 & 12.9 & --- & 12.443 & 0.012 & 13.017 & 0.031 & 11.678 & 0.015 & 10.711 & 0.026 & 10.119 & 0.043 & 0.25 & 0.003 & a,c\\
  NGC6779 & V1 & 289.16386 & 30.20461 & 10.43 & 0.14 & 1.510 & 15.46 & --- & 15.250 & 0.004 & 15.594 & 0.005 & 14.689 & 0.003 & 13.993 & 0.016 & 13.553 & 0.058 & 0.25 & 0.003 & a,c\\
  NGC7078 & V86 & 322.49647 & 12.16863 & 10.71 & 0.1 & 16.825 & 13.659 & 12.646 & 12.991 & 0.011 & 12.969 & 0.019 & 11.984 & 0.009 & 11.568 & 0.04 & 11.05 & 0.041 & 0.11 & 0.001 & a,e\\
  NGC7078 & V1 & 322.45908 & 12.17401 & 10.71 & 0.1 & 1.438 & 14.954 & 14.362 & 14.839 & 0.005 & 15.033 & 0.005 & 14.389 & 0.006 & 13.943 & 0.031 & 13.649 & 0.032 & 0.11 & 0.001 & a,e\\
  NGC7089 & V6 & 323.36471 & $-$0.83324 & 11.69 & 0.11 & 19.349 & 13.14 & --- & 12.946 & 0.023 & 13.085 & 0.190 & 12.020 & 0.094 & 11.665 & 0.03 & 11.204 & 0.042 & 0.04 & 0.00 & a,c\\
  NGC7089 & V5 & 323.34935 & $-$0.82026 & 11.69 & 0.11 & 17.547 & 13.28 & --- & 13.074 & 0.010 & 13.406 & 0.007 & 12.420 & 0.011 & 11.803 & 0.032 & 11.31 & 0.03 & 0.04 & 0.00 & a,c\\
  NGC7089 & V11 & 323.38507 & $-$0.81828 & 11.69 & 0.11 & 33.400 & 12.11 & --- & 11.975 & 0.013 & 12.257 & 0.041 & 11.405 & 0.030 & 10.86 & 0.06 & 10.401 & 0.042 & 0.04 & 0.00 & b,c\\
  NGC7089 & V1 & 323.36859 & $-$0.79874 & 11.69 & 0.11 & 15.565 & 13.36 & --- & 13.145 & 0.007 & 13.514 & 0.015 & 12.573 & 0.003 & 11.939 & 0.018 & 11.446 & 0.024 & 0.04 & 0.00 & a,c\\
  Terzan1 & V5 & 263.94223 & $-$30.48428 & 6.7 & 0.17 & 18.849 & --- & 14.576 & 16.030 & 0.012 & 18.133 & 0.065 & 14.464 & 0.021 & 11.96 & 0.026 & 10.578 & 0.035 & 1.99 & 0.05 & a,c\\
\noalign{\smallskip}
\hline  
\noalign{\smallskip}
\end{tabular}
\tablefoot{Columns: $(1)$ Identification of the GGCs; $(2)$ Identification of the T2C: $(3)$-$(4)$ Ra and Dec; $(5)$-$(6)$ Distance of the GGC and his relative uncertainty from \citet{ngeow2022,baumgardt2021accurate}; $(7)$ Periods; $(8)$-$(9)$ Magnitudes in $V\,I$ band from \citet{ngeow2022}; $(10)$-$(11)$ Magnitude in $G$ and relative uncertainty; $(12)$-$(13)$ As for column $(10)$ and $(11)$ but for the $G_{BP}$; $(14)$-$(15)$ As for column $(10)$ and $(11)$ but for the $G_{RP}$; $(16)$-$(17)$ As for column $(10)$ and $(11)$ but for the $J$; $(18)$-$(19)$ As for column $(11)$ and $(12)$ but for the $K_s$; $(20)$-$(21)$ Extinction and relative uncertainty from \citet{harris2010new}; $(22)$ Notes: a = Gaia data from SOS; b = Gaia data from Gaia source; c = $J\,K_s$ data and periods from \citet{Bhardwaj2017_Bulge}; d = $J\,K_s$ data and periods from \citet{braga2020separation}; e = $J\,K_s$ data and periods from \citet{bhar2021}.}    
\end{sidewaystable*}

\FloatBarrier


\section{Calibration with Galactic T2Cs parallaxes} \label{calplx}

\subsection{Calibration with the Photometric parallax}

We first defined the photometric parallax (in mas):

\begin{equation}
    \varpi_{phot}= 10^{-2(m-M-10)} ,
\end{equation}

\noindent
where $m$ is the apparent magnitude (or apparent Wesenheit magnitude); $M$ is the absolute magnitude (or absolute Wesenheit magnitude), defined as:

\begin{equation}
    M = \alpha + \beta_{LMC} \times \log_{10}P .
\end{equation}
\noindent where $\beta_{LMC}$ are the slopes obtained for the LMC and listed in Table~\ref{tab:lmcT2}.
Given the scarcity of the sample of stars with known metallicity \citep{Wielgorski2022}, we did not include a metallicity term in the $PL/PW$ relations. However, this is not a major issue, as several papers reported a negligible metallicity dependence of these relations for the T2Cs \citep[][]{Matsunaga2009,ripepi2015vmc,ngeow2022} at odds with \citet{Wielgorski2022} results, which are, however, based on a very small sample of T2Cs with metallicity measurement 
 and must be confirmed. 

The value of $\alpha$ is thus obtained by minimising the following $\chi^2$ expression:

\begin{equation} 
    \chi^2 = \sum \frac{(\varpi_{EDR3}-\varpi_{phot})^2}{\sigma^2},
\end{equation}
\noindent
where $\varpi_{EDR3}$ are the parallaxes from $Gaia$ EDR3 corrected individually with the recipe provided by \citet{Lindegren2021}. The fitting procedure is described in detail by \citet{ripepi2022classical} and is not repeated here. The interested reader is referred to the quoted paper.

\subsection{Calibration with the ABL}

In addition to the photometric parallax method, we also used the ABL with a different fitting technique to estimate the zero points of T2C $PL/PW$. By using two different methods, we can compare the obtained results. The ABL is defined as: 

\begin{equation} \label{eq:abl}
    ABL= 10^{0.2W}= 10^{0.2(\alpha + \beta_{LMC} \times \log P)} = \varpi 10^{0.2w-2} ,
\end{equation}

where, as above, $W$ and $w$ are the absolute and apparent Wesenheit magnitudes and $\varpi$ is the parallax, while, as in the previous case, the slope ($\beta_{LMC}$) is fixed to that of the LMC.  The $\chi^2$ expression in this case is:
\begin{equation} 
\chi^2 = \sum \frac{(\varpi 10^{0.2w-2}-10^{0.2(\alpha + \beta_{LMC} \times \log P)})^2}{\sigma^2}
\end{equation}
\noindent
where the different quantities have the same meaning as in Eq.~\ref{eq:abl} and $\sigma$ includes all the errors. The results for the unknown $\alpha$ are listed in the bottom part of Table~\ref{tab:0pointsphotpar}.

\end{appendix}

\end{document}